\documentclass[acmsmall]{acmart}

\usepackage{etoolbox}
\makeatletter
\expandafter\patchcmd\csname\string\maketitle\endcsname
  {\vskip\z@\@plus3fill}
  {\vskip\z@\@plus2fill\box\abstractbox\vskip\z@\@plus1fill}
  {}{}
\makeatother

\usepackage[utf8]{inputenc}

\usepackage[T1]{fontenc}

\usepackage[american]{babel}

\usepackage{amsthm}
\usepackage{centernot}

\usepackage{graphicx} %
\usepackage{xspace} % to get the spacing after macros right

  \usepackage{comment}
  \usepackage{mdframed}
  \usepackage{pdflscape}

  \usepackage{geometry}

  \usepackage[color=blue!20]{todonotes}

  \usepackage{tikz}
  \usetikzlibrary{positioning}
  \usetikzlibrary{backgrounds,fit,calc,decorations,decorations.pathmorphing,decorations.pathreplacing}

  \usepackage{booktabs}

\usepackage{subcaption}
\definecolor{CTsemi}{gray}{0.55} % chapter numbers will be semi transparent .5 .55 .6 .0
\definecolor{CTcitation}{rgb}{0,0.5,0} % WebGreen
\definecolor{CTurl}{named}{Maroon} % Maroon
\definecolor{CTtitle}{named}{Maroon} % Maroon {cmyk}{0, 0.87, 0.68, 0.32}
\definecolor{CTlink}{named}{RoyalBlue} % RoyalBlue {cmyk}{1, 0.50, 0, 0}
\definecolor{halfgray}{gray}{0.55} % chapter numbers will be semi transparent .5 .55 .6 .0
\definecolor{webgreen}{rgb}{0,0.5,0}
\definecolor{webbrown}{rgb}{0.6,0,0}

\usepackage{microtype}
\usepackage{hyperref}

\hypersetup{%
  colorlinks=true, linktocpage=true, pdfstartpage=3, pdfstartview=FitV,%
  breaklinks=true, pageanchor=true,%
  pdfpagemode=UseNone, %
  plainpages=false, bookmarksnumbered, bookmarksopen=true, bookmarksopenlevel=1,%
  hypertexnames=true, pdfhighlight=/O,%nesting=true,%frenchlinks,%
  urlcolor=CTurl, linkcolor=CTlink, citecolor=CTcitation, %pagecolor=RoyalBlue,%
  pdfcreator={pdfLaTeX},%
  pdfproducer={LaTeX}%
}

\usepackage[symbols,nogroupskip]{glossaries-extra}
\glsdisablehyper % Don't link back to glossary
\makeglossaries
\newcommand{\GlsSymbol}[3]{\glsxtrnewsymbol[description={#3}]{#1}{#2}}
\newcommand{\GlsChildSymbol}[4]{\glsxtrnewsymbol[description={#4},parent={#2}]{#1}{#3}}

\usepackage[capitalise]{cleveref}

\newtheorem{theorem}{Theorem}[section]

\newtheorem{lemma}[theorem]{Lemma}

\newtheorem{corollary}[theorem]{Corollary}
\newtheorem{definition}[theorem]{Definition}

\newtheorem{example}[theorem]{Example}

\usepackage{bussproofs}

\input{math-macros}

\DefineNamedColor{named}{azul2}{RGB}{40,96,127}
\DefineNamedColor{named}{azul1}{RGB}{4,151,229}
\DefineNamedColor{named}{azul3}{RGB}{1,41,64}
\DefineNamedColor{named}{laranja1}{RGB}{255,146,5}

\newmdenv[roundcorner=5pt,backgroundcolor=RoyalBlue!10,linecolor=RoyalBlue]{infobox}
\newmdenv[roundcorner=5pt,backgroundcolor=Maroon!10,linecolor=Maroon,frametitle=Warning]{warningbox}

\makeatletter
\let\orig@infobox=\infobox
\def\infobox#1{
  \orig@infobox[frametitle={#1}]
}
\makeatother

\newcommand{\tagA}{\ensuremath{(\dag)}}
\newcommand{\tagB}{\ensuremath{(\ddag)}}
\newcommand{\tagC}{\ensuremath{(\clubsuit)}}

\DefineNamedColor{named}{high}{RGB}{255,0,0}
\DefineNamedColor{named}{medium}{RGB}{128,128,0}
\DefineNamedColor{named}{low}{RGB}{0,255,0}

\tikzset{modelnode/.style={draw,rounded corners,minimum size=14pt,inner
    sep=2pt}}
\tikzset{stack/.style={fill=green!20}}
\tikzset{utnode/.style={draw,inner sep=2pt}}

\tikzstyle{internal} = [circle, draw, fill=white, text=black, minimum
size = 8mm]
\tikzstyle{external} = [circle, draw, fill=gray!50, text=black, minimum size = 8mm,inner sep=0.5pt]
\tikzstyle{free} = [double, circle, draw, fill=white, text=black, minimum size = 8mm]
\tikzstyle{call} = [rectangle, draw, fill=white, text=black, minimum size = 8mm]

\tikzset{tll/.style={draw,circle,fill=azul2!30,minimum size=16pt}}
\tikzset{tlls/.style={draw,circle,fill=azul2!30,minimum
    size=12pt,font={\footnotesize},inner sep=1pt}}
\tikzset{letter/.style={draw,circle,fill=laranja1!30,minimum
    size=8pt}}
\tikzset{nil/.style={}}
\tikzset{fnode/.style={draw,rounded corners,font={\footnotesize}}}

\tikzset{loc/.style={draw,rounded corners,minimum size=14pt,inner
    sep=2pt}}
\tikzset{dangling/.style={fill=orange!50}}

\tikzset{hnode/.style={draw,circle,minimum size=14pt,inner
    sep=2pt}}
\tikzset{stable/.style={fill=blue!20}}
\tikzset{hedge/.style={->}}

\tikzset{rgnode/.style={draw,thick,rounded corners,minimum size=14pt,inner
    sep=2pt}}
\tikzset{datanode/.style={minimum size=14pt,inner sep=2pt}}
\tikzset{rgedge/.style={->,thick}}

\tikzset{essential/.style={rounded corners,thick,fill=green!15,inner sep=2pt,minimum size=14pt,draw}}
\tikzset{triangle/.style={fill = blue!5,isosceles triangle,draw=,thick,shape border rotate=90,isosceles triangle stretches=true, minimum height=20mm,minimum width=15mm,inner sep=0,yshift={-10mm}}}

\tikzset{legendedge/.style={->, thick, >=stealth, gray, dashed, line width = 1pt }}
\tikzset{legendnode/.style={align = center, minimum height = 10pt, fill= green!20}}

\GlsSymbol{---bot}{\ensuremath{\bot}}{undefined; $f(x)=\bot$ iff
  $x\notin \dom(f)$}
\GlsSymbol{---concat}{\ensuremath{\concat}}{sequence concatenation}
\GlsSymbol{---funcomp}{\ensuremath{\funcomp}}{function composition}
\GlsSymbol{--eq-Iso}{\ensuremath{\Iso}}{isomorphism}
\GlsSymbol{---pfun}{\ensuremath{\pfun}}{partial function}
\GlsSymbol{--pto}{\ensuremath{\ptoArrow}}{points-to assertion}
\GlsSymbol{--bin-oursunion}{\ensuremath{\oursunion}}{strong disjoint union of
  heaps w.r.t.~stack $\S$}
\GlsSymbol{--bin-stdunion}{\ensuremath{\stdunion}}{disjoint union of
  functions}
\GlsSymbol{--rel-ltAP}{\ensuremath{\ltAP}}{access path ordering on
  locations in directed graph}
\GlsSymbol{--rel-sleq}{\ensuremath{\sleqSymbol}}{equality between variables}
\GlsSymbol{--rel-slneq}{\ensuremath{\slneqSymbol}}{dis-equality
  between variables}
\GlsSymbol{--rel-subfun}{\ensuremath{\subfun}}{sub-function relation;
  $f\subfun g$ holds iff for all $x\in\dom(f)$, $f(x)\subseteq g(x)$}
\GlsSymbol{--sl-sep}{\ensuremath{\sep}}{separating conjunction}
\GlsSymbol{--sl-sep-iterated}{\ensuremath{\IteratedStar}}{iterated separating conjunction}
\GlsSymbol{--sl-wsep}{\ensuremath{\wsep}}{weak separating conjunction}
\GlsSymbol{--slwand-mw}{\ensuremath{\mw}}{magic wand / separating implication}
\GlsSymbol{--slwand-sept}{\ensuremath{\protect\sept}}{septraction /
  existential wand}
\GlsSymbol{--zz-size}{\ensuremath{\size{\cdot}}}{size / cardinality}
\GlsSymbol{--zz-seq}{\ensuremath{\tuple{x_1,\ldots,x_k}}}{ordered
  sequence consisting of the elements $x_1,\ldots,x_k$}
\GlsSymbol{--zz-ssl-ex1}{\ensuremath{\exudp{\dform}}}{existential
  unary data predicate}
\GlsSymbol{--zz-ssl-ex2}{\ensuremath{\exbdp{\fany}{\dform}}}{existential
  binary data predicate for field $\fany$}
\GlsSymbol{--zz-ssl-fa1}{\ensuremath{\faudp{\dform}}}{universal
  unary data predicate}
\GlsSymbol{--zz-ssl-fa2}{\ensuremath{\fabdp{\fany}{\dform}}}{universal
  binary data predicate for field $\fany$}
\GlsSymbol{--zz-x-Hoaretriple}{\ensuremath{\partialC{ \phi }{ \proga
    }{\psi}}}{Hoare triple with precondition $\phi$, program $c$, and
  postcondition $\psi$}
\GlsSymbol{--zz-eqclass}{\ensuremath{\eqclass{\S}{x}}}{equivalence
  class of $x$ w.r.t.~stack $\S$}
\GlsSymbol{-E-emptyseq}{\ensuremath{\emptyseq}}{the empty sequence}
\GlsSymbol{-L-lambda}{\ensuremath{\lambda}}{notation for defining
  anonymous functions}
\GlsSymbol{-P-Pi}{\ensuremath{\Pure}}{a pure constraint}
\GlsSymbol{-S-renfun}{\ensuremath{\renfun}}{a function renaming
  variables and/or locations}
\GlsSymbol{-S-scls}{\ensuremath{\scls}}{a stack-aliasing constraint}
\GlsSymbol{-S-scls-restr}{\ensuremath{\eqrestr{\scls}{\vec{y}}}}{restriction
  of stack-aliasing constraint $\scls$ to $\vec{y}$}
\GlsSymbol{-V-pinst}{\ensuremath{\pinst{\fa}{\vec{y}}{\vec{z}}}}{instantiate
  (free or bound) variables $\vec{y}$ with $\vec{z}$ in $\fa$}
\GlsSymbol{-V-phiapp}{\ensuremath{\phiapp{\phi}{\vec{z}}}}{replace
  free variables of $\phi$ with $\vec{z}$, equivalent to $\pinst{\phi}{\fvs{\phi}}{\vec{z}}$}
\GlsSymbol{ACx}{\ensuremath{\EqClasses{\vec{x}}}}{the set of all stack-aliasing constraints over
    variables $\vec{x}$}
\GlsSymbol{aliasing}{\ensuremath{\eqclasses{\S}}}{the stack-aliasing
  constraint of stack $\S$}
\GlsSymbol{alloced}{\ensuremath{\valloc{\S,\H}}}{set of allocated
  variables of the model}
\GlsSymbol{classes-s}{\ensuremath{\eqclassesAMS{\S}}}{the equivalence
  classes of $\eqclasses{\S}$}
\GlsSymbol{danglinglocs}{\ensuremath{\danglinglocs{\H}}}{set of
  dangling locations of the heap $\H$}
\GlsSymbol{dom-f}{\ensuremath{\dom(f)}}{domain of (partial) function $f$}
\GlsSymbol{dom--scls}{\ensuremath{\dom(\scls)}}{domain of
stack-aliasing constraint $\scls$}
\GlsSymbol{E-G}{\ensuremath{\EDG}}{the edge relation of a directed (indexed) graph}
\GlsSymbol{emp}{\ensuremath{\emp}}{empty-heap predicate}
\GlsSymbol{f}{\ensuremath{\false}}{false}
\GlsSymbol{fvars-phi}{\ensuremath{\fvsKW(\phi)}}{the set of free variables of formula $\phi$}
\GlsSymbol{fvars-pred}{\ensuremath{\fvsKW(\pred)}}{the set of
  parameters of predicate $\pred$}
\GlsSymbol{G}{\ensuremath{\DG}}{a directed (indexed) graph}
\GlsSymbol{graph-h}{\ensuremath{\ugraphof{\H}}}{induced graph of heap $\H$}
\GlsSymbol{h}{\ensuremath{\H}}{a heap}
\GlsSymbol{Heaps}{\ensuremath{\HH}}{the set of all heaps}
\GlsSymbol{igraph-h}{\ensuremath{\igraphof{\H}}}{induced indexed graph of heap
  $\H$}
\GlsSymbol{img-f}{\ensuremath{\img(f)}}{image of (partial) function $f$}
\GlsSymbol{Loc}{\ensuremath{\Loc}}{the set of memory locations}
\GlsSymbol{locs-h}{\ensuremath{\locsKW(\H)}}{the locations
  $\dom(\H)\cup\bigcup\img(\H)$ for heap $\H$}
\GlsSymbol{locs-phi}{\ensuremath{\locsKW(\phi)}}{locations that occur
  as terms in formula $\phi$}
\GlsSymbol{max-v}{\ensuremath{\max}}{returns the maximal variable
  among a set or sequence of variables}
\GlsSymbol{nil}{\ensuremath{\nil}}{a variable representing the null pointer}
\GlsSymbol{refed}{\ensuremath{\vrefed{\S,\H}}}{set of referenced
  variables of the model}
\GlsSymbol{s}{\ensuremath{\S}}{a stack}
\GlsSymbol{s-1}{\ensuremath{\stkinv}}{the inverse of stack
  $\S$}
\GlsSymbol{s-1-max}{\ensuremath{\stkchc}}{the \stkchoicefun{} of stack $\S$}
\GlsSymbol{SLbase}{\ensuremath{\SLbase}}{a basic first-order
  separation logic}
\GlsSymbol{sinkseq-h}{\ensuremath{\sinkseq{\H}}}{ordered
  (w.r.t.~$\ltAP$) sink sequence of directed tree $\H$}
\GlsSymbol{sinkvars-s-h}{\ensuremath{\sinkvars{\S}{\H}}}{the
  stack-equivalence classes w.r.t.~$\S$ corresponding to the sinks
  of $\H$, in the same order as the sink sequence}
\GlsSymbol{sourcevars-s-h}{\ensuremath{\sourcevars{\S}{\H}}}{the
  stack-equivalence classes w.r.t.~$\S$ corresponding to the sources
  of $\H$}
\GlsSymbol{Stacks}{\ensuremath{\SS}}{the set of all stacks}
\GlsSymbol{t}{\ensuremath{\true}}{true}
\GlsSymbol{V-G}{\ensuremath{\VDG}}{the nodes of a directed
  (indexed) graph}
\GlsSymbol{Val}{\ensuremath{\Val}}{the set of values that can be
  stored in the stack and heap}
\GlsSymbol{Var}{\ensuremath{\Var}}{the set of (program and logical) variables}
\GlsSymbol{--Compose-aCompose}{\ensuremath{\Compose}}{composition of
  AMS or sets of AMS}
\GlsSymbol{--ComposeAdjoint-gmw}{\ensuremath{\gmw}}{abstract magic
  wand in AMS computation}
\GlsSymbol{--Compose-DCompose}{\ensuremath{\DCompose}}{composition of
  EAMS}
\GlsSymbol{--ComposeAdjoint-gmwD}{\ensuremath{\Dgmw}}{abstract magic
  wand in EAMS computation}
\GlsSymbol{--models-datamodels}
{\ensuremath{\datamodels}} {model relationship of the data
  theory $\Tdata$}
\GlsSymbol{--models-xmodels}
{\ensuremath{\entailsx}} {entailment on models with
  $\dom(\S)=\vec{x}$}
\GlsSymbol{--zz-chunksize}{\ensuremath{\csize{\phi}}}{the chunk size of
formula $\phi$}
\GlsSymbol{--pto-lr}{\ensuremath{\ptoArrow_{\fleft,\fright}}}{points-to assertion
allocating a tree node (no data field specified)}
\GlsSymbol{--pto-ls}{\ensuremath{\ptoArrow_\lsDS}}{points-to assertion
allocating a list node}
\GlsSymbol{--pto-n}{\ensuremath{\ptoArrow_\fnext}}{points-to assertion
allocating a list node (no data field specified)}
\GlsSymbol{--pto-tree}{\ensuremath{\ptoArrow_\treeDS}}{points-to assertion
allocating a tree node}
\GlsSymbol{-A-predvara}{\ensuremath{\predvara}}{first free variable of a unary or binary data
  predicate}
\GlsSymbol{-B-predvarb}{\ensuremath{\predvarb}}{second free variable of a binary data predicate}
\GlsSymbol{A}{\ensuremath{\A}}{an abstract memory state (AMS)}
\GlsSymbol{abst-s-phi}{\ensuremath{\abst{\S}{\phi}}}{algorithm for
  computing the set $\phiabsts{\S}{\phi} \cap \AMSks$, for
  $k=\csize{\phi}$}
\GlsSymbol{AbstLists-S-x-y}{\ensuremath{\AbstLists{\S}{x}{\vec{y}}}}{set of
  abstract lists w.r.t.~$\S$ with head $x$ and holes $\vec{y}$}
\GlsSymbol{AbstLists2-S-x-y}{\ensuremath{\AbstGeqtwoLists{\S}{x}{\vec{y}}}}{set
  of abstract lists of size at least $2$ w.r.t.~$\S$ with head $x$ and
  holes $\vec{y}$}
\GlsSymbol{AbstTrees-S-x-y}{\ensuremath{\AbstTrees{\S}{x}{\vec{y}}}}{set of
  abstract trees w.r.t.~$\S$ with root $x$ and holes $\vec{y}$}
\GlsSymbol{AbstTrees2-S-x-y}{\ensuremath{\AbstGeqtwoTrees{\S}{x}{\vec{y}}}}{set
  of abstract trees of size at least $2$ w.r.t.~$\S$ with root $x$ and
  holes $\vec{y}$}
\GlsSymbol{alloc-A}{\ensuremath{\allocAMS{\A}}}{the allocated
  variables of an AMS}
\GlsSymbol{alloc-x}{\ensuremath{\isalloced{x}}}{derived SSL formula
  expressing that $x$ is allocated}
\GlsSymbol{allocn-sh}{\ensuremath{\mfalloc{\S,\H}}}{the sets of
  variables allocated in negative chunks of $\SH$}
\GlsSymbol{ams-sh}{\ensuremath{\amsof{\S,\H}}}{the induced AMS of
  $\SH$}
\GlsSymbol{amss-phi}{\ensuremath{\phiabsts{\S}{\phi}}}{the abstraction
of $\SSL$ formula $\phi$ by the set of all AMS of models with stack
$\S$ that satisfy $\phi$}
\GlsSymbol{amsx-phi}{\ensuremath{\phiabstx{\vec{x}}{\phi}}}{the abstraction
of $\SSL$ formula $\phi$ by the set of all AMS of models with
$\dom(\S)=\vec{x}$ that satisfy $\phi$}
\GlsSymbol{AMS}{\ensuremath{\AMS}}{the set of all AMS}
\GlsSymbol{AMSks}{\ensuremath{\AMSks}}{the set of all AMS with nodes
  $\eqclassesAMS{\S}$ and garbage-chunk count at most $k$}
\GlsSymbol{AMSs}{\ensuremath{\AMSs}}{the set of all AMS with nodes
  $\eqclassesAMS{\S}$}
\GlsSymbol{array-s1-s2}{\ensuremath{\ArraySort{s_1}{s_2}}}{the sort of
arrays with indices of sort $\mathsf{s_1}$ and elements of sort $\mathsf{s_2}$}
\GlsSymbol{Bool}{\ensuremath{\sbool}}{Boolean sort in SMT}
\GlsSymbol{bound-phi}{\ensuremath{\sslpdatabound{\phi}}}{upper bound
  on the size of the minimal model of $\phi\in\SSLpdata$}
\GlsSymbol{char-S-E}{\ensuremath{\charform{\sdc,\EA}}}{characteristic
  formula for stack-data constraint $\sdc$ and EAMS $\EA$}
\GlsSymbol{chunks-sh}{\ensuremath{\chunks\SH}}{all chunks of the model
  $\SH$}
\GlsSymbol{chunksn-sh}{\ensuremath{\badchunks{\S,\H}}}{the negative chunks of the model
$\SH$}
\GlsSymbol{chunksp-sh}{\ensuremath{\goodchunks{\S,\H}}}{the positive chunks of the model
  $\SH$}
\GlsSymbol{classes-S}{\ensuremath{\eqclassesAMS{\sdc}}}{the
  stack-equivalence classes implied by stack-data constraint $\sdc$}
\GlsSymbol{d-fdata-l}{\ensuremath{\fdata(\ell)}}{data stored at
  location $\ell$}
\GlsSymbol{D}{\ensuremath{\DS}}{the set of built-in data structures of
  $\SSL$}
\GlsSymbol{Data}{\ensuremath{\sdata}}{the set of data values that can
  be stored in the heap}
\GlsSymbol{dconstraint-tau}{\ensuremath{\dconstraintphi{\tau}}}{the
  syntactic data constraint of $\tau$}
\GlsSymbol{decide+}{\ensuremath{\decidesslpKW}}{oracle for deciding
  $\SSLp$ satisfiability}
\GlsSymbol{dropdata-phi}{\ensuremath{\dropdp{\phi}}}{drop all data
  predicates, data fields, and data formula from $\phi$}
\GlsSymbol{ds}{\ensuremath{\anyDS}}{an arbitrary data structure from $\DS$}
\GlsSymbol{eams}{\ensuremath{\eamsof{\S,\H}}}{the induced
  EAMS of $\SH$ w.r.t.~$\Iset$ and $\Pset$}
\GlsSymbol{eamsphi}{\ensuremath{\phieabsts{\S}{\phi}}}{the set of
  induced EAMS of the models of formula $\phi$ with stack $\S$
  w.r.t.~predicate calls $\Iset$ and data predicates $\Pset$}
\GlsSymbol{edges-s-h}{\ensuremath{\iedge(\H_c)}}{the induced
  hyperedges of chunk $\H_c$ w.r.t.~stack $\S$}
\GlsSymbol{FData}{\ensuremath{\FData}}{the set of quantifier-free
  $\Tdata$ formulas}
\GlsSymbol{I-phi}{\ensuremath{\callsof{\phi}}}{the $\phi$-relevant
  predicate calls}
\GlsSymbol{K}{\ensuremath{\aKKW}}{constant combinator of array theory}
\GlsSymbol{l-fleft-l}{\ensuremath{\fleft(\ell)}}{left child of tree
  location $\ell$}
\GlsSymbol{lift-mn-A}{\ensuremath{\lift{m}{n}{\A}}}{bound-lifting of AMS $\A$ from $m\in\N$ to $n\in\N$}
\GlsSymbol{loc-ls-H}{\ensuremath{\slistloc{\H}}}{allocated list
  locations of heap $\H$}
\GlsSymbol{loc-tree-H}{\ensuremath{\streeloc{\H}}}{allocated tree
  locations of heap $\H$}
\GlsSymbol{ls}{\ensuremath{\ls}}{singly-linked list predicate in
  $\SSL$}
\GlsSymbol{ls-two}{\ensuremath{\lstwo}}{restriction of $\ls$ to lists of
  length at least $2$}
\GlsSymbol{map}{\ensuremath{\amapKW}}{map combinator of array theory}
\GlsSymbol{n-fnext-l}{\ensuremath{\fnext(\ell)}}{successor of list location $\ell$}
\GlsSymbol{P-phi}{\ensuremath{\dpsof{\phi}}}{$\phi$-implied data
  predicates, i.e., data predicates that occur explicitly in $\phi$}
\GlsSymbol{r-fright-l}{\ensuremath{\fright(\ell)}}{right child of tree
  location $\ell$}
\GlsSymbol{rsize-A}{\ensuremath{\rsize{\A}}}{the realizability size of
AMS $\A$}
\GlsSymbol{rsize-phi}{\ensuremath{\rsize{\phi}}}{the realizability size of
formula $\phi \in \SSLpdata$}
\GlsSymbol{S}{\ensuremath{\sdc}}{a stack-data constraint}
\GlsSymbol{SDC-phi-x}{\ensuremath{\SDCphix}}{the set of all stack-data
constraints w.r.t.~$\phi$ and $\vec{x}$}
\GlsSymbol{sig-ds}{\ensuremath{\nodeSig{\anyDS}}}{node signature of
  data structure $\anyDS$}
\GlsSymbol{sort-f}{\ensuremath{\sortof{\fany}}}{sort associated with the
  field $\fany$}
\GlsSymbol{SSL}{\ensuremath{\SSL}}{strong-separation logic}
\GlsSymbol{SSL-p}{\ensuremath{\SSLp}}{positive strong-separation
  logic}
\GlsSymbol{SSLdata}{\ensuremath{\SSLdata}}{strong-separation logic
  with data predicates}
\GlsSymbol{SSLdata-p}{\ensuremath{\SSLpdata}}{positive
  strong-separation logic with data predicates}
\GlsSymbol{store-A-i-v}{\ensuremath{\astore{\Arr}{i}{v}}}{store value
  $v$ at index $i$ of array $\Arr$}
\GlsSymbol{subsumed-I-A}{\ensuremath{\subsumedAMS{\Iset}{\A}}}{all
  predicate calls in $\Iset$ that hold in sub-models of the models of
  AMS $\A$}
\GlsSymbol{Tarray}{\ensuremath{\Tarray}}{array theory}
\GlsSymbol{Tdata}{\ensuremath{\Tdata}}{background data theory}
\GlsSymbol{tree}{\ensuremath{\tree}}{binary tree predicate in
  $\SSL$}
\GlsSymbol{tree-two}{\ensuremath{\treetwo}}{restriction of $\tree$ to trees of
  depth at least $2$}
\GlsSymbol{WSL}{\ensuremath{\WSL}}{weak-separation logic}
\GlsSymbol{WSL-p}{\ensuremath{\WSLp}}{positive weak-separation logic}
\GlsSymbol{--eq-AEQ}{\ensuremath{\AEQ}}{$\alpha$-equivalence on
  $\Sid$-trees and $\Sid$-forests}
\GlsSymbol{--Compose-ACompose}{\ensuremath{\ACompose}}{composition of $\Sid$-types}
\GlsSymbol{--Compose-FCompose}{\ensuremath{\FCompose}}{composition of
  $\Sid$-forests}
\GlsSymbol{--Compose-PCompose}{\ensuremath{\PCompose}}{composition of
  unfolded symbolic heaps}
\GlsSymbol{--derive-cfgderive}{\ensuremath{\RuleStep}}{one-step
  derivability relation between words induced by CFG}
\GlsSymbol{--derive-cfgderivestar}{\ensuremath{\RuleSteps}}{transitive
closure of $\RuleStep$}
\GlsSymbol{--derive-fderive}{\ensuremath{\fderive}}{one-step derivability
  relation on $\Sid$-forests}
\GlsSymbol{--derive-fderivestar}{\ensuremath{\fderivestar}}{reflexive--transitive
  closure of $\fderive$}
\GlsSymbol{--derive-pderive}{\ensuremath{\deriveqf}}{one-step derivability
  relation on unfolded symbolic heaps}
\GlsSymbol{--derive-pderivestar}{\ensuremath{\derivestar}}{reflexive--transitive
  closure of $\deriveqf$}
\GlsSymbol{--euinst}{\ensuremath{\xrightarrow{\protect\eexistsSMALL
      \vec{e} / \protect\fforallSMALL \vec{u}}}}{partial instantiation
  of universals with existentials from $\vec{e}$ or universals
  from $\vec{u}$}
\GlsSymbol{--sl-zrescope}{\ensuremath{\rescopeOP}}{separating conjunction
  with re-scoping (pushing existentials out)}
\GlsSymbol{--eq-rewreq}{\ensuremath{\rewreq}}{rewrite equivalence on
  formulas with guarded quantifiers}
\GlsSymbol{--models}{\ensuremath{\models}}{model relationship}
\GlsChildSymbol{--models-entailment}{--models}{\ensuremath{\models}}{entailment}
\GlsSymbol{--models-sidmodels}
{\ensuremath{\sidmodels}} {model relationship or entailment w.r.t.~SID
  $\Sid$}
\GlsSymbol{--zzall}{\ensuremath{\protect\fforall}}{guarded universal quantifier}
\GlsSymbol{--zzexists}{\ensuremath{\protect\eexists}}{guarded existential quantifier}
\GlsSymbol{-V-Phi}
{\ensuremath{\Sid}}
{system of inductive definitions}
\GlsSymbol{allholepredst}{\ensuremath{\tallholepredsKW(\ftree)}}{set of all
  hole predicates of $\Sid$-tree $\ftree$}
\GlsSymbol{allholesf}{\ensuremath{\fallholesKW(\frst)}}{set of all
  holes across all trees of $\Sid$-forest $\frst$}
\GlsSymbol{allholest}{\ensuremath{\tallholesKW(\ftree)}}{set of all
  holes of $\Sid$-tree $\ftree$}
\GlsSymbol{arity-pred}
{\ensuremath{\arityKW(\pred)}}
{arity of predicate identifier $\pred$}
\GlsSymbol{callst}{\ensuremath{\tcallsKW_\ftree(\la)}}{the recursive
  calls at location $\la$ in $\Sid$-tree $\ftree$}
\GlsSymbol{CFG}{\ensuremath{\CFG}}{the set of all context-free grammars}
\GlsSymbol{dom-frst}{\ensuremath{\dom(\frst)}}{the domain of
  the $\Sid$-forest $\frst$}
\GlsSymbol{DUSH-Phi}{\ensuremath{\DUSH{\Sid}}}{set of all DUSHs over
  SID $\Sid$}
\GlsSymbol{DUSHx-Phi}{\ensuremath{\DUSHx{\Sid}{\vec{x}}}}{set of all
  DUSHs $\fa$ over SID $\Sid$ with $\fvs{\fa}\subseteq\vec{x}$}
\GlsSymbol{dushroots-phi}{\ensuremath{\dushroots{\phi}}}{roots (i.e.,
  root variables of the predicates on the right-hand side of magic
  wands) of DUSH $\phi$}
\GlsSymbol{even}{\ensuremath{\even}}{singly-linked lists of even length}
\GlsSymbol{f-frst}{\ensuremath{\frst}}{a $\Sid$-forest}
\GlsSymbol{forestsphi}{\ensuremath{\sidfrstsof{\H}}}{set of
  $\Sid$-forests $\frst$ with $\fheapof{\frst}=\H$}
\GlsSymbol{ForgetPhi}{\ensuremath{\Forget{\scls,y}{\fa}}}{forget variable
  $y$ in $\fa$, assuming stack-aliasing constraint $\scls$}
\GlsSymbol{ForgetT}{\ensuremath{\Forget{\scls,y}{\typ}}}{forget variable
$y$ in $\Sid$-type $\typ$, assuming stack-aliasing constraint $\scls$}
\GlsSymbol{G-cfg}{\ensuremath{\G}}{a context-free grammar}
\GlsSymbol{graphf}{\ensuremath{\tgraphKW(\frst)}}{induced graph of
  $\Sid$-forest $\frst$}
\GlsSymbol{grapht}{\ensuremath{\tgraphKW(\ftree)}}{induced graph of
  $\Sid$-tree $\ftree$}
\GlsSymbol{headt}{\ensuremath{\theadKW_\ftree(\la)}}{the head
  predicate at location $\la$ in $\Sid$-tree $\ftree$}
\GlsSymbol{heapf}{\ensuremath{\fheapofKW(\frst)}}{the induced
  heap of $\Sid$-forest $\frst$}
\GlsSymbol{heapt}{\ensuremath{\fheapofKW(\ftree)}}{the induced
  heap of $\Sid$-tree $\ftree$}
\GlsSymbol{heaptl}{\ensuremath{\fheapofKW_\ftree(\la)}}{the induced
  heap of location $\la$ in $\Sid$-tree $\ftree$}
\GlsSymbol{heightt}{\ensuremath{\theight{\ftree}}}{the height of
  $\Sid$-tree $\ftree$}
\GlsSymbol{holepredst}{\ensuremath{\tholepredsKW_\ftree(\la)}}{the
  hole predicates of location $\la$ in $\Sid$-tree $\ftree$}
\GlsSymbol{holest}{\ensuremath{\tholesKW_\ftree(\la)}}{the holes of
  location $\la$ in $\Sid$-tree $\ftree$}
\GlsSymbol{IDbtw}{\ensuremath{\IDbtw}}{SIDs that satisfy
  progress, connectivity, and establishment}
\GlsSymbol{interface}{\ensuremath{\finterface{\frst}}}{the interface,
  i.e., the roots and holes, of the $\Sid$-forest $\frst$}
\GlsSymbol{L-G}{\ensuremath{\Lang{\G}}}{the language of context-free
  grammar $\G$}
\GlsSymbol{lalloc-phi}{\ensuremath{\lallocKW(\phi)}}{variables allocated
  locally, i.e., not considering recursive calls}
\GlsSymbol{lref-phi}{\ensuremath{\lrefKW(\phi)}}{variables referenced
  locally, i.e., not considering recursive calls}
\GlsChildSymbol{lsSID}{ls}{\ensuremath{\ls}}{user-defined
  singly-linked list predicate in $\SLID$}
\GlsSymbol{lsSIDlseg}{\ensuremath{\lseg}}{user-defined
  singly-linked list segment in $\SLID$}
\GlsSymbol{ModelsgSid}{\ensuremath{\Mpos{\Sid}}}{the set of guarded
  models w.r.t.~SID $\Sid$}
\GlsSymbol{odd}{\ensuremath{\odd}}{singly-linked lists of odd length}
\GlsSymbol{predroot}{\ensuremath{\prootKW}}{root parameter of a predicate call}
\GlsSymbol{Preds}
{\ensuremath{\PredsKW}}
{set of predicate identifiers}
\GlsSymbol{PredsPhi}
{\ensuremath{\Preds{\Sid}}}
{the set of predicate identifiers that occur in SID $\Sid$}
\GlsSymbol{project-Sf}{\ensuremath{\sfproj{\S}{\frst}}}{stack--forest projection}
\GlsSymbol{projectLocf}{\ensuremath{\lfproj{\vec{v}}{\frst}}}{forest
  projection of $\Sid$-forest $\frst$ w.r.t.~locations $\vec{v}$}
\GlsSymbol{projectLoct}{\ensuremath{\ltproj{\vec{v}}{\ftree}}}{tree
  projection of $\Sid$-tree $\ftree$}
\GlsSymbol{ptrlocst}{\ensuremath{\tptrlocsKW(\ftree)}}{Set of all
  locations that occur in a points-to assertion in $\Sid$-tree $\ftree$}
\GlsSymbol{ptrpred_k}
{\ensuremath{\ptrpred_k}}
{a predicate defined by the single rule $\ptrpred_k(\vec{x}) \Rule
  \pto{x_1}{\tuple{x_2,\ldots,x_{k+1}}}$}
\GlsSymbol{ptypes-phi-Sigma}{\ensuremath{\ptypesof{\phi}{\scls}}}{the
  $\vec{x}$-types of $\phi$ for aliasing-constraint $\scls$ that can
  be computed by looking up the types of predicate calls
  using the function $\thefun$}
\GlsSymbol{root-t}{\ensuremath{\trootKW(\ftree)}}{the root location of
  $\Sid$-tree $\ftree$}
\GlsSymbol{rootpred}{\ensuremath{\trootpred{\ftree}}}{the predicate
  call at the root of the $\Sid$-tree $\ftree$}
\GlsSymbol{roots-f}{\ensuremath{\froots{\frst}}}{the set of all roots
  of the $\Sid$-forest $\frst$}
\GlsSymbol{ruleinstf}{\ensuremath{\fruleinst{\frst}{l}}}{the rule instance
  at location $l$ in one of the $\Sid$-trees in $\Sid$-forest $\frst$}
\GlsSymbol{ruleinstt}{\ensuremath{\truleinstKW_\ftree(\la)}}{the rule instance
  at location $\la$ in $\Sid$-tree $\ftree$}
\GlsSymbol{SHe}
{\ensuremath{\SymHeap}}
{set of existentially-quantified symbolic heaps}
\GlsSymbol{SLIDa}{\ensuremath{\SLIDa}}{SLID formulas with guarded
  universals}
\GlsSymbol{SLIDea}{\ensuremath{\SLIDea}}{SLID formulas with guarded
  existentials and guarded universals}
\GlsSymbol{SLIDguarded}
{\ensuremath{\AllSLIDguarded}}
{guarded quantifier-free separation logic}
\GlsSymbol{SLIDguardedbtw}
{\ensuremath{\SLIDguarded}}
{guarded quantifier-free separation logic}
\GlsSymbol{SLIDqf}
{\ensuremath{\AllSLIDqf}}
{quantifier-free separation logic}
\GlsSymbol{SLIDqfbtw}
{\ensuremath{\SLIDqf}}
{quantifier-free separation logic}
\GlsSymbol{SLIDxtension}{\ensuremath{\SLop{\cdot_1,\ldots,\cdot_k}}}{the
SLID variant with additional atoms and operators $\cdot_1,\ldots,\cdot_k$.}
\GlsSymbol{splitfl}{\ensuremath{\fsplit{\frst}{\vec{l}}}}{the unique
  $\vec{l}$-split of $\Sid$-forest $\frst$}
\GlsSymbol{succt}
{\ensuremath{\tsuccKW_\ftree(\la)}}
{successors of location $\la$ in $\Sid$-tree $\ftree$}
\GlsSymbol{t-ftree}{\ensuremath{\ftree}}{a $\Sid$-tree}
\GlsSymbol{tll}{\ensuremath{\tll}}{user-defined predicate for trees
  with linked leaves}
\GlsChildSymbol{treeSID}{tree}{\ensuremath{\tree}}{user-defined
  binary tree predicate in $\SLID$}
\GlsSymbol{type-SH}{\ensuremath{\sidtypeof{\S,\H}}}{$\Sid$-type of
  $\SH$}
\GlsSymbol{types-phi-Sigma}{\ensuremath{\ttypesof{\phi}{\scls}}}{algorithm
that computes the $\Sid$-types of $\SLIDguarded$ formula $\phi$
for fixed stack-aliasing constraint $\scls$}
\GlsSymbol{Types-Phi}{\ensuremath{\allsidtypes{\Sid}}}{all
  $\Sid$-types over SID $\Sid$}
\GlsSymbol{Types-sPhi}{\ensuremath{\ssidtypes{\S}{\Sid}}}{restriction of
  $\allsidtypes{\Sid}$ to types of models with stack $\S$}
\GlsSymbol{TypesAliasing-phi}{\ensuremath{\phitypes{\scls}{\phi}}}{all
  $\Sid$-types of the models of $\phi$ with stack-aliasing constraint $\scls$}
\GlsSymbol{Typess-phi}{\ensuremath{\phitypes{\S}{\phi}}}{all
  $\Sid$-types of the models of $\phi$ with stack $\S$}
\GlsSymbol{Typesx-phi}{\ensuremath{\phitypes{\vec{x}}{\phi}}}{all
  $\Sid$-types of the models of $\phi$ with $\dom(\S)\subseteq\vec{x}$}
\GlsSymbol{USHPhi}{\ensuremath{\USH{\Sid}}}{set of all unfolded
  symbolic heaps w.r.t.~SID $\Sid$}
\let\oursunion\undefined

\setcopyright{acmcopyright}
\copyrightyear{2018}
\acmYear{2018}
\acmDOI{10.1145/1122445.1122456}

\acmJournal{JACM}
\acmVolume{37}
\acmNumber{4}
\acmArticle{111}
\acmMonth{8}

\begin{document}
\selectlanguage{american} % american ngerman

\title{A Decision Procedure for Guarded Separation Logic}
\subtitle{Complete Entailment Checking for Separation Logic with Inductive Definitions}


\author{Christoph Matheja}
\affiliation{%
  \institution{ETH Zurich}
  \city{Zurich}
  \country{Switzerland}}
\email{cmatheja@inf.ethz.ch}

\author{Jens Pagel}
\email{pagel@forsyte.at}
\affiliation{%
  \institution{TU Wien}
  \city{Vienna}
  \country{Austria}
}

\author{Florian Zuleger}
\affiliation{%
  \institution{TU Wien}
  \city{Vienna}
  \country{Austria}
}
\email{zuleger@forsyte.at}

\renewcommand{\shortauthors}{Matheja, Pagel and Zuleger}

\begin{abstract}%
  We develop a doubly-exponential decision procedure for the satisfiability problem
of \emph{guarded separation logic}---a novel fragment
of separation logic featuring user-supplied inductive predicates,
Boolean connectives, and separating connectives, including restricted (guarded) versions
of negation, magic wand, and septraction.
Moreover, we show that dropping the guards for any of the above connectives leads to an undecidable fragment.

We further apply our decision procedure to reason about \emph{entailments} in the popular symbolic heap fragment of separation logic.
In particular, we obtain a doubly-exponential decision procedure for entailments between (quantifier-free) symbolic heaps with inductive predicate definitions of bounded treewidth ($\SLIDbtw$)---one of the most expressive decidable fragments of separation logic.
Together with the recently shown $\TwoExpTime$-hardness for entailments in said fragment, we conclude that the entailment problem for $\SLIDbtw$
is $\TwoExpTime$-complete---thereby closing a previously open complexity gap. 
% \noindent{}In \cite{katelaan2019effective}, we proposed a novel decision
%  procedure for entailment checking in the symbolic-heap segment of
%  separation logic with user-defined inductive definitions of bounded
%  treewidth.
%  %
%  In the meantime, we discovered that the decision procedure
%  in~\cite{katelaan2019effective} is incomplete.
%  %
%  In this article, we fix the incompleteness issues while retaining
%  the double-exponential asymptotic complexity bound.
%  %
%  In doing so, we also remove several of the simplifying assumptions
%  made in \cite{katelaan2019effective}.
%  %
%  Furthermore, we generalize our decision procedure to the fragment of
%  \emph{positive formulas}, in which conjunction, disjunction, and
%  guarded occurrences of negation, septraction and the magic wand can
%  be freely combined with the separating conjunction.

\end{abstract}

%\begin{CCSXML}
%<ccs2012>
%  <concept_id>10010520.10010553.10010562</concept_id>
%  <concept_desc>Computer systems organization~Embedded systems</concept_desc>
%  <concept_significance>500</concept_significance>
%  <concept_id>10010520.10010575.10010755</concept_id>
%  <concept_desc>Computer systems organization~Redundancy</concept_desc>
%  <concept_significance>300</concept_significance>
%  <concept_id>10010520.10010553.10010554</concept_id>
%  <concept_desc>Computer systems organization~Robotics</concept_desc>
%  <concept_significance>100</concept_significance>
%  <concept_id>10003033.10003083.10003095</concept_id>
%  <concept_desc>Networks~Network reliability</concept_desc>
%  <concept_significance>100</concept_significance>
%</ccs2012>
%\end{CCSXML}
%
%\ccsdesc[500]{Computer systems organization~Embedded systems}

%\keywords{datasets, neural networks, gaze detection, text tagging}

\maketitle

\section{Introduction}\label{sec:intro}
%
%\paragraph{Separation logic}
Separation Logic (SL) \cite{ishtiaq2001bi,reynolds2002separation} is a popular formalism for Hoare-style verification of imperative, heap-manipulating programs.
At its core, SL extends first-order logic with two connectives---the \emph{separating conjunction} $\sep$ and the \emph{separating implication} $\mw$ (aka magic wand)---for concisely specifying how resources, such as program memory, can be split-up and extended, respectively.
Based on these connectives, SL enables \emph{local reasoning}, i.e., sound verification of program parts in isolation, about the resources employed by a program---a key property responsible for SL's broad adoption in
static analysis~\cite{berdine2007shape,gotsman2007threadmodular,calcagno2011compositional,calcagno2011infer,calcagno2015moving}, automated verification~\cite{berdine2005smallfoot,berdine2011slayer,chin2012automated,jacobs2011verifast,piskac2014grasshopper,muller2017viper,ta2018automated}, and interactive theorem proving~\cite{appel2014program,jung2018iris}.

Regardless of the flavor of formal reasoning, any \emph{automated} approach based on SL ultimately relies on a solver for discharging either the \emph{satisfiability problem}---does the SL formula $\phi$ have a model?---or the \emph{entailment problem}---is every model of $\phi$ also a model of $\psi$, or, equivalently, is $\phi \wedge \neg \psi$ unsatisfiable?
While both problems are undecidable (and equivalent) in general~\cite{calcagno2001computability}, various decidable SL fragments, in which entailments cannot be reduced to the (un)satisfiability problem because negation is forbidden, have been proposed in the literature, e.g.,~\cite{berdine2004decidable,cook2011tractable,iosif2013tree,echenim2020bernays}.

In particular, the \emph{symbolic heap} fragment---an idiomatic form of SL formulas with $\sep$ but without $\mw$ that is often encountered when manually writing program proofs~\cite{berdine2005symbolic}---has received a lot of attention.
Symbolic heaps appear, for instance, in the automated tools \textsc{Infer}~\cite{calcagno2011infer}, \textsc{Sleek}~\cite{chin2012automated}, \textsc{Songbird}~\cite{ta2016automated}, \textsc{Grasshopper}~\cite{piskac2014grasshopper}, \textsc{Verifast}~\cite{jacobs2011verifast}, \textsc{SLS}~\cite{ta2018automated}, and \textsc{Spen}~\cite{enea2017spen}.
To support complex data structure specifications,
%, e.g., linked lists or trees,
symbolic heaps are often enriched with \emph{systems of inductive predicate definitions} (SIDs).
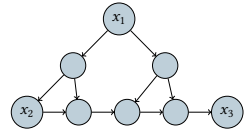
\begin{figure}[t!]
  \begin{subfigure}[b]{0.6\textwidth}
      $\begin{array}{lll}
        \gls{tll}(x_1,x_2,x_3) &\Rule& \ppto{x_1}{\tuple{\nil,\nil,x_3}} \sep (\sleq{x_1}{x_2})
        \\
    \tll(x_1,x_2,x_3)& \Rule& \SHEX{\tuple{l,r,m}} \ppto{x_1}{\tuple{l,r,\nil}}
      \\&&\phantom{\SHEX{\tuple{l,r,m}}}\sep \tll(l,x_2,m)\\&&\phantom{\SHEX{\tuple{l,r,m}}} \sep \tll(r,m,x_3)
      \end{array}$
      \caption{An SID specifying trees with linked leaves.}
      \label{fig:ex:sid:tll}
    \end{subfigure}\hfill
    \begin{subfigure}[b]{0.3\textwidth}
      \scalebox{0.6}{\begin{tikzpicture}
        \node[tll] (x) {$x_1$};
        \node[tll,below left=8mm of x] (l) {};
        \node[tll,below right=8mm of x] (r) {};
        \node[tll,below left=8mm of l] (ll) {$x_2$};
        \node[tll,right=5mm of ll] (lr) {};
        \node[tll,right=5mm of lr] (rl) {};
        \node[tll,right=5mm of rl] (rr) {};
        \node[tll,right=5mm of rr] (x3) {$x_3$};
        \draw (x) edge[->] (l);
        \draw (x) edge[->] (r);
        \draw (l) edge[->] (ll);
        \draw (l) edge[->] (lr);
        \draw (r) edge[->] (rl);
        \draw (r) edge[->] (rr);
        \draw (ll) edge[->] (lr);
        \draw (lr) edge[->] (rl);
        \draw (rl) edge[->] (rr);
        \draw (rr) edge[->] (x3);
      \end{tikzpicture}

%%% Local Variables:
%%% mode: latex
%%% TeX-master: "../main"
%%% End:
}
      \caption{A model of $\tll(x_1,x_2,x_3)$.}
      \label{fig:ex:sid:heap}
    \end{subfigure}
    \caption[Example: SID defining trees with linked leaves]{
        The SID of \citet{iosif2013tree} defining trees with linked lives and an illustration of a model.}\label{fig:tll-example}
\end{figure}
For example, \cref{fig:tll-example} depicts an SID specifying trees with linked leaves as well as an illustration of a model ($\nil$-pointers have been omitted for readability).

The precise form of permitted SIDs has a significant impact on the decidability and complexity of reasoning about symbolic heaps:
\citet{brotherston2014decision} showed that \emph{satisfiability} is $\ExpTime$-complete for symbolic heaps over arbitrary SIDs, whereas the \emph{entailment problem} is undecidable in general (cf.~\cite{antonopoulos2014foundations,iosif2014deciding}).
To deal with entailments, tools rely on specialized methods for fixed predicates~\cite{berdine2004decidable,cook2011tractable,piskac2013automating,piskac2014automating}, decision procedures for restricted classes of SIDs~\cite{iosif2013tree,iosif2014deciding}, or incomplete approaches, e.g., fold/unfold reasoning~\cite{chin2012automated} or cyclic proofs~\cite{brotherston2011automated}.

Among the largest decidable classes of symbolic heaps with user-supplied SIDs is the fragment of \emph{symbolic heaps with bounded treewidth} ($\SLIDbtw$) developed by \citet{iosif2013tree}, which supports rich data structure definitions, such as the one in~\cref{fig:tll-example}.
Further examples include doubly-linked lists and binary trees with parent pointers.
Decidability is achieved by imposing three syntactic conditions on SIDs, which allow reducing the entailment problem for $\SLIDbtw$ to the (decidable) satisfiability problem for monadic second-order logic (MSO) over graphs of bounded treewidth (cf.~\cite{courcelle2012graph}).
This reduction yields an elementary decision procedure (by analyzing the resulting quantifier depth, it is in $4\ExpTime$).
However, it is infeasible in practice.
Furthermore, there is a ``complexity gap'' between the above decision procedure and a recent result proving that the entailment problem for $\SLIDbtw$ is at least $\TwoExpTime$-hard~\cite{echenim2019lower}

The goal of this article is twofold:
First, we look beyond symbolic heaps and study
\emph{guarded separation logic} ($\SLIDguarded$)---a novel SL fragment featuring
both standard Boolean and separating connectives (including restricted forms of negation and magic wand) as well as SIDs supported by $\SLIDbtw$.
In particular, we develop a doubly-exponential decision procedure for the \emph{satisfiability problem} of $\SLIDguarded$.
Second, we show that the \emph{entailment problem} for $\SLIDbtw$
can be reduced to the satisfiability problem for $\SLIDguarded$ because an $\SLIDbtw$ formula $\phi$ entails an $\SLIDbtw$ formula $\psi$ iff the $\SLIDguarded$ formula $\phi \wedge \neg \psi$ is unsatisfiable.
Consequently, we close the aforementioned complexity gap and conclude that the entailment problem for $\SLIDbtw$ is \emph{$\TwoExpTime$-complete}.
\paragraph{Guarded separation logic}
Inspired by work on first-order logic with \emph{guarded negation}, we propose the fragment $\SLIDguarded$ of \emph{guarded separation logic}.
$\SLIDguarded$ supports negation $\neg$, magic wand $\mw$ and septraction $\sept$~\cite{brochenin2012almighty}, but requires each of these connectives to appear in conjunction with another $\SLIDguarded$ formula $\phi$, i.e., $\phi \wedge \neg \psi$, $\phi \wedge (\psi \mw \vartheta)$, or $\phi \wedge (\psi \sept \vartheta)$, acting as its \emph{guard}; hence, the name.
By construction, a guard is never equivalent to true and thus cannot be dropped.

While we consider the satisfiability problem of quantifier-free $\SLIDguarded$ formulas, we admit arbitrary inductive predicates as long as they can be defined in $\SLIDbtw$, which supports existential quantifiers.
Hence, the formulas below belong to $\SLIDguarded$ and are thus covered by our decision procedure.
\begin{align*}
& \tll(x,y,z) \wedge \neg\pto{x}{\tuple{\nil,\nil,z}}
\tag{a tree with linked leaves and at least three nodes}
\\
& \left(\pto{x}{\tuple{y,z}} \sep \tll(y,\ell,r) \sep \tll(z,\ell,\nil)\right)
\wedge \neg\tll(x,\ell,\nil)
\tag{encoding of an entailment in $\SLIDbtw$}
\\
& \left(\pto{x}{\tuple{y,z}} \sep \tll(z,\ell,\nil)\right)
       \wedge \left( \tll(y,\ell,r) \mw \tll(x,\ell,\nil) \right)
\tag{tll where the root's left subtree is missing}
\end{align*}
\paragraph{Abstraction-based satisfiability checking}
Our decision procedure for $\SLIDguarded$ satisfiability---and thus also for $\SLIDbtw$ entailments---is
based on the \emph{compositional computation} of an \emph{abstraction} of program states, i.e., the universe of potential models, that \emph{refines} the satisfaction relation $\models$ of $\SLIDguarded$.\footnote{We will properly formalize all notions mentioned in this section in the remainder of this article.}
That is, we will develop an abstract domain $\AD$ and an abstraction function
$\genabst\colon\States \to \AD$ with the following three key properties:
\begin{enumerate}
    \item \emph{Refinement.}
    Whenever
    $\genabst(\sigma)=\genabst(\sigma')$ holds for two states $\sigma$ and $\sigma'$, then $\sigma$ and $\sigma'$ satisfy the same $\SLIDguarded$ formulas, i.e.,
  the equivalence relation induced by our abstraction function,
  \[
    \sigma \equiv_{\genabst} \sigma'
    \quad\text{iff}\quad
    \genabst(\sigma)=\genabst(\sigma'),
  \]
  refines the satisfaction relation $\models$ of $\SLIDguarded$.
\item \emph{Compositionality.}
  For each logical connective supported by $\SLIDguarded$, the abstraction function $\genabst$ can be computed compositionally from already known abstractions.
  For example, for the separating conjunction $\sep$, this means that there exists an effectively computable operation $\Compose\colon \AD \times \AD \pfun \AD$ such that, for all states $\sigma$ and $\sigma'$,
  \[ \genabst(\sigma \stdunion \sigma') = \genabst(\sigma) \Compose \genabst(\sigma'), \]
   where $\sigma \stdunion \sigma'$ denotes the ``disjoint union'' of two states used to assign semantics to the separating conjunction.
\item \emph{Finiteness.}
    The abstract domain $\AD$ has only finitely many elements.
\end{enumerate}
Put together, refinement and compositionality allow lifting the abstraction function over states
\[ \genabst\colon \States \to \AD \]
to a function over models of $\SLIDguarded$ formulas
\begin{align*}
   \genabst_\SLIDguarded\colon & \SLIDguarded \to \AD,
   \qquad
   \phi \mapsto \set{ \genabst(\sigma) \mid \sigma \in \States, \sigma \models \phi },
\end{align*}
Provided we can compute the abstraction $\genabst_{\SLIDguarded}$ of every atomic $\SLIDguarded$ formula, we can then use $\genabst_{\SLIDguarded}$ for satisfiability checking: The $\SLIDguarded$ formula $\phi$ is satisfiable iff $\genabst_{\SLIDguarded}(\phi) \neq \emptyset$.
Finiteness then ensures that the set $\genabst_{\SLIDguarded}(\phi)$ is finite; it can thus be computed and checked for emptiness.

We will provide a more detailed overview in \cref{ch:towards} of our abstraction once we have precisely defined the semantics of guarded separation logic formulas.
\paragraph{Contributions}
The main contributions of this article can be summarized as follows:
\begin{itemize}
\item%
We study the decidability of (quantifier-free) \emph{guarded} separation logic ($\SLIDguarded$)---a novel separation logic fragment that goes beyond symbolic heaps with user-defined inductive definitions by featuring restricted (guarded) versions of the magic wand, septraction, and negation.
\item
We show that omitting the guards for \emph{any} of the  three operators --- magic wand, septraction, and negation --- leads to an undecidable logic.
Together with our decidability results, this yields an almost tight decidability delineation between for separation logics that admit user-defined inductive predicate definitions.
\item%
We present a decision procedure for the satisfiability problem of $\SLIDguarded$ based on
the compositional computation of finite abstractions, called $\Sid$-types, of potential models.
\item%
We analyze the complexity of the above decision procedure and show that satisfiability of
$\SLIDguarded$ is decidable in $\TwoExpTime$.
\item%
We apply our decision procedure for $\SLIDguarded$ to decide, again in $\TwoExpTime$, the \emph{entailment} problem for (quantifier-free) symbolic heaps with user-defined inductive definitions of bounded-tree; in light of the recently shown $\TwoExpTime$-hardness by \citet{echenim2019lower}, we obtain that said entailment problem is
$\TwoExpTime$-\emph{complete}---thereby closing an existing complexity gap.
\end{itemize}
This article unifies and revises the results of two conference papers \cite{katelaan2019effective,katelaan2020beyond}.
We note that both papers only sketch the main ideas and most of the proofs were omitted.
In this article, we dedicate a whole section to the careful motivation of our abstraction (see \cref{ch:towards}) and present all the proofs
(an early version of this article was put on arXiv~\cite{journals/corr/abs-2002-01202} in order to convince the reviewers of \cite{katelaan2020beyond} of the correctness of our results).
We remark that we have significantly improved the presentation and reworked all the technical details in comparison to our earlier technical report \cite{journals/corr/abs-2002-01202}.

\paragraph{Organization of the article}
%
%In \cref{sec:related-work}, we present related work and discuss how our fragment of guarded separation logic fits into the landscape of (decidable) fragments of separation logic considered in the literature.
After agreeing on basic notational conventions in \cref{sec:notation}, we briefly recap separation logic in \cref{sec:sl-basics}.
In particular, we consider user-defined inductive definitions and the bounded treewidth fragment upon which our own SL fragments are based.
We introduce the novel fragment of guarded separation logic in \cref{sec:sl-guarded}.
\cref{ch:undec} shows that even small extensions of guarded separation logic lead to an unsatisfiable satisfiability problem.
The remainder of this article is concerned with developing a decision procedure for guarded separation logic and, by extension, the entailment problem for symbolic heaps with inductive definitions of bounded treewidth.
\cref{ch:towards} informally discusses the main ideas underlying our decision procedure.
The formal details are worked out in \cref{ch:forests,ch:types,ch:deciding-btw}.
In particular, in \cref{ch:deciding-btw}, we present the decision procedure itself and analyze its complexity.
Finally, we conclude in \cref{sec:conclusion}.
To improve readability, some technical proofs have been moved to the Appendix at the end of this article.
\paragraph{Acknowledgments}
We thank Mnacho Echenim, Radu Iosif, and
Nicolas Peltier for their outstandingly thorough study of \cite{katelaan2019effective}, which presented the originally proposed abstraction-based decision procedure, and their help in discovering an incompleteness issue, which we were able to fix in our follow-up work \cite{katelaan2020beyond,journals/corr/abs-2002-01202}.

\section{Notation}
\label{sec:notation}
Throughout this article, we adhere to the following basic notational conventions.
\paragraph{Sequences}
Finite sequences are denoted either in boldface, e.g., $\vec{x}$, or by explicitly listing their elements, e.g., $\tuple{x_1,\ldots,x_k}$; the \emph{empty sequence} is $\emptyseq$.
The \emph{length} of the sequence $\vec{x}$ is $\size{\vec{x}}$.
We call $\vec{x}$ \emph{repetition-free} if its elements are pairwise different. 
To reduce notational clutter, we often omit the brackets around sequences of length one, i.e., we write $x$ instead of $\tuple{x}$.
The sequence $\vec{x} \concat \vec{y}$ is obtained from \emph{concatenating} the sequences $\vec{x}$ and $\vec{y}$.
$A^{*}$ is the \emph{set of all finite sequences} over some set $A$; $A^{+}$ is the set of all \emph{non-empty} finite sequences over $A$.
\paragraph{Sets from sequences}
We frequently treat sequences as sets if the ordering of elements is irrelevant.
For example, $x \in \vec{x}$ states that the sequence $\vec{x}$ contains the element $x$,
$\vec{x} \cup \vec{y}$ is the set consisting of all elements of the sequences $\vec{x}$ and $\vec{y}$, etc.
\paragraph{Partial functions}
We denote by $f\colon A \pfun B$ a \emph{(partial) function} with \emph{domain} $\dom(f) \defn A$ and \emph{image} $\img(f) \defn B$.
If $f$ is undefined on $x$, i.e., $x \notin \dom(f)$, we write $f(x) = \bot$.
Moreover, $f \funcomp g$ is the \emph{composition} of the functions $f$ and $g$  mapping every $x$ to $f(g(x))$.
We interpret the \emph{size} $\size{f}$ of a partial function $f$ as the cardinality of its domain, i.e., $\size{f} \defn \size{\dom(f)}$;
$f$ is \emph{finite} if $\size{f}$ is finite.
We often describe finite partial functions as sets of mappings.
The set $\set{x_1 \mapsto y_1, \ldots, x_k \mapsto y_k}$, for example, 
represents the partial function that, for every $i \in [1,k]$, maps $x_i$ to $y_i$; it is undefined for all other values.
%In particular, $\empty$ is the everywhere-undefined function.
%We write $f \cap g$ as a shortcut for the common domain $\dom(f) \cap \dom(g)$ of the partial function $f$ and $g$.
Furthermore, $f \cup g$ denotes the (not necessarily disjoint) union of $f$ and $g$; it is defined iff $f(x) = g(x)$ holds for all $x \in \dom(f) \cap \dom(g)$.
Formally:
\begin{align*}
    (f \cup g)(x) \defn \begin{cases} f(x), & \text{if}~ x \in \dom(f), \\ g(x), & \text{if}~ x \in \dom(g) \setminus \dom(f), \\ \bot, & \text{otherwise.} \end{cases}
\end{align*}
We write $f \stdunion g$ instead of $f \cup g$ whenever we additionally require that $\dom(f) \cap \dom(g) = \emptyset$.
Furthermore, we denote by $\pinst{f}{x}{v}$ the \emph{updated} partial function in which $x$ maps to $v$, i.e.,
\begin{align*}
    %\pinst{f}{x}{v}\colon \dom(f) \cup \{x\} \to \img(f) \cup \{v\}, \qquad
    \pinst{f}{x}{v}(y) \defn \begin{cases} v & \text{if}~ y = x, \\ f(y), & \text{otherwise.} \end{cases}
\end{align*}
In particular, if $f(x)$ is undefined, $\pinst{f}{x}{v}$ adds $x$ to the domain of the resulting function. By slight abuse of notation, we write $\pinst{f}{x}{\bot}$ to denote the function in which $x$ is removed from the domain of $f$.
To compare partial functions, we say that function $g$ \emph{subsumes} function $f$, written $f \subseteq g$, if (1) $g$ is at least as defined as $g$ and (2) $g$ agrees with $f$ on their common domain. Formally:
\[
    f \subseteq g \quad\text{iff}\quad \dom(f) \subseteq \dom(g) \quad\text{and}\quad \forall x \in \dom(f)\colon f(x) = g(x).
\]
Whenever $f$ and $g$ map to sets or functions, we also use a weaker ordering. 
The relation $f \subfun g$ is defined as $f \subseteq g$ but only requires that $g(x)$ subsumes $f(x)$ if both are defined on $x$, i.e.,
\[
    f \subfun g \quad\text{iff}\quad \dom(f) \subseteq \dom(g) \quad\text{and}\quad \forall x \in \dom(f)\colon f(x) \subseteq g(x).
\]
\paragraph{Functions over sequences}
We implicitly lift partial functions $f\colon A \pfun B$ to functions $f\colon A^{*} \pfun B^{*}$ over sequences 
by pointwise application. That is, for a sequence $\tuple{a_1,\ldots,a_k} \in A^{*}$, we define 
\[
    f(\tuple{a_1,\ldots,a_k})
    \defn
    f(a_1,\ldots,a_k)
    \defn
    \tuple{f(a_1), \ldots, f(a_k)},
\]
where, as indicated above, we omit the brackets indicating sequences to improve readability.

Finally, we lift the update $\pinst{f}{x}{v}$ of a single value to sequences of values $\vec{x} = \tuple{x_1,\ldots,x_k}$ and $\vec{v} = \tuple{v_1,\ldots,v_k}$
by setting $\pinst{f}{\vec{x}}{\vec{v}} \defn \pinst{\pinst{\pinst{f}{x_1}{v_1}}{x_2}{v_2}\ldots}{x_k}{v_k}$.

\section{Separation Logic with Inductive Definitions}
\label{sec:sl-basics}
We briefly recapitulate the basics of first-order separation logic with user-defined predicates.
That is, we introduce the syntax and semantics of both separation logic
and systems of inductive definitions, the symbolic heap fragment, and the bounded treewidth fragment originally
studied by \citet{iosif2013tree}.
Most of the presented material is fairly standard (cf., among others,~\cite{ishtiaq2001bi,reynolds2002separation,brotherston2014decision,iosif2014deciding})
with the notable exception
that \emph{our semantics of pure formulas enforces the heap to be empty}.
A reader familiar with separation logic may skim over this section to familiarize herself with our notation.

\subsection{The Syntax of Separation Logic}
\label{sec:sl-basics:syntax}
\begin{figure}
  \centering
  $
  \begin{array}{lll}
               &     & x \in \Var,~ u, v \in \Var \cup \Val,~ \vec{w} \in (\Var \cup \Val)^+,~ \pred \in \PredsKW \\
      \phiAtom & ::= & \emp \mid \sleq{u}{v} \mid \slneq{u}{v} \mid
                       \pto{u}{\vec{w}} \mid \pred(\vec{w}) \\
  \phi & ::= & \phiAtom \mid \phi \sep \phi \mid \phi \mw \phi \mid
               \phi \wedge \phi \mid \phi \vee \phi \mid \neg \phi
               \mid \EXO{x}\phi \mid \FAO{x}\phi\\
    %
    % \phiSH & ::= & \EXO{\underbrace{x_1,\ldots,x_k}_{k \geq 0}} \underbrace{\phiAtom \sep \cdots
    %                \sep \phiAtom}_{\text{1 or more atoms}}
  \end{array}
  $
  \caption[Syntax of $\SLgeneric$]{The syntax of a
    first-order separation logic with user-defined predicates ($\SLgeneric$)}
  \label{fig:sl:generic-syntax}
\end{figure}
\Cref{fig:sl:generic-syntax} defines the syntax of first-order separation logic with user-defined predicates  ($\SLgeneric$ for short),
where $x$ is drawn from a countably infinite set $\Var$ of \emph{variables},
$u$ and $v$ are either variables or \emph{values} drawn from the countably infinite set $\Val$,
and $\vec{w}$ is a finite sequence whose elements can be both variables and values.
In particular, notice that any value in $\Val$ may appear as a constant in formulas.
Moreover, $\pred$ is taken from a finite set $\PredsKW$ of \emph{predicate identifiers}; each predicate $\pred$ is equipped with an \emph{arity} $\arity{\pred} \in \mathbb{N}$ that determines its number of parameters. %passed to the predicate, i.e., $\pred(\vec{z})$ implies $\arity{\pred} = \size{\vec{z}}$.

Informally, the meaning of the atomic formulas is as follows:
\begin{itemize}
\item The \emph{empty-heap predicate} $\emp$ denotes the empty heap.
\item The \emph{equality} $\sleq{u}{v}$ and the \emph{disequality} $\slneq{u}{v}$ express that $u$ and $v$ alias and that they do not alias in the current program state (whose heap needs to be empty), respectively.
\item The \emph{points-to assertion} $\pto{u}{\vec{w}}$ states that the address $u$ points to a heap-allocated object consisting of $\size{\vec{w}} > 0$ fields, where the $i$-th field stores the $i$-th value of the sequence $\vec{w}$.
\item The \emph{predicate call} $\pred(\vec{w})$ allows to refer to user-defined data structures, e.g., lists and trees.
  %
  %
% \item The predicate $\ls(x,y)$ states that the heap contains a
%   singly-linked list segment from $x$ to $y$ of arbitrary length.
\end{itemize}
The $\SLgeneric$ formulas $\emp$ and $\pto{u}{v}$ are called
\emph{spatial} atoms because they describe the spatial layout of the
heap, whereas (dis-)equalities are called \emph{pure}
atoms~\cite{ishtiaq2001bi} because they do not depend on the heap.
Apart from atoms, $\SLgeneric$ formulas are built from
\begin{itemize}
\item classical propositional connectives, i.e., conjunction ($\wedge$), disjunction ($\vee$), and negation ($\neg$),
\item existential ($\exists$) and universal ($\forall$) quantifiers, and
\item \emph{separating connectives}, i.e., \emph{separating conjunction} $\sep$ and \emph{implication} (or magic wand) $\mw$.
\end{itemize}
As usual, one can derive additional operators such as standard implication
$\phi\Rightarrow\psi \defn \neg \phi \vee \psi$ and \emph{septraction}
$\phi \gls{--slwand-sept} \psi \defn \neg (\phi \mw \neg
\psi)$~(cf.~\cite{brochenin2012almighty,thakur2014satisfiability}).

The semantics of the classical connectives is standard.
Let us briefly compare their meaning with the intuition underlying the separating connectives.
While $\phi \wedge \psi$ means that the program state satisfies both
$\phi$ and $\psi$ simultaneously, $\phi \sep \psi$ denotes that
(the heap component of) the program state can be split into two disjoint parts which separately
satisfy $\phi$ and $\psi$.
Similarly, while $\phi \Rightarrow \psi$ means that every program
state satisfying $\phi$ also satisfies $\psi$, $\phi \mw \psi$
means that the \emph{extension} of the program state with any program
state that satisfies $\phi$ yields a program state that satisfies
$\psi$.

The magic wand is useful for weakest-precondition
reasoning, e.g., to express
memory allocation~\cite{ishtiaq2001bi,reynolds2002separation,batz2019quantitative}.
However, automated verification tools often do not or only partially support the magic wand,
%In much of the separation logic literature as well as in the bulk of
%the tools, the magic wand is not considered or only partially
%supported, however,
because its inclusion quickly leads to
undecidability~\cite{appel2014program,blom2015witnessing,schwerhoff2015lightweight}.
%
%
%\paragraph{Substitution}
\subsubsection{Substitution}
Various constructions throughout this article involve syntactically replacing variables and values---we thus give a generic definition that allows performing multiple substitutions at once.
Let $\vec{y},\vec{z} \in (\Var \cup \Val)^*$ be sequences of the same length, where $\vec{y}$ is repetition-free.
We denote by $\pinst{\phi}{\vec{y}}{\vec{z}}$ the formula obtained from $\phi$ through (simultaneous) \emph{substitution} of each element in $\vec{y}$ by the element in $\vec{z}$ at the same position; %\Cref{fig:substitution} 
\Cref{app:substitution} provides a formal definition.
%the substitution is undefined if the result is not an $\SLgeneric$ formula (e.g., we cannot replace a quantified variable by a value). 
%For example, %we have
E.g.,
\[
    \pinst{\left(\SHEX{x} \pto{x}{\tuple{y, 7}} \sep \ls(y,x)\right)}{\tuple{y,7}}{\tuple{3,x}}
    ~=~
    \SHEX{x} \ppto{x}{\tuple{3, x}} \sep \ls(3,x).
\]
%For example, for $\vec{y}=\tuple{x}$ and $\vec{z}=\tuple{w}$, we have $\pinst{\ppto{x}{v}}{\vec{y}}{\vec{z}} = \pto{w}{v}$
%and $\pinst{(\SHEX{x} \ls(z,x))}{\vec{y}}{\vec{z}} = \SHEX{w} \ls(z,w)$.
%
Moreover, we write $\phi(\vec{z})$ as a shortcut for the substitution $\pinst{\phi}{\fvs{\phi}}{\vec{z}}$.
\subsection{The Stack-Heap Model}
\label{sec:sl-basics:stack-heap}
We interpret $\SLgeneric$ in terms of the widely-used \emph{stack-heap model}, which already appears in the seminal
papers of~\citet{ishtiaq2001bi} and~\citet{reynolds2002separation}.
A stack-heap pair $\SH$ consists of a stack $\S$ assigning values to variables
and a heap $\H$ assigning values to allocated memory locations.

Towards a formal definition, we fix the set $\Val \defn \mathbb{Z}$ of \emph{values}
and the set
$\Loc \defn \mathbb{N}_{> 0} \subseteq \Val$ of \emph{addressable memory locations}; the \emph{null pointer} $0$ is a value but \emph{not} a location.
The set $\SS$ of \emph{stacks} then consists of all finite partial functions mapping variables to values, i.e.,
\begin{align*}
   \SS \defn \set{ \S \mid \S\colon V \pfun \Val,\,  V \subseteq \Var,\, \size{V} < \infty, },
\end{align*}
In order to treat both evaluations of variables and constant values uniformly, we slightly abuse notation and
set $\S(v) \defn v$ for all values $v \in \Val$.\footnote{This convention does not affect the formal definition of stacks; in particular, their domain and image remains unchanged.}
The set $\HH$ of \emph{heaps} consists of all finite partial functions mapping allocated memory locations to sequences of values, i.e.,
\begin{align*}
    \HH \defn \set{ \H \mid \H\colon L \pfun \Val^{+},\, L \subseteq \Loc,\, \size{L} < \infty }.
\end{align*}
By mapping locations to sequences rather than single values, the heap assigns every allocated memory location to the entire structure allocated at this location.
This is a fairly standard---but far from ubiquitous~\cite{reynolds2002separation,calcagno2006beyond}---abstraction of the actual memory layout; it simplifies the memory model without losing precision as long as we do not use pointer arithmetic.
We frequently refer to stack-heap pairs $\SH$ as (program) \emph{states}.
%
%\paragraph{Location terminology}
\subsubsection{Value \& location terminology}\label{sec:sl-basics:loc-terminology}
We denote by $\values{\fa}$ the set of all values in $\Val$ that explicitly appear as constants symbols in $\SLgeneric$ formula $\fa$. Similarly, the set of all values appearing in heap $\H$ is $\values{\H} \defn \dom(\H) \cup \bigcup_{\vec{v} \in \img(\H)} \vec{v}$; we lift this set to states $\SH$ by setting $\values{\SH} \defn \img(\S) \cup \values{\H}$.
The restriction of values in a heap to \emph{locations} is $\locs{\H} = \values{\H} \cap \Loc$ (analogously for states).
We often distinguish between allocated, referenced, and dangling values $v \in \Val$:
$v$ is \emph{allocated} in heap $\H$ if $v \in \dom(\H)$; it is \emph{referenced} if $v \in \img(\H)$.
%Moreover, $\ell$ is \emph{labeled} in stack $\S$ iff $\ell \in \img(\S)$.
%We collect all locations appearing in heap $\H$ in the set
%$\gls{locs-h} \defn \dom(\H) \cup ((\bigcup\img(\H))\cap\sbothloc)$.
Finally, $v$ is \emph{dangling} if it appears in $\H$ but is neither allocated nor a null pointer, i.e.,
$v \in \values{\H} \setminus (\dom(\H) \cup \{\nil\})$.

%In the same spirit, w
We call a variable $x \in \Var$ allocated, referenced, or dangling if the value $\S(x)$ is allocated, referenced, or dangling, respectively.
We collect all allocated variables and all referenced variables in state $\SH$
in the sets
$%\[ 
   \gls{alloced} \defn \set{x \mid \S(x)\in\dom(\H)}$
   and
   %\quad\text{and}\quad
   $\gls{refed} \defn \set{x \mid \S(x)\in\img(\H) }$. %~.
%\]
%
\subsection{The Semantics of Separation Logic}
\label{sec:sl-basics:semantics}
\begin{figure}[tb!]
  \centering
  \begin{tabular}{l@{\qquad\qquad}l}
  \hline
  \hline
  $\phi$ & $\SH \sidmodels \phi$ \quad iff \\
  \hline
  \hline
  $\emp$ & $\H = \emptyset$ \\
  $\sleq{u}{v}$ & $\S(u) = \S(v)$ and $\H = \emptyset$ \\
  $\slneq{u}{v}$ & $\S(u) \neq \S(v)$ and $\H = \emptyset$ \\
  $\pto{u}{\vec{y}}$ & $\H = \set{\S(u) \mapsto \S(\vec{y})}$ \\
      $\pred(\vec{y})$ & $\SH \sidmodels \psi(\vec{y}) \text{ for some } (\pred(\vec{x}) \Rule \psi)\in\Sid$ \\
  $\psi \sep \theta$ &
      exists $\H_1,\H_2$ such that $\H = \H_1 \stdunion \H_2$,
      %$\dom(\H_1)\cap\dom(\H_2)=\emptyset$, \\
      %& \qquad $\H=\H_1\cup\H_2$,
        $\SHi{1}\sidmodels\psi$, and $\SHi{2}\sidmodels \theta$\\
  $\psi \mw \theta$ &
      for all $\H_0$, if
      $\dom(\H_0)\cap\dom(\H)=\emptyset$ and $\SHi{0} \sidmodels \psi$ %\\
      %& \qquad
      then $\SHpair{\S}{\H_0\stdunion\H}\sidmodels \theta$ \\
  $\psi \wedge \theta$ & $\SH \sidmodels \psi$ and $\SH \sidmodels \theta$ \\
  $\psi \vee \theta$ & $\SH \sidmodels \psi$ or $\SH \sidmodels \theta$ \\
  $\neg \psi$ & not $\SH \sidmodels \psi$ \\
  $\EXO{x} \psi$ & exists $v\in\Val$ such that $\SHpair{\pinst{\S}{x}{v}}{\H} \sidmodels \psi$ \\
  $\FAO{x} \psi$ & for all $v\in\Val$, $\SHpair{\pinst{\S}{x}{v}}{\H} \sidmodels \psi$\\
  \hline
  \hline
  \end{tabular}
  \caption{Semantics of $\SLgeneric$}
  \label{fig:sl:generic-semantics}
\end{figure}
\Cref{fig:sl:generic-semantics} defines the semantics of $\SLgeneric$ in terms of a satisfaction relation $\sidmodels$, where
the sole purpose of $\Sid$---explained in \cref{sec:sl-basics:sid} below---is to assign semantics to user-defined predicate calls.
A state $\SH$ that satisfies an $\SLgeneric$ formula $\fa$, i.e., $\SH \sidmodels \fa$, is called a \emph{model} of $\fa$.
The empty-heap predicate $\emp$ holds iff the heap is empty;
equalities and disequalities between variables hold iff the stack maps
the variables to identical and different values, respectively.
For (dis-)equalities, we additionally require that the heap is
empty. This is non-standard, but not
unprecedented~\cite{piskac2013automating}, and will simplify the technical development.
A points-to assertion $\pto{x}{\tuple{y_1,\ldots,y_k}}$ holds in the
singleton heap that allocates exactly the location $\S(x)$ and stores the
values $\S(y_1),\ldots,\S(y_k)$ at this location.
This interpretation of points-to assertions is often called
a \emph{precise}~\cite{yang2001local,calcagno2007local} semantics, because the heap
contains precisely the object described by the points-to assertion,
and nothing else. %---in contrast to an \emph{intuitionistic} semantics,
In particular, any points-to assertion of the form $\pto{\nil}{\ldots}$ is always false because $\nil$ is not a location.
For the separating conjunction,
$\SH \models \psi \sep \theta$ holds if and only if there exist
domain-disjoint heaps $\H_1$, $\H_2$ such that their
union ($\stdunion$, see \cref{sec:notation})
%point-wise union
is $\H$ and both $\SHi{1}\sidmodels\psi$ and
$\SHi{2}\sidmodels\theta$ hold.

While the separating conjunction is about splitting the heap, the
magic wand is about extending it: $\SH \sidmodels \phi \mw \psi$ holds
iff all ways to extend $\H$ with a disjoint model of $\phi$ yields a
model of $\psi$.

The semantics of the Boolean connectives and the quantifiers is standard.
In particular, as justified by the lemma below, the semantics of quantifiers
can also be interpreted in terms of syntactic substitution rather than updating the stack
(which is formally defined in \cref{sec:notation}).
\begin{lemma}[Substitution Lemma]\label{lem:existential-variable-stack-extension}
    For all $\SLgeneric$ formulas $\fa$, states $\SH$, variables $x \in \fvs{\fa}$, and values $v \in \Val$, we have
    $(\pinst{\S}{x}{v}, \H) \sidmodels \fa$ iff
    $\SH \sidmodels \pinst{\fa}{x}{v}$.
\end{lemma}
\begin{proof}
  By induction on the structure of $\SLgeneric$ formulas.
\end{proof}
%
%\begin{lemma}\label{lem:existential-variable-stack-extension}
%  Let $\fa \in \SymHeap$ and $e\in\Var$. Then
%  $\SH \sidmodels \SHEX{e}\fa$ if and only if there exists a location
%  $v\in\Loc$ such that
%  $\SHpair{\S \cup \set{e\mapsto v}}{\H} \sidmodels \fa$.
%\end{lemma}
%
%; for simplicity I only allow quantifying over locations, not
%over arbitrary values. This is necessary because variables in
%$\SLbase$ (as well as the other logics in this thesis) are not
%sorted---their interpretation is not constrained to either $\Loc$ or
%$\Val\setminus\Loc$---and it rarely makes sense for a single variable
%to range both over memory locations and over the domain of data values
%stored in the heap.
%
\begin{example}
  \begin{enumerate}
  \item $\ppto{x}{y} \sep \ppto{y}{\nil}$ states that the heap consists
    of exactly two objects, one pointed to by $x$, the other pointed
    to by $y$; that the object pointed to by $x$ contains a pointer to
    the object pointed to by $y$; and that the object pointed to by
    $y$ contains a null pointer.
    Put less precisely but more concisely, $x$ points to $y$, $y$
    points to $\nil$, and $x$ and $y$ are separate objects on the
    heap.
    The precise semantics of assertions guarantees that
    there are no other objects in the heap.
  \item $\ppto{x}{y}\wedge\ppto{z}{y}$ states that (a) the heap consists
    of a single object $x$ that points to $y$ and that simultaneously
    (b) the heap consists of a single object $z$ that points to $y$.
    This formula is only satisfiable for stacks $\S$ with
    $\S(x)=\S(z)$.
  \item $\ppto{x}{y} \mw \ppto{z}{y}$ states that after adding a pointer
    from $x$ to $y$ to the heap, we obtain a heap that contains a
    single pointer from $z$ to $y$.
    This formula is only satisfiable for the empty heap and for stacks
    $\S$ with $\S(x)=\S(z)$.
  \item
    $\FAO{x} \ppto{x}{\nil} \mw ((\neg \emp) \sep (\neg \emp))$
    states that the heap contains at least one pointer: no matter
    which variable we additionally allocate, the resulting heap can be
    split into two nonempty parts, so the original heap must itself
    have been nonempty---the formula is equivalent to $\neg \emp$.
  \end{enumerate}
\end{example}
%
%\paragraph{Systems of inductive definitions}
\subsubsection{Systems of inductive definitions}\label{sec:sl-basics:sid}
Predicates are interpreted in terms of a user-supplied \emph{system of inductive definitions} (SID).
An SID is a finite set $\Sid$ of \emph{rules} of the form $\pred(\vec{x}) \Rule \fa(\vec{x})$, where $\pred \in \PredsKW$ is
a predicate symbol, $\fvs{\pred} = \vec{x} \in \Var^*$ are the formal
\emph{parameters} of $\pred$ with $\size{\vec{x}} = \arity{\pred}$, and $\fa$ is an $\SLgeneric$ formula with free variables $\vec{x}$;
the \emph{size} $\size{\Sid}$ of $\Sid$ is the sum of the sizes of the formulas in its rules.
We collect all predicates that occur in $\Sid$ in the
set \gls{PredsPhi}.
Moreover, we assume that all rules with the same predicate $\pred$ on the left-hand side have the same
(repetition-free) sequence of parameters $\vec{x}$.
A stack-heap pair $\SH$ satisfies the predicate call $\pred(\vec{y})$ with respect to SID $\Sid$
iff $\Sid$ contains a rule $\pred(\vec{x}) \Rule \fa$ such that $\SH$ satisfies the rule's right-hand side
once we instantiate its formal parameters with the arguments passed to the predicate call, i.e., $\SH \sidmodels \fa(\vec{y})$.

Notice that rules involving arbitrary $\SLgeneric$ formulas---e.g., $\pred(x) \Rule \neg \pred(x)$---do \emph{not} necessarily lead to
a well-defined semantics of predicate calls.
We will restrict the formulas allowed to appear in SIDs in \Cref{sec:sl-basics:slid} to
ensure that our semantics is always well-defined.
\begin{example}[Inductive Definitions]\label{ex:sids}
  \begin{enumerate}
      \item Let $\SidLs$ be the SID given by the following rules:
    \[
      \begin{array}{lllllll}
        \lseg(x_1,x_2) &\Rule& \pto{x_1}{x_2}
        &\quad& 
        \ls(x_1) &\Rule& \pto{x_1}{\nil} \\
        \lseg(x_1,x_2) &\Rule& \SHEX{y} \pto{x_1}{y} \sep \lseg(y,x_2) 
        &\quad& 
        \ls(x_1) &\Rule& \SHEX{y} \ppto{x_1}{y} \sep \ls(y)
      \end{array}
    \]
    The predicate $\lseg(x_1,x_2)$ describes non-empty singly-linked list segments
    with head $x_1$ and tail $x_2$; the predicate $\ls(x_1)$ describes those list segments
    that are terminated by a null pointer.
    Hence, the formulas $\lseg(x_1,\nil)$ and $\ls(x_1)$ are
    equivalent with respect to the SID $\SidLs$.
  \item The SID $\SidOddEven$ below defines all non-empty list segments of odd and even length, respectively.
    \[
      \begin{array}{lllllll}
        \gls{odd}(x_1,x_2) &\Rule& \pto{x_1}{x_2}
        &&
        \gls{even}(x_1,x_2) &\Rule& \SHEX{y} \ppto{x_1}{y} \sep
                               \odd(y,x_2) \\
        \odd(x_1,x_2) &\Rule& \SHEX{y} \ppto{x_1}{y} \sep
                              \even(y,x_2)
      \end{array}
    \]
  \item The SID $\SidTree$ below defines null-terminated binary trees with root $x_1$.
    \index{binary tree!SID defining}
        \[
          \begin{array}{lllllll}
            \gls{tree}(x_1) &\Rule& \pto{x_1}{\tuple{\nil,\nil}}
            &\quad&
            \tree(x_1) &\Rule& \SHEX{\tuple{l,r}} \ppto{x_1}{\tuple{l,r}} \sep \tree(l) \sep
                               \tree(r)
          \end{array}
        \]
  \end{enumerate}
\end{example}
%
%\paragraph{Satisfiability and entailment}
\subsubsection{Satisfiability and entailment}
An $\SLgeneric$ formula $\fa$ is \emph{satisfiable} with respect to $\Sid$
iff there exists a state $\SH$ such that $\SH \sidmodels \fa$.
Moreover, the $\SLgeneric$ formula $\fa$ \emph{entails} the $\SLgeneric$ formula $\fb$ given SID $\Sid$,
written $\fa \sidmodels\fb$, iff for all states $\SH$, we have $\SH \sidmodels \fa$ implies $\SH \sidmodels \fb$.
\subsubsection{Isomorphic states}
Our decision procedure will exploit that $\SLgeneric$ formulas cannot---at least as long as we do not explicitly use constant values other than the null pointer---distinguish between individual values.
More formally, they cannot distinguish isomorphic states.
\begin{definition}[Isomorphic States]\label{def:isomorphic-states}
  Two states $\SH$ and $\SHpair{\S'}{\H'}$ are
  \emph{isomorphic}, written $\SH \Iso \SHpair{\S'}{\H'}$, iff there
  exists a bijection
    $\renfun\colon \values{\SH} \to \values{\SHpprime}$ such that
  \begin{enumerate}
    \item for all $x$, $\S'(x) = \renfun(\S(x))$,
    \item $\H' = \{ \renfun(l) \mapsto \renfun(\H(l)) \mid l \in \dom(\H)\}$, and
    \item the null pointer cannot be renamed, i.e., if $0 \in \values{\SH}$, then $\renfun(\nil) = \nil \in \values{\SHpprime}$.
  \end{enumerate}
\end{definition}

\begin{lemma}\label{lem:sid:iso-models-same-formulas}
    Let $\fa$ be an $\SLgeneric$ formula with $\values{\fa} \subseteq \{\nil\}$.
  Then, for all states $\SH$ and $\SHpprime$, %we have
    \[ \SH \Iso \SHpprime \quad\text{implies}\quad
       \SH \sidmodels \fa \text{ iff } \SHpprime \sidmodels \fa.
    \]
\end{lemma}
\begin{proof}
 By induction on the structure of $\SLgeneric$ formulas.
\end{proof}
\subsection{The Bounded Treewidth Fragment}
\label{sec:sl-prelim:bounded-treewidth}
Our main goal is to develop a decision procedure for entailments in an $\SLgeneric$ fragment that extends the so-called bounded treewidth fragment ($\SLIDbtw$) of \citet{iosif2013tree}.
In this section, we briefly recapitulate that fragment as we will rely on similar restrictions for SIDs. %---to ensure decidability.
%
%\paragraph{Symbolic heaps}
\subsubsection{Symbolic heaps}
\label{sec:sl-prelim:symbolic-heaps}
Formulas in $\SLIDbtw$ are restricted to symbolic heaps with user-supplied predicates---a popular fragment of $\SLgeneric$ that is both expressive enough for specifying complex heap shapes and suitable to serve as an abstract domain for
program analyses (cf.~\cite{berdine2005symbolic,berdine2007shape,calcagno2011compositional}).
A \emph{symbolic heap} is a formula of the form
  \[
\EXO{\underbrace{x_1,\ldots,x_k}_{k \geq 0}} \underbrace{\phiAtom \sep \cdots
                    \sep \phiAtom}_{\text{1 or more atoms}}.
  \]
Notice that negation, disjunction, universal quantifiers, and magic wands are \emph{not} allowed in symbolic heaps.
In particular, this means---since pure formulas are evaluated in the empty heap---that
there is \emph{no} symbolic heap that is always satisfied, i.e., equivalent to $\mathsf{true}$.

When working with symbolic heaps, it is convenient to group the atoms into (1) a spatial part collecting all points-to assertions, (2) a part collecting all predicate calls, and (3) a pure part collecting all equalities and disequalities (in that order).
Hence, the set $\SymHeap$ of \emph{symbolic heaps} $\phiSH$ is given by
\begin{align*}
    \phiSH ::= & \underbrace{\SHEX{\vec{e}}}_{\vec{e} \in \Var^*}
                ~ \underbrace{
                    \ppto{u_1}{\veci{v}{1}} \sep \cdots
                   \sep
                    \ppto{u_k}{\veci{v}{k}}
                }_{\text{spatial part}, \emp \text{ for } k = 0}
                ~\sep~
                \underbrace{
                  \pred_1(\veci{w}{1}) \sep \cdots \sep \pred_l(\veci{w}{l})
                }_{\text{predicate calls},  \emp \text{ for } l = 0}
                \\&
                 \phantom{\SHEX{\vec{e}} \ppto{u_1}{\veci{v}{1}}}
                ~\sep~
                \underbrace{
                  \sleq{u_1}{v_1} \sep \cdots \sep \sleq{u_m}{v_m} \sep
                  \slneq{u_1'}{v_1'} \sep \cdots \sep \slneq{u_n'}{v_n'}
              }_{\text{pure part}, \emp \text{ for } m = 0, n = 0 \text{ respectively}}.
\end{align*}
%
%\paragraph{Symbolic heap SIDs}
\subsubsection{Symbolic heap SIDs}
\label{sec:sl-basics:slid}
Our semantics of predicate calls (see \Cref{fig:sl:generic-semantics}) is well-defined as long as all formulas appearing in the underlying SID are symbolic heaps---a requirement that we impose throughout the remainder of this article.
For instance, all SIDs in \Cref{ex:sids} only use symbolic heaps in their rules.
The restriction of SID rules to symbolic heaps is standard. % in the literature.
In fact, our semantics coincides with other semantics from the separation logic literature  that---instead of replacing predicates by rules step by step---are based on least fixed points~\cite{brotherston2007formalised,brotherston2014decision} or derivation trees~\cite{iosif2013tree,iosif2014deciding,jansen2017unified,matheja2020automated}.
%
%\paragraph{The bounded treewidth fragment}
\subsubsection{The bounded treewidth fragment}\label{sec:sl-basics:btw}
 Since negation is not available,
 the entailment problem for symbolic heaps is genuinely different from the
  satisfiability problem: it is impossible to solve an entailment
  $\phi \sidmodels \psi$ by checking the unsatisfiability of
  $\phi \wedge \neg \psi$, because the latter formula is not a
  symbolic heap.
  In fact, the satisfiability problem for symbolic heaps is decidable in general~\cite{brotherston2014decision},
  whereas the entailment problem is not~\cite{antonopoulos2014foundations,iosif2014deciding}.
  However, various subclasses of symbolic heaps with a decidable---and even tractable~\cite{cook2011tractable}---entailment problem have been studied in the literature (e.g.,~\cite{berdine2004decidable,iosif2013tree,iosif2014deciding,le2017decidable}); as such, the symbolic heap fragment has been the main focus of a recent competition of entailment solvers (SL-COMP)~\cite{sighireanu2019slcomp}.
 The largest of these fragments has been developed by \citet{iosif2013tree}; it achieves decidability by imposing three restrictions---progress, connectivity, and establishment---on SIDs to ensure that all models of predicates are of \emph{bounded treewidth}.\footnote{More precisely, when viewed as graphs, all
  models of the SIDs satisfying the three aforementioned restrictions have bounded treewidth. For a formal
  definition of treewidth, we refer to~\cite{diestel2016graph}.}
\paragraph{Local allocation and references}
%\subsubsection{Local allocation and references}
To formalize the above three assumptions for SIDs, we need two auxiliary definitions:
we collect all variables and locations that appear on the left-hand side of points-to assertions in formula $\phi$
in the \emph{local allocation} set $\lallocKW(\phi)$; analogously, the \emph{local references} set $\lrefKW(\phi)$ collects all variables and values appearing on the right-hand side of points-to assertions.
We now present the three aforementioned conditions imposed on SIDs $\Sid$ to ensure decidability of the entailment problem for symbolic heaps.
\paragraph{Progress}
%\subsubsection{Progress}
A predicate $\pred$ satisfies
\emph{progress} iff there exists a free variable $x \in \fvs{\pred}$ such that, for all rules
$(\pred(\vec{x}) \Rule \phi) \in \Sid$,
(1) $\phi$ contains exactly one point-to assertion, and
(2) $x$ is allocated in $\phi$, i.e., $\lalloc{\phi} =\set{x}$.
In this case, we call $x$ the \emph{root} of $\pred$.
%and
%define $\lalloc{\pred(\vec{y})} \defn \set{ \pinst{x}{\vec{x}}{\vec{y}}}$.
%\fztodo{The notation $\lalloc{\pred(\vec{y})} \defn \set{ \pinst{x}{\vec{x}}{\vec{y}}}$ is cumbersome. Do we need it?}
%
Moreover, if the $i$-th parameter of $\pred$, say $x_i$, is its root,
then we set $\proot{\pred(\vec{x})} \defn x_i$.
\paragraph{Connectivity}
%\subsubsection{Connectivity}
A predicate $\pred$ satisfies
\emph{connectivity} iff for all rules of
$\pred$, all variables that are allocated in the recursive calls of
  the rule are also referenced in the rule.
  Formally, for all rules $(\pred \Rule \phi) \in \Sid$ and for all
  calls $\pred'(\vec{y})$ appearing in $\phi$, we have
  $\proot{\phiapp{\pred'}{\vec{y}}}
    \subseteq \lref{\phi}$.

\paragraph{Establishment}
%\subsubsection{Establishment}
%
A predicate $\pred$ is
\emph{established} iff all existentially
quantified variables across all rules of $\pred$ are eventually
allocated, or equal to a parameter (or the null pointer).
  Formally, for all rules $(\pred(\xx) \Rule \SHEX{\vec{y}}\phi)\in\Sid$
  and for all states$\SH$, if $\SH \sidmodels \phi$ then
  $\S(\vec{y}) \subseteq \dom(\H) \cup \S(\xx) \cup \set{0}$.
%
%\paragraph{SIDs of bounded treewidth}
\subsubsection{SIDs of bounded treewidth}
We denote by $\IDbtw$ the set of all SIDs in which all predicates satisfy \emph{progress},
\emph{connectivity}, and \emph{establishment}.
For instance, the SIDs belong to \cref{ex:sids}.
\begin{theorem}[\cite{iosif2013tree,echenim2019lower}]
 The entailment problem for symbolic heaps over SIDs in $\IDbtw$ is decidable, of elementary complexity, and \textsc{2-EXPTIME} hard.
\end{theorem}
\noindent
In the remainder of this article, we strengthen the above theorem in two ways: First, we give a larger decidable $\SLgeneric$ fragment, and, second, we develop a \textsc{2-EXPTIME} decision procedure which, by the above lower bounds, is of optimal asymptotic complexity. Moreover, we show that even small extensions of our fragments lead to an undecidable entailment problem.

\subsubsection{Global Assumptions about SIDs}\label{sec:sl-basics:sid-assumptions}
Unless stated otherwise, we assume that all SIDs considered in this article belong to $\IDbtw$. Moreover, to avoid notational clutter, $\Sid$ always refers to an arbitrary, but fixed SID in $\IDbtw$ unless it is explicitly given.
Without loss of generality, we make two further assumptions about the rules in SIDs to simplify the technical development.
First, we assume that non-recursive
rules do \emph{not} contain existential quantifiers because they can always be eliminated:
due to progress and establishment, all existentially-quantified variables in a non-recursive rule must be
provably equal to either a constant or a parameter of the predicate.
Second, to avoid dedicated reasoning about points-to assertions, we may add dedicated
predicates simulating points-to assertions to every SID; we call the
resulting SIDs \emph{\ptrclosed{}}:
\begin{definition}[Pointer-closed SID]\label{def:ptrclosed}
  An SID $\Sid$ is \emph{\ptrclosed{}} w.r.t.~$\phi$ iff
  it contains
  a predicate \gls{ptrpred_k} and a single rule
  $\ptrpred_k(\tuple{x_1,\ldots,x_{k+1}}) \Rule \pto{x_1}{\tuple{x_2,\ldots,x_{k+1}}}$
  for all points-to assertions mapping to structures of length $k$ in~$\phi$.
\end{definition}
\noindent
Since all predicates introduced by transforming an SID into a pointer-closed one satisfy progress, connectivity, and establishment,
we can safely assume that SIDs in $\IDbtw$ are pointer-closed.
As a consequence of this assumption, we consider the number of formal parameters of predicates to be at least as large as the number of fields of points-to assertions whenever we analyze complexities.

\section{The Guarded Fragment of Separation Logic}
\label{sec:sl-guarded}
To obtain fragments of $\SLgeneric$ with both support for complex data structure predicates and a decidable entailment problem,
we rely on the same restrictions on user-defined predicates as \citet{iosif2013tree}: the semantics of predicate calls needs to be determined by SIDs taken from the bounded treewidth fragment $\IDbtw$.
In contrast to \citet{iosif2013tree}, our work is, however, not limited to entailments between symbolic heaps over the predicates at hand.
Rather, we additionally consider reasoning about a novel (quantifier-free) \emph{guarded} %quantifier-free fragment of
fragment of separation logic ($\SLIDguarded$) featuring restricted variants of negation $\neg$, magic wand $\mw$, and septraction $\sept$. %; this fragment is introduced in this section.

Intuitively, the guarded fragment enforces that the aforementioned connectives $\neg$, $\mw$, and $\sept$ only appear in conjunction with another formula restricting the possible shapes of the heap.
We will show in \Cref{ch:undec} that this restriction is crucial: lifting it for any of these connectives yields an undecidable entailment problem---even if the remaining connectives are removed.

\subsection{Guarded Formulas}
The set $\SLIDguarded$ of formulas in (quantifier-free) \emph{guarded separation logic} is given by the grammar
\begin{align*}
    \phiGuarded ::=~ & \phiAtom \quad\left(~::=~ \emp \mid \sleq{u}{v} \mid \slneq{u}{v} \mid \pto{u}{\vec{w}} \mid \pred(\vec{w})~\right) \tag{same atoms as $\SLgeneric$} \\
                    & \mid \phiGuarded \sep \phiGuarded 
                    \mid \phiGuarded \wedge \phiGuarded 
                    \mid \phiGuarded \vee \phiGuarded \tag{standard connectives} \\
                    & \mid \phiGuarded \wedge \neg \phiGuarded \tag{\emph{guarded} connectives} 
                    \mid \phiGuarded \wedge (\phiGuarded \mw \phiGuarded) 
                    \mid \phiGuarded \wedge (\phiGuarded \sept \phiGuarded). 
\end{align*}
%\begin{align*}
%    \phiGuarded ::=~ & \phiAtom \quad\left(~::=~ \emp \mid \sleq{u}{v} \mid \slneq{u}{v} \mid \pto{u}{\vec{w}} \mid \pred(\vec{w})~\right) \tag{same atoms as $\SLgeneric$} \\
%                    & \mid \phiGuarded \sep \phiGuarded \tag{separating conjunction} \\
%                    & \mid \phiGuarded \wedge \phiGuarded \tag{standard conjunction} \\
%                    & \mid \phiGuarded \vee \phiGuarded \tag{standard disjunction} \\
%                    & \mid \phiGuarded \wedge \neg \phiGuarded \tag{\emph{guarded} negation} \\
%                    & \mid \phiGuarded \wedge (\phiGuarded \mw \phiGuarded) \tag{\emph{guarded} magic wand} \\
%                    & \mid \phiGuarded \wedge (\phiGuarded \sept \phiGuarded). \tag{\emph{guarded} septraction}
%\end{align*}
%
The atoms as well as the connectives $\sep$, $\wedge$, and $\vee$ are the same as for the full logic $\SLgeneric$ introduced in \Cref{sec:sl-basics:syntax}.
Moreover, negation $\neg$, magic wand $\mw$, and septraction $\sept$ may only appear in \emph{guarded} form, i.e., in conjunction with another guarded formula $\phiGuarded$.
Since $\SLIDguarded$ is a syntactic fragment of $\SLgeneric$, the semantics of $\SLIDguarded$ is given by the semantics of $\SLgeneric$ presented in \Cref{sec:sl-basics:semantics}.
\begin{example}
  Assume the predicate $\lseg(x_1,x_2)$ represents all non-empty list segments
  from $x_1$ to $x_2$; a formal definition is found in \Cref{ex:sids}.
  Moreover, consider the following guarded formulas:
  \begin{enumerate}
  \item
    $\lseg(x,y) \wedge \neg \pto{x}{y}$
    states that the heap consists of a list of length at least two.
  \item
    $\lseg(x,y) \wedge (\lseg(y,z) \sept \lseg(x,x))$
    states that the heap consists of a list segment from $x$ to $y$
    that can be extended to a cyclic list by adding a list
    from $y$ to $z$;
    it entails that $x$ and $z$ are aliases.
  \end{enumerate}
\end{example}
\noindent
In contrast to variants of separation logic in the literature (cf.~\cite{reynolds2002separation,calcagno2011compositional}),
our separation logic $\SLgeneric$ does not contain an atom $\mathsf{true}$, which is always satisfied.
While $\mathsf{true}$ is, of course, definable in $\SLgeneric$, e.g., $\emp \vee \neg \emp$,
it is \emph{not} definable in $\SLIDguarded$.
In particular, $\sleq{x}{x}$ is not equivalent to $\mathsf{true}$
as our semantics of equalities and disequalities requires the heap to be empty.
This is crucial: If $\mathsf{true}$ were definable, $\SLIDguarded$ would coincide with the set of all
quantifier-free $\SLgeneric$ formulas, because we could choose $\mathsf{true}$ for all guards.
\subsection{Guarded States and Dangling Pointers}\label{sec:sid:dangling}
The decision procedure developed in this article exploits that all models of guarded formulas are themselves guarded in the sense that they have only a limited amount of dangling pointers.
We recall from \cref{sec:sl-basics:loc-terminology} that a dangling pointer is a value that is neither allocated nor equal to the null pointer; the set of all dangling values in heap $\H$ is
%thus given by
%\[ 
$\danglinglocs{\H} \defn \values{\H} \setminus (\dom(\H) \cup \{\nil\})$.
%~. \]

In the following, we first define guarded states, then show that establishment implies a models of atomic predicates to be guarded, and finally lift this result to arbitrary guarded formulas.
\begin{definition}[Guarded State]\label{def:positive-model}
  The set $\Mpos{\Sid}$ of \emph{guarded states} is given by
  \[ \Mpos{\Sid} ~\defn~ \{ \SH \mid \S \in \SS, \H \in \HH, \danglinglocs{\H} \subseteq \img(\S) \}~.\]
%  A state $\SH$ is \emph{guarded} if
%  $\danglinglocs{\H} \subseteq \img(\S)$.
%  %
%  We denote by $\Mpos{\Sid}$ the set of all guarded states.
\end{definition}
\noindent
Guarded states are well-behaved with regard to taking the union of heaps:
\begin{lemma}\label{lem:union-pos-model}
  Let $\SHpair{\S}{\H_1}, \SHpair{\S}{\H_2} \in \Mpos{\Sid}$ with $\H_1 \stdunion \H_2 \neq \bot$.
  Then, $\SHpair{\S}{\H_1 \stdunion \H_2} \in \Mpos{\Sid}$.
\end{lemma}
\begin{proof}
  We observe that $\danglinglocs{\H_1 \stdunion \H_2} \subseteq \danglinglocs{\H_1} \cup \danglinglocs{\H_2} \subseteq \img(\S)$
\end{proof}
\noindent
Furthermore, due to establishment (cf. \cref{sec:sl-basics:sid-assumptions}), models of predicate calls are guarded:

\begin{lemma}\label{lem:predicate-pos-model}
  %Let $\Sid \in \IDbtw$.
  For all predicates $\pred\in\Preds{\Sid}$ and all states $\SH$, we have
  \[ \SH \sidmodels \pred(\vec{x})  \quad\text{implies}\quad \SH \in \Mpos{\Sid}. \]
\end{lemma}
\begin{proof}
 By induction on the number of rule applications needed to establish $\SH \sidmodels \pred(\vec{x})$; a detailed proof is found in \cref{app:predicate-pos-model}.
\end{proof}
\noindent
We now lift the result from~\cref{lem:predicate-pos-model} from atomic predicates to arbitrary guarded formulas.
We will use the following result that every model of a guarded formula satisfies a finite number of predicates conjoined by the separating conjunction:

\begin{lemma}\label{lem:guarded-iterated-star-predicates}
  Let $\phi \in \SLIDguarded$ be a guarded formula with $\fvs{\phi} = \xx$.
  Then, for every state $\SH \sidmodels \phi$, there are predicates $\pred_i \in \Preds{\Sid}$ and variables $\vec{z_i} \subseteq \xx$ such that $\SH \sidmodels \IteratedStar_{1\leq i\leq k}\pred_i(\vec{z_i})$.
\end{lemma}
\begin{proof}
    By structural induction on $\phi$; see \cref{app:guarded-iterated-star-predicates} for details.
\end{proof}

\begin{corollary}\label{lem:pos-formula-pos-model}
  %Let $\Sid \in \IDbtw$.
  For all $\phi \in \SLIDguarded$ and all states $\SH \sidmodels \phi$, we have
  $\SH \in \Mpos{\Sid}$.
  %\[ \SH \sidmodels \phi \quad\text{implies}\quad \SH \in \Mpos{\Sid}. \]
\end{corollary}
\begin{proof}
  Immediate from \cref{lem:predicate-pos-model} and \cref{lem:guarded-iterated-star-predicates}.
\end{proof}

\paragraph{On the importance of guardedness}
The fact that all appearances of negation, magic wand, and septraction in $\SLIDguarded$ are guarded by a conjunction with
another $\SLIDguarded$ formula is crucial for limiting the number of dangling pointers in the above lemma.

For the negation and the magic wand this is straightforward: without guards, both can be used to define $\mathsf{true}$,
e.g., $\emp \vee \neg \emp$ and $(\ppto{x}{\nil}\sep\ppto{x}{\nil})\mw\emp$).
Since $\mathsf{true}$ is satisfied by all states, the number of dangling locations is unbounded. %including non-guarded ones.
For the septraction $\sept$, consider the following SID:
\[
  \begin{array}{ll}
      \gls{tll}(r,l,t)& \Rule \pto{l}{t} \sep \sleq{r}{l} 
    \qquad
    \qquad
    \qquad
    \qquad
    \qquad
    \qquad
    \lseg(l, t) \Rule  \SHEX{n} \ppto{l}{n} \sep \lseg(n, t) \\
    \tll(r,l,t)& \Rule  \SHEX{\tuple{u,v,m}} \pbpto{r}{u,v} \sep
                 \tll(u,l,m) \sep \tll(s_2,m,r) 
    \qquad
    \qquad
    \lseg(l, t) \Rule  \pto{l}{t} 
  \end{array}
\]
The $\tll$ predicate encodes a binary tree with root $r$ and leftmost
leaf $l$ overlaid with a singly-linked list segment from $l$ to $t$ whose nodes
are the leaves of the tree.
Now, assume a state $\SH$ satisfying the unguarded formula $\lseg(l,t) \sept \tll(r,l,t)$.
In other words, there exists a heap $\H_1$ with $\SHi{1} \sidmodels \lseg(l,t)$ and
$\SHpair{\S}{\H\cup\H_1} \sidmodels \tll(r,l,t)$.
Since each list element is a leaf of the tree in heap $\H$, we have
$\danglinglocs{\H} = \dom(\H_1)$---a finite, but unbounded set of dangling pointers.
%that contradicts
%\Cref{lem:btw:pos-non-dangling}.
%Hence, $\lseg(h,t) \sept \tll(r,h,t)$ is satisfied by non-guarded states.

\section{Beyond Guarded Separation Logic: Undecidability Proofs}\label{ch:undec}

Before we develop our decision procedure for the fragment $\SLIDguarded$
of guarded separation logic with inductive definitions of bounded treewidth,
we further justify the need for guarding negation, magic wand, and septraction.
More precisely, we show in this section that omitting the guards for \emph{any} of the above three operators
leads to an undecidable logic.
Together with our decidability results presented afterwards, %in \cref{ch:deciding-btw},
this yields an almost tight delineation between undecidability of separation logics that allow arbitrary SIDs in $\IDbtw$.
\subsection{Encoding Context-Free Language in SIDs}\label{sec:unguarded-undec}
%\subsection{Undecidability of Unguarded SLID}\label{sec:unguarded-undec}
%

%\paragraph*{Context-free grammars.}
All of our undecidability results, which are presented in \cref{sec:unguarded-undec-proofs}, rely
on a novel encoding of the language-intersection problem for
context-free grammars---a well-known undecidable problem.

\begin{definition}[Context-free grammar]\index{context-free
    grammar}\index{CFG|see{context-free grammar}}\index{terminal}\index{nonterminal}\index{production rule}
  A \emph{context-free grammar} (CFG) in Chomsky normal form is a 4-tuple
  $\gls{G-cfg}=\tuple{\NTerm,\Term,\Rules,\Start}$, where $\NTerm$ is a finite
  set of \emph{nonterminals}; $\Term$ is a finite set of terminals, which is
  disjoint from $\NTerm$;
  $\Rules \subseteq \NTerm \times (\NTerm^2 \cup \Term)$ is a finite
  set of \emph{production rules} mapping nonterminals to two nonterminals or a single terminal; and $\Start\in\NTerm$ is the
  \emph{start symbol}. $\gls{CFG}$ is the set of all CFGs.
\end{definition}
\noindent
We often denote production rules $\tuple{a,b}$ by $\GRule{a}{b}$ to improve
readability.
Since we assume all CFGs to be in Chomsky normal form, all rules
are either of the form
$\GRule{N}{AB}$ or $\GRule{N}{a}$, where $N,A,B$ are nonterminals in $\NTerm$ and $a$ is a terminal in $\Term$.

\begin{definition}
  Let $\G=\tuple{\NTerm,\Term,\Rules,\Start}\in\CFG$ and let
  $v,w \in \NTwords$. We write $v \gls{--derive-cfgderive} w$ if there
    exist strings $u_1,u_2 \in \NTwords$ and a rule $\GRule{a}{b}$ such that
  $v=u_1\concat a \concat u_2$ and $w=u_1\concat b \concat u_2$.
  We write $\gls{--derive-cfgderivestar}$ for the transitive closure of $\RuleStep$.
  The
  \emph{language} of $\G$ is given by
  $\gls{L-G}\defn\set{w \in \Term^{*} \mid \Start \RuleSteps w }$.
\end{definition}
\noindent
In the following, we exploit the following undecidability classic result (cf.~\cite{bar1961formal}):
\begin{theorem}[Undecidability of language intersection]
  Given two CFGs $\GA$ and $\GB$, it is undecidable whether
  $\Lang{\GA}\cap\Lang{\GB}\neq\emptyset$ holds---even if neither $\GA$ nor $\GA$ accept the empty string $\emptyseq$.
\end{theorem}

\paragraph*{Encoding CFGs as SIDs.}
%Our undecidability proofs rely on a common encoding of CFGs.
Throughout the remainder of this section, we fix a set $\Term = \set{a_1,\ldots,a_n}$ of terminals and two CFGs $\G_1$ and $\G_2$,
where we assume that
their sets of nonterminals do not overlap, i.e., $\NTerm_1 \cap \NTerm_2=\emptyset$.
\begin{figure}[tb!]
  \centering
%   $\begin{array}{llll}
%       \letter{i}(a) &\Rule&
%                             \pto{a}{\underbrace{\tuple{\nil,\ldots,\nil}}_{\text{length }i}}, & \ForAllLetters
%       \\
% %
%       N(x_1,x_2,x_3) &\Rule& \SHEX{l,r,m} \ppto{x_1}{\tuple{l,r}} \sep
%                              A(l,x_2,m) \sep B(r,m,x_3), & j \in \set{1,2},\GRule{N}{AB}\in\Rules_j
%       \\
% %
%       N(x_1,x_2,x_3) &\Rule& \SHEX{a}\ppto{x_1}{\tuple{x_3,a}} \sep
%                              \letter{k}(a) \sep \sleq{x_1}{x_2}, &
%                j
%                                                                   \in
%                                                                   \set{1,2},\GRule{N}{a_k}\in\Rules_j
%       \\
%   %
%       \word(x,y) &\Rule& \SHEX{a} \ppto{x}{y,a}
%                          \sep \letter{i}(a), & \ForAllLetters \\
% %
%       \word(x,y) &\Rule& \SHEX{\tuple{n,a}}
%                          \ppto{x}{n,a} \sep \letter{i}(a) \sep \word(n,y), & \ForAllLetters
%       \\
       %      \end{array}$
  %
  $\begin{array}{lllr}
      \letter{i}(a) &\Rule&
                            \pto{a}{\underbrace{\tuple{\nil,\ldots,\nil}}_{\text{length }i}},\; \ForAllLetters
      \\
      N(x_1,x_2,x_3) &\Rule& \SHEX{l,r,m} \ppto{x_1}{\tuple{l,r}} \sep
                             A(l,x_2,m) \sep B(r,m,x_3),
     %\\\multicolumn{3}{l}{\quad
      &
      j \in \set{1,2},(\GRule{N}{AB})\in\Rules_j
      %}
      \\
      N(x_1,x_2,x_3) &\Rule& \SHEX{a}\ppto{x_1}{\tuple{x_3,a}} \sep
                             \letter{k}(a) \sep \sleq{x_1}{x_2},
     %\\\multicolumn{3}{l}{\quad
      &
      j \in \set{1,2},(\GRule{N}{a_k})\in\Rules_j%}
      \\
      \word(x,y) &\Rule& \SHEX{a} \ppto{x}{\tuple{y,a}}
                         \sep \letter{i}(a), %\\\multicolumn{3}{l}{\quad
                         & \ForAllLetters%}
                         \\
      \word(x,y) &\Rule& \SHEX{\tuple{n,a}}
                         \ppto{x}{\tuple{n,a}} \sep \letter{i}(a) \sep \word(n,y), %\\\multicolumn{3}{l}{\quad
                         & \ForAllLetters %}
      \\
    \end{array}$
  \caption[An SID encoding of CFG derivations]{The SID $\Sid$ encoding derivations of the
    CFGs $\GA=\theGi{1}$ and $\GB=\theGi{2}$.}
  \label{fig:sid-encoding}
\end{figure}
\cref{fig:sid-encoding} depicts the SID $\Sid$ encoding $\G_1$ and $\G_2$:
For each terminal symbol $a_i$, we introduce a predicate $\letter{i}(a)$.\footnote{While it is convenient to model each terminal $a_i$ through a single points-to assertion mapping $a$ to $i$ null pointers, it is noteworthy, that, in principle, points-to assertions mapping to at most two values suffice.}
Moreover, for each nonterminal $N \in \NTerm_1\cup\NTerm_2$, there is a corresponding predicate encoding
the derivations of $\G_1$ and $\G_2$ as trees with linked leaves (TLL), similar to the SID in \cref{fig:tll-example} on
p.~\pageref{fig:tll-example}.
The predicate $\word$ overapproximates the possible \emph{front}, i.e.,
the list of linked leaves of the TLL; we will need it later to prove undecidability of individual fragments.

By construction, every word in $\Lang{\G_i}$ corresponds to at least one state $\SH$
with $\SH \sidmodels \Start_i(x_1,x_2,x_3)$.
Furthermore, every model $\SH$ of  $\Start_i(x_1,x_2,x_3)$ corresponds to both
a \emph{derivation tree}\footnote{We do not formally define derivation trees for CFGs as they are not required for the formal development; we refer to \cite{hopcroft2007introduction} for a thorough introduction of CFGs.} and a word in $\Lang{\G_i}$, where the inner nodes of the TLL
correspond to the derivation tree and its front corresponds to the word in $\Lang{\G_i}$.
\begin{example}
    \cref{fig:word-to-heap} illustrates both a derivation tree (\Cref{fig:word-to-heap:deriv}) and a model of our encoding (\cref{fig:word-to-heap:heap})
    for the CFG $\G=\tuple{\set{S,A,B,C}, \set{a_1,a_2}, \Rules, S}$
    whose rules are provided in
    \Cref{fig:word-to-heap:cfg}.
    We observe that the
    depicted model
    encodes the aforementioned derivation tree: Every nonterminal is
  translated to a node in a binary tree (blue). The leaves of the tree
  are linked. They each have a successor that encodes a terminal
  symbol of the derivation (orange): The node contains $k$
  pointers to $\nil$ to represent
  terminal $a_k$.
  The list of linked leaves and orange nodes together form the
  \emph{induced word}, i.e., $a_2a_2a_1a_1a_1$.
    %as defined later in
  %\cref{def:inducedN}.

\begin{figure}[tb!]
  \begin{subfigure}[b]{0.22\textwidth}
    \begin{align*}
      & \GRule{S}{AB}\\
      & \GRule{A}{CC}\\
      & \GRule{B}{BB}\\
      & \GRule{B}{a_1}\\
      & \GRule{C}{a_2}
    \end{align*}
    \caption{Production rules of a CFG $\G$.}\label{fig:word-to-heap:cfg}
  \end{subfigure}\hfill
  \begin{subfigure}[b]{0.37\textwidth}
    \begin{tikzpicture}
  % Layer 3
  \node (c1) {$a_2$};
  \node[right=3mm of c1] (c2) {$a_2$};
  \node[right=3mm of c2] (B31) {$B$};
  \node[right=3mm of B31] (B32) {$B$};
  \node[right=3mm of B32] (b31) {$a_1$};
  % Layer 4
  \node[below=3mm of B31] (b41) {$a_1$};
  \node[below=3mm of B32] (b42) {$a_1$};
  % Layer 2
  \node[above=3mm of c1] (C1) {$C$};
  \node[above=3mm of c2] (C2) {$C$};
  \node[above=3mm of B31,xshift=4.5mm] (B21) {$B$};
  \node[above=3mm of b31] (B22) {$B$};
  % Layer 1
  \node[above=3mm of C1,xshift=4.5mm] (A) {$A$};
  \node[above=3mm of B21,xshift=4.5mm] (B1) {$B$};
  % Layer 0
  \node[above=3mm of A,xshift=10mm] (s) {$S$};

  % Edges
  \draw (s) edge[->] (A);
  \draw (s) edge[->] (B1);
  \draw (A) edge[->] (C1);
  \draw (A) edge[->] (C2);
  \draw (B1) edge[->] (B21);
  \draw (B1) edge[->] (B22);
  \draw (C1) edge[->] (c1);
  \draw (C2) edge[->] (c2);
  \draw (B21) edge[->] (B31);
  \draw (B22) edge[->] (b31);
  \draw (B21) edge[->] (B32);
  \draw (B31) edge[->] (b41);
  \draw (B32) edge[->] (b42);
\end{tikzpicture}

%%% Local Variables:
%%% mode: latex
%%% TeX-master: "../main"
%%% End:
    \caption{A derivation tree for the word $a_2a_2a_1a_1a_1 \in \Lang{\G}$.}\label{fig:word-to-heap:deriv}
  \end{subfigure}\hfill
  \begin{subfigure}[b]{0.37\textwidth}
    \begin{tikzpicture}
  % Layer 3
  \node[tlls] (c1) {\footnotesize $x_2$};
  \node[tlls,right=3mm of c1] (c2) {};
  \node[tlls,right=3mm of c2] (B31) {};
  \node[tlls,right=3mm of B31] (B32) {};
  \node[tlls,right=3mm of B32] (b31) {};
  \node[tlls,right=3mm of b31] (x3) {\footnotesize $x_3$};
  % Layer 4
  \node[tlls,below=2mm of B31] (b41) {};
  \node[tlls,below=2mm of B32] (b42) {};
  % Layer 2
  \node[tlls,above=2mm of c1] (C1) {};
  \node[tlls,above=2mm of c2] (C2) {};
  \node[tlls,above=2mm of B31,xshift=4.5mm] (B21) {};
  \node[tlls,above=2mm of b31] (B22) {};
  % Layer 1
  \node[tlls,above=2mm of C1,xshift=4.5mm] (A) {};
  \node[tlls,above=2mm of B21,xshift=4.5mm] (B1) {};
  % Layer 0
  \node[tlls,above=0mm of A,xshift=10mm] (s) {\footnotesize $x_1$};

  % Edges
  \draw (s) edge[->] (A);
  \draw (s) edge[->] (B1);
  \draw (A) edge[->] (C1);
  \draw (A) edge[->] (C2);
  \draw (B1) edge[->] (B21);
  \draw (B1) edge[->] (B22);
  \draw (C1) edge[->] (c1);
  \draw (C2) edge[->] (c2);
  \draw (B21) edge[->] (B31);
  \draw (B22) edge[->] (b31);
  \draw (B21) edge[->] (B32);
  \draw (B31) edge[->] (b41);
  \draw (B32) edge[->] (b42);

  % Linked leaves
  \draw (c1) edge[->] (c2);
  \draw (c2) edge[->] (b41);
  \draw (b41) edge[->] (b42);
  \draw (b42) edge[->] (b31);
  \draw (b31) edge[->] (x3);

  % Letter encoding
  \node[letter,below=3mm of c1] (l1) {};
  \node[letter,below=3mm of c2] (l2) {};
  \node[letter,below=3mm of b41] (l3) {};
  \node[letter,below=3mm of b42] (l4) {};
  \node[letter,below=3mm of b31] (l5) {};

  \node[nil,below=3mm of l1,xshift=-1mm] (n11) {\tiny $\nil$};
  \node[nil,below=3mm of l1,xshift=1mm] (n12) {\tiny $\nil$};
  \node[nil,below=3mm of l2,xshift=-1mm] (n21) {\tiny $\nil$};
  \node[nil,below=3mm of l2,xshift=1mm] (n22) {\tiny $\nil$};
  \node[nil,below=3mm of l3] (n3) {\tiny $\nil$};
  \node[nil,below=3mm of l4] (n4) {\tiny $\nil$};
  \node[nil,below=3mm of l5] (n5) {\tiny $\nil$};

  \draw (c1) edge[->] (l1);
  \draw (c2) edge[->] (l2);
  \draw (b41) edge[->] (l3);
  \draw (b42) edge[->] (l4);
  \draw (b31) edge[->] (l5);

  \draw (l1) edge[->] (n11);
  \draw (l1) edge[->] (n12);
  \draw (l2) edge[->] (n21);
  \draw (l2) edge[->] (n22);
  \draw (l3) edge[->] (n3);
  \draw (l4) edge[->] (n4);
  \draw (l5) edge[->] (n5);

  %\draw[laranja1,thick]\convexpath{c1,c2,b41.north,b31,x3,n5,n4,n3,n22.south,n11}{12pt};
\end{tikzpicture}

%%% Local Variables:
%%% mode: latex
%%% TeX-master: "../main"
%%% End:
    \caption{The corresponding model of the predicate call $S(x_1,x_2,x_3)$.}\label{fig:word-to-heap:heap}
  \end{subfigure}
  \centering
  \caption[Example: Encoding a CFG derivation as stack--heap
  model]{Encoding a derivation of a context-free grammar as a
    stack--heap model.}
  \label{fig:word-to-heap}
\end{figure}
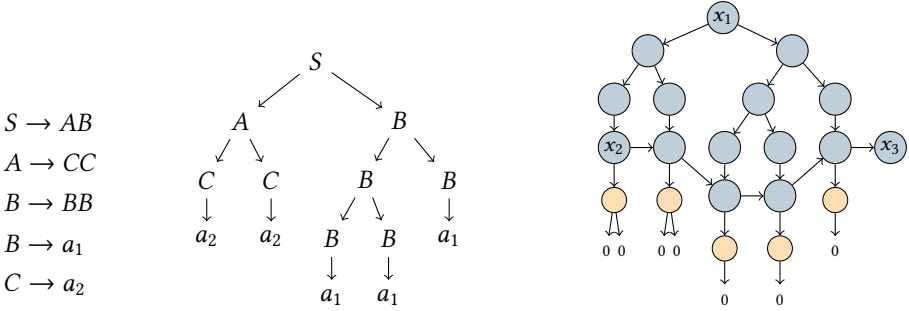

\end{example}
\noindent
To show that our encoding is correct, i.e., it adequately captures the language of a given CFG, we need to refer to the word induced by a given model.
We first define the terminals of such a word, which are given by the letter predicates in the models' list of linked leaves.
\begin{definition}[Induced letters]\label{def:inducedW}\index{induced letters}
  Let $\G=\theG$ and let $\Sid$ be the corresponding SID encoding. Let
  $\SH \sidmodels \word(x,y)$ and let
  $j_1,\ldots,j_m \in \set{1,\ldots,n}$ be such that
%
% \begin{align*}
%   \SH \sidmodels \EX{\tuple{n_1,\ldots,n_{m-1},b_1,\ldots,b_m}}&
%                                                                  (\ppto{x}{\tuple{n_1,b_1}} \sep \letter{j_1}(b_1)) \\\sep&
%                                                                                                                             (\ppto{n_1}{\tuple{n_2,b_2}} \sep \letter{j_2}(b_2)) \\\sep \cdots \sep& (\ppto{n_{m-1}}{\tuple{y,b_m}} \sep \letter{j_m}(b_m)).
% \end{align*}
% %
\begin{align*}
  \SH \sidmodels& \SHEX{{n_1,\ldots,n_{m-1},b_1,\ldots,b_m}}
                                                                 (\ppto{x}{\tuple{n_1,b_1}} \sep \letter{j_1}(b_1)) \\&\phantom{\exists}\sep
                                                                                                                            (\ppto{n_1}{\tuple{n_2,b_2}} \sep \letter{j_2}(b_2))\\&\phantom{\exists} \sep \cdots \sep (\ppto{n_{m-1}}{\tuple{y,b_m}} \sep \letter{j_m}(b_m)).
\end{align*}
We define the \emph{induced letters} of $\SH$ and $x,y$ as
$\inducedW{\S,\H,x,y} \defn a_{j_1}a_{j_2}\cdots a_{j_m}$.
\end{definition}
\noindent
Every model of the predicate $N(x_1,x_2,x_3)$ contains a sub-heap satisfying the $\word$ predicate:
\begin{lemma}\label{lem:g-heap-has-word-heap}
  Let $\G=\theG$ and let $\Sid$ be its %the corresponding 
    SID encoding.
  Let $x_1,x_2,x_3\in\Var$, $N \in \NTerm$ and let
  $\SH \sidmodels N(x_1,x_2,x_3) \sep \true$. Then there exists a
  unique heap $\H_w \subseteq\H$ with
  $\SHi{w}\sidmodels \word(x_2,x_3)$.
\end{lemma}
\begin{proof}
  A straightforward induction shows that the models of the predicate
  call $N(x_1,x_2,x_3)$ are trees with linked leaves with root
  $\S(x_1)$, leftmost leaf $\S(x_2)$, and successor of rightmost leaf
  $\S(x_3)$.
  We pick as $\H_w$ the heap that contains $\S(x_2)$ as well as all
  values reachable from $\S(x_2)$ in $\H$.
  This gives us precisely the list from $\S(x_2)$ to $\S(x_3)$.
  Moreover, every leaf satisfies a formula of the form $\SHEX{a}\ppto{y}{\tuple{z,a}} \sep
  \letter{k}(a)$.
  Consequently, $\SHi{w} \sidmodels \word(x_2,x_3)$.
\end{proof}
\noindent
\Cref{lem:g-heap-has-word-heap} ensures that models of our encoding of CFGs
induce a word over the given alphabet.

\begin{definition}[Induced word]\label{def:inducedN}
  Let $\G=\theG$ and let $\Sid$ be its %the corresponding 
    SID encoding.
  Let $x_1,x_2,x_3\in\Var$, $N \in \NTerm$ and let
  $\SH \sidmodels N(x_1,x_2,x_3)$. Let $\H_w\subseteq\H$ be the unique
  heap with $\SHi{w}\sidmodels \word(x_2,x_3)$.  We define the
  \emph{induced word} of $\SH$ and $N$ as
  $\inducedN{N}{\S,\H,x_2,x_3} \defn \inducedW{\S,\H_w,x_2,x_3}$.
\end{definition}
\noindent
Every word $w \in \Lang{\G}$ is then the induced word of a model of the
corresponding SID encoding.
\begin{lemma}[Completeness of the encoding]\label{lem:g-to-sid}
  Let $\G=\theG$ and let $\Sid$ be the corresponding SID encoding. Let
  $1 \leq i \leq 2$, $x_1,x_2,x_3\in\Var$, and let
  $w \in \Lang{\G}$. Then there exists a model $\SH$ of
  $\Start(x_1,x_2,x_3)$ with
  $\inducedN{\Start}{\S,\H,x_2,x_3}=w$.
\end{lemma}
\begin{proof}
  By mathematical induction on the number of
  $\RuleStep$ steps; see \Cref{app:g-to-sid}.
\end{proof}

\noindent
Likewise, every induced word of a model of the corresponding SID
encoding is in $\Lang{\G}$.

\begin{lemma}[Soundness of the encoding]\label{lem:sid-to-g}
  Let $\G=\theG$ and let $\Sid$ be the corresponding SID encoding. Let
  $x_1,x_2,x_3\in\Var$ and let
  $\SH\sidmodels \Start(x_1,x_2,x_3)$. Then
  $\inducedN{\Start}{\S,\H,x_2,x_3} \in \Lang{\G}$.
\end{lemma}
\begin{proof}
  By induction on the height $h$ of the tree contained in $\H$; see \Cref{app:sid-to-g}.
\end{proof}
\subsection{Undecidability of Unguarded Fragments}\label{sec:unguarded-undec-proofs}
We are now ready to prove that omitting guards leads to undecidable SL fragments.
To conveniently describe these fragments, we write $\gls{SLIDxtension}$ for the
restriction of quantifier-free formulas in $\SLIDbtw$ to formulas built from all atoms as well as the additional symbols and connectives
$\cdot_1,\ldots,\cdot_k$. For example, formulas in $\SLop{\wedge,\sep,\true}$
are built from atomic predicates, the predicate $\true$ (true), and the binary connectives $\sep$, $\wedge$.
As usual, $\true$ holds in all models, i.e., $\SH \sidmodels \true$ for all states $\SH$.

First, we show that allowing $\true$ as well as both standard and separating conjunction
$\wedge$ and $\sep$ (i.e., $\SLIDguarded$ without
disjunction or any of the guarded connectives) immediately leads to
undecidability.

\begin{theorem}\label{thm:undec-true}\index{undecidability!of SLID1@of $\SLop{\wedge,\sep,\true}$}
  The satisfiability problem for the fragment $\SLop{\wedge,\sep,\true}$ is undecidable.
\end{theorem}
\begin{proof}
  Let $\Sid$ be the encoding of the CFGs
  $\GA=\theGi{1}$ and $\GB=\theGi{2}$ as described in \cref{sec:unguarded-undec}.
  Moreover, consider the $\SLop{\wedge,\sep,\true}$ formula
  %\[
     $\phi \defn (\cfgA(a,x,y) \sep \true) \wedge (\cfgB(b,x,y) \sep \true)$.
  %\]
  %
  Then $\phi$ is satisfiable iff $\Lang{\GA}\cap\Lang{\GB}\neq\emptyset$; 
  see \Cref{app:undec-true} for details.
\end{proof}

\begin{corollary}\label{cor:undec-neg}\index{undecidability!of SLID2@of $\SLop{\wedge,\sep,\neg}$}
   The satisfiability problem of $\SLop{\wedge,\sep,\neg}$ is undecidable.
\end{corollary}
\begin{proof}
  This follows directly from
    the undecidability of $\SLop{\wedge,\sep,\true}$ (\cref{thm:undec-true}), because $\true$
    is definable in $\SLop{\wedge,\sep,\neg}$; for example,
    $\true \defn \neg(\emp\wedge\neg\emp)$.
\end{proof}

\begin{corollary}\label{cor:undec-mw}\index{undecidability!of SLID3@of $\SLop{\wedge,\sep,\mw}$}
   The satisfiability problem of $\SLop{\wedge,\sep,\mw}$ is undecidable.
\end{corollary}
\begin{proof}
  Follows directly from the undecidability of
  $\SLop{\wedge,\sep,\true}$ (\cref{thm:undec-true}), because $\true$
  is definable in $\SLop{\wedge,\sep,\mw}$; for example,
  $\true \defn (\slneq{x}{x}) \mw \emp$.
\end{proof}

Our final undecidability proof concerns unguarded septractions. We
need one more auxiliary result before we can prove this result.

\begin{lemma}\label{lem:wordb-correct}
  Let $\G_2=\theGi{2}$ be the CFG fixed in \cref{sec:unguarded-undec}.
  Moreover, let $\Sid$ be the corresponding SID encoding,
  $\wordB(x,y) \defn (\word(x,y) \sept \cfgB(a,x,y)) \sept \cfgB(a,x,y)$,
  and let $\SH$ be a state. Then  $\SH \sidmodels \wordB(x,y)$ iff
  $\SH\sidmodels\word(x,y)$ and $\inducedW{\S,\H,x,y} \in \Lang{\GB}$.
\end{lemma}
\begin{proof}
    See \Cref{app:wordb-correct}.
\end{proof}
%\begin{proof}
%  Assume $\SH \sidmodels \wordB(x,y)$.
%  %
%  By the semantics of $\sept$, there exists a heap $\H_1$ with
%  $\SHi{1} \sidmodels \word(x,y) \sept \cfgB(a,x,y)$ such that
%  $\SHpair{\S}{\H\stdunion\H_1} \sidmodels \cfgB(a,x,y)$.
%  %
%  %
%  Observe that $\H_1$ contains precisely the inner nodes of
%  $\SHpair{\S}{\H\stdunion\H_1}$, i.e., everything \emph{except} the
%  part of the state that induces the word.
%  %
%  Consequently, $\H$ is the part of the state that induces the word,
%  i.e.,
%  $\inducedN{\cfgB}{\S,\H\stdunion\H_1,x,y}=\inducedW{\S,\H,x,y}$ and
%  $\SH \sidmodels \word(x,y)$.
%  %
%  \cref{lem:sid-to-g} then yields
%  $\inducedW{\S,\H,x,y}\in\Lang{\GB}$.
%
%  Conversely, assume a state $\SH$ be such that
%  $w \defn \inducedW{\S,\H,x,y} \in \Lang{\GB}$.
%  %
%  As a consequence of \cref{lem:g-to-sid}, there exists a heap $\H_1$
%  with $\SHpair{\S}{\H\stdunion\H_1} \sidmodels \cfgB(a,x,y)$.
%  %
%  Because $\SH \sidmodels\word(x,y)$
%  by assumption, the semantics of $\sept$ yields that
%  $\SHi{1}\sidmodels \word(x,y) \sept \cfgB(a,x,y)$.
%  %
%  Because $\SHpair{\S}{\H\stdunion\H_1} \sidmodels \cfgB(a,x,y)$, we
%  obtain by the semantics of $\sept$ that
%  $\SH \sidmodels (\word(x,y) \sept \cfgB(a,x,y)) \sept \cfgB(a,x,y)$.
%\end{proof}
%
\noindent
To prove the undecidability of separation logic in the presence of unguarded septractions,
we show that
$\psi \defn \wordB(x,y) \sept \cfgA(a,x,y)$ is satisfiable iff
$\Lang{\GA}\cap\Lang{\GB}\neq\emptyset$.
Intuitively, this holds because $\psi$ is satisfiable iff it is
possible to replace the ``word part'' of models of $\cfgA(a,x,y)$
with the ``word part'' of models of $\cfgB(b,x,y)$.
\begin{theorem}\label{thm:undec-sept}\index{undecidability!of SLID4@of
  $\SLop{\sept}$}
  The satisfiability problem of $\SLop{\sept}$ is undecidable.
\end{theorem}
\begin{proof}
    $\psi \defn \wordB(x,y) \sept \cfgA(a,x,y)$ is satisfiable iff
    $\Lang{\GA}\cap\Lang{\GB}\neq\emptyset$; see \Cref{app:undec-sept}.
\end{proof}
\noindent
We have shown that all extensions of the guarded fragment $\SLIDguarded$, in which one of the guards is dropped, lead to an undecidable satisfiability problem.
In the remainder of this paper, we develop a decision procedure for $\SLIDguarded$; keeping all guards thus indeed ensures decidability.

%%% Local Variables:
%%% mode: latex
%%% TeX-master: "../Thesis"
%%% End:

\section{Towards a Compositional Abstraction for GSL}\label{ch:towards}

\newcommand{\abstsh}{\genabst_{1}}
\newcommand{\abstpuf}{\genabst_{2}}
\newcommand{\abstproj}{\genabst_{3}}

In \cref{sec:intro}, we sketched our goal of using a finite \emph{compositional abstraction} that \emph{refines} the satisfaction relation in order to decide the satisfiability problem for the separation logic fragment $\SLIDguarded$.
the same procedure then also allows deciding entailments between (quantifier-free) symbolic heaps in $\SymHeap$ with user-defined predicates (defined by rules that may, of course, contain quantifiers).
The key challenge is to develop an abstraction mechanism that can deal with arbitrary user-defined predicates from the $\IDbtw$ fragment.
To get an abstraction that satisfies refinement, we need to be able to deduce from the abstraction which predicate calls hold in the underlying model.
To this end, we will abstract every state by a set of formulas that relates the state to predicates of the SID.

In the following we introduce our abstraction, starting with a simple but insufficient idea, and then incrementally improve on it.
\paragraph*{Purpose of this section}
This section serves as a roadmap; it outlines the main concepts underlying our decision procedure and explains them informally by means of examples.
We will formalize all of these concepts in follow-up sections---references to the formal details are provided where appropriate.
Similarly, the remaining sections will frequently refer back to this section to either give further details on the examples, or to pin-point the progress of our formalization.
\subsection{First Attempt: Abstracting States by Symbolic Heaps}
Our first idea is to abstract a state by the
quantifier-free symbolic heaps that it satisfies:
\[
    \abstsh(\S,\H) \defn \set{ \phi \in \SymHeap \mid \phi \text{ quantifier-free},
     \SH \sidmodels \phi }.
\]
Let us analyze the properties of this abstraction function.
(For the moment, we ignore whether we can actually compute this abstraction.)

\subsubsection{Finiteness}
The abstraction $\abstsh$ is finite, because there are only finitely many quantifier-free symbolic heaps up to logical equivalence:
Since $\Sid\in\IDbtw$, every predicate call in a symbolic heap $\phi$ has to allocate at least one free variable due to the \emph{progress} property (cf. \cref{sec:sl-basics:btw});
the same holds for every points-to assertion. Consequently, every satisfiable quantifier-free formula can contain at most $\size{\fvs{\phi}}$ many predicate calls and points-to assertions.
In principle, we can, of course, ``blow up'' a satisfiable formula
$\phi$ to arbitrary size by adding $\emp$ atoms and
\mbox{(dis-)equalities}, but
% it is never necessary to add more than
% $\bigO({\size{\fvs{\phi}}}^2)$ such atoms, because this suffices to
% express any fixed \emph{aliasing constraint}
% (cf.~\cref{sec:sl-intro:other}) over $\fvs{\phi}$.
any fixed \emph{aliasing constraint} over $\fvs{\phi}$, i.e., any fixed relationship between the free variables of formula $\phi$,
can be expressed with at most ${\size{\fvs{\phi}}}^2)$ such atoms.
Hence, it suffices to consider only symbolic heaps up to that size in the abstraction.
\subsubsection{Refinement}\label{sec:towards:refinement}
Recall that our abstraction satisfies refinement iff states leading to the same abstraction satisfy the same formulas.
This immediately holds for $\abstsh$---at least on the quantifier-free symbolic-heap fragment of $\SLIDguarded$.
\subsubsection{Compositionality}\label{sec:towards:compositionality}
Can we also \emph{compose} abstractions,
i.e., can we find a (computable) operator $\Compose$ with
$\abstsh(\S,\H_1)\Compose\abstsh(\S,\H_2)=\abstsh(\S,\H_1\stdunion\H_2)$?
Unfortunately, the example below demonstrates that finding such an operator is quite challenging.
Assume that $\Sid$ defines the list-segment predicate $\lseg$
(cf.~\cref{ex:sids}).
Moreover, consider a state $(\S,\H_1\stdunion\H_2)$ such that
\begin{itemize}
\item $\S(u)\neq\S(v)$ for all $u,v \in \{x,y,z\}$ with $u \neq v$,
\item $\abstsh(\S,\H_1) = \set{ \lseg(x,y) }$,
%\item 
$\abstsh(\S,\H_2) = \set{ \lseg(y,z) }$, and
%\item 
$\abstsh(\S,\H_1\stdunion\H_2) = \set{ \lseg(x,z) }$,
\end{itemize}
where we omit pure constraints in the sets $\abstsh(\cdots)$ for readability.
We highlight that it is a-priori unclear how to infer that $\lseg(x,z)$ holds in the composed state, i.e., that $\abstsh(\S,\H_1\stdunion\H_2) = \set{ \lseg(x,z) }$---at least by relying solely on 
%the abstractions of the individual model, i.e., using only 
the assumptions $\abstsh(\S,\H_1) = \set{ \lseg(x,y) }$ and $\abstsh(\S,\H_2) = \set{ \lseg(y,z) }$.
In particular, it is unclear how to derive this fact by a syntactic argument.
One might turn towards an argument based on the semantics, i.e., considering the definition of $\lseg$ in $\Sid$.
However, then the above composition operation $\Compose$ boils down to an entailment check
$%\[
  \lseg(x,y) \sep \lseg(y,z) \sidmodels \lseg(x,z)$.
%\]
%
Hence, we end up with a chicken-and-egg problem:
we need an entailment checker to implement the composition operator $\Compose$ that we would like to
use in the implementation of our abstraction-based (satisfiability and) entailment checker.

\subsection{Second Attempt: Unfolding Predicates into Forests}\label{sec:towards:snd-attempt}

Next, we attempt to extend the abstraction $\abstsh$ to get a ``more syntactic'' composition operation.
To this end, we need to take a step back and reflect on the semantics of SIDs.

\subsubsection{Unfolding predicate calls}
%
%Assume $\vec{x}$ are the parameters of a predicate $\pred \in
%\Preds{\Sid}$.
%
According to the $\SLgeneric$ semantics (\cref{sec:sl-basics:semantics}),
$\SH \sidmodels \pred(\vec{z})$ holds iff there exists a rule $\pred(\vec{x}) \Rule \psi(\vec{x}) \in \Sid$ such that $\SH \sidmodels \phiapp{\psi}{\vec{z}}$.
We say that we have \emph{unfolded} the predicate $\pred$ by the above rule.
In general, $\psi$ may itself contain predicate calls. To prove $\SH \sidmodels \psi(\vec{z})$, we must continue unfolding the remaining predicate calls according to rules of the respective predicates until no predicate calls remain.

It is natural to visualize an unfolding process as a tree.
In fact, defining the semantics of inductive predicates based on such \emph{unfolding trees} is a common approach in the literature (cf.~\cite{iosif2013tree,iosif2014deciding,jansen2017unified,matheja2020automated}).
In this article, we use a variant of unfolding trees, called \emph{$\Sid$-trees}, which we will formally introduce in \cref{def:types:tree}.

\begin{example}[$\Sid$-tree]\label{ex:towards-sid-tree}
  Recall the SID $\Sid$ from \cref{fig:tll-example}, which defines the predicate $\tll$.
  \Cref{fig:ex:sid-tree:model} depicts a state $\SH$ with
  $\SH \sidmodels \tll(x,y,z)$.
  Each node is labeled with a location and the stack variable evaluating to the location (if any).
  The depicted state $\SH$ thus corresponds to
  \begin{align*}
      \S ~=~ & \set{x \mapsto 1, y \mapsto 4, z \mapsto 8, a \mapsto 5,b \mapsto 6,c \mapsto 7},~\text{and} \\
      \H ~=~ & \{1 \mapsto \tuple{2,3,0}, \allowbreak 2\mapsto \tuple{4,5,0}, \allowbreak 5 \mapsto \tuple{6,7,0},
  \allowbreak 4 \mapsto \tuple{0,0,6}, \allowbreak \\
             & \qquad 6 \mapsto \tuple{0,0,7}, \allowbreak 7 \mapsto \tuple{0,0,3}, \allowbreak 3 \mapsto \tuple{0,0,8}\}.
  \end{align*}

  \Cref{fig:ex:sid-tree:forest} shows a $\Sid$-tree $\ftree$ corresponding to this state.
  \begin{figure}[tb!]
    \centering
    \begin{subfigure}[b]{0.25\textwidth}
      \begin{tikzpicture}[state/.style={minimum size=2cm}]
        \node[tlls] (x) {$1\colon x$};
        \node[tlls,below left=6mm of x] (l) {2};
        \node[tlls,below right=6mm of x] (r) {3};
        \node[tlls,below left=6mm of l] (y) {$4\colon y$};
        \node[tlls,below right=6mm of l] (a) {$5\colon a$};
        \node[tlls,below left=6mm of a] (b) {$6\colon b$};
        \node[tlls,below right=6mm of a] (c) {$7\colon c$};
        \node[tlls,right=3mm of r] (z) {$8\colon z$};
        \draw (x) edge[->] (l);
        \draw (x) edge[->] (r);
        \draw (l) edge[->] (y);
        \draw (l) edge[->] (a);
        \draw (a) edge[->] (b);
        \draw (a) edge[->] (c);
        \draw (y) edge[->] (b);
        \draw (b) edge[->] (c);
        \draw (c) edge[->] (r);
        \draw (r) edge[->] (z);
      \end{tikzpicture}
      \caption{A model $\SH$ of the predicate $\tll(x,y,z)$.}
      \label{fig:ex:sid-tree:model}
    \end{subfigure} \hfill
    \begin{subfigure}[b]{0.7\textwidth}
      \scalebox{0.9}{\begin{tikzpicture}
  \node[fnode] (t0) {$\tll(1,4,8) \Rule \ppto{1}{\tuple{2,3,0}}\sep \tll(2,4,3) \sep \tll(3,3,4)$};
  \node[fnode,below=9mm of t0,xshift=-1.8cm] (t1) {$\tll(2,4,3) \Rule \ppto{2}{\tuple{4,5,0}}\sep \tll(4,4,6) \sep \tll(5,6,3)$};
  \node[fnode,below=2mm of t0,xshift=1.8cm] (t2) {$\tll(3,3,4) \Rule \ppto{3}{\tuple{\nil,\nil,4}} \sep (\sleq{3}{3})$};;

  \node[fnode,below=3mm of t1,xshift=-2.8cm] (t3) {$\tll(4,4,6) \Rule \ppto{4}{\tuple{\nil,\nil,6}} \sep (\sleq{4}{4})$};

  \node[fnode,below=3mm of t1,xshift=2.8cm] (t4) {$\tll(5,6,3) \Rule \ppto{5}{\tuple{6,7,0}}\sep \tll(6,6,7) \sep \tll(7,7,3)$};

  \node[fnode,below=3mm of t4,xshift=-4cm] (t5) {$\tll(6,6,7) \Rule \ppto{6}{\tuple{\nil,\nil,7}} \sep (\sleq{6}{6})$};

  \node[fnode,below=3mm of t4,xshift=0.8cm] (t6) {$\tll(7,7,3) \Rule \ppto{7}{\tuple{\nil,\nil,3}} \sep (\sleq{7}{7})$};
  
  \draw (t0) edge[->] (t1);
  \draw (t0) edge[->] (t2);
  \draw (t1) edge[->] (t3);
  \draw (t1) edge[->] (t4);
  \draw (t4) edge[->] (t5);
  \draw (t4) edge[->] (t6);
\end{tikzpicture}
%%% Local Variables:
%%% mode: latex
%%% TeX-master: "../../Thesis"
%%% End:
}
      \caption{A corresponding $\Sid$-tree $\ftree$.}
      \label{fig:ex:sid-tree:forest}
    \end{subfigure}

    \caption[Example: A model and its $\Sid$-tree]{A model $\SH\sidmodels\tll(x,y,z)$ and the $\Sid$-tree $\ftree$ corresponding to this model.}
    \label{fig:ex:sid-tree}
  \end{figure}
  Each node of $\ftree$ is labeled with a \emph{rule instance}, i.e., a rule of the SID in which all variables---both formal parameters and existentially-quantified variables---have been instantiated with the locations and values of the state.
  This is different from other notions of unfolding trees, e.g.,~\cite{iosif2013tree},  in which nodes are labeled by rules, not rule
  instances.
  Note that $\ftree$ induces the heap $\H$: % in a very direct way:
  $\H$ is the union of all the points-to assertions that occur in the node labels of $\ftree$.
  We denote this heap by $\theapof{\ftree}$.
\end{example}

\noindent
$\Sid$-trees enable an alternative reading of the entailment
$\lseg(x,y) \sep \lseg(y,z) \sidmodels \lseg(x,z)$:
The entailment is valid iff whenever $\SH \sidmodels \lseg(x,y)\sep\lseg(y,z)$ holds, it is possible to find a $\Sid$-tree $\ftree$ with root $\lseg(\S(x),\S(z))$
such that $\theapof{\ftree}=\H$.
\subsubsection{Abstracting states by forests}
Our next abstraction attempt is to encode the existence of a suitable $\Sid$-tree.
More precisely, we encode that the models of $\lseg(x,y)$ and $\lseg(y,z)$ each correspond to \emph{partial} $\Sid$-trees of $\lseg(x,z)$ that can be combined into an unfolding tree of $\lseg(x,z)$.
A partial $\Sid$-tree is obtained by prematurely stopping the unfolding process.
Consequently, such a tree may contain \emph{holes}---predicate calls that have not been unfolded.
\begin{example}[$\Sid$-trees with holes]\label{ex:towards:tree-with-holes}
  Recall the entailment $\lseg(x,y) \sep \lseg(y,z) \sidmodels \lseg(x,z)$ from above.
  \Cref{fig:ex:partial-trees:model} shows states $\SHi{1},\SHi{2}$ with $\SHi{1}\sidmodels\lseg(x,y)$ and $\SHi{2}\sidmodels\lseg(y,z)$.
  By the semantics of $\sep$, it holds for $\H\defn \H_1\stdunion\H_2$ that $\SH \sidmodels \lseg(x,y) \sep \lseg(y,z)$.

  How can $\Sid$-trees be used to argue that $\SH \sidmodels \lseg(x,z)$?
  \cref{fig:ex:partial-trees:forest} shows a \emph{$\Sid$-forest}---a set of $\Sid$-trees---consisting of the partial $\Sid$-trees $\ftree_1$ and $\ftree_2$
  with $\theapof{\ftree_1} = \H_1$ and $\theapof{\ftree_2} = \H_2$, respectively.
  Notice that $\ftree_1$ contains a \emph{hole}:
  Since the predicate call $\lseg(4,6)$ is not unfolded,
  the hole, location $4$, is not allocated in the tree, even though it is the root parameter of the \emph{hole predicate} $\lseg(4,6)$.
  We can \emph{merge} $\ftree_1$ and $\ftree_2$ into a larger tree by
  plugging $\ftree_2$ into the hole of $\ftree_1$, i.e., we add an
  an edge from the hole of $\ftree_1$ to the root of $\ftree_2$.
  This is possible because the root of  $\ftree_2$ is labeled with the aforementioned hole predicate, $\lseg(4,6)$.
  The resulting tree is a $\Sid$-tree for $\lseg(x,z)$. That is, it is a tree without holes whose root is labeled with a rule instance of the predicate call $\lseg(\S(x),\S(z))$.
  By merging the two trees, we verified the above entailment: every model of $\lseg(x,y)\sep\lseg(y,z)$ is also a model of $\lseg(x,z)$.
  %
  % There is a different $\Sid$-tree for $\SHi{1}$, with root
  % $\lseg(1,4)$ and without any holes.
  %
\end{example}

\begin{figure}[tb!]
    \centering
    \begin{subfigure}[b]{0.40\textwidth}
      \centering
      \begin{tikzpicture}
        \node[tlls] (1) {$1\colon x$};
        \node[tlls,below=3mm of 1] (2) {$2$};
        \node[tlls,below=3mm of 2] (3) {$3$};
        \node[tlls,below=3mm of 3] (4) {$4\colon y$};

        \node[tlls,right=of 1] (4b) {$4\colon y$};
        \node[tlls,below=3mm of 4b] (5b) {$5$};
        \node[tlls,below=3mm of 5b] (6b) {$6\colon z$};

        \draw (1) edge[->] (2);
        \draw (2) edge[->] (3);
        \draw (3) edge[->] (4);
        \draw (4b) edge[->] (5b);
        \draw (5b) edge[->] (6b);
\end{tikzpicture}
%%% Local Variables:
%%% mode: latex
%%% TeX-master: "../../Thesis"
%%% End:
      \caption{The states $\SHi{1}$, $\SHi{2}$.}\label{fig:ex:partial-trees:model}
    \end{subfigure} \hfill
    \begin{subfigure}[b]{0.55\textwidth}
      \centering
      \scalebox{0.9}{\begin{tikzpicture}
  \node[fnode] (l1-1) {$\lseg(1, 6) \Rule \ppto{1}{2} \sep
    \lseg(2,6)$};
  \node[fnode,below=3mm of l1-1] (l1-2) {$\lseg(2, 6) \Rule
    \ppto{2}{3} \sep \lseg(3,6) $};
  \node[fnode,below=3mm of l1-2] (l1-3) {$\lseg(3, 6) \Rule
    \ppto{3}{4} \sep \lseg(4,6) $};
  \draw (l1-1) edge[->] (l1-2);
  \draw (l1-2) edge[->] (l1-3);

  \begin{scope}[on background layer]
    \node[fit=(l1-1)(l1-3),fill=laranja1!30] (tree1) {};
    \node[right=2mm of tree1] {$\ftree_1$};
  \end{scope}

  \node[fnode,below=15mm of l1-3] (l2-1) {$\lseg(4, 6) \Rule
    \ppto{4}{5} \sep \lseg(5,6) $};
  \node[fnode,below=3mm of l2-1] (l2-2) {$\lseg(5, 6) \Rule
    \ppto{5}{6}$};
  \draw (l2-1) edge[->] (l2-2);

    \begin{scope}[on background layer]
    \node[fit=(l2-1)(l2-2),fill=red!20] (tree2) {};
    \node[right=2mm of tree2] {$\ftree_2$};
  \end{scope}

  \node[right=0mm of l1-3,xshift=-8mm] (hole) {};
  \node[legendnode,below=5mm of l1-3,xshift=13mm] (holelbl) {hole of $\ftree_1$};
  \draw (holelbl) edge[legendedge] (hole);
\end{tikzpicture}
%%% Local Variables:
%%% mode: latex
%%% TeX-master: "../../Thesis"
%%% End:
}
      \caption{$\Sid$-trees corresponding to the states.}
      \label{fig:ex:partial-trees:forest}
    \end{subfigure}

    \caption[Example: $\Sid$-trees with holes]{States
      $\SHi{1}\sidmodels\lseg(x,y)$ and $\SHi{2}\sidmodels\lseg(y,z)$ and
      $\Sid$-trees $\ftree_1,\ftree_2$ corresponding to the states.
      The tree $\ftree_1$ contains one predicate call that is not unfolded, $\lseg(4,6)$.
      We say that $4$, the root of this folded predicate call, is a hole of the tree.}
    \label{fig:ex:partial-trees}
  \end{figure}
%We formalize
%partial unfolding trees w.r.t.~SID $\Sid$ as
%\emph{$\Sid$-trees} in \cref{def:types:tree} in the next chapter.
%
\noindent
In fact, we go one step further and consider \emph{$\Sid$-forests}
(cf.~\cref{def:types:forest}), i.e., sets of partial $\Sid$-trees.
For example, the set $\set{\ftree_1,\ftree_2}$ illustrated in \cref{fig:ex:partial-trees} is a $\Sid$-forest.
%This makes sense once we go beyond lists or trees.
%
\begin{example}[$\Sid$-forest]\label{ex:treerp-forest}
  Continuing \cref{ex:towards-sid-tree},
  \cref{fig:ex:towards:forest} depicts a \emph{$\Sid$-forest} $\frst=\set{\ftree_1,\ftree_2,\ftree_3,\ftree_4}$ that encodes one way to obtain the state $\SH$ through iterative unfolding of predicate calls.
  Both $\ftree_1$ and $\ftree_2$ only \emph{partially} unfold the predicates at their roots, leaving locations $5$ resp. $6$ and $7$ as holes.
  By merging the four trees, we get the tree $\ftree$ from \cref{ex:towards-sid-tree}.
  \begin{figure}[tb!]
      \begin{tikzpicture}

  % t1
  \node[fnode] (t0) {$\tll(1,4,8) \Rule \ppto{1}{\tuple{2,3,0}}\sep \tll(2,4,3) \sep \tll(3,3,8)$};
  \node[fnode,below=9mm of t0,xshift=-1.8cm] (t1) {$\tll(2,4,3) \Rule \ppto{2}{\tuple{4,5,0}}\sep \tll(4,4,6) \sep \tll(5,6,3)$};
  \node[fnode,below=2mm of t0,xshift=1.8cm] (t2) {$\tll(3,3,8) \Rule \ppto{3}{\tuple{\nil,\nil,8}} \sep (\sleq{3}{3})$};;
  \node[fnode,below=3mm of t1,xshift=-2.8cm] (t3) {$\tll(4,4,6) \Rule \ppto{4}{\tuple{\nil,\nil,6}} \sep (\sleq{4}{4})$};
  \draw (t0) edge[->] (t1);
  \draw (t0) edge[->] (t2);
  \draw (t1) edge[->] (t3);
  \begin{scope}[on background layer]
    \node[fit=(t0)(t1)(t2)(t3),fill=laranja1!30] (tree1) {};
    \node[right=2mm of tree1] {$\ftree_1$};
  \end{scope}

  % t2
  \node[fnode,below=12mm of t1,xshift=3cm] (t4) {$\tll(5,6,3) \Rule \ppto{5}{\tuple{6,7,0}}\sep \tll(6,6,7) \sep \tll(7,7,3)$};
  \begin{scope}[on background layer]
    \node[fit=(t4),fill=red!20] (tree2) {};
    \node[right=2mm of tree2] {$\ftree_2$};
  \end{scope}

  % t3
  \node[fnode,below=4mm of t4,xshift=-4.7cm] (t5) {$\tll(6,6,7) \Rule \ppto{6}{\tuple{\nil,\nil,7}} \sep (\sleq{6}{6})$};
  \begin{scope}[on background layer]
    \node[fit=(t5),fill=green!10] (tree3) {};
    \node[right=2mm of tree3] {$\ftree_3$};
  \end{scope}

  % t4
  \node[fnode,below=4mm of t4,xshift=1.2cm] (t6) {$\tll(7,7,3) \Rule \ppto{7}{\tuple{\nil,\nil,3}} \sep (\sleq{7}{7})$};
  \begin{scope}[on background layer]
    \node[fit=(t6),fill=green!10] (tree4) {};
    \node[right=2mm of tree4] {$\ftree_4$};
  \end{scope}

%  % t1
%  \node[fnode] (t11) {$\tree(1) \Rule \ppto{1}{2}\sep\treerp(2,1)$};
%  \begin{scope}[on background layer]
%    \node[fit=(t11),fill=laranja1!30] (tree1) {};
%    \node[right=2mm of tree1] {$\ftree_1$};
%  \end{scope}
%
%  % t2
%  \node[fnode,below=5mm of t11] (t21) {$\treerp(2,1) \Rule
%    \ppto{2}{\tuple{3,4,1}}\sep\treerp(3,1)\sep\treerp(4,1)$};
%  \node[fnode,below=3mm of t21] (t22) {$\treerp(3,1) \Rule
%    \ppto{3}{\tuple{0,0,1}}$};
%  \draw (t21) edge[->] (t22);
%  \begin{scope}[on background layer]
%    \node[fit=(t21)(t22),fill=red!20] (tree2) {};
%    \node[right=2mm of tree2] {$\ftree_2$};
%  \end{scope}
%
%  % t3
%  \node[fnode,below=5mm of t22] (t31) {$\treerp(4,1) \Rule
%    \ppto{4}{\tuple{5,6,1}}\sep\treerp(5,1)\sep\treerp(6,1)$};
%  \node[fnode,below=3mm of t31,xshift=-2.5cm] (t32) {$\treerp(5,1) \Rule
%    \ppto{5}{\tuple{0,0,1}}$};
%  \node[fnode,below=3mm of t31,xshift=2.5cm] (t33) {$\treerp(6,1) \Rule
%    \ppto{6}{\tuple{0,0,1}}$};
%  \draw (t31) edge[->] (t32);
%  \draw (t31) edge[->] (t33);
%  \begin{scope}[on background layer]
%    \node[fit=(t31)(t32)(t33),fill=green!10] (tree3) {};
%    \node[right=2mm of tree3] {$\ftree_3$};
%  \end{scope}

\end{tikzpicture}
%%% Local Variables:
%%% mode: latex
%%% TeX-master: "../../Thesis"
%%% End:
      \caption[Example: A $\Sid$-forest for trees with parent pointers]{A $\Sid$-forest
        $\frst=\set{\ftree_1,\ftree_2,\ftree_3,\ftree_4}$ for the state from
        \cref{ex:towards-sid-tree}, used in
        \cref{ex:treerp-forest}.}
        \label{fig:ex:towards:forest}
  \end{figure}
\end{example}
\noindent
Our second idea towards a suitable abstraction is
to abstract a state by computing all $\Sid$-forests consisting of trees whose combined heap matches the heap of the state:
\[
  \abstpuf(\S,\H) \defn \set{ \frst ~\middle|~
   \frst \text{ is a } \Sid \text{ -forest of } (\S,\H) \text{ with } \H = \bigcup_{\ftree\in\frst}\theapof{\ftree} }.
\]

\subsubsection{Compositionality}\label{sec:towards:compo2}
We can define a suitable composition operation $\abstpuf(\S,\H_1)\Compose\abstpuf(\S,\H_2)$ by computing all ways to merge the $\Sid$-forests of $\abstpuf(\S,\H_1)$ and $\abstpuf(\S,\H_2)$.
This approach yields precisely the set of all $\Sid$-forests of
$\SHpair{\S}{\H_1\stdunion\H_2}$, i.e., the set
$\abstpuf(\S,\H_1\stdunion\H_2)$ from above, as required. % by the compositionality property.

\subsubsection{Finiteness} Unfortunately, $\abstpuf$ results in an \emph{infinite} abstraction due to two main issues:
\begin{description}
\item[Issue 1] The tree nodes are labeled with concrete locations, so the result of $\abstpuf$ differs even if the states are isomorphic, i.e., identical up to renaming of locations.
    Apart from leading to an infinite abstraction, distinguishing such states is undesirable as they satisfy the same $\SLIDguarded$ formulas as long as we do not explicitly use constant values other than the null pointer.
\item[Issue 2] If we keep track of all $\Sid$-forests, the size of $\abstpuf(\S,\H)$ grows with the size of $\H$---it is unbounded.
    For example, the abstraction of a list-segment of size $n$ contains the forest that consist of a single tree with $n$ nodes, the forest that consists of $n$ one-node trees as well as all possibilities in between.
\end{description}
%
%
% \begin{example}[There are infinitely many
%   $\Sid$-forests]\label{ex:infinite-forests}
%   \TODO{Write this up properly.}
%   For example, a list of length $n$ gives rise to a partial unfolding
%   forest that consists of $n$ trees: one tree per location allocated
%   in the list.
% \end{example}
%
%This part of the thesis is to a large extent concerned with overcoming these two issues to obtain a viable abstraction for $\SLIDguarded$.

\subsection{Third Attempt: Forest Projections}\label{sec:towards:projection}
Our first attempt yields a finite abstraction that is not compositional,
whereas our second attempt is compositional but not not finite.
We now construct a finite \emph{and} compositional abstraction by
considering an additional abstraction---called the \emph{projection}---
on top of $\Sid$-forests with holes.
To this end, we denote by $\trootpred{\ftree}$ the root and by $\tallholepreds{\ftree}$ the hole predicates of a $\Sid$-tree $\ftree$.
% In the previous section, we introduced $\Sid$-trees as unfolding
% trees
\begin{example}
  \begin{enumerate}
  \item Let $\ftree_1,\ftree_2$ be the $\Sid$-trees from \cref{ex:towards:tree-with-holes}, which are illustrated in \cref{fig:ex:partial-trees:forest}.
      Then,
      \begin{itemize}
        \item $\trootpred{\ftree_1}=\lseg(1,6)$,
        $\tallholepreds{\ftree_1}=\set{\lseg(4,6)}$, and
        \item
        $\trootpred{\ftree_1}=\lseg(4,6)$,
        $\tallholepreds{\ftree_2}=\emptyset$.
      \end{itemize}

  \item Let $\ftree_1,\ftree_2,\ftree_3,\ftree_4$ be the $\Sid$-trees from \cref{ex:treerp-forest}, which are illustrated in \cref{fig:ex:towards:forest}.
      Then,
        \begin{itemize}
        \item $\trootpred{\ftree_1}=\tll(1,4,8)$,
        $\tallholepreds{\ftree_1}=\set{\tll(5,6,3)}$,
        \item $\trootpred{\ftree_2}=\tll(5,6,3)$,
        $\tallholepreds{\ftree_2}=\set{\tll(6,6,7),\tll(7,7,3)}$,
        \item $\trootpred{\ftree_3}=\tll(6,6,7)$,
        $\tallholepreds{\ftree_3}=\emptyset$, and
        \item $\trootpred{\ftree_4}=\tll(7,7,3)$, $\tallholepreds{\ftree_4}=\emptyset$.
        \end{itemize}
  \end{enumerate}
\end{example}
\noindent
The \emph{projection} of a $\Sid$-forest $\frst$ can be viewed as a $\SLIDguarded$ formula encoding, for each tree $\ftree \in \frst$, a model of $\trootpred{\ftree}$ from which models of  $\tallholepreds{\ftree}$ have been subtracted, i.e.,\footnote{Recall that $\IteratedStar \{\phi_1, \ldots \phi_n\}$ is a shortcut for $\phi_1 \sep \ldots \sep \phi_n$.}
%The main insight behind the \emph{$\Sid$-type} abstraction is that every $\Sid$-tree $\ftree$ can be viewed as encoding a model of $\trootpred{\ftree}$ from which models of $\tallholepreds{\ftree}$ have been subtracted.
%
%The tree $\ftree$ can thus be \emph{projected} onto the formula
%
\[
\IteratedStar_{\ftree\in \frst} \left[ \left(\IteratedStar\tallholepreds{\ftree}\right) \mw
  \trootpred{\ftree} \tag*{\tagA} \right].
\]
\noindent
The goal of the projection operation is to combat Issue 2 identified above:
to restore finiteness, we keep only limited information about each unfolding tree, and remember only its root predicate and its hole predicates.
The magic wand introduced by the projection operation in the formula \tagA{} allows us to maintain the compositionality of the abstraction.
\begin{example}[Forest projection---with locations]\label{ex:lseg-projection-loc}
  Recall from \cref{ex:towards:tree-with-holes},
  the states $\SHi{1}\sidmodels\lseg(x,y)$, $\SHi{2}\sidmodels\lseg(y,z)$, and the corresponding $\Sid$-trees $\ftree_1,\ftree_2$.
  The projection of stack $\S$ and $\Sid$-forest $\set{\ftree_1,\ftree_2}$ is then the formula      $(\lseg(4,6) \mw \lseg(1,6)) \sep (\emp \mw \lseg(4,6))$.
\end{example}
\subsubsection{Abstracting from locations}
\label{sec:towards:replacing}
%The formula \tagA{} still contains locations rather than variables, because the parameters of predicate calls in our $\Sid$-forests are locations, not variables.
Our goal, which we will soon complete, has been to define a compositional abstraction over states.
To this end, we introduced (partial) unfolding trees.
These trees are naturally defined through the instantiation of SID rules with locations.
Unfortunately, locations present an obstacle towards obtaining a finite abstraction (Issue 1):
we get a different abstraction even for $\Sid$-forests that encode the same model up to isomorphism!
However, after having projected unfolding trees to formulas, we are able to reverse the instantiation of variables with locations.
We in fact define the projection operation \tagA{} to output variables instead of locations:
Say $\ftree$ is a $\Sid$-tree of the state $\SH$.
Then we replace the locations in the formula \tagA{} as follows:
\begin{enumerate}
    \item Every location $v\in\img(\S)$ is replaced by a variable $x$ satisfying $\S(x) = v$. %an arbitrary variable in $\stkinv(v)$.
    \item Every location in $\dom(\H)\setminus\img(\S)$ is replaced by an \emph{existentially-quantified} variable, because there exists a location in the heap $\H$ that corresponds to the location in the formula  \tagA{}.

    \item All other locations are replaced by a \emph{universally-quantified} variable, because these locations do not occur in the heap $\H$ (this holds because we will always assume  $\SHpair{\S}{\fheapof{\ftree}} \in \Mpos{\Sid}$)
        and can thus be picked in an arbitrary way.
\end{enumerate}
We remark that the formal definition of projection uses non-standard quantifiers $\eexists$ and $\fforall$ in projections (further discussed below).
    For the moment, this difference does not matter; it is safe to replace them with the usual first-order quantifiers $\exists$ and $\forall$ for intuition.
\begin{example}[Forest projection---with variables (without quantifiers)]
  \label{ex:lseg-projection}
  Continuing \cref{ex:lseg-projection-loc},
  %We recall the models $\SHi{1}\sidmodels\lseg(x,y)$ and $\SHi{2}\sidmodels\lseg(y,z)$ from \cref{ex:towards:tree-with-holes}, and the corresponding $\Sid$-trees $\ftree_1,\ftree_2$.
  the projection of $\S$ and $\Sid$-forest $\set{\ftree_1,\ftree_2}$ using the above replacement is
      \[
        (\lseg(y,z) \mw \lseg(x,z)) \sep (\emp \mw \lseg(y,z)).
      \]
  We will later prove that the projection operation is \emph{sound}
  (Lemma~\ref{lem:sf-projection-sound}), i.e., for the given example,
  \[
  (\S,\H_1\stdunion\H_2) \sidmodels (\lseg(y,z) \mw \lseg(x,z)) \sep (\emp \mw \lseg(y,z)).
  \]
  We further note that the idea of connecting holes with corresponding roots in $\Sid$-trees is mirrored on the level of formulas:
    since the magic wand $\mw$ is the left-adjoint of the separating conjunction $\sep$, an application of \emph{modus ponens}\footnote{i.e., $\phi \sep (\phi \mw \psi) ~\Rightarrow~ \psi$} (formalized in Lemma~\ref{lem:gmp}) suffices to establish that
  \[
  (\S,\H_1\stdunion\H_2) \sidmodels \emp \mw \lseg(x,z).
  \]
\end{example}
\begin{figure}[tb!]
    \centering
    \begin{subfigure}[b]{0.30\textwidth}
      \centering
      \begin{tikzpicture}
        \node[tlls] (x) {$0\colon x$};
        \node[tlls,right=3mm of x] (a) {$1$};
        \node[tlls,right=3mm of a] (ay) {$2$};
        \node[tlls,right=3mm of ay] (y) {$3\colon y$};
        \node[tlls,right=3mm of y] (yz) {$4$};
        \node[tlls,right=3mm of yz] (z) {$5\colon z$};
        \node[tlls,above=3mm of y] (za) {$6$};
        
        \draw (x) edge[->] (a);
        \draw (a) edge[->] (ay);
        \draw (ay) edge[->] (y);
        \draw (y) edge[->] (yz);
        \draw (yz) edge[->] (z);        
        \draw (z) edge[->,bend right] (za);
        \draw (za) edge[->,bend right] (a);
\end{tikzpicture}
%%% Local Variables:
%%% mode: latex
%%% TeX-master: "../../Thesis"
%%% End:
      \caption{A model $\SH$ of $\topCyclic(x,y,z)$}
      \label{fig:ex-cyclic:partial-trees:model}
    \end{subfigure} \hfill
    \begin{subfigure}[b]{0.65\textwidth}
      \centering
      \scalebox{0.9}{}
      {\begin{tikzpicture}
  \node[fnode] (l1-1) {$\topCyclic(0,3,5) \Rule \ppto{0}{\tuple{1,3,5}} \sep \lseg(1,1)$};
  \node[fnode,below=3mm of l1-1] (l1-2) {$\lseg(1,1) \Rule \ppto{1}{2} \sep \lseg(2,1) $};
  \node[fnode,below=3mm of l1-2] (l1-3) {$\lseg(2,1) \Rule \ppto{2}{3} \sep \lseg(3,1) $};
  \draw (l1-1) edge[->] (l1-2);
  \draw (l1-2) edge[->] (l1-3);

  \begin{scope}[on background layer]
    \node[fit=(l1-1)(l1-3),fill=laranja1!30] (tree1) {};
    \node[right=2mm of tree1] {$\ftree_1$};
  \end{scope}

  \node[fnode,below=5mm of l1-3,xshift=-2cm] (l2-1) {$\lseg(3, 1) \Rule \ppto{3}{4} \sep \lseg(4,1)$};
  \node[fnode,below=3mm of l2-1] (l2-2) {$\lseg(4, 1) \Rule \ppto{4}{5} \sep \lseg(5,1)$};
  \draw (l2-1) edge[->] (l2-2);

  \begin{scope}[on background layer]
    \node[fit=(l2-1)(l2-2),fill=red!20] (tree2) {};
    \node[right=2mm of tree2] {$\ftree_2$};
  \end{scope}

  \node[fnode,right=10mm of l2-1] (l3-1) {$\lseg(5, 1) \Rule \ppto{5}{6} \sep \lseg(6,1) $};
  \node[fnode,below=3mm of l3-1] (l3-2) {$\lseg(6, 1) \Rule \ppto{6}{1}$};
  \draw (l3-1) edge[->] (l3-2);

  \begin{scope}[on background layer]
    \node[fit=(l3-1)(l3-2),fill=green!10] (tree3) {};
    \node[right=2mm of tree3] {$\ftree_3$};
  \end{scope}
\end{tikzpicture}
%%% Local Variables:
%%% mode: latex
%%% TeX-master: "../../Thesis"
%%% End:
}
      \caption{$\Sid$-trees corresponding to the state $\SH$.}
      \label{fig:ex-cyclic:partial-trees:forest}
    \end{subfigure}

    \caption[Example: $\Sid$-trees with holes]{A state $\SH\sidmodels\topCyclic(x,y,z)$ and $\Sid$-trees corresponding to this state.}
    \label{fig:ex-cyclic:partial-trees}
  \end{figure}
\begin{example}[Forest projection---with variables (and quantifiers)]
  \label{ex:cyclic-projection}
    We consider an $\Sid$, consisting of the list segment predicate $\lseg$ (see \cref{ex:sids}) and the following additional predicate:
  \[
    \begin{array}{lll}
      \topCyclic(x,y,z) & \Rule & \SHEX{a} \pto{x}{\tuple{a,y,z}} \sep \ls(a,a)
    \end{array}
  \]
  \cref{fig:ex-cyclic:partial-trees:forest} depicts a model $\SH$ of $\topCyclic(x,y,z)$ and $\Sid$-trees $\ftree_1,\ftree_2,\ftree_3$ with $\H = \theapof{\ftree_1} \cup \theapof{\ftree_2} \cup \theapof{\ftree_3}$.
  The projection of stack $\S$ and $\Sid$-forest $\set{\ftree_1,\ftree_2,\ftree_3}$ is
  \[
      \EEX{a} (\lseg(y,a) \mw \topCyclic(x,y,z)) \sep
      (\lseg(z,a) \mw \lseg(y,a)) \sep \lseg(z,a).
  \]
  We will later prove that the projection operation is \emph{sound} (Lemma~\ref{lem:sf-projection-sound}), i.e., for the given example,
  \[
  \SH \sidmodels \EEX{a} (\lseg(y,a) \mw \topCyclic(x,y,z)) \sep
      (\lseg(z,a) \mw \lseg(y,a)) \sep \lseg(z,a).
  \]
\end{example}
\noindent
Equipped with this extended projection operation, we are now in a position to specify the third (and almost final) abstraction function:
\[
  \abstproj(\S,\H) \defn \set{ \phi ~\middle|~
   \phi \text{ is the projection of a } \Sid \text{ -forest } \frst \text{ of } (\S,\H) \text{ with } \H = \bigcup_{\ftree\in\frst}\theapof{\ftree} }.
\]

\subsubsection{Compositionality}
As already hinted at in Example~\ref{ex:lseg-projection},
the projection of formulas allows us to define a (computable) operator $\Compose$ with
$\abstproj(\S,\H_1)\Compose\abstproj(\S,\H_2)=\abstproj(\S,\H_1\stdunion\H_2)$ such that:
\begin{enumerate}
  \item We have $\phi \sep \psi \in \abstproj(\S,\H_1)\Compose\abstproj(\S,\H_2)$ for all $\phi \in \abstproj(\S,\H_1)$ and $\psi \in \abstproj(\S,\H_2)$ .
  \item The set of formulas $\abstproj(\S,\H_1)\Compose\abstproj(\S,\H_2)$ is closed under application of modus ponens.
  \item The set $\abstproj(\S,\H_1)\Compose\abstproj(\S,\H_2)$ is closed under certain rules for manipulating quantifiers.
\end{enumerate}

\begin{example}[Composition Operation on Projections]
\label{ex:composition-projection}
  \begin{enumerate}
  \item  We recall the states $\SHi{1}\sidmodels\lseg(x,y)$ and $\SHi{2}\sidmodels\lseg(y,z)$ from \cref{ex:towards:tree-with-holes}, and the %corresponding 
      $\Sid$-trees $\ftree_1,\ftree_2$.
      The projection of $\S$ and $\set{\ftree_1}$ is
      %\[
        $\lseg(y,z) \mw \lseg(x,z)$,
      %\]
      and the projection of $\S$ and $\set{\ftree_2}$ is
      $%\[
       \emp \mw \lseg(y,z)$.
      %\]
      Hence, %we have
      \[
        (\lseg(y,z) \mw \lseg(x,z)) \sep (\emp \mw \lseg(y,z)) \in \abstproj(\S,\H_1)\Compose\abstproj(\S,\H_2).
      \]
      By applying modus ponens, we get %for the magic wand, we get
      %\[
        $\emp \mw \lseg(x,z) \in \abstproj(\S,\H_1)\Compose\abstproj(\S,\H_2)$.
      %\]
      The above reasoning approach will lead to a compositional proof of the entailment
      \[
          \lseg(x,y) \sep \lseg(y,z) \sidmodels \lseg(x,z).
      \]

  \item Let $\SH$ be the model and let $\ftree_1,\ftree_2,\ftree_3$ be the $\Sid$-trees from \cref{ex:cyclic-projection}.
      We set $\H_1 = \theapof{\ftree_1} \cup \theapof{\ftree_3}$ and $\H_2 = \theapof{\ftree_2}$.
      The projection of $\S$ and $\set{\ftree_1,\ftree_3}$ is
      \[
        \EEX{a} (\lseg(y,a) \mw \topCyclic(x,y,z)) \sep \lseg(z,a),
      \]
      and the projection of $\S$ and $\set{\ftree_2}$ is
      %\[
        $\FFA{a'} \lseg(z,a') \mw \lseg(y,a')$.
      %\]
      Hence, we have
      \begin{align*}
          & \left[ \EEX{a} (\lseg(y,a) \mw \topCyclic(x,y,z)) \sep \lseg(z,a) \right] \sep \\
          & \qquad \qquad \left[ \FFA{a'} \lseg(z,a') \mw \lseg(y,a') \right] \in \abstproj(\S,\H_1)\Compose\abstproj(\S,\H_2).
      \end{align*}
      By instantiating $a'$ with $a$ and moving $\EEX{a}$ to the front of the formula, we get that
      \begin{align*}
          & \EEX{a} (\lseg(y,a) \mw \topCyclic(x,y,z)) \sep \lseg(z,a) \sep \\
          & \qquad\qquad (\lseg(z,a) \mw \lseg(y,a)) \in \abstproj(\S,\H_1)\Compose\abstproj(\S,\H_2).
      \end{align*}
      By applying modus ponens (twice), we get %for the magic wand, we get
      %By applying modus ponens for the magic wand  (twice), we get that
      %\[
        $\EEX{a} \topCyclic(x,y,z) \in \abstproj(\S,\H_1)\Compose\abstproj(\S,\H_2)$.
      %\]
      As the variable $a$ does not appear free anymore, the quantifier can be dropped, and we get
      \[
        \topCyclic(x,y,z) \in \abstproj(\S,\H_1)\Compose\abstproj(\S,\H_2).
      \]

      %We note that using the above reasoning our approach will enable a compositional proof of the entailment
      The above reasoning our approach will lead to a compositional proof of the entailment
      \[
          \fork(x,y,z) \sep \lseg(y,z) \sidmodels \topCyclic(x,y,z),
      \]
      where $\Sid$ extends the SID from \cref{ex:cyclic-projection} by the predicate
      \[
        \begin{array}{lll}
        \fork(x,y,z) & \Rule & \SHEX{a} \ppto{x}{\tuple{a,y,z}} \sep \ls(a,y)\sep \ls(z,a).
      \end{array}
      \]
  \item Let $\ftree_1,\ftree_2,\ftree_3,\ftree_4$ be the $\Sid$-trees from \cref{ex:treerp-forest} for the state $\SH$ of \cref{ex:towards-sid-tree}.
      We set $\H_1 = \theapof{\ftree_1} \cup \theapof{\ftree_3} \cup \theapof{\ftree_4}$ and $\H_2 = \theapof{\ftree_2}$.
      The projection of $\S$ and $\set{\ftree_1,\ftree_3,\ftree_4}$ is
      \[
        \EEX{r} (\tll(a,b,c) \mw \tll(x,y,z)) \sep \ppto{b}{\tuple{0,0,c}} \sep \ppto{c}{\tuple{0,0,r}},
      \]
      and the projection of $\S$ and $\set{\ftree_2}$ is
      %\[
        $\FFA{r'} (\ppto{b}{\tuple{0,0,c}} \sep \ppto{c}{\tuple{0,0,r'}}) \mw \tll(a,b,c)$.
      %\]
      Hence, %we have
      \begin{multline*}
        \left[ \EEX{r} (\tll(a,b,c) \mw \tll(x,y,z)) \sep \ppto{b}{\tuple{0,0,c}} \sep \ppto{c}{\tuple{0,0,r}} \right] \sep\\
        \left[        \FFA{r'} (\ppto{b}{\tuple{0,0,c}} \sep \ppto{c}{\tuple{0,0,r'}}) \mw \tll(a,b,c) \right] \in \abstproj(\S,\H_1)\Compose\abstproj(\S,\H_2).
      \end{multline*}
      By instantiating $r'$ with $r$ and moving $\EEX{r}$ to the front of the formula, we get that
      \begin{multline*}
         \EEX{r} (\tll(a,b,c) \mw \tll(x,y,z)) \sep \ppto{b}{\tuple{0,0,c}} \sep \ppto{c}{\tuple{0,0,r}} \sep \\
         ((\ppto{b}{\tuple{0,0,c}} \sep \ppto{c}{\tuple{0,0,r}}) \mw \tll(a,b,c)) \in \abstproj(\S,\H_1)\Compose\abstproj(\S,\H_2).
      \end{multline*}
      By applying modus ponens for the magic wand (twice), we get that
      \[
        \EEX{r} \tll(x,y,z) \in \abstproj(\S,\H_1)\Compose\abstproj(\S,\H_2).
      \]
      As the variable $r$ does not appear free anymore, the quantifier can be dropped and we get
      \[
        \tll(x,y,z) \in \abstproj(\S,\H_1)\Compose\abstproj(\S,\H_2).
      \]

       The above reasoning our approach will lead to a compositional proof of the entailment
      \[
          \tllHole(x,y,z,a,b,c) \sep \ppto{a}{\tuple{b,c,0}} \sidmodels \tll(x,y,z),
      \]
      where $\Sid$ extends the TLL SID from \cref{fig:tll-example} by the predicates
      \[
        \begin{array}{lll}
        \tllHole(x,y,z,a,b,c)& \Rule & \SHEX{l,r} \ppto{x}{\tuple{l,r,0}} \sep \tllHoleHelper(l,r,y,z,a,b,c)\\
        \tllHoleHelper(l,r,y,z,a,b,c) & \Rule & \ppto{l}{\tuple{y,a,0}} \sep \tllListFour(y,b,c,r,z)\\
        \tllListFour(y,b,c,r,z)& \Rule & \ppto{y}{\tuple{0,0,b}} \sep \tllListThree(b,c,r,z)\\
        \tllListThree(b,c,r,z)& \Rule & \ppto{b}{\tuple{0,0,c}} \sep \tllListTwo(c,r,z)\\
        \tllListTwo(c,r,z)& \Rule & \ppto{c}{\tuple{0,0,r}} \sep \ptrpred(r,z).
        \end{array}
      \]

  \end{enumerate}
\end{example}

\subsubsection{Guarded Quantifiers.}
\label{sec:towards:guarded-quantifiers}
We now discuss the semantics of the special quantifiers $\eexists$ and $\fforall$ used in the projection operation.
We rely on a non-standard semantics because we want our approach to directly support SIDs with (dis-)equalities.
If one would disallow (dis-)equalities in SIDs, one could use the usual $\exists$ and $\forall$ quantifiers instead of $\eexists$ and $\fforall$ (which is sufficient for the SIDs in \cref{ex:composition-projection}).
We motivate our non-standard semantics with the SID $\Sid$ given by the following predicates:
\[
    \begin{array}{lllllll}
    p(x,a,b)& \Rule & \SHEX{y}\ppto{x}{y} \sep q(y,a) \sep x \neq a \sep a \neq b
    &\quad&
    q(y,a) & \Rule & \ppto{y}{\nil} \sep y \neq z
    \end{array}
\]
We consider the stack $\S = \set{x \mapsto 1}$ and the heap $\H = \set{1 \mapsto 2, \allowbreak 2 \mapsto 3}$.
We further consider the unfolding tree $\ftree$ consisting of the root $p(1,4,5) \Rule \ppto{1}{2}\sep 1\neq4 \sep 4\neq5$ with the single child $q(2,4) \Rule \ppto{2}{3} \sep 2\neq4$;
note that $\theapof{\ftree} = \H$.
The projection of this unfolding tree is the formula
%\[
    $\FFA{a,b} p(x,a,b)$.
%\]
As discussed earlier, we want that the projection operation is sound, i.e.,
%we want that the following holds:
$\SH \sidmodels \FFA{a,b} p(x,a,b)$.
However, using the standard quantifier $\forall$ instead of $\fforall$ does not work:
\begin{align*}
    \text{(1)} 
    \SH \not\sidmodels p(1,5,5)
    \qquad
    \text{(3)} 
    \SH \not\sidmodels p(1,1,5)
    \qquad
    \text{(3)} 
    \SH \not\sidmodels p(1,2,5)
\end{align*}
%\begin{enumerate}
%\item $\SH \not\sidmodels p(1,5,5)$
%\item $\SH \not\sidmodels p(1,1,5)$
%\item $\SH \not\sidmodels p(1,2,5)$
%\end{enumerate}
The above example shows that we need to prevent instantiating universally quantified 
\mbox{variables with}
\begin{enumerate}
\item identical locations, see $\SH \not\sidmodels p(1,5,5)$,
\item locations that are in the image of the stack, see $\SH \not\sidmodels p(1,1,5)$, and
\item locations that are existentially quantified, see $\SH \not\sidmodels p(1,2,5)$.
\end{enumerate}
For the semantics of $\fforall$ we use that in $\SLIDbtw$ all existentially quantified variables (that are not equal to a parameter or the null pointer) are allocated because of the establishment requirement, and set
\begin{multline*}
  \SH \sidmodels \FFA{\tuple{a_1,\ldots,a_k}}\phi \text{ iff for all pairwise different locations}  \\
  v_1,\ldots,v_k \in \Loc\setminus(\dom(\H) \cup \img(\S)) \text{ it holds that } \SHpair{\S \cup \set{a_1\mapsto v_1,\ldots,a_k\mapsto v_k}}{\H} \sidmodels \phi
\end{multline*}
Our main requirement for giving semantics to the
$\eexists$ quantifier is the correctness of the following entailment, which we already used in \cref{ex:composition-projection}:
%\begin{align*}
    $(\EEX{e}\phi) \sep (\FFA{a}\psi) \sidmodels \EEX{e} \phi \sep \psi[a/e]$ \hfill \tagA %\tag*{\tagA}
%\end{align*}

This is ensured by the following semantics for $\eexists$:
\begin{multline*}
\SH \sidmodels \EEX{\tuple{e_1,\ldots,e_k}}\phi \text{ iff for all pairwise different locations}  \\
v_1,\ldots,v_k\in\dom(\H) \setminus \img(\S)  \text{ such that } \SHpair{\S \cup \set{e_1\mapsto v_1,\ldots,e_k\mapsto v_k}}{\H} \sidmodels \phi.
\end{multline*}
We call our quantifiers $\eexists$ and $\fforall$  \emph{guarded} because they exclude the instantiation of variables with certain locations.
We note that our quantifiers are \emph{not} dual, i.e., $\EEX{\vec{x}}{\phi}$ is \emph{not} equivalent to $\neg\FFA{\vec{x}}{\neg\phi}$.
However, we believe that our semantics is sufficiently motivated by our considerations on the soundness of the projection and the entailment \tagA.

\subsubsection{Finiteness}
\label{sec:towards:finiteness}
Did we solve Issues 1 and 2 from the second attempt?
Unfortunately not completely.
However, one additional restriction on unfolding forests will be sufficient to guarantee the finiteness of the abstraction.
We first explain the issue by means of an example:

\begin{example}
\label{ex:motivation-delimited}
  Let $\Sid$ be an SID that defines the list-segment predicate $\lseg$.
  Let $\SH \sidmodels \lseg(x,\nil)$ with $\size{\H}>n$. Then, there exists a forest $\frst$
  %$\frst = \{\ftree_1,\ldots,\ftree_n\}$
  such that $\H = \bigcup_{\ftree \in \frst} \theapof{\ftree}$ and whose projection is
  \begin{align*}
    \EEX{y_1,\ldots,y_n}& \lseg(y_n,\nil) \sep
    (\lseg(y_{n},\nil)\mw\lseg(y_{n-1},\nil)) \\&\sep \cdots \sep
    (\lseg(y_{2},\nil)\mw\lseg(y_{1},\nil)) \sep
    (\lseg(y_1,\nil)\mw\lseg(x,\nil))
  \end{align*}
  As there exist such models $\SH$ for arbitrary $n\in\N$,
  there are infinitely many (non-equivalent) formulas resulting from projections of unfolding forests.
\end{example}

\noindent
Fortunately, we do not need to consider all unfolding forests for deciding the satisfiability of the considered separation logic $\SLIDguarded$.
%We will make use of our observation from Section~\ref{ch:seplog} that for all guarded models the targets of dangling pointers are in the image of the stack (Lemma~\ref{lem:btw:pos-non-dangling}).
%
We recall that our goal is to define a compositional abstraction:
$$\abstproj(\S,\H_1)\Compose\abstproj(\S,\H_2)=\abstproj(\S,\H_1\stdunion\H_2)$$
Hence, we need to ensure that every unfolding tree of $\SHpair{\S}{\H_1\stdunion\H_2}$ can be composed via $\Compose$ from unfolding trees of $\SHpair{\S}{\H_1}$ and $\SHpair{\S}{\H_2}$.
In our approach we will have the guarantee that $\SHpair{\S}{\H_1}$ and $\SHpair{\S}{\H_2}$ are guarded (cf.~\cref{lem:types:homo:compose}).
With this in mind, let us consider an unfolding tree of $\SHpair{\S}{\H_1\stdunion\H_2}$ that is composed of some trees of $\SHpair{\S}{\H_1}$ and $\SHpair{\S}{\H_2}$, i.e., without loss of generality there is a pointer that is allocated in $\SHpair{\S}{\H_1}$ and points to a value in $\SHpair{\S}{\H_2}$.
Then, this pointer is dangling for the state $\SHpair{\S}{\H_1}$ and the target of this pointer is in the image of the stack.
Now, we recall from the definition of composition that the target of the pointer is a hole of an unfolding tree of $\SHpair{\S}{\H_1}$ and the root of an unfolding tree of $\SHpair{\S}{\H_2}$.
Thus, we can restrict our attention to unfolding trees whose roots and holes are in the image of the stack!
This motivates the following definition:
\begin{center}
  An unfolding tree $\ftree$ is $\S$-\emph{delimited}, if the root and holes of $\ftree$ are in the image of the stack $\S$.
\end{center}
Equipped with this definition (which we will formalize in \cref{def:types:frst:delimited}),
we restrict the abstraction function $\abstproj$ to forests of delimited unfolding trees.
This guarantees the finiteness of the abstraction:
The formulas resulting from the projection of such forests have the property that (1) all root parameters of predicate calls are free variables and every variable occurs at most once as a root parameter, and (2) all root parameters of predicates calls on the left-hand side of a magic wand are free variables and every variable occurs at most once as a root parameter.
Because the number of free variables is bounded, finiteness easily follows.
%\emph{Delimited $\Sid$-forests.}

\subsection{Summary of Overview}\label{sec:towards:summary}
To sum up, we propose abstracting the state $\SH$ in the following way:
\begin{enumerate}
\item We compute all $\S$-delimited $\Sid$-forests of $\SH$.
\item We project these forests onto formulas.
\item The abstraction of $\SH$ is the set of all these formulas; we call this set %of formulas 
    the \emph{type} of $\SH$.
\end{enumerate}
The resulting abstraction is
(1) finite (the set of types is finite),
(2) compositional (we have $\abstproj(\S,\H_1)\Compose\abstproj(\S,\H_2)=\abstproj(\S,\H_1\stdunion\H_2)$), and
(3) computable (we only need to apply rules for modus ponens and for manipulating quantifiers as illustrated in Example~\ref{ex:composition-projection}).
%\end{enumerate}

\paragraph*{Outline of the following sections.}
In the remainder of this article, we give the technical details for the material overviewed in this section.
In \cref{ch:forests}, we formalize $\Sid$-forests (\cref{sec:forests:forests-def}), their projections (\cref{sec:forests:projections}), and how to compose forest projections (\cref{sec:forests:composing}).
The type abstraction is introduced in \cref{ch:types}.
We discuss how the satisfiability problem for guarded $\SLgeneric$ formulas can be reduced to computing types in \cref{sec:types:understanding}.
We formalize $\S$-delimited forests in \cref{sec:types:delimited}
and discuss how types can be computed compositionally in \cref{sec:types:abstraction,sec:using-types:refinement}.
Finally, in \cref{ch:deciding-btw}, we present algorithms for computing the types of $\SLIDguarded$ formulas, summarize our overall decision procedure,
and discuss our decidability and complexity results.
%
%In \cref{ch:forests}, we formalize $\Sid$-forests and their projections.
%%%
%\Cref{ch:types} defines the \emph{$\Sid$-type} abstraction as the set of projections of all \emph{delimited $\Sid$-forests}.
%%%
%In \cref{ch:deciding-btw}, we show the \emph{refinement theorem} for the $\Sid$-type abstraction and $\SLIDguarded$ formulas, and develop a decision procedure for $\SLIDguarded$ based on computing $\Sid$-types.

%%% Local Variables:
%%% mode: latex
%%% TeX-master: "../Thesis"
%%% End:

\section{Forests and Their Projections}\label{ch:forests}

We now start formalizing the concepts that have been informally introduced
in \cref{ch:towards}: \emph{$\Sid$-forests}
(\cref{sec:forests:forests-def}), their projection onto formulas
(\cref{sec:forests:projections}), and how to compose them
(\cref{sec:forests:composing}).
%
%The \emph{$\Sid$-type} abstraction is introduced afterward in \cref{ch:types}. 
%Moreover, our decision procedures for $\SLIDguarded$ based on
%$\Sid$-types follow in \cref{ch:deciding-btw}.

\subsection{Forests}\label{sec:forests:forests-def}
Our main objects of study in this section are
\emph{$\Sid$-forests}
(\cref{def:types:forest}) made up of \emph{$\Sid$-trees}
(\cref{def:types:tree}). As motivated in \cref{ch:towards}, a
$\Sid$-tree encodes one fixed way to unfold a predicate call by means
of the rules of the SID $\Sid$.
The differences between the unfolding trees of
\citet{iosif2013tree,iosif2014deciding,jansen2017unified} and our
$\Sid$-trees are that (1) we instantiate variables with locations, and
(2) $\Sid$-trees can have \emph{holes}, i.e., we allow that one
or more of the predicate calls introduced (by means of recursive
rules) in the unfolding process remain folded.
%
%As explained in \cref{ch:towards}, we allow holes to obtain a
%compositional abstraction.
%

\subsubsection{Rule instances.}\index{rule instance}
We annotate every node of a $\Sid$-tree with a \emph{rule instance} of
the SID $\Sid$, i.e., a formula obtained from a rule of the
SID by instantiating both the formal arguments of the predicates and
the existentially quantified variables of the rule with locations:
\begin{align*}
  \RuleInst{\Sid} \defn \{&\pred(\sla) \Rule
    \pinst{\fa}{\svx \cdot \svy}{\sla \cdot \slb}
  \mid \;(\pred(\svx) \Rule \SHEX{\svy} \fa) \in \Sid, \\&\;
  \quad
  \sla\in\Loc^{\arity{\pred}}, \slb \in
  \Loc^{\size{\svy}}, \text{ and }%\\&\;
    \text{all (dis-)equalities in }
  \pinst{\fa}{\svx \cdot \svy}{\sla \cdot \slb}
  \text{ are valid}\}
\end{align*}
In the above definition, we refer only to those
(dis-)equalities that occur explicitly in the
formula, not those implied by recursive calls or by the separating conjunction.
Validity of these (dis-)equalities is straightforward to check because all variables have been instantiated with concrete locations.
Moreover, we remark that the null pointer $\nil$, which is a value but not a location, remains untouched.

The notion of a rule instance is motivated as follows: whenever $\SH \sidmodels \pred(\sla)$, there is at
least one rule instance
$(\pred(\sla) \Rule \fb) \in \RuleInst{\Sid}$ such that
$\SH \sidmodels \fb$.

\subsubsection{$\Sid$-trees.} We represent a
$\Sid$-tree\index{tree!$\Sid$-trees} as a partial function
%
%\[
$\gls{t-ftree}\colon \WitnessType$, %\]
where the set $\Loc$ of locations serves as the nodes of the tree; every
node is mapped to its successors in the (directed) tree and to
its label, a rule instance.
Moreover, for $\ftree$ to be a $\Sid$-tree, it must satisfy additional
consistency criteria.
To formalize these criteria, we fix some SID $\Sid$ and a node
\[\ftree(l) = \tuple{\sla, (\pred(\vec{v}) \Rule \ppto{a}{\vec{b}}
  \sep \pred_1(\vec{v_1}) \sep\cdots\sep\pred_m(\vec{v_m}) \sep \Pure
  )},\]
where $\Pure$ is a set of equalities and disequalities.
Notice that all rule instances are of the above form because---by our global assumptions in \cref{sec:sl-basics:sid-assumptions}---$\Sid$ satisfies the progress property.
We introduce the following shortcuts for the node at location $l$ to simplify working with $\Sid$-trees:
\begin{align*}
\gls{succt} ~\defn~ & \sla
\tag{locations corresponding to the successors of node $l$}\\
\gls{headt} ~\defn~ & \pred(\vec{v})
\tag{the predicate on the lhs of the rule instance} \\
\gls{heaptl} ~\defn~ & \set{a \mapsto \vec{b}}
\tag{the unique heap satisfying the points-to assertion in the rule instance} \\
\gls{callst} ~\defn~ & \set{\pred_1(\vec{v_1}),\ldots,\pred_m(\vec{v_m})}
\tag{the predicate calls in the rule instance} \\
\gls{ruleinstt} ~\defn~ & \pred(\vec{v}) \Rule \ppto{a}{\vec{b}}
\sep \pred_1(\vec{v_1}) \sep\cdots\sep\pred_m(\vec{v_m}) \sep \Pure
\tag{the rule instance}
\end{align*}
Moreover, we define the \emph{hole predicates}\index{hole predicate}
of $l$ as those predicate calls in $\tcalls{\ftree}{l}$ whose root
does not occur in $\tsucc{\ftree}{l}$; the
\emph{holes} of $l$ are the corresponding locations:
\begin{itemize}
\item
  $\gls{holepredst} \defn \{\pred'(\vec{z'}) \in \tcalls{\ftree}{l}
    \mid \forall c \in \tsucc{\ftree}{l} \ldotp \thead{\ftree}{c} \neq
    \pred'(\vec{z'})\}$, and
\item
  $\gls{holest} \defn \set{\proot{\pred'(\vec{z'})} \mid
    \pred'(\vec{z'}) \in \tholepreds{\ftree}{l}}$.
\end{itemize}
We lift some of the above definitions from individual locations and values to
entire trees $\ftree$:
\begin{align*}
    \gls{heapt} ~\defn~ & \bigcup_{c \in \dom(\ftree)} \locheapof{\ftree}{c}
    \tag{the heap satisfying exactly the points-to assertions in $\ftree$} \\
    \gls{ptrlocst} ~\defn~ & \bigcup_{\ppto{c}{\vec{d}} \in \theapof{\ftree}}\set{c}\cup\vec{d}
    \tag{all values that appear in points-to assertions in $\ftree$} \\
    \gls{allholepredst} ~\defn~ & \bigcup_{c \in \dom(\ftree)} \tholepreds{\ftree}{c}
    \tag{all hole predicates in $\ftree$} \\
    \gls{allholest} ~\defn~ & \bigcup_{l \in \dom(\ftree)} \tholes{\ftree}{l}
    \tag{all holes in $\ftree$}
\end{align*}
We denote by $\gls{grapht}$ the directed graph 
%corresponding to
%$\ftree$ that is 
induced by the successors of locations in $\ftree$.
That is,
\[\tgraph{\ftree} \defn \tuple{\dom(\ftree), \set{ \tuple{x,y} \mid x \in \dom(\ftree), y \in
      \tsucc{\ftree}{x}}}.\]
The \emph{height} of $\ftree$ is the length of the longest path in
the directed graph $\tgraph{\ftree}$.
\begin{definition}[$\Sid$-Tree]\label{def:types:tree}\index{Phi-tree@$\Sid$-tree}
  A partial function
  $\ftree\colon \WitnessType$ is a \emph{$\Sid$-tree} iff
  \begin{enumerate}
      \item $\Sid$ is in the fragment of SIDs of bounded treewidth, i.e., $\Sid\in\IDbtw$,
      \item $\tgraph{\ftree}$ is a directed tree, and
      \item $\ftree$ is \emph{$\Sid$-consistent}, i.e., for all locations $l \in \dom(\ftree)$, we have:
      \begin{itemize}
        \item $l$ is the single allocated location in its rule instance, i.e.,
              $\locheapof{\ftree}{l}=\set{l \mapsto \ldots}$,
        \item $l$ points to its successors in $\ftree$, i.e.,
              $\locheapof{\ftree}{l}=\set{l \mapsto \vec{b}}$ implies
              $\tsucc{\ftree}{l} \subseteq \vec{b}$, and
        \item the predicate calls associated with the successors
              $\tsucc{\ftree}{l} = \tuple{v_1,\ldots,v_k}$,
            of $l$ appear in the rule instance at location $l$, i.e.,
            $\{\thead{\ftree}{v_1}, \ldots,
              \thead{\ftree}{v_k}\} \subseteq \tcalls{\ftree}{l}$.
      \end{itemize}
  \end{enumerate}
  Since every $\Sid$-tree $\ftree$ is a directed tree, it has a root, which
    we denote by $\gls{root-t}$; the corresponding predicate call is
  $\trootpred{\ftree} \defn
  \thead{\ftree}{\troot{\ftree}}$.
  % We
  % call a tree $\ftree$ with $\size{\dom(\ftree)}=1$ a \emph{singleton
  %   tree}\index{singleton tree}.
\end{definition}

\begin{example}[$\Sid$-Tree]\label{ex:types:tree}
  \begin{enumerate}
      \item A $\Sid$-tree over the SID $\SidOddEven$ (cf. \cref{ex:sids}) is given by
    \[\ftree(l) \defn
      \begin{cases}
        \tuple{b, \even(l_1,a) \Rule \ppto{l_1}{b} \sep \odd(b, a)}&
        \text{if } l = l_1 \\
        \tuple{\emptyset, \odd(b,a) \Rule \ppto{b}{l_2} \sep \even(l_2, a)}& \text{if } l = b \\
        \bot & \text{otherwise.}
      \end{cases}
    \]
    Formally, $\ftree$ is defined over the locations
    $\dom(\ftree)=\set{l_1,b}$.
    Moreover, we have
    $\tsucc{\ftree}{l_1}=b$, $\thead{\ftree}{l_1}=\even(l_1,a)$,
    $\tcalls{\ftree}{l_1} = \set{\odd(b, a)}$,
    $\theapof{\ftree}=\set{l_1 \mapsto b, b \mapsto l_2}$,
    $\locheapof{\ftree}{l_1}=\set{l_1 \mapsto b}$,
    $\tptrlocs{\ftree}=\set{l_1,b,l_2}$,
    $\tallholes{\ftree}=\set{l_2}$, and
    $\tallholepreds{\ftree}=\set{\even(l_2,a)}$.
    \item All of the trees considered in \Cref{ch:towards} are $\Sid$-trees.
  \end{enumerate}
\end{example}
\noindent
We remark that the above definition of $\Sid$-trees does not account for rule instances in which the same predicate call appears multiple times. Similarly, we do not account for multiple predicate calls with the same root parameter.
As we will see in \cref{sec:types:delimited}, such cases do not need to be considered.
We can thus ignore these cases in favor of a simpler formalization.
Our main motivation for considering $\Sid$-trees is that they give a more structured view on models of predicate calls.
In particular, every such model corresponds to (at least one) $\Sid$-tree without holes:

\begin{lemma}\label{lem:model-of-pred-to-tree}
  %Let $\Sid \in \IDbtw$, 
  Let $\SH$ be a state 
  and $\pred \in \Preds{\Sid}$. %be a predicate.
  Then, $\SH \sidmodels \pred(z_1,\ldots,z_k)$ iff there exists a $\Sid$-tree $\ftree$ with
  $\trootpred{\ftree}=\pred(\S(z_1),\ldots,\S(z_k))$,
  $\tallholepreds{\ftree}=\emptyset$, and $\fheapof{\set{\ftree}}=\H$.
\end{lemma}
\begin{proof}
  The statement directly follows by induction on the number of rules applied to derive $\SH \sidmodels \pred(z_1,\ldots,z_k)$ resp. the height of the tree $\ftree$.
\end{proof}

\subsubsection{$\Sid$-Forests}
We combine zero or more $\Sid$-trees into \emph{$\Sid$-forests}.
\begin{definition}[$\Sid$-Forest]\label{def:types:forest}
  A \emph{$\Sid$-forest} $\frst$ %over some SID
  %$\Sid \in \IDbtw$ 
  is a finite set of
    $\Sid$-trees $\frst = \set{\ftree_1,\ldots,\ftree_k}$ with pairwise disjoint locations, i.e.,
  $\dom(\ftree_i) \cap \dom(\ftree_j) = \emptyset$ for $i \neq
  j$.
\end{definition}
\noindent
We assume that all definitions are lifted from $\Sid$-trees to
$\Sid$-forests, i.e., %  in the obvious way. 
%In particular, 
for %a forest
$\frst=\set{\ftree_1,\ldots,\ftree_k}$,
we define
\begin{itemize}
\item the \emph{induced heap of $\frst$}\index{induced heap!of
    Phi-forest@of $\Sid$-forest} as
  $\fheapof{\frst} \defn \bigcup_{\ftree \in \frst}
  \theapof{\ftree}$;
      if $l \in \dom(\ftree_i)$ then $\gls{ruleinstf}=\truleinst{\ftree_i}{l}$;
\item
  $\begin{aligned}[t]
    \gls{graphf} \defn \big\langle &\dom(\frst), \{ \tuple{x,y} \mid 1
      \leq i \leq k, x \in \dom(\ftree_i), y \in
      \tsucc{\ftree_i}{x}\} \big\rangle;~\text{and}
  \end{aligned}$
\item $\gls{dom-frst} \defn \bigcup_i \dom(\ftree_i)$;
%\item 
%\item 
      $\gls{roots-f} \defn \set{\troot{\ftree_i} \mid 1 \leq i \leq k}$;
%\item 
      $\gls{allholesf} \defn \bigcup_{1\leq i \leq k}\tallholes{\ftree_i}$.
% \item $\gls{graphf} \defn \big\langle {\dom(\frst), \{ \tuple{x,y} \mid
%       1 \leq i \leq k, x \in \dom(\ftree_i), y \in
%       \tsucc{\ftree_i}{x}}\} \big\rangle$;
 \end{itemize}

%We combine the roots and holes of a forest into its interface.

\begin{example}[$\Sid$-Forest]\label{ex:types:forest}
  Both
  \Cref{ex:treerp-forest} and %defines a $\Sid$-forest.
  \Cref{ex:lseg-projection} define a $\Sid$-forest.

\end{example}
\subsubsection{Composing Forests}\label{sec:forests-def:composing}

As motivated in \cref{sec:towards:compo2}, $\Sid$-forests are composed by (1) taking
their disjoint union and (2) optionally merging pairs of trees of
the resulting forest by identifying the root of one tree with a hole
of another tree.

\paragraph*{Disjoint union of forests.} The union of two
$\Sid$-forests corresponds to ordinary set union, provided no location
is in the domain of both forests; otherwise, it is undefined.

\begin{definition}[Union of
  $\Sid$-forests]\label{def:types:funion}
  \index{union!of forests}\index{forests!union of}
  Let $\frst_1, \frst_2$ be $\Sid$-forests. The \emph{union} of
  $\frst_1$, $\frst_2$ is given by
  \[ \frst_1\funion\frst_2 \defn
    \begin{cases}
      \frst_1\cup\frst_2&\text{if } \dom(\frst_1)\cap\dom(\frst_2)=\emptyset,\\
      \bot, & \text{otherwise.}
    \end{cases}
  \]
\end{definition}

\begin{lemma}\label{lem:forests:funion-heap}
  Let $\frst = \frst_1\funion\frst_2$. Then
  $\fheapof{\frst} = \fheapof{\frst_1} \stdunion \fheapof{\frst_2}$.
\end{lemma}
\begin{proof}
$\fheapof{\frst} = \bigcup_{\ftree \in \frst} \theapof{\ftree} =
    (\bigcup_{\ftree \in \frst_1} \theapof{\ftree}) \cup
    (\bigcup_{\ftree \in \frst_2} \theapof{\ftree}) =
    \fheapof{\frst_1} \stdunion \fheapof{\frst_2}$. (Where we have
    $\stdunion$ rather than $\cup$ because $\frst_1\funion\frst_2$ is
    defined.)
\end{proof}

\paragraph*{Splitting forests.}
%
% Our next goal is to formalize the process of merging $\Sid$-trees. We
% do so in a roundabout way:
We formalize the process of merging $\Sid$-trees in a roundabout way:
we first define a way to \emph{split} the
trees of a forest into sub-trees at a fixed set of locations---the
inverse of merging forests.
This may seem like an arbitrary choice, but will simplify the technical
development in follow-up sections.
We first consider two examples of splitting before
formalizing it in
\cref{def:forest:fsplit}.

\begin{example}[Splitting forests]
  \begin{enumerate}
  \item Let $\ftree$ be the $\Sid$-tree from \cref{ex:types:tree}.
    The $\set{b}$-split of $\set{\ftree}$ is given by
    $\set{\ftree_1,\ftree_2}$,
    for the trees
    %\begin{itemize}
    %\item 
            $\ftree_1=\set{l_1 \mapsto \tuple{\emptyset, \even(l_1,a) \Rule
          \ppto{l_1}{b} \sep \odd(b, a)}}$ and
    %\item 
            $\ftree_2=\set{b \mapsto \tuple{\emptyset, \odd(b,a) \Rule \ppto{b}{l_2} \sep \even(l_2, a)}}$.
    %\end{itemize}
    %In fact, 
    $\set{\ftree_1,\ftree_2}$ is the $\vec{l}$-split of
    $\set{\ftree}$ for all $\vec{l}\supseteq\set{b}$: in our
    definition of $\vec{l}$-split we will not require for the
    locations in $\vec{l}$ to actually occur in the forest.

  \item Recall the forest $\frst=\set{\ftree_1,\ftree_2,\ftree_3}$
    from \cref{ex:treerp-forest} and the tree $\ftree$ from
    \cref{ex:towards-sid-tree}.
    Then $\frst$ is the $\set{2,4}$-split of $\set{\ftree}$.
    Likewise, $\frst$ is the $\set{1,2,4,7}$-split of $\set{\ftree}$,
        because $1$ is %already 
          the root of a tree and $7$ does not occur
        in the forest.
        In contrast, $\frst$ is \emph{not} the $\set{1,2,5}$-split of
        $\set{\ftree}$, because
        $5 \in \dom(\frst) \setminus \froots{\frst}$.
      \end{enumerate}
    \end{example}
    \begin{definition}[$\vec{l}$-split]\label{def:forest:fsplit}\index{l-split@$\vec{l}$-split}
      Let $\frst,\barfrst$ be $\Sid$-forests and
      $\vec{l}\subseteq\Loc$.
      Then $\barfrst$ is an \emph{$\vec{l}$-split} of $\frst$
      if
      \begin{enumerate}
          \item both forests cover the same locations, i.e., $\dom(\frst)=\dom(\barfrst)$,
          \item both forests contain the same rule instances, i.e., $\fruleinst{\frst}{d}=\fruleinst{\barfrst}{d}$ for all
      $d\in\dom(\frst)$, and
          \item the graph of $\barfrst$ is obtained from the graph of $\frst$ by removing edges leading to locations in $\vec{l}$, i.e., $\fgraph{\barfrst}= \fgraph{\frst} \setminus \set{ (a,b) \mid a \in \Loc, b \in \vec{l}}$.
      \end{enumerate}
    \end{definition}

    \begin{lemma}[Uniqueness of $\vec{l}$-split]\label{lem:fsplit-unique}
        For all $\vec{l} \subseteq \Loc$, every $\Sid$-forest has a unique $\vec{l}$-split
        $\gls{splitfl}$.
    \end{lemma}
    \begin{proof} See \Cref{app:fsplit-unique}. \end{proof}
    \noindent
    To formalize how we merge trees, we define a derivation relation $\fderivestar$ between
    forests in which we iteratively split trees at suitable locations.
    Intuitively, $\frst_1 \fderivestar \frst_2$ holds if splitting the
    trees in $\frst_2$ at zero or more locations yields $\frst_1$; or,
    equivalently, if ``merging'' zero or more trees of $\frst_1$ yields
    $\frst_2$.

    \begin{definition}[Forest derivation]\index{forests!derivation between}
      The forest $\frst_2$ is
      \emph{one-step derivable} from the forest $\frst_1$, denoted
      $\frst_1 \gls{--derive-fderive} \frst_2$ iff there exists a location
      $l \in \dom(\frst)$ such that $\frst_1 = \fsplit{\frst_2}{\set{l}}$.

      The reflexive-transitive closure of $\fderive$ is denoted by $\fderivestar$.
      %if $\frst_1 \fderivestar \frst_2$, then we call $\frst_1$ \emph{derivable}
      %from $\frst_2$.
    \end{definition}

    \begin{example}\label{ex:types:fderivestar}
      Let $\bot$ denote the everywhere undefined partial function.
      Then, consider the forests
      $\frst \defn \{\ftree_1,\ftree_2\}$
      and $\barfrst \defn \set{\bartree}$
      given by the trees below.
      Then $\frst \fderive \barfrst$ because $\frst = \fsplit{\barfrst}{l_2}$.
    \begin{align*}
       \ftree_1 \defn \{& \tentry{l_1}{\bot}{\odd(l_1,l_{4})}{
                      \ppto{l_1}{l_2} \sep \even(l_2,l_{4})} \} 
                  \\
    \ftree_2 \defn \{& \tentry{l_2}{l_3}{\even(l_2,l_4)}{\ppto{l_2}{l_3}
      \sep \odd(l_3,l_4)},\\&
    \tentry{l_3}{\bot}{\odd(l_3,l_4)}{\ppto{l_3}{l_4}} \}\\
    \bartree \defn \{&
                    \tentry{l_1}{l_2}{\odd(l_1,l_{4})}{
                  \ppto{l_1}{l_2} \sep \even(l_2,l_{4})}, \\
                  &\tentry{l_2}{l_3}{\even(l_2,l_4)}{\ppto{l_2}{l_3}
      \sep \odd(l_3,l_4)},\\&
    \tentry{l_3}{\bot}{\odd(l_3,l_4)}{\ppto{l_3}{l_4}}\}.
\end{align*}
\end{example}
\noindent
We note that multiple steps of $\fderive$ correspond to splitting at multiple locations, because
 \[\fsplit{\frst}{\set{\la_1,\ldots,\la_k}} =
   \fsplit{\ldots\fsplit{\fsplit{\frst}{\set{\la_1}}}{\set{\la_2}}}{\ldots,\set{\la_k}}. \qedhere\]
\begin{lemma}\label{lem:fderivestar-is-split}
  $\frst_1 \fderivestar \frst_2$ iff there exists a set of locations $\vec{l}$ with
  $\frst_1 = \fsplit{\frst_2}{\vec{l}}$.
\end{lemma}
\noindent
Moreover, forests in the $\fderivestar$ relation describe the same states:
\begin{lemma}\label{lem:fderive:samemodel}
  Let $\frst$ be a $\Sid$-forest and $\barfrst \fderivestar
  \frst$. Then $\fheapof{\barfrst}=\fheapof{\frst}$.
\end{lemma}
\begin{proof}
  Since $\barfrst \fderivestar \frst$, there exists---by
  \cref{lem:fderivestar-is-split}---a set of locations $\vec{l}$ with
  $\barfrst=\fsplit{\frst}{\vec{l}}$.
  By definition of $\vec{l}$-splits, we have (1)
  $\dom(\barfrst) = \dom(\frst)$ and (2)
  $\fruleinst{\barfrst}{\la}=\fruleinst{\frst}{\la}$
  for every location $\la \in \dom(\barfrst)$.
  Consequently, $\fheapof{\barfrst}=\fheapof{\frst}$.
\end{proof}
\noindent
Based on the $\fderivestar$, we define the \emph{composition} operation on pairs of forests
%that we motivated informally in \cref{sec:towards:snd-attempt}:
as motivated in \cref{sec:towards:snd-attempt}:
\begin{definition}[Forest composition]\label{def:forest-composition}
  %Let $\frst_1,\frst_2$ be $\Sid$-forests.
 % with $\frst_1\funion\frst_2\neq \bot$.
%
The \emph{composition} of $\frst_1$ and $\frst_2$ is %given by
$\frst_1 \gls{--Compose-FCompose} \frst_2 \defn \set{ \frst \mid
  \frst_1 \funion \frst_2 \fderivestar \frst }$.
%
% \TODO{Isn't it okay that this is $\emptyset$ if
%   $\frst_1\funion\frst_2=\bot$? Then we wouldn't need a case
%   distinction.}
\end{definition}

%%% Local Variables:
%%% mode: latex
%%% TeX-master: "../Thesis"
%%% End:

\subsection{Forest Projections}\label{sec:forests:projections}
In \cref{sec:towards:projection}, we informally presented the \emph{projection} of $\Sid$-forests onto $\SLIDguarded$ formulas, and discussed the need for using \emph{guarded quantifiers}.
As a reminder, we repeat here the informal definition of the projection:
Given a stack $\S$ and a $\Sid$-forest $\frst=\ktrees$,
\begin{enumerate}
\item we compute the formula
$\phi \defn \IteratedStar_{1 \leq i \leq k}\left(\IteratedStar\tallholepreds{\ftree_i}\right) \mw
  \trootpred{\ftree_i}$, in which all parameters of all predicate calls are locations;

\item we replace in $\phi$ every %location 
      $v\in\img(\S)$ by an arbitrary but fixed variable $x$ with
      %for which 
      $\S(x)=v$ holds;
\item we replace every location $v\in\dom(\ftree)\setminus\img(\S)$ by a \emph{guarded existential};
\item we replace every other location by a \emph{guarded universal}.
\end{enumerate}
\noindent
We now make these definitions precise.
First, we introduce the \emph{projection} of trees and forests (\cref{sec:projection:forests}).
Then, we state the definition of \emph{guarded quantifiers} (in
\cref{sec:projection:guarded});
Finally, we introduce the \emph{stack-projection} (in
\cref{sec:projection:stack}).

\subsubsection{Tree and Forest Projections}\label{sec:projection:forests}

We are now ready to define the forest projection outlined in \cref{sec:towards:projection}.
We begin with defining the projection of a tree:

\begin{definition}[Projection of a Tree]\label{def:types:t-projection}\index{tree projection}
  The \emph{projection} $\gls{projectLoct}$
  of a $\Sid$-tree $\ftree$ is given by
  \begin{displaymath}
    \ltproj{\vec{v}}{\ftree} \defn
                     \left(\IteratedStar\tallholepreds{\ftree}\right) \mw \trootpred{\ftree}.
  \end{displaymath}
\end{definition}

%\begin{definition}[Tree projection]\label{def:types:t-projection}\index{tree projection}
%  The \emph{tree projection} $\gls{projectLoct}$
%  of a $\Sid$-tree $\ftree$ is given by
%  \begin{align*}
%    \ltproj{\vec{v}}{\ftree} \defn&
%                     \pinst{\big(\underbrace{\left(\IteratedStar\tallholepreds{\ftree}\right) \mw \trootpred{\ftree}\big)}_{~\defn~\psi}}{\vec{w}}{\vec{a}},
%  \end{align*}
%  where $\vec{w} \defn \locs{\psi} \setminus (\dom(\theapof{\ftree}) \cup \vec{v})$ collects all locations that are neither allocated in $\ftree$ nor ``blocked'' by $\vec{v}$, and where $\vec{a} \defn \tuple{a_{1},\ldots,a_{\size{\vec{w}}}}$ contains a fresh variable for each location in $\vec{w}$.
%\end{definition}

\begin{example}
  Recall from \cref{ex:types:tree} the $\Sid$-tree $\ftree$ over an SID describing lists of even and odd length.
  This tree admits the tree projection
  %\begin{displaymath}
    $\ltproj{\set{l_1,l_2,a}}{\ftree} = \even(l_2, a)
    \mw \even(l_1, a)$.
  %\end{displaymath}
%
%  \begin{enumerate}
%  \item $\ltproj{\emptyset}{\ftree} = \FFA{a_1} \even(l_2, a_1)
%    \mw \even(l_1, a_1)$.
%  \item $\ltproj{\set{l_1,l_2}}{\ftree} = \FFA{a_1} \even(l_2, a_1)
%    \mw \even(l_1, a_1)$.
%  \item $\ltproj{\set{l_1,l_2,a}}{\ftree} = \even(l_2, a)
%    \mw \even(l_1, a)$.
%  \end{enumerate}
\end{example}
\noindent
Tree projections are sound in the sense that the induced heap of a tree satisfies its tree projection.
To prove this result, we need the following variant of modus ponens (cf.~\cite{reynolds2002separation}):
\begin{lemma}[Generalized modus
  ponens]\label{lem:gmp}\index{generalized modus ponens}
  \index{modus ponens}
  \begin{align*}
    ((\pred_2(\vec{x_2}) \sep \psi) \mw \pred_1(\vec{x_1})) \sep
    (\psi' \mw \pred_2(\vec{x_2})) \quad\text{implies}\quad (\psi \sep \psi')
    \mw \pred_1(\vec{x_1}).
  \end{align*}
\end{lemma}
\begin{lemma}[Soundness of Tree Projections]\label{lem:t-projection-sound}
  %Let $\Sid\in\IDbtw$ and
  Let $\ftree$ be a $\Sid$-tree- with $\fheapof{\ftree} = \H$.
  Then, $\SHpair{\_}{\fheapof{\ftree}} \sidmodels \ltproj{\img(\S)}{\ftree}$ (where $\_$ denotes an arbitrary stack).
\end{lemma}
\begin{proof}
  By mathematical induction on the height of $\ftree$;
  see \cref{app:t-projection-sound} for details.
\end{proof}

\subsubsection{Guarded Quantifiers}\label{sec:projection:guarded}
As motivated in \cref{sec:towards:guarded-quantifiers}, we introduce \emph{guarded} quantifiers,
%versions of both existential and universal quantifiers,
which we denote by \gls{--zzexists} and \gls{--zzall}, respectively.
Specifically, we consider formulas %of the form
%\[
  $\EEX{\vec{e}} (\FFA{\vec{a}} (\phiQf \sep \cdots                \sep \phiQf))$,
%\]
%
where $\phiQf$ denotes \emph{quantifier-free} $\SLgeneric$ formulas (cf.~\cref{sec:sl-basics:syntax}). %over SIDs of bounded treewidth (cf.~\cref{sec:sl-basics:btw}).
We collect all formulas of the above form in the set
$\gls{SLIDea}$.
%$\SLIDea$ formulas without guarded existentials are collected in $\gls{SLIDa}$.
%
Our guarded quantifiers have the following semantics:
\begin{itemize}
\item $\SH \sidmodels \EEX{\tuple{e_1,\ldots,e_k}}\phi$ iff there
  exist pairwise different locations
  \[
    \qquad\quad v_1,\ldots,v_k\in\dom(\H) \setminus \img(\S)
    \quad\text{such that}\quad
    \SHpair{\S \cup \set{e_1\mapsto v_1,\ldots,e_k\mapsto v_k}}{\H}
  \sidmodels \phi.\]
\item $\SH \sidmodels \FFA{\tuple{a_1,\ldots,a_k}}\phi$ iff for all pairwise different locations
 \[ \qquad\quad v_1,\ldots,v_k\in\Loc\setminus(\dom(\H) \cup \img(\S)),
    \quad\text{we have}\quad
    \SHpair{\S \cup \set{a_1\mapsto v_1,\ldots,a_k\mapsto v_k}}{\H}
  \sidmodels \phi.\]
\end{itemize}
Notice that our guarded quantifiers differ from the standard ones in three aspects:
first, guarded quantifiers cannot be
instantiated with locations that are already in the stack.
Second, we require that the quantified locations are pairwise different.
Third, our quantifiers are \emph{not} dual, i.e.,
$\EEX{\vec{x}}{\phi}$ is \emph{not} equivalent to
$\neg\FFA{\vec{x}}{\neg\phi}$.
%
% In particular, $\eexists$ \emph{cannot} be instantiated with a
% location $v \in \img(\H)\setminus\dom(\H)$.
%
%In particular, $\fforall$ ranges over \emph{all} locations that are not in
%$\locs{\H}$, whereas $\eexists$ ranges only over \emph{some} locations
%that are in $\locs{\H}$, namely $\dom(\H)$.
%
For guarded states, location terms in a formula can be replaced by a guarded
universal quantifier as long as they do not appear in the state in question:
\begin{lemma}\label{lem:guarded-universal-intro}
  Let %$\Sid\in\IDbtw$ and
  $\SH \in \Mpos{\Sid}$
  and $\fa$ be a quantifier free $\SLgeneric$ formula with
  $\SH \sidmodels \fa$.
  Moreover, let
  $\vec{v} \in {(\Loc \setminus (\dom(\H)\cup\img(\S)))}^{*}$ be
  a repetition-free sequence of locations.
  Then, for every set
  $\vec{a}\defn\set{a_1,\ldots,a_{\size{\vec{v}}}}$
  of fresh variables (i.e., $\vec{a}\cap\dom(\S)=\emptyset$),
  we have
  $\SH \sidmodels \FFA{\vec{a}} \pinst{\fa}{\vec{v}}{\vec{a}}$.
\end{lemma}
\begin{proof}
  See \Cref{app:guarded-universal-intro}.
\end{proof}
%%\noindent
%%The semi-distributivity of quantifiers in separation
%%logic\footnote{See, for example,~\cite{reynolds2002separation}.} also
%%holds for our guarded quantifiers:
%%%
%%\begin{lemma}\label{lem:guarded-semi-distributive}
%%  For all $\SLgeneric$ formulas $\phi,\psi$
%%    with $\fvs{\psi}\cap\vec{z}=\emptyset$
%%    and $\mathbb{Q}\in \set{\eexists,\fforall}$, we have
%%  \[
%%    \SH \sidmodels (\mathbb{Q}\vec{z}\ldotp \phi) \sep \psi
%%    \quad\text{implies}\quad
%%    \SH \sidmodels \mathbb{Q}\vec{z}\ldotp (\phi \sep \psi).
%%  \]
%%%  Let $\phi,\psi$ be formulas with
%%%    $\fvs{\psi}\cap\vec{z}=\emptyset$.
%%%  Let $\mathbb{Q}\in \set{\eexists,\fforall}$ and
%%%  $\SH \sidmodels (\mathbb{Q}\vec{z}\ldotp \phi) \sep \psi$. Then
%%%  $\SH \sidmodels \mathbb{Q}\vec{z}\ldotp (\phi \sep \psi)$.
%%\end{lemma}
%%%
%%\begin{proof}
%%  See \cref{app:guarded-semi-distributive}.
%%\end{proof}
%%%
%%\noindent
%%Notice that, in contrast to the ordinary quantifiers, the above implication is \emph{not} an equivalence:
%%moving any guarded quantifier out of a separating conjunction results in a
%%\emph{strictly} weaker formula. That is,
%%$\EEX{\vec{z}} (\phi \sep \psi)$ does not entail $(\EEX{\vec{z}} \phi) \sep
%%\psi$ whereas
%%%. % even if $\vec{z}\cap\fvs{\psi}=\emptyset$.
%%%
%%%In contrast,
%%$\EXO{\vec{z}} (\phi \sep \psi)$ entails $(\EXO{\vec{z}} \phi) \sep
%%\psi$ if $\vec{z}\cap\fvs{\psi}=\emptyset$.

% whereas this
% entailment is valid for unguarded existentials.

\noindent
Many standard equivalences of separation logic continue to hold for formulas with guarded quantifiers;
we list corresponding \emph{rewriting rules}
in \cref{fig:sl:rewreq}.
These rules establish the \emph{rewriting equivalence} $\equiv$, which preserves logical equivalence---we will only consider formulas up to $\equiv$.
\begin{figure}[tb!]
  \centering
    \begin{tabular}{c}
    \AxiomC{$\phi_1 \rewreq \phi_2$}
    %\AxiomC{$\oplus \in \set{\sep,\mw}$}
    \RightLabel{\RuleMono}
    \UnaryInfC{$\phi_1 \sep \psi \rewreq \phi_2 \sep \psi$}
    \DisplayProof{} 
    \AxiomC{}
    \RightLabel{\RuleEmp}
\UnaryInfC{$\phi_1 \sep \emp \rewreq \phi_1$}
                      \DisplayProof{}
    %%%
    %%%
    %%%
    \AxiomC{$\phi_1 \rewreq \phi_2$}
   \RightLabel{\RuleAnti}
   \UnaryInfC{$\phi_2 \mw \psi \rewreq \phi_1 \mw \psi$}
                      \DisplayProof{}
    \AxiomC{}
    \RightLabel{\RuleId}
    \UnaryInfC{$\phi_1 \rewreq \phi_1$}
                      \DisplayProof{} 
    \\[14pt]
    \AxiomC{}
    \RightLabel{\RuleAssoc}
    \UnaryInfC{$\phi_1 \sep (\phi_2 \sep \phi_3) \rewreq (\phi_1 \sep \phi_2) \sep \phi_3$}
      \DisplayProof{}
    \AxiomC{}
    \RightLabel{\RuleComm}
    \UnaryInfC{$\phi_1 \sep \phi_2 \rewreq \phi_2 \sep \phi_1$}
      \DisplayProof{}
    \AxiomC{$\phi_1\rewreq \phi_2$}
    \RightLabel{\RuleSym}
                      \UnaryInfC{$\phi_2\rewreq \phi_1$}
    \DisplayProof{} 
    %%%
    \\[14pt]
    %%%
    %%%
          \AxiomC{$\mathbb{Q} \in \set{\fforall,\eexists}$}
    \AxiomC{$z \notin \allvs{\phi}$}
                      \RightLabel{\RuleRen}
                      \BinaryInfC{$\mathbb{Q} y\ldotp\phi \rewreq \mathbb{Q}
                                        z\ldotp \pinst{\phi}{y}{z}$}
                                        \DisplayProof{} 
    \AxiomC{$\phi_1 \rewreq \phi_2$}
    \RightLabel{\RuleEIntro}
    \UnaryInfC{$\EEX{\vec{y}}\phi_1 \rewreq \EEX{\vec{y}}\phi_2$}
    \DisplayProof{}
    \AxiomC{$\phi_1 \rewreq \phi_2$}
    \RightLabel{\RuleAIntro}
    \UnaryInfC{$\FFA{\vec{y}}\phi_1 \rewreq \FFA{\vec{y}}\phi_2$}
    \DisplayProof{}
    \\[14pt]
        \AxiomC{$\phi_1 \rewreq \phi_3$}
                      \AxiomC{$\phi_3 \rewreq \phi_2$}
                      \RightLabel{\RuleTrans}
                      \BinaryInfC{$\phi_1 \rewreq \phi_2$}
    \DisplayProof{}
    %%%
    %%%
    %%%
    \AxiomC{$\mathbb{Q} \in \set{\fforall,\eexists}$}
    \AxiomC{$z \notin \allvs{\phi}$}
                      \RightLabel{\RuleDrop}
                      \BinaryInfC{$\mathbb{Q} z\ldotp\phi \rewreq \phi$}
                                        \DisplayProof{}
    %%%
  \end{tabular}
%%% Local Variables:
%%% mode: latex
%%% TeX-master: "../../Thesis"
%%% End:

  \caption[Rewriting rules for $\SLIDea$]{A set of rules for rewriting $\SLIDea$ formulas into equivalent formulas.
  }
  \label{fig:sl:rewreq}
\end{figure}

\begin{lemma}[Soundness of rewriting equivalence]\label{lem:rewreq-sound}
  If $\fa_1 \rewreq \fa_2$ then $\fa_1 \sidmodels \fa_2$.
\end{lemma}
% \begin{proof}
%   All rules are standard, see e.g.~\cite{reynolds2002separation}.
%   %
%   \TODO{Not true now that we have the nonstandard quantifiers. Add
%     proper proof?}
% \end{proof}

\subsubsection{Stack-Projection}
\label{sec:projection:stack}
We now abstract from locations in projections (cf. \cref{sec:towards:replacing}),
replacing every location $l$ in the projection of a forest $\frst$ by a variable:
a stack variable,
if $l$ is in the image of the stack,
an existentially-quantified
one if $l \in \dom(\frst)$, and a universally-quantified one otherwise.
%and a universally-quantified variable otherwise.

\paragraph{Aliasing and Variable Order.}
In case of aliasing, i.e., if there are multiple variables that are mapped to the same location $l$, there are multiple choices for replacing $l$ by a stack variable $x$ with $\S(x) = l$.
This has the consequence that the projection would not be unique.
In order to avoid this problem, we assume an arbitrary, but fixed, linear ordering of the variables $\Var$.
We then choose the variable among all the aliases of a variable that is maximal according to this variable ordering.
Formally:
%We formally state our choice function below:

\begin{definition}[\stkchoicefun{} $\stkchc$]
\label{def:sid:stack-inverse}
  Let $\S$ be a stack.
  Then, the \stkchoicefun{} of $\S$ maps a location $l \in \img(\S)$ to
  $\gls{s-1-max}(l) = \max \{ x\in\dom(\S) \mid \S(x)=l \}$.
%
%  \begin{enumerate}
%   \item The \emph{stack inverse} of $\S$ is the partial function given by $\gls{s-1} \defn \set{ v \mapsto \set{x \mid \S(x)=v} \mid v \in \img(\S)}$.
%    %
%  \item The \emph{\stkchoicefun{}} of $\S$ is given by
%      $\gls{s-1-max} \defn \set{ v \mapsto \max(\stkinv(v)) \mid v \in \img(\S)}$.
%  \end{enumerate}
\end{definition}

\paragraph{Quantified Variables.}
We (mostly) maintain the convention that we denote (guarded) universally resp. existentially quantified variables by $a_1,a_2,\ldots$ resp. $e_1,e_2,\ldots$.
We will always assume that $\set{a_{1},a_{2},\ldots}\cap\set{e_1,e_2,\ldots} = \emptyset$ and that $\dom(\S) \cap
(\set{a_{1},a_{2},\ldots}\cup\set{e_1,e_2,\ldots})
= \emptyset$ for any stack $\S$.
%
%We also need the \emph{\stkchoicefun{}}, $\stkchc$, defined in \cref{def:sid:stack-inverse} on page~\pageref{def:sid:stack-inverse}.
%
We are now ready to give the main definition of this subsection:

\begin{definition}[Stack-projection]\label{def:types:sf-projection}
  Let $\frst=\ktrees$ be a $\Sid$-forest, %and let 
  $\S$ be a stack, and
  %Further,
  \begin{itemize}
    \item let $\phi = \IteratedStar_{1 \leq i \leq k}\; \ltproj{\vec{v}}{\ftree_i}$ be the projection of trees of $\frst$ conjoined by $\sep$,
    \item let $\vec{w} = \locs{\phi} \cap (\dom(\frst) \setminus \img(\S))$ be some  (arbitrarily ordered) sequence of locations that occur in the formula $\phi$ and are allocated in $\fheapof{\frst}$ but are not the value of any stack variable,
    \item and let $\vec{v} = \locs{\phi} \setminus (\img(\S) \cup \dom(\frst))$ be some  (arbitrarily ordered) sequence of locations that occur in the formula $\phi$ and are neither allocated nor the value of any stack variable.
    \end{itemize}
  Then, we define the \emph{stack-projection} of $\S$ and $\frst$ as
  \begin{displaymath}
      \gls{project-Sf} ~\defn~ \EEX{\vec{e}} \FFA{\vec{a}} \pinst{\phi}{\dom(\stkchc)\concat\vec{v}\concat\vec{w}}{\img(\stkchc)\concat\vec{a}\concat\vec{e}},
  \end{displaymath}
  where $\vec{e} \defn \tuple{e_1,e_2,\ldots,e_{\size{\vec{w}}}}$ and $\vec{a} \defn \tuple{a_1,a_2,\ldots,a_{\size{\vec{v}}}}$ denote disjoint sets of fresh variables.
\end{definition}
\noindent
%We note that 
The stack-projection is well-defined because $\dom(\stkchc)$, $\vec{w}$ and $\vec{v}$ form a partitioning of $\locs{\phi}$. Furthermore, the null pointer $\nil$ is not a location; it thus remains untouched.
%\fztodo{should be values($\phi$) here, right?
%check that in the proofs values vs locs is correct.
%adapt explanation here and earlier that 0 is left as location in the formulas.}
Notice that the stack-projection is \emph{unique} (w.r.t. the rewriting equivalence $\rewreq$ defined in \cref{fig:sl:rewreq}):
while the stack-projection involves picking an (arbitrary) order on the trees $\ftree_1,\ldots,\ftree_k$ and a choice of the fresh variables $\vec{e}$ and $\vec{a}$, this does not matter because of the commutativity and associativity of $\sep$ and the possibility to rename quantified variables, which is allowed for by the rules of the rewriting equivalence $\rewreq$.

%
%Further, I am aware that the notation
%$\pinst{\phi}{\dom(\stkchc)\concat\vec{w}}{\img(\stkchc)\concat\vec{e}}$
%is not fully formal, as $\dom(\stkchc)$ and $\img(\stkchc)$ are sets,
%not sequences. Just like in earlier chapters, I assume that a suitable
%order is imposed on these sets to instantiate every location
%$v\in\dom(\stkchc)$ with the variable $\stkchc(v)$.
%Intuitively, the stack-projection replaces all locations
%\todo{Use
% labeled location terminology in this part of the thesis?}~
%in the image of the stack with stack variables;
%all other locations are replaced by existentially-quantified variables (if they are in $\fheapof{\frst}$) or by universally-quantified variables (otherwise).
%Consequently, the resulting formula no longer contains any locations.

%

%%We use guarded existential quantifiers instead of standard ones (i.e., $\exists$)
%%to ensure compatibility with guarded universal quantifiers.
%%That is, while the implication
%%\[
%%  (\FFA{a} \phi) \sep (\EEX{e} \psi) \implies \EEX{e}
%%  \pinst{\phi}{a}{e} \sep \psi
%%\]
%%is valid, it does \emph{not} hold when using a standard existential quantifier instead:
%%\[
%%  (\FFA{a} \phi) \sep (\EXO{e} \psi) \centernot\implies \EXO{e}
%%  \pinst{\phi}{a}{e} \sep \psi.
%%\]
%%We will rely on this kind of reasoning when \emph{composing} projections later on (cf.~\cref{def:rescope}).
%
\begin{example}[Stack-projection]\label{ex:sf-projection}
  We consider three examples of stack-projections:
  \begin{enumerate}
  \item Let $\ftree$ be the $\Sid$-tree from
    \cref{ex:types:tree}.
    Then, for $\frst = \set{\ftree}$ and $\S = \set{x_1 \mapsto l_1, x_2 \mapsto l_2}$,
    we have
    \begin{align*}
    \fheapof{\frst} = \set{l_1 \mapsto l_2}
    \quad\text{and}\quad
    \lfproj{\img(\S)}{\frst} = \FFA{a_1} \odd(l_2,a_1) \mw
    \even(l_1,a_1).
    \end{align*}
    As all locations in this formula are in the image of the stack, we
    have
    \begin{align*}
    \sfproj{\S}{\frst} = %&=
    \pinst{\lfproj{\set{l_1,l_2}}{\frst}}{\dom(\stkchc)}{\img(\stkchc)} %\\
    %&= 
    = \FFA{a_1} \odd(x_2,a_1) \mw \even(x_1,a_1).
    \end{align*}
  \item Let $\SH$ be the model and let $\ftree_1,\ftree_2,\ftree_3$ be the $\Sid$-trees from \cref{ex:cyclic-projection}.
    Then,
    \begin{align*}
        \sfproj{\S}{\set{\ftree_1,\ftree_3}} &= \EEX{a} (\lseg(y,a) \mw \topCyclic(x,y,z)) \sep \lseg(z,a), ~\text{and} \\
     %   & \text{and}\\
        \sfproj{\S}{\set{\ftree_2}} &= \FFA{a'} \lseg(z,a') \mw \lseg(y,a').
    \end{align*}

  \item Let $\ftree_1,\ftree_2,\ftree_3,\ftree_4$ be the $\Sid$-trees from \cref{ex:treerp-forest} for the state $\SH$ of \cref{ex:towards-sid-tree}.
      Then,
    \begin{align*}
        \sfproj{\S}{\set{\ftree_1,\ftree_3,\ftree_4}} &= \EEX{r} (\tll(a,b,c) \mw \tll(x,y,z)) \sep \ppto{b}{\tuple{0,0,c}} \sep \ppto{c}{\tuple{0,0,r}},~\text{and}\\
        %& \text{and}\\
        \sfproj{\S}{\set{\ftree_2}} &= \FFA{r'} (\ppto{b}{\tuple{0,0,c}} \sep \ppto{c}{\tuple{0,0,r'}}) \mw \tll(a,b,c).
    \end{align*}

\end{enumerate}
\end{example}
\noindent
In each of the above examples,
we observe that
$\SHpair{\S}{\fheapof{\frst}} \sidmodels \sfproj{\S}{\frst}$.
This is not a coincidence as eliminating locations preserves the soundness of forest projections:
\begin{lemma}[Soundness of stack-projection]\label{lem:sf-projection-sound}
  Let %$\Sid\in\IDbtw$ and
  $\SH \in \Mpos{\Sid}$.
  Moreover, let $\frst$ be a $\Sid$-forest with $\fheapof{\frst} = \H$.
  Then, we have $\SH \sidmodels \sfproj{\S}{\frst}$.
\end{lemma}
\begin{proof}
   See \Cref{app:sf-projection-sound}.
\end{proof}

\begin{example}[Why we need guarded quantifiers]
  We now have the machinery available
  to discuss why guarded quantifiers are needed.
  To this end, let us revisit the motivating example
  in \cref{sec:towards:guarded-quantifiers}:
  We consider the state $\SH$, given by the stack $\S = \set{x \mapsto 1}$ and the heap $\H = \set{1 \mapsto 2, \allowbreak 2 \mapsto 3}$, and the SID $\Sid$ given by the following predicates:
\[
    \begin{array}{lllllll}
    p(x,a,b)& \Rule & \SHEX{y}\ppto{x}{y} \sep q(y,a) \sep x \neq a \sep a \neq b
    &\quad&
    q(y,a) & \Rule & \ppto{y}{\nil} \sep y \neq z
    \end{array}
\]
We further consider the unfolding tree $\ftree$ consisting of the rule instance $p(1,4,5) \Rule \ppto{1}{2}\sep 1\neq4 \sep 4\neq5$ at the root with a single child for the rule instance $q(2,4) \Rule \ppto{2}{3} \sep 2\neq4$.
Hence, we have $\trootpred{\ftree} = p(1,4,5)$, $\tallholepreds{\ftree} = \emptyset$, and $\locs{p(1,4,5)} \setminus (\dom(\H) \cup \img(\S)) = \set{4,5}$.
By \cref{def:types:t-projection}, we then obtain the tree projection
\begin{align*}
    \sfproj{\S}{\ftree} ~=~ &
    \FFA{a,b} \pinst{
        \big((\IteratedStar \tallholepreds{p(1,a,b)}) \mw \trootpred{\ftree}\big)
    }{
        \tuple{1,4,5}
    }{
        \tuple{x,a,b}
    }
    \\
    ~=~ &
    \FFA{a,b} \pinst{
        \emp \mw p(x,4,5)
    }{
        \tuple{4,5}
    }{
        \tuple{a,b}
    } 
    ~=~ 
    \FFA{a,b} p(x,a,b).
\end{align*}
By \cref{lem:t-projection-sound}, we have $\SH \sidmodels \FFA{a,b} p(x,a,b)$.
In particular, the semantics of $\fforall$ guarantees that $a$ and $b$ refer to distinct locations that are not allocated and that are not the value of any stack variable.
This is crucial to ensure soundness of the stack-projection:
if we would use a standard universal quantifier instead of a guarded one, $\SH$ would not be a model $\ltproj{\img(\S)}{\ftree}$ as neither $\SH \sidmodels p(x,1,5)$, $\SH \sidmodels p(x,2,5)$ nor
$\SH \sidmodels p(x,5,5)$ holds.
\end{example}

\subsection{Composing Projections}\label{sec:forests:composing}
\subsubsection{Motivation}\label{sec:composing:motivation}
Recall from \cref{sec:towards:compositionality} that
our goal is the definition of a \emph{composition} operator for the projections of forests.
This operation should collect exactly those projections
of forests $\frst \in \frst_1 \FCompose \frst_2$ (see \cref{def:forest-composition})
that can be derived from $\sfproj{\S}{\frst_1}$ and
$\sfproj{\S}{\frst_2}$, i.e.,
\begin{align*}
  \sfproj{\S}{\frst_1}\PCompose\sfproj{\S}{\frst_2} \overset{?}{=}
  \set{\sfproj{\S}{\frst} \mid \frst \in \frst_1 \FCompose \frst_2}.
\end{align*}
Put differently, we are looking for an operation $\PCompose$ such that $\sfproj{\S}{\cdot}$ is a homomorphism from the set of $\Sid$-forests and $\FCompose$ to the set of
% unfolded symbolic heaps
projections and $\PCompose$.

How can we define such an operation $\PCompose$?
Intuitively, we need to conjoin the projections via $\sep$ in order to simulate the operation $\frst_1\funion\frst_2$, and apply the generalized modus ponens rule (see~\cref{lem:gmp}) in order to simulate the operation $\fderive$ on trees.
There is, however, one complication:
our forest projections contain quantifiers.
In particular,
$\sfproj{\S}{\frst_1} \sep \sfproj{\S}{\frst_2}$ is of the form
$(\EEX{\vec{e_1}} \FFA{\vec{a_1}}  \phi_1) \sep (\EEX{\vec{e_2}} \FFA{\vec{a_2}} \phi_2)$,
whereas $\sfproj{\S}{\frst_1\funion\frst_2}$ is of the form
$\EEX{\vec{e}} \FFA{\vec{a}} \phi$, where $\phi_1$, $\phi_2$, and $\phi$ do not contain guarded
quantifiers.
In other words, $\PCompose$ has to push the guarded quantifiers to the front before the modus ponens rule can be applied.

\subsubsection{Definition of the Composition Operation}
We will define $\PCompose$ in terms of two operations:
(1) An operator $\rescopeOP$ that captures all sound ways to move the guarded quantifiers to the front of the formula $\sfproj{\S}{\frst_1}\sep\sfproj{\S}{\frst_2}$ (i.e., ``re-scopes the guarded quantifiers'').
(2) A derivation operator $\deriveqf$ that rewrites formulas based on the generalized modus ponens rule (\cref{lem:gmp}).

\begin{definition}[Re-scoping]
We say $\chi$ is a \emph{re-scoping} of $\EEX{\vec{e_1}} \FFA{\vec{a_1}} \phi_1$ and $\EEX{\vec{e_2}} \FFA{\vec{a_2}} \phi_2$, in signs $\chi \in \left(\EEX{\vec{e_1}} \FFA{\vec{a_1}} \phi_1\right) \rescopeOP \left(\EEX{\vec{e_2}} \FFA{\vec{a_2}} \phi_2\right)$,
if there are repetition-free sequences of variables $\vec{a}$, $\vec{d_i}$ and $\vec{u_i} \subseteq \vec{a} \cup \vec{d_{3-i}}$,
%(of not necessarily pairwise different elements)
for  $i=1,2$,  such that
%\begin{itemize}
  %\item 
        (1) $\vec{a}$, $\vec{d_1}$ and $\vec{d_2}$ are pairwise disjoint, and \\
  %\item 
        (2) $\chi \rewreq \EEX{\vec{d_1} \concat \vec{d_2}} \FFA{\vec{a}} \pinst{\phi_1}{\vec{e_1}\concat \vec{a_1}}{\vec{d_1}\concat\vec{u_1}} \sep \pinst{\phi_2}{\vec{e_2}\concat\vec{a_2}}{\vec{d_2}\concat\vec{u_2}}$.
%\end{itemize}
\end{definition}

\noindent
The re-scoping operation is sound with regard to the semantics of separation logic:

\begin{lemma}[Soundness of Re-scoping]
\label{lem:ex:rescoping:sound}
  Let $\EEX{\vec{e_1}} \FFA{\vec{a_1}} \phi_1$ and $\EEX{\vec{e_2}} \FFA{\vec{a_2}} \phi_2$ be some formulas whose predicates are defined by some SID $\Sid$.
  Then,
     \[
        \chi \in \left(\EEX{\vec{e_1}} \FFA{\vec{a_1}} \phi_1\right) \rescopeOP \left(\EEX{\vec{e_2}} \FFA{\vec{a_2}} \phi_2\right)
        \quad\text{implies}\quad
        \left(\EEX{\vec{e_1}} \FFA{\vec{a_1}} \phi_1\right) \sep \left(\EEX{\vec{e_2}} \FFA{\vec{a_2}} \phi_2\right)
        \sidmodels \chi.
   \]
\end{lemma}
\begin{proof}
  Follows directly from the semantics of the guarded quantifiers $\eexists$ and $\fforall$.
\end{proof}

\begin{definition}[Derivability]
We say $\chi$ can be \emph{derived} from $\EEX{\vec{e}} \FFA{\vec{a}} \phi$, in signs
$\EEX{\vec{e}} \FFA{\vec{a}} \phi \deriveqf \chi$, if
$\chi$ can be obtained from $\EEX{\vec{e}} \FFA{\vec{a}} \phi$ by applying %the modus ponens rule (see~\cref{lem:gmp}) 
    \cref{lem:gmp}
    and the rewriting equivalence $\rewreq$ (see~\cref{fig:sl:rewreq}), formally,
if there are predicates $\pred_1(\vec{x_1}), \pred_2(\vec{x_2})$, and formulas
$\psi,\psi',\zeta$ such that
\begin{enumerate}
  \item $\phi \rewreq (\pred_2(\vec{x_2}) \sep \psi) \mw \pred_1(\vec{x_1}))) \sep
    (\psi' \mw \pred_2(\vec{x_2})) \sep \zeta$, and
  \item $\chi \rewreq \EEX{\vec{e}} \FFA{\vec{a}} (\psi \sep \psi') \mw \pred_1(\vec{x_1}) \sep \zeta$.
\end{enumerate}
\end{definition}

\noindent
The derivability relation is sound with regard to the semantics of separation logic:

\begin{lemma}[Soundness of Derivability]
\label{lem:ex:derivability:sound}
  Let $\EEX{\vec{e}} \FFA{\vec{a}} \phi_1$ be some formula whose predicates are defined by some SID $\Sid$.
  Then,
     $%\[
        \EEX{\vec{e}} \FFA{\vec{a}} \phi_1 \deriveqf \chi
        \quad\text{implies}\quad
        \EEX{\vec{e}} \FFA{\vec{a}} \phi_1
        \sidmodels
        \chi%.
   $.%\]
\end{lemma}
\begin{proof}
  Follows directly from the soundness of the generalized modus ponens rule (see \cref{lem:gmp}) and the soundness of the rewriting equivalence $\rewreq$.
\end{proof}

\noindent
We now define composition based on the re-scoping and derivation operations:

\begin{definition}[Composition Operation]\label{def:pcompose}
  We define the \emph{composition} of $\phi_1$ and $\phi_2$ by
  \[
      \phi_1 \PCompose \phi_2
      \defn
      \{ \phi \mid \zeta \derivestar \phi ~\text{for some}~\zeta \in \rescope{\phi_1}{\phi_2} \}.
  \]
\end{definition}

\begin{corollary}[Soundness of $\PCompose$]\label{lem:pcompose:sound}
  %For two formulas $\phi_1$ and $\phi_2$, we have
  %\[
    $\phi \in \phi_1 \PCompose \phi_2$ implies
    %\quad\text{implies}\quad
    $\phi_1 \sep \phi_2 \sidmodels \phi$.
  %\]
\end{corollary}
\begin{proof}
  Follows immediately from \cref{lem:ex:rescoping:sound,lem:ex:derivability:sound}.
\end{proof}
%
%We now illustrate the composition operation:
%
\begin{example}
  \begin{itemize}
  \item For $\phi_1=\ls(x_2,x_3) \mw \ls(x_1,x_3)$ and
    $\phi_2=\emp \mw \ls(x_2,x_3)$, it holds that
    $\phi_1 \sep \phi_2 \deriveqf \emp \mw \ls(x_1,x_3)$.
    Hence, $(\emp \mw \ls(x_1,x_3)) \in \phi_1\PCompose\phi_2$.
  \item For $\phi_1=\FFA{a}\ls(x_2,a) \mw \ls(x_1,a)$
    and $\phi_2=\FFA{b}\ls(x_3,b) \mw \ls(x_2,b)$,
    we have $\FFA{c} (\ls(x_2,c) \mw \ls(x_1,c)) \sep (\ls(x_3,c) \mw \ls(x_2,c)) \in \phi_1 \rescopeOP \phi_2$.
    With $\FFA{c} (\ls(x_2,c) \mw \ls(x_1,c)) \sep (\ls(x_3,c) \mw \ls(x_2,c)) \deriveqf \FFA{c} (\ls(x_3,c) \mw \ls(x_1,c))$,
    we have $\FFA{c} (\ls(x_3,c) \mw \ls(x_1,c)) \in \phi_1 \PCompose \phi_2$.
  \end{itemize}
\end{example}

\noindent
Let us also revisit our informal exposition in~\cref{ex:composition-projection} and make it precise:

\begin{example}[Composition Operation on Projections]
\label{ex:composition-projection-precise}
  \qquad
  \begin{itemize}
  \item Let %$\S$ be the stack and let 
      $\ftree_1,\ftree_2,\ftree_3$ be the $\Sid$-trees from \cref{ex:cyclic-projection}.
      We set $\frst_1 = \set{\ftree_1,\ftree_3}$ and $\frst_1 = \set{\ftree_2}$.
      We then have:
      \begin{align*}
          & \sfproj{\S}{\frst_1} = \EEX{a} (\lseg(y,a) \mw \topCyclic(x,y,z)) \sep \lseg(z,a),~\text{and}\\
          & \sfproj{\S}{\frst_2} = \FFA{a'} \lseg(z,a') \mw \lseg(y,a').~\text{Then,} \\
      %\end{align*}
      %Then,
      %\begin{multline*}
          & \EEX{a} (\lseg(y,a) \mw \topCyclic(x,y,z)) \sep \lseg(z,a) \sep (\lseg(z,a) \mw \lseg(y,a)) 
          \\ & \qquad \in \sfproj{\S}{\frst_1} \rescopeOP \sfproj{\S}{\frst_2}~\text{and further} \\%~\text{Further,} \\
      %\end{multline*}
      %\end{multline*}
      %Further,
      %\begin{multline*}
         %\quad \quad \quad \quad 
          & \EEX{a} (\lseg(y,a) \mw \topCyclic(x,y,z)) \sep \lseg(z,a) \sep (\lseg(z,a) \mw \lseg(y,a)) 
        \derivestar \topCyclic(x,y,z).
      %\end{multline*}
      \end{align*}
      Hence, we have $\topCyclic(x,y,z) \in \sfproj{\S}{\frst_1} \PCompose \sfproj{\S}{\frst_2}$.
  \item %Let $S$ be the stack and l
        Let $\ftree_1,\ftree_2,\ftree_3,\ftree_4$ be the $\Sid$-trees from \cref{ex:treerp-forest}.
      We set $\frst_1 = \set{\ftree_1,\ftree_3,\ftree_4}$ and $\frst_2 = \set{\ftree_2}$.
      We have
      \begin{align*}
          & \sfproj{\S}{\frst_1} = \EEX{r} (\tll(a,b,c) \mw \tll(x,y,z)) \sep \ppto{b}{\tuple{0,0,c}} \sep \ppto{c}{\tuple{0,0,r}},~\text{and}\\
          & \sfproj{\S}{\frst_2} = \FFA{r'} (\ppto{b}{\tuple{0,0,c}} \sep \ppto{c}{\tuple{0,0,r'}}) \mw \tll(a,b,c).~\text{Then,} \\
      %\end{align*}
     %Then,
      %\begin{multline*}
         %\quad \quad \quad \quad 
          & \EEX{r} (\tll(a,b,c) \mw \tll(x,y,z)) \sep \ppto{b}{\tuple{0,0,c}} \sep \ppto{c}{\tuple{0,0,r}} \sep \\
          & \qquad ((\ppto{b}{\tuple{0,0,c}} \sep \ppto{c}{\tuple{0,0,r}}) \mw \tll(a,b,c)) \in        \sfproj{\S}{\frst_1} \rescopeOP \sfproj{\S}{\frst_2}.~\text{Further,} \\
      %\end{multline*}
      %\end{align*}
      %Further,
      %\begin{multline*}
         %\quad \quad \quad \quad
          & \EEX{r} (\tll(a,b,c) \mw \tll(x,y,z)) \sep \ppto{b}{\tuple{0,0,c}} \sep \ppto{c}{\tuple{0,0,r}} \sep \\
          & \qquad ((\ppto{b}{\tuple{0,0,c}} \sep \ppto{c}{\tuple{0,0,r}}) \mw \tll(a,b,c))
        \EEX{r} \tll(x,y,z)\derivestar \tll(x,y,z).
      %\end{multline*}
      \end{align*}
      Hence, we have $\tll(x,y,z) \in \sfproj{\S}{\frst_1} \PCompose \sfproj{\S}{\frst_2}$.
  \end{itemize}
\end{example}

\subsubsection{Relating the Composition of Forests and of Projections}
Recall from \cref{sec:composing:motivation}
our design goal that the projection function $\sfproj{\S}{\cdot}$ should be a homomorphism from forests and forest composition $\FCompose$ (\cref{def:forest-composition}) to projections
and projection composition $\PCompose$ (\cref{def:pcompose}), i.e.,
\begin{align*}
  \sfproj{\S}{\frst_1}\PCompose\sfproj{\S}{\frst_2} \overset{?}{=}
  \set{\sfproj{\S}{\frst} \mid \frst \in \frst_1 \FCompose \frst_2}. %\tag*{\tagA}
\end{align*}
\noindent
Indeed, in one direction our composition operation achieves this:
\begin{lemma}\label{cor:fcompose-to-pcompose}
  Let $\S$ be a stack and let $\frst_1,\frst_2$ be $\Sid$-forests such that $\frst_1\funion\frst_2\neq\bot$.
  Then,
  \[
    \frst \in \frst_1\FCompose\frst_2
    \quad\text{implies}\quad
    \sfproj{\S}{\frst} \in \sfproj{\S}{\frst_1} \PCompose
    \sfproj{\S}{\frst_2}.
  \]
\end{lemma}
\begin{proof}
  See~\cref{app:fcompose-to-pcompose}.
\end{proof}
\noindent
Unfortunately, as demonstrated below, the homomorphism breaks in the other direction:
\begin{example}[Projection is not homomorphic]\label{ex:proj-no-homo}
  Consider
  the $\Sid$-forests $\frst_1 = \set{\ftree_1}$ and $\frst_2 = \set{\ftree_2}$
  and the stack $\S \defn \set{x_1 \mapsto l_1, x_2 \mapsto l_2, x_3 \mapsto l_3}$, where
  \begin{align*}
    & \ftree_1 = \set{l_1 \mapsto \tuple{\emptyset, (\odd(l_1,m_1)
      \Rule \ppto{l_1}{l_2} \sep \even(l_2,m_1)) }},~\text{and}\\
    & \ftree_2 = \set{l_2 \mapsto \tuple{\emptyset, (\even(l_2,m_2)
      \Rule \ppto{l_2}{l_3} \sep \odd(l_3,m_2)) }}.
  \end{align*}
  The corresponding projections are
  \begin{align*}
    \sfproj{\S}{\frst_1} = \FFA{a} \even(x_2,a) \mw
      \odd(x_1,a)~\text{and}~
    \sfproj{\S}{\frst_2} = \FFA{a} \odd(x_3,a) \mw
      \even(x_2,a).
  \end{align*}
  Moreover, we have
  %\begin{align*}
    %& 
      $\FFA{a} \odd(x_3,a) \mw \odd(x_1,a) \in \sfproj{\S}{\frst_1} \PCompose \sfproj{\S}{\frst_2}$.
  %\end{align*}

  %
  However, since different locations, namely $m_1$ and $m_2$, are unused in the two forests,
  there is only one forest in $\frst_1 \FCompose \frst_2$:
  $\set{\ftree_1,\ftree_2}$.
  It is not possible to merge the trees, because the hole predicate of the first tree, $\even(l_2,m_1)$, is different from the root of the second tree, $\even(l_2,m_2)$.
  In particular, there does \emph{not} exist a forest $\frst$ with $\frst \in \frst_1\FCompose\frst_2$ and
  $\sfproj{\S}{\frst} \rewreq \FFA{a} \odd(x_3,a) \mw \odd(x_1,a)$.
\end{example}
\noindent
The essence of \cref{ex:proj-no-homo} is that while $\PCompose$ allows renaming quantified universals, $\FCompose$ does not allow renaming locations, breaking the homomorphism.
To get a correspondence between the two notions of composition,  we therefore allow renaming all locations that do \emph{not} occur as the value of any stack variable.
We capture this in the notion of \emph{$\S$-equivalence}:

\begin{definition}[$\S$-equivalence]\label{def:alpha-eq}
  %Let $\S$ be a stack.
  Two $\Sid$-forests $\frst_1,\frst_2$ are \emph{$\S$-equivalent},
  denoted $\frst_1\gls{--eq-AEQ}\frst_2$, iff
  there is a bijective function $\renfun\colon \Val \to \Val$ such that
    $\renfun(l) = l$ for all $l\in \img(\S)$,
    $\renfun(\nil) = \nil$, and
    %$l \in \locs{\fheapof{\frst_1}}$,
    $\renfun(\frst_1) = \frst_2$, where
  %\end{itemize}
  %where
  \begin{itemize}
    \item $\renfun(\{\ftree_1,\ldots,\ftree_k\}) \defn \{\renfun(\ftree_1),\ldots,\renfun(\ftree_k)\}$,
    \item $\renfun(\ftree) \defn \set{ \renfun(l) \mapsto \tuple{\renfun(\tsucc{\ftree_1}{l}),    \pinst{\truleinst{\ftree_1}{l}}{\dom(\renfun)}{\img(\renfun)} \mid l \in \dom(\ftree)}}$, and
    \item $\pinst{(\pred(\vec{l}) \Rule \phi)}{\vec{v}}{\vec{w}} \defn
        \pred(\pinst{\vec{l}}{\vec{v}}{\vec{w}}) \Rule \pinst{\phi}{\vec{v}}{\vec{w}}$ for sequences of locations $\vec{v}$ and $\vec{w}$.
  \end{itemize}
\end{definition}

\noindent
Note that $\frst_1\AEQ\frst_2$ implies that  $\SHpair{\S}{\fheapof{\frst_1}}$ and $\SHpair{\S}{\fheapof{\frst_2}}$ are isomorphic.
In fact, %$\S$-equivalent forests have the same projections:
\begin{lemma}\label{lem:stack-equivalence}
  If $\frst_1$ and $\frst_2$ are $\Sid$-forests with $\frst_1\AEQ\frst_2$, then
  $\sfproj{\S}{\frst_1} \rewreq \sfproj{\S}{\frst_2}$.
\end{lemma}
\begin{proof}
  Direct from the definition of $\S$-equivalence and the stack-projection.
\end{proof}
\noindent
With the definition of $\S$-equivalence in place, we indeed obtain the desired composition:

\begin{theorem}\label{thm:fcompose-eq-pcompose}
  If $\frst_1,\frst_2$ be $\Sid$-forests with
  $\frst_1\funion\frst_2\neq\bot$.
  Then,
  \begin{align*}
  \sfproj{\S}{\frst_1}\PCompose\sfproj{\S}{\frst_2} ~=~
  \set{\sfproj{\S}{\frst} \mid
  \frst \in \barfrst_1 \FCompose \barfrst_2,~
  \barfrst_1\AEQ\frst_1,~
  \barfrst_2\AEQ\frst_2
  }.
  \end{align*}
\end{theorem}
\begin{proof}
  See~\cref{app:fcompose-eq-pcompose}.
\end{proof}
%\begin{proof}
%  Immediate from
%  \cref{cor:fcompose-to-pcompose,lem:rescope-to-funion,lem:derive-to-fderive}.
%\end{proof}

%%% Local Variables:
%%% mode: latex
%%% TeX-master: "../Thesis"
%%% End:

%%% Local Variables:
%%% mode: latex
%%% TeX-master: "../Thesis"
%%% End:

\section{The Type Abstraction}\label{ch:types}
We now formally introduce the abstraction on which our decision procedure for $\SLIDguarded$ will be built.
As motivated in \cref{sec:towards:summary},
we abstract every (guarded) state to a \emph{$\Sid$-type}, which is a \emph{set} of stack-forest projections.
In order to ensure the \emph{finiteness} of the abstraction we need to restrict $\Sid$-types to certain kinds of stack-forest projections.
%
%We now give the details of our abstraction.
Let us denote by $\gls{forestsphi} \defn \set{\frst \mid \fheapof{\frst} = \H}$ the set of all $\Sid$-forests whose induced heap is $\H$.
We will then abstract a state $\SH$ to a \emph{subset} of the stack-forest projections whose induced heap is $\H$, i.e.,
$%\[
  \set{\sfproj{\S}{\frst} \mid \frst \in \sidfrstsof{\H}}$.
%\]

We call the formulas $\sfproj{\S}{\frst}$ \emph{unfolded symbolic heaps}\index{unfolded symbolic heap}\index{USH} (USHs) with respect to SID $\Sid$ because any such stack-forest projection can be obtained by ``partially unfolding'' a symbolic heap (which might require adding appropriate (guarded) quantifiers).
Intuitively, the USHs satisfied by a state $\SH$ capture all ways in which $\SH$ relates to the predicates in SID $\Sid$.
%
%\begin{infobox}{Unfolded symbolic heaps vs.~inductive wands}
%  Note that USHs are a generalization of the \emph{inductive
%    wands}\index{inductive wand} proposed
%  in~\cite{tatsuta2019completeness}: Inductive wands also correspond
%  to partial unfolding of predicates, but Tatsuta et al.~only consider
%  quantifier-free formulas with inductive wands, whereas we allow a
%  (guarded) exists--forall prefix.
%\end{infobox}
%
While the entire set of USHs is finite for every fixed state $\SH$, the set of all USHs w.r.t.~an SID $\Sid$ is infinite in general: %, as illustrated by the example below.

\begin{example}
  Assume the SID $\Sid$ defines the list-segment predicate
  $\lseg$ (see \cref{ex:sids}).
  Moreover, let $\SH$ be a state with $\size{\H} > n \in \N$
  such that $\SH \sidmodels \lseg(x,\nil)$.
  Then there exists a forest $\frst$ with $\fheapof{\frst}=\H$ whose projection
  consists of $n$ components, i.e.,
    \begin{align*}
    \sfproj{\S}{\frst}=\EEX{y_1,\ldots,y_n}& \lseg(y_n,\nil) \sep
    (\lseg(y_{n},\nil)\mw\lseg(y_{n-1},\nil)) \\&\sep \cdots \sep
    (\lseg(y_{2},\nil)\mw\lseg(y_{1},\nil)) \sep
    (\lseg(y_1,\nil)\mw\lseg(x,\nil)).
    \end{align*}
  As there exist such states $\SH$ for arbitrary natural numbers $n$, there are
  infinitely many USHs w.r.t.~$\Sid$.
\end{example}
%
%
%
%The $\Sid$-type of a model $\SH$, formally introduced in
%\cref{sec:types:abstraction}, is the set of all DUSHs that can be
%obtained via stack--forest projection of the forests
%$\sidfrstsof{\H}$.
\noindent
To obtain a \emph{finite} abstraction, we restrict ourselves to \emph{delimited USHs} (DUSHs), in which (1) all root parameters of predicate calls are free
variables and (2) every variable occurs at most once as a root parameter on the left-hand side of a magic wand:

\begin{definition}\label{def:types:ush:delimited}
  An unfolded symbolic heap $\fa$ is \emph{delimited} iff
  \begin{enumerate}
  \item for all $\pred(\vec{z}) \in \fa$,
    $\proot{\pred(\vec{z})} \in \fvs{\fa}$, and
  \item for every variable $x$ there exists at most one predicate call
    $\pred(\vec{z}) \in \fa$ such that
    $\pred(\vec{z})$ occurs on the left-hand side of a magic wand
    and $x=\proot{\pred(\vec{z})}$.
  \end{enumerate}
\end{definition}

The notion of delimited unfolded symbolic heaps is motivated as follows:
(1) For every guarded state, the targets of dangling pointers are in the image of the stack.
(2) That every variable occurs at most once as the root of a predicate on the left-hand side of a magic wand is a prerequisite for ``eliminating'' the magic wand through the generalized modus ponens rule.

\begin{example}
\label{ex:dush-formulas}
  Recall from \cref{ex:sids} the SID $\SidTree$ that defines binary trees.
  We consider the stack $\S = \set{x \mapsto l_1, y \mapsto l_2, z \mapsto l_3}$.
  \begin{enumerate}
  \item
    We consider the following trees and corresponding forest-projections:
    \begin{align*}
      & \ftree_1 \defn \{ l_1 \mapsto \tuple{\emptyset, \tree(l_1)
        \Rule \ppto{l_1}{\tuple{l_2,l_3}} \sep \tree(l_2) \sep
        \tree(l_3) } \} \\
      & \ftree_2 \defn  \{ l_2 \mapsto \tuple{\emptyset, \tree(l_2)
                     \Rule \pto{l_2}{\tuple{\nil,\nil}}} \},
      \qquad
      \ftree_3 \defn \{l_3 \mapsto \tuple{\emptyset, \tree(l_3)
                          \Rule \pto{l_3}{\tuple{\nil,\nil}}} \},\\
                        %
%    \end{align*}
%    %
%    Then,
%    \begin{align*}
      & \sfproj{\S}{\{\ftree_1,\ftree_2,\ftree_3\}} = ((\tree(y) \sep \tree(z)) \mw \tree(x)) \sep \tree(y) \sep \tree(z)\\
      & \bartree \defn \{l_1 \mapsto \tuple{\tuple{l_2,l_3}, \ftree_1(l_1)}\} \cup \ftree_2\cup\ftree_3 \qquad
      \sfproj{\S}{\{\bartree\}} = \tree(x)
    \end{align*}
    We observe that $\sfproj{\S}{\{\ftree_1,\ftree_2,\ftree_3\}}$ and $\sfproj{\S}{\{\bartree\}}$ are delimited.
    $\sfproj{\S}{\{\bartree\}}$ can be obtained from $\sfproj{\S}{\{\ftree_1,\ftree_2,\ftree_3\}}$ by two applications of modus ponens.
  \item  We consider the following trees and the projection of the corresponding forest:
    \begin{align*}
    \ftree_1 \defn \{& l_1 \mapsto \tuple{\emptyset, \tree(l_1)
        \Rule \ppto{l_1}{\tuple{l_2,l_2}} \sep \tree(l_2) \sep
        \tree(l_2) } \}\\
    \ftree_2 \defn  \{& l_2 \mapsto \tuple{\emptyset, \tree(l_2)
                     \Rule \pto{l_2}{\tuple{\nil,\nil}}} \} \\
%    \end{align*}
%    %
%    Then,
%    \begin{align*}
    \sfproj{\S}{\{\ftree_1,\ftree_2\}} & = ((\tree(y) \sep \tree(y)) \mw \tree(x)) \sep \tree(y)
    \end{align*}
    We note that $\sfproj{\S}{\{\ftree_1,\ftree_2\}}$ is not delimited because the variable $y$ appears twice on the LHS of a magic wand; 
    at most one occurrence of $y$ can be eliminated using modus ponens.
  \end{enumerate}
\end{example}

\noindent
We collect the set of all delimited unfolded symbolic heaps (DUSH) over the SID $\Sid$ in
  \begin{align*}
\gls{DUSH-Phi} \defn \{\sfproj{\S}{\frst} \mid \S \in \SS, \frst
    \text{ is a $\Sid$-forest}, \sfproj{\S}{\frst} \text{ is delimited}\}.
  \end{align*}

We are now ready to introduce the type abstraction.
Given a state $\SH$ and an SID $\Sid$, we call the set of all projections of $\Sid$-forests
capturing the heap $\H$ in the DUSH fragment the \emph{$\Sid$-type} of $\SH$:
\begin{definition}[$\Sid$-Type]\label{def:sid-type}
The \emph{$\Sid$-type} (type for short) of a state $\SH$ and an SID $\Sid$ is given by
\[
\gls{type-SH} \defn \set{\sfproj{\S}{\frst}
    \mid \frst \in \sidfrstsof{\H} } \cap \DUSH{\Sid}.
\]
\end{definition}
\noindent
In the remainder of this section, we discuss the main results and building blocks required for turning the type abstraction into a decision procedure for guarded separation logic ($\SLIDguarded$).
%; a corresponding algorithm is discussed afterward in \cref{ch:deciding-btw}.
%
%
\subsection{Understanding Satisfiability as Computing Types}\label{sec:types:understanding}
The main idea underlying our decision procedure is to the reduce the satisfiability problem for $\SLIDguarded$---``given a $\SLIDguarded$ formula $\fa$, does $\fa$ have a model $\SH \sidmodels \fa$?''---to the question of whether some type $\typ$ can be computed from a model of $\fa$, i.e., $\typ = \sidtypeof{\S,\H}$ should hold for some $\SH \sidmodels \fa$; the set of all such types will be formally defined further below.
\subsubsection{Aliasing Constraints}\label{sec:aliasing-constraint}
To conveniently reason about sets of types, we require that types in the same set have the same free variables and the same aliases, i.e., we will group types by \emph{aliasing constraint}---an equivalence relation $\scls \subseteq \Var \times \Var$ representing all aliases under consideration.
More formally, every stack $\S$ induces an aliasing constraint $\eqclasses{\S}$ given by
\[
    \eqclasses{\S} \defn \set{(x,y) \mid x,y\in\dom(\S) \text{ and } \S(x)=\S(y)}.
\]
We denote the \emph{domain} of an aliasing constraint $\scls$ by $\dom(\scls) \defn \set{x \mid (x,x)\in\scls}$.
Furthermore, we write $\scls(x)$ for the set of aliases of $x$ given by the aliasing constraint $\scls$, i.e.,
the equivalence class $\scls(x) \defn \{ y \mid (x,y) \in \scls \}$ of $\scls$ that contains $x$.
To obtain a canonical formalization, we frequently represent the equivalence class of $x$ by its largest\footnote{w.r.t. the linear ordering over variables we assume throughout this article; notice that the maximum is  well-defined as long as the set of aliases of a variable is finite.}
member; formally, $\eqclass{\scls}{x} \defn \max \scls(x)$.
\subsubsection{From $\SLIDguarded$ satisfiability to types}
As outlined at the beginning of \cref{sec:types:understanding}, our decision procedure will be based on computing sets of types of the following form:
\begin{definition}[$\scls$-Types]\label{def:types:x-phi-types}
 Let $\scls$ be an aliasing constraint (cf. \cref{sec:aliasing-constraint}).
 Then the set
 $\phitypes{\scls}{\phi}$ of \emph{$\scls$-types} of $\SLIDguarded$ formula $\phi$
 is defined as
\begin{align*}
  \phitypes{\scls}{\phi}
  \defn
    \set{ \sidtypeof{\S,\H} \mid \S \in \SS,~ \H \in \HH,~ \eqclasses{\S} = \scls,~ \SH \sidmodels \phi }.
\end{align*}
\end{definition}
\noindent
By the above definition, a $\SLIDguarded$ formula $\fa$ with at least
one non-empty set of $\scls$-types is satisfiable: some type coincides with $\sidtypeof{\S,\H}$, where
$\SH \sidmodels \fa$.
Conversely, if $\fa$ is satisfiable, then there exists a model $\SH \sidmodels \fa$ and the set
$\phitypes{\eqclasses{\S}}{\fa}$ is non-empty. In summary:
\[
  \fa~\text{is satisfiable} \quad\text{iff}\quad
  \exists \scls.~\phitypes{\scls}{\fa} \neq \emptyset.
\]
On a first glance, finding a suitable aliasing constraint $\scls$ and proving non-emptiness of $\phitypes{\scls}{\fa}$
might appear as difficult as finding a state $\SH$ such that $\SH \sidmodels \fa$ holds due to
three concerns:
\begin{enumerate}
\item\label{type-issue-a} There are, in general, both infinitely many aliasing constraints and infinitely many $\Sid$-types, because the size of stacks---and thus the number of free variables to consider---is unbounded.
\item\label{type-issue-b} Even if the set $\phitypes{\scls}{\fa}$ is finite, effectively computing it is non-trivial.
\item\label{type-issue-c} Deciding whether a type $\typ$ belongs to $\phitypes{\scls}{\fa}$ is non-trivial:
assume that $\SH \sidmodels \fa$, $\SHpprime \not\sidmodels \fa$,
and both states yield the same type, i.e., $\typ = \sidtypeof{\S,\H} = \sidtypeof{\S',\H'}$.
Determining that $\typ \in \phitypes{\scls}{\fa}$ would then require us to know that $\typ$ can be computed from a specific state, namely $\SH$.
\end{enumerate}
As informally motivated in \cref{ch:towards}, our type abstraction can deal with each of the above concerns; we provide the formal details addressing each concern in the remainder of this section:
Regarding (\ref{type-issue-a}), we discuss in \cref{sec:using-types:finiteness} how both aliasing constraints and types can safely be restricted to finite subsets.
%we note that we cboth aliasing constraints and types can be safely restricted to a finite set $\vec{x}$ of variables as long as $\vec{x}$ covers at least the free variables of the formula $\fa$ under consideration. %, i.e., $\vec{x} \supseteq \fvs{\fa}$.
%The number of aliasing constraints to consider---those in $\EqClasses{\vec{x}}$---then equals the $\size{\vec{x}}$-th Bell number, bounded by $n^n \in \bigO(2^{n \log(n)})$, where $n = \size{\Sid} + \size{\vec{x}}$.
%We will show in \cref{sec:using-types:finiteness} that the total number of $\scls$-types over all $\SLIDguarded$ formulas is bounded, too.
Determining whether $\exists \scls.~\phitypes{\scls}{\fa} \neq \emptyset$ holds thus amounts to computing finitely many finite sets.
This corresponds to achieving \emph{finiteness} in \cref{ch:towards}.
Regarding (\ref{type-issue-b}), we introduce operations for effectively computing $\Sid$-types from existing ones in \cref{sec:types:delimited,sec:types:abstraction}; they will be the building blocks of our decision procedure.
This corresponds to achieving \emph{compositionality} in \cref{ch:towards}.

Regarding (\ref{type-issue-c}), we show in \cref{sec:using-types:refinement} that one can decide whether a type $\typ$ belongs
to $\phitypes{\scls}{\fa}$ \emph{without reverting to any state underlying $\typ$}.
In particular, 
we will show %a \emph{refinement theorem} ensuring 
that states yielding the same $\Sid$-type satisfy the same $\SLIDguarded$ formulas.
This corresponds to achieving \emph{refinement} in \cref{ch:towards}.

%---thus ruling out the issue raised in (\ref{type-issue-c}).
%
%In the remainder of the section we argue the finiteness of the abstraction domain and introduce operations on $\Sid$-Type, which will be the building blocks of our decision procedure for $\SLIDguarded$.
%For the latter we first need to precisely capture the forests that correspond to DUSH and state a result on the (de-)composition of such forests.
%This result will then allow us introduce our operations on $\Sid$-Types.
%
\subsection{Finiteness}\label{sec:using-types:finiteness}
To ensure finiteness of the type abstraction, we only consider stacks with variables taken from some arbitrary, but fixed, finite set $\vec{x}$ of variables.
In particular, we denote by $\gls{DUSHx-Phi}$ the restriction of delimited unfolded symbolic heaps ($\DUSH{\Sid}$) to formulas $\fa$ with free variables in $\vec{x}$, i.e.,  $\fvs{\fa}\subseteq\vec{x}$.
With this restriction in place, we are immediately able to argue the finiteness of the DUSH Fragment
based on the following observation:
Every variable can appear at most twice (once as the projection of a hole and once as the projection of a tree).
\begin{lemma}\label{lem:dush-finite}
  Let $n \defn \size{\Sid}+\size{\vec{x}}$, where $\vec{x}$ is a finite set of variables.
  Then %the number of DUSHs over $\Sid$ and $\vec{x}$
  %is exponentially bounded in $n$; more precisely,
  $\size{\DUSHx{\Sid}{\vec{x}}} \in 2^{\bigO(n^2 \log(n))}$.
\end{lemma}
\begin{proof}
   See \Cref{app:dush-finite}.
\end{proof}
\noindent
Analogously to $\DUSHx{\Sid}{\vec{x}}$, we only consider aliasing constraints and
types over the finite set %of variables 
$\vec{x}$, i.e.,
%That is, 
we consider the finite set of aliasing constraints
%\[
    $\EqClasses{\vec{x}} \defn \set{ \eqclasses{\S} \mid \S \in \SS, \dom(\S) = \vec{x}}$.
%.\]
We note that the number of aliasing constraints in $\EqClasses{\vec{x}}$ equals the $|\vec{x}|$-th Bell number, bounded by $n^n \in \bigO(2^{n \log(n)})$, where $n = \size{\xx}$.
Furthermore, we collect in $\allxtypes$ all $\scls$-types over $\Sid$ and $\vec{x}$, i.e.,
\[
  \allxtypes
  \defn
  \bigcup_{\scls \in \EqClasses{\vec{x}}} \bigcup_{\phi \in \SLIDguarded} \phitypes{\scls}{\phi}.
\]
The above restriction of types to variables in $\vec{x}$ indeed ensures finiteness:
\begin{theorem}\label{thm:types-finite}
 Let $\vec{x} \subseteq \Var$ be finite and $n \defn \size{\Sid} + \size{\vec{x}}$.
 Then $\size{\allxtypes} \in 2^{2^{\bigO(n^2 \log(n))}}$.
\end{theorem}
\begin{proof}
  Recall from \cref{lem:dush-finite} that
  the set $\DUSHx{\Sid}{\vec{x}}$ of DUSHs over $\Sid$
  with free variables taken from $\vec{x}$ is of size
  $2^{\bigO(n^2 \log(n))}$.
  We show below that every type $\typ \in \allxtypes$ is a subset of $\DUSHx{\Sid}{\vec{x}}$.
  Hence, the size of $\allxtypes$ is bounded by the number of subsets of $\DUSHx{\Sid}{\vec{x}}$, i.e.,
  $\size{\allxtypes} \in 2^{2^{\bigO(n^2 \log(n))}}$.

  It remains to show that, for every $\typ \in \allxtypes$, we have $\typ \subseteq \DUSHx{\Sid}{\vec{x}}$:
  %\begin{itemize}
      %\item 
          By definition of $\allxtypes$, there exists an aliasing constraint $\scls \in \EqClasses{\vec{x}}$  and a $\SLIDguarded$ formula $\fa$ such that $\typ \in \phitypes{\scls}{\fa}$.
    %\item 
          By \cref{def:types:x-phi-types}, there exists a state $\SH$ such that          $\dom(\S) = \dom(\scls) \subseteq \vec{x}$, $\SH \sidmodels \fa$, and $\typ = \sidtypeof{\S,\H}$.
    %\item 
          By \cref{def:sid-type},
          $\typ = \set{\sfproj{\S}{\frst} \mid \frst \in \sidfrstsof{\H} } \cap \DUSH{\Sid} \subseteq \DUSHx{\Sid}{\vec{x}}$. %\qedhere
  %\end{itemize}
\end{proof}

\subsection{$\S$-Delimited Forests}\label{sec:types:delimited}

To introduce the forests that correspond to DUSHs we make use of the notions of an \emph{interface} of a $\Sid$-forest, which is the set of locations that appear in some tree either as the root or as a hole:
\begin{definition}[Interface]\label{def:types:forest:interface}\index{interface}
The \emph{interface} of a $\Sid$-forest $\frst = \ktrees$ is given by
  \[\gls{interface} \defn \bigcup_{1\leq i \leq
    k}(\set{\troot{\ftree_i}}\cup \tallholes{\ftree_i}).\]
\end{definition}

\begin{example}[Interface]\label{ex:forest:interface}
  Recall the forest $\frst$ from \cref{ex:types:forest}.
We have $\finterface{\frst}=\set{l_1,l_2,l_3}$: the locations
$l_1,l_2,l_3$ all occur as the roots of a tree; $l_2$ and $l_3$
additionally occur as holes (of $\ftree_3$ and $\ftree_1$,
respectively); $l_4$ occurs neither as root nor as hole of a tree and is thus not part of the interface.
\end{example}
\noindent
An \emph{$\S$-delimited forest} is a $\Sid$-forest
whose interface consists only of locations covered by some stack variable and which does not have any duplicate holes:
\begin{definition}[$\S$-delimited $\Phi$-Forest]\label{def:types:frst:delimited}
  A $\Sid$-forest $\frst$ is
  \emph{$\S$-delimited}\index{s-delimited@$\S$-delimited} iff
  %\begin{enumerate}
      %\item 
      (1) $\finterface{\frst}\subseteq\img(\S)$, and
      %\item 
      (2) for every $l \in \fallholes{\frst}$ in some tree $\ftree \in \frst$, there
          is exactly one rule instance (for $l' \in \dom(\ftree)$)
            \[ \truleinstKW_\ftree(l') = \pred(\vec{v}) \Rule \ppto{a}{\vec{b}} \sep \pred_1(\vec{v_1}) \sep\cdots\sep\pred_m(\vec{v_m}) \sep \Pure \]
            and exactly one index $i \in [1,m]$
            such that $\proot{\pred_i(\vec{v_i})} = l$.
  %\end{enumerate}
\end{definition}
%  \fztodo{It is a big ugly to go over the rule instance here but before it was imprecise.
%  An alternative would be to define trees with no holes first and then define partial trees as the trees which can result from splits.
%  These partial trees would automatically have the desired property, but this would be a much bigger change... We can add this to the list (with the lowest priority though ;)
%  \cmtodo{agreed; I'm also not eager to introduce trees without holes}
%    }
%
% Note that by requiring that every hole occurs at most once in a
% delimited forest $\frst$, we have the guarantee that we can ``remove''
% all holes of a delimited forest via forest composition and
% $\fderivestar$, i.e., there exist forests $\frst',\barfrst$ such that
% $\frst\funion\frst'\fderivestar\barfrst$ and
% $\fallholes{\barfrst}=\emptyset$.
\begin{example}
  We consider the forests from \cref{ex:dush-formulas}:
  We note that $\{\ftree_1,\ftree_2,\ftree_3\}$ from (1) is $\S$-delimited,
  while $\{\ftree_1,\ftree_2\}$ from (2) is not.
\end{example}
\noindent
A $\Sid$-forest is $\S$-delimited precisely when its projection is delimited (see \cref{app:delimited-forest--delimited-projection} for a proof):
\begin{lemma}\label{lem:delimited-forest--delimited-projection}
  Let $\frst$ be a forest and let $\S$ be a stack. Then $\frst$ is
  $\S$-delimited iff $\sfproj{\S}{\frst}$ is delimited.
\end{lemma}

\noindent
We now state that the $\S$-delimitedness of forests is preserved under decomposition;
this result will allows to lift the composition of DUSH formulas (resp. $\S$-delimited forests) to types.
\begin{theorem}\label{lem:delimited-forest-delimited-sub-forests}
  Let $\SHi{1},\SHi{2}\in\Mpos{\Sid}$ be guarded states, and let $\frst$ be a $\S$-delimited forest with $\frst \in \sidfrstsof{\H_1\stdunion\H_2}$.
  Then, there exist $\S$-delimited forests $\frst_1,\frst_2$ with $\fheapof{\frst_i}=\H_i$ and $\frst \in \frst_1 \FCompose \frst_2$.
\end{theorem}
\begin{proof}
  See~\cref{app:delimited-forest-delimited-sub-forests}.
\end{proof}

%%% Local Variables:
%%% mode: latex
%%% TeX-master: "../Thesis"
%%% End:

\subsection{Operations on Types}\label{sec:types:abstraction}

\emph{type composition}, \emph{renaming of variables}, \emph{forgetting variables}, and \emph{type extension}.
These operations will be the building blocks of our decision procedure for $\SLIDguarded$.
\subsubsection{Type Composition}
%
%Our goal is to 
We define an operation $\ACompose$ on the level of $\Sid$-types such that $\sidtypeof{\S, \H_1 \stdunion \H_2} = \sidtypeof{\S,\H_1} \ACompose  \sidtypeof{\S,\H_2}$, i.e., $\sidtypeof{\S,\cdot}$ is a homomorphism w.r.t. to the operation $\stdunion$ on heaps and the operation $\ACompose$ on types.
As justified below, we can define $\ACompose$ by applying our composition operation for forest projections, $\PCompose$ (cf. \cref{def:pcompose}), to all elements of the involved types.
\begin{theorem}[Compositionality of $\Sid$-types]\label{thm:oursunion-to-derive}
  For all guarded states $\SHi{1}$ and $\SHi{2}$ with
  $\H_1\stdunion\H_2\neq\bot$,
  $\sidtypeof{\S, \H_1 \stdunion \H_2}$ can be computed from
  $\sidtypeof{\S, \H_1}$ and $\sidtypeof{\S,\H_2}$ as follows:
  \begin{align*}
    \sidtypeof{\S,\H_1 \stdunion \H_2} = \{
        \fa \in \DUSH{\Sid} \mid
        \text{ex. }
        \fb_1 \in \sidtypeof{\S,\H_1},
        \fb_2 \in \sidtypeof{\S,\H_2}
        \text{ such that }
        \fa\in\fb_1 \PCompose \fb_2
      \}.
  \end{align*}
\end{theorem}
\begin{proof}
  See \Cref{app:oursunion-to-derive}.
\end{proof}
\noindent
Our second consideration for defining the composition operation $\ACompose$ on $\Sid$-types is that the operation $\stdunion$ is only defined on disjoint heaps.
In order to be able to express a corresponding condition on the level of types, we will make use of the following notion:
\begin{definition}[Allocated variables of a type]
  The set of \emph{allocated variables} of $\Sid$-type $\typ$ is % are given by the set
  \begin{align*}
    \typealloc{\typ} \defn \{x \mid \text{there ex.~} \fa \in \typ
      \text{ and } (\fb\mw\pred(\vec{z})) \text{ in } \fa ~\text{s.t.~} x = \proot{\pred(\vec{z})}\}\}.
  \end{align*}
\end{definition}

\noindent
The above notion is motivated by the fact that,
for each non-empty type, the allocated variables of the type agree with the allocated variables of every state having that type.

\begin{lemma}\label{lem:typealloc-definable}
  Let $\SH$ be a state with $\sidtypeof{\S,\H} \neq \emptyset$.
  Then,
  $\valloc{\S,\H} = \typealloc{\sidtypeof{\S,\H}}$.
\end{lemma}
\begin{proof}
  See \cref{app:typealloc-definable}.
\end{proof}
\noindent
We note that, for every model $\SH$ of some predicate call $\pred(z_1,\ldots,z_k)$, there is at least one tree $\ftree$ with $\fheapof{\set{\ftree}}=\H$ (see~\cref{lem:model-of-pred-to-tree});
hence, $\sfproj{\S}{\set{\ftree}} \in \sidtypeof{\S,\H}$ and the non-emptiness requirement of~\cref{lem:typealloc-definable} is fulfilled---a fact that generalizes to all models of guarded formulas:
%
%This fact can be generalized to models of guarded formulas:
%
\begin{lemma}
  \label{cor:guarded-formula-non-empty-type}
  Let $\phi \in \SLIDguarded$ and let $\SH$ be a state with $\SH \sidmodels \phi$.
  Then, $\sidtypeof{\S,\H} \neq \emptyset$.
\end{lemma}
\begin{proof}
  See \cref{app:guarded-formula-non-empty-type}.
  %Follows directly from~\cref{lem:guarded-iterated-star-predicates} and~\cref{lem:typealloc-definable}.
  %\cmtodo{The proof refers to the wrong lemma}
%%
%  Since $\SH$ is guarded,
%  \cref{lem:guarded-iterated-star-predicates} yields that
%  there exist predicate calls
%  such that
%  $\SH \sidmodels \IteratedStar_{1\leq i\leq k}\pred_i(\vec{x_i})$.
%  %
%  We split $\H$ into
%  $\H_1\stdunion\cdots\stdunion\H_k$ such that
%  $\SHi{i}\sidmodels\pred_i(\vec{x_i})$ for each $i \in [1,k]$.
%  Next, consider the forest
%  $\frst\defn\set{\ftree_1,\ldots,\ftree_k}$,
%  where each $\ftree_i$ is a
%  $\Sid$-tree with $\fheapof{\ftree_i}=\H_i$; such trees
%  exist by \cref{lem:model-of-pred-to-tree}. Observe further that
%  each of these trees is delimited, because they do not have holes and
%  their root is in $\S(\vec{x_i})$.
%  By \Cref{lem:forests:funion-heap}, we have $\fheapof{\frst}=\H$.
%  %
%  Finally, \cref{lem:sf-projection-sound} yields
%  $\SH \sidmodels \sfproj{\S}{\frst}$ and thus
%  $\sfproj{\S}{\frst}\in\sidtypeof{\S,\H}$.
\end{proof}
\noindent
We are now ready to state our composition operation $\ACompose$ on $\Sid$-types:
\begin{definition}[Type composition]\label{def:type:composition}
  The \emph{composition} $\typi{1} \ACompose \typi{2}$ of $\Sid$-types $\typi{1}$ and $\typi{2}$ is given by
  \[
        \typi{1} \ACompose \typi{2} \defn
    \begin{cases}
      \bot, &\text{if } \typealloc{\typi{1}} \cap \typealloc{\typi{2}} \neq \emptyset, \\
      %\phi_1 \PCompose \phi_2,& \text{otherwise.}
        \left(\bigcup_{\phi_1 \in \typi{1}, \phi_2 \in \typi{2}} \phi_1 \PCompose \phi_2\right) \cap \DUSH{\Sid}, & \text{otherwise.}
    \end{cases}
  \]
\end{definition}
\noindent
We now state two results that $\ACompose$ indeed has the desired properties, i.e., that $\sidtypeof{\S,\cdot}$ is a homomorphism w.r.t. to the operation $\stdunion$ on heaps and the operation $\ACompose$ on types:
\begin{corollary}[Compositionality of type abstraction]\label{lem:types:homo:compose}
  %
%  Let $\S$ be a stack and let $\H_1,\H_2$ be heaps with $\SHi{1},\SHi{2}\in\Mpos{\Sid}$ and
%  $\H_1\stdunion\H_2\neq\bot$.
  For guarded states $\SHi{1}$ and $\SHi{2}$ with $\H_1 \stdunion \H_2 \neq \bot$, we have
  $\sidtypeof{\S, \H_1 \stdunion \H_2} = \sidtypeof{\S,\H_1} \ACompose \sidtypeof{\S,\H_2}$.
\end{corollary}
\begin{proof}
See \cref{app:types:homo:compose}.
%  We need to show that $\sidtypeof{\S,\H_1} \ACompose \sidtypeof{\S,\H_2} \neq \bot$.
%  Assume that $\sidtypeof{\S,\H_i} = \emptyset$ for $i=1$ or $i=2$.
%  Then, $\typealloc{\typi{i}} = \emptyset$ and we get that $\typealloc{\typi{1}} \cap \typealloc{\typi{2}} = \emptyset$,
%  Otherwise, we have $\sidtypeof{\S,\H_i} \neq \emptyset$ for $i=1,2$.
%  Then, $\valloc{\S,\H_i} = \typealloc{\sidtypeof{\S,\H_i}}$.
%  $\H_1\stdunion\H_2\neq\bot$ then implies that
%  $\typealloc{\typi{1}} \cap \typealloc{\typi{2}} = \emptyset$.
%  The claim then follows from~\cref{thm:oursunion-to-derive}.
\end{proof}

\begin{lemma}
  \label{lem:types:homo:existence}
    For $i \in \{1,2\}$, let $\SHi{i}$ be states with $\sidtypeof{\S,\H_i} = \typi{i} \neq \emptyset$ and $\typi{1} \ACompose \typi{2} \neq \bot$.
  Then, there are states $\SHpair{\S}{\H'_i}$ such that $\sidtypeof{\S,\H'_i} = \typi{i}$ and
  $\sidtypeof{\S, \H_1' \stdunion \H'_2} = \typi{1} \ACompose \typi{2}$.
\end{lemma}
\begin{proof}
See \Cref{app:types:homo:existence}.
\end{proof}
%\begin{proof}
%  We choose some states $\SHpair{\S}{\H'_i}$ that are isomorphic to $\SHi{i}$ such that $\locs{\H'_1} \cap \locs{\H'_2} \subseteq \img(\S)$.
%  We have that $\SHpair{\S}{\H'_i} = \sidtypeof{\S,\H_i} = \typi{i}$ because isomorphic states have the same types (observe that the stack-projection replaces location that are not in the image of the stack by quantified variables).
%  By~\cref{lem:typealloc-definable}, we have    $\typealloc{\typi{i}} = \typealloc{\SHpair{\S}{\H'_i}} = \valloc{\S,\H'_i}$.
%  Thus, we get $\H'_1\stdunion\H'_2 \neq \bot$ from    $\typi{1} \ACompose \typi{2} \neq \bot$ and $\locs{\H'_1} \cap \locs{\H'_2} \subseteq \img(\S)$.
%  Then, \cref{thm:oursunion-to-derive} yields that $\sidtypeof{\S, \H'_1 \stdunion \H'_2} = \sidtypeof{\S,\H'_1} \ACompose \sidtypeof{\S,\H'_2}$.
%\end{proof}

\subsubsection{Renaming Variables}
To compute the types of predicate calls $\pred(\vec{y})$ compositionally, we need a mechanism to rename variables in $\Sid$-types:
Once we know the types of a predicate call $\pred(\vec{x})$ over the formal arguments $\vec{x} = \fvs{\pred}$, we can compute the types of $\pred(\vec{y})$ by renaming $\vec{x}$ to $\vec{y}$.
Such a renaming amounts to a simple variable substitution:
\begin{definition}[Variable Renaming]\label{def:type-instantiation}
  Let $\vec{x}$ be a sequence of pairwise distinct variables, let $\vec{y}$ be an arbitrary sequence of variables with $|\vec{y}| = |\vec{x}|$, and let $\scls$ be an aliasing constraint with and $\vec{y} \subseteq \dom(\scls)$.
  Moreover, let $\vec{y'}$ be the sequence obtained by replacing every variable in $y \in \vec{y}$ by the maximal variable in its equivalence class, i.e., by $\eqclass{\scls}{y}$.
  Then, the \emph{$\pinstINST{\vec{x}}{\vec{y}}$-renaming} of type $\typ$ w.r.t. aliasing constraint $\scls$ is given by
  $%\begin{align*}
    \tinst{\typ}{\scls}{\vec{x}}{\vec{y}} \defn \set{ \pinst{\fa}{\vec{x}}{\vec{y'}} \mid \fa \in \typ}$.
  %\end{align*}
\end{definition}
\noindent
Variable renaming is compositional as it corresponds to first renaming variables at the level of stacks and then computing the type of the resulting state.
More formally, assume a state $\SH$, where we already renamed $\vec{x}$ to $\vec{y}$ in stack $\S$; in particular, $\vec{x} \cap \dom(\S) = \emptyset$.
Computing $\sidtypeof{\S,\H}$ then coincides with the $\pinstINST{\vec{x}}{\vec{y}}$-renaming of
$\sidtypeof{\S',\H}$, where $\S' = \pinst{\S}{\vec{x}}{\vec{y}} \defn \pinst{\S}{\vec{x}}{\S(\vec{y})}$
is the stack $\S$ in which the variables $\vec{x}$ have not been renamed to $\vec{y}$ yet.\footnote{Recall %from \cref{sec:notation}  
that $\pinst{\S}{\vec{u}}{\vec{v}}$ denotes a stack \emph{update} in which variables in $\vec{u}$ are added to the domain of stack $\S$ if necessary.}
\begin{lemma}\label{lem:types:homo:rename}
  For $\vec{x}$, $\vec{y}$ as above and a stack $\S$
  with $\vec{y} \subseteq \dom(\S)$ and $\vec{x} \cap \dom(\S) = \emptyset$, we have
  \[
    \tinst{\sidtypeof{\pinst{\S}{\vec{x}}{\vec{y}},\H}}{\eqclasses{\S}}{\vec{x}}{\vec{y}}
    =
    \sidtypeof{\S,\H}
    .
  \]
\end{lemma}
\begin{proof}
   See \Cref{app:types:homo:rename}.
\end{proof}
\subsubsection{Forgetting Variables}
Our third operation on types removes a free variable $x$ from a type $\typ$.
Intuitively, for every formula $\fa \in \typ$, there are two cases:
\begin{enumerate}
\item If $x$ aliases with some free variable, then we replace $x$ by its largest alias.
\item If $x$ does not alias with any free variable, then we remove it from the set of free variables by introducing a (guarded) existential quantifier.
%\item we universally quantify over $y$ in case $y$ otherwise (i.e., $y$ is not allocated and does not alias with any free variable).
\end{enumerate}
Formally, we fix an aliasing constraint $\scls$ (cf. \cref{sec:aliasing-constraint}) characterizing which free variables are aliases.
Forgetting a variable $x$ in a formula $\fa$ with respect to $\scls$ is then defined as follows:

%formula $\EEX{\vec{e}} \FFA{\vec{a}} \psi \in \typ$ w.r.t. a stack-aliasing constraint $\scls$ as follows:
%Formally, we define an operation for removing a variable $y$ from a
%formula given stack-aliasing
%constraint $\scls\defn\eqclasses{\S}$.

\begin{align*}
  \Forget{\scls,x}{\fa} \defn
\begin{cases}
    \pinst{\fa}{x}{\max (\scls(x) \setminus  \{x\})},
    & \text{if}~ x \in \fvs{\fa} ~\text{and}~ \scls(x) \neq  \{x\}, \\
    \EEX{x} \fa,
    & \text{if}~ x \in \fvs{\fa} ~\text{and}~ \scls(x) = \{x\}, \\
    \fa, & \text{if}~ x \notin \fvs{\fa}~.
\end{cases}
\end{align*}
\noindent
Forgetting a variable in a \emph{type} $\typ$ %then 
corresponds to applying the above operation to all $\fa \in \typ$.
However, $\Forget{\scls,x}{\fa}$ does---in general---not belong to the fragment $\DUSH{\Sid}$ because we might existentially quantify over a root variable of $\fa$.
Hence, we additionally intersect the result %of our forget operation 
with $\DUSH{\Sid}$:

\begin{definition}[Forgetting a variable]\label{def:types:forget}
  The $\Sid$-type obtained from \emph{forgetting} variable $x$ in $\Sid$-type $\typ$ w.r.t. aliasing constraint $\scls$ is defined by
  ~$%\begin{align*}
      \Forget{\scls,x}{\typ}
      \defn
      \set{ \Forget{\scls,x}{\fa} \mid \fa \in \typ } \cap \DUSH{\Sid}
  $.%\end{align*}
\end{definition}
\noindent
The above operation is compositional
as forgetting an allocated variable in the type of a guarded state coincides with first removing the variable from the state and then computing its type:
\begin{lemma}\label{lem:types:homo:forget}
  Let $\SH$ be a guarded state such that $\S(x) \in \dom(\H)$ holds for some variable $x$.
  Then,
  \[ \Forget{\eqclasses{\S},x}{\sidtypeof{\S,\H}} = \sidtypeof{\pinst{\S}{x}{\bot},\H}~. \]
\end{lemma}
\begin{proof}
See~\Cref{app:types:homo:forget}.
\end{proof}

\subsubsection{Type Extension}
Our fourth and final operation is concerned with extending types to stacks over larger domains.
To this end, we instantiate universally quantified variables with free variables that do not appear in the type so far.
Formally, let
$\phi = \EEX{\vec{e}} \FFA{(\vec{a} \cdot u \cdot \vec{b})} \psi$ be a formula and let $x$ be a fresh variable, i.e., $x \not\in \fvs{\phi}$.
We call the formula
$\EEX{\vec{e}} \FFA{(\vec{a} \cdot \vec{b})} \pinst{\psi}{u}{x}$
an \emph{$x$-instantiation} of $\phi$.
Extending a type by variable $x$ then corresponds to adding all $x$-instantiations of its members:

\begin{definition}[$x$-extension of a Typ]\label{def:x-extension}
The \emph{$x$-extension} of a $\Sid$-type $\typ$ by a fresh variable $x$ is
\[ \Extend{x}{\typ} \defn \typ \cup \{ \phi' \text{ is an } x \text{-instantiation of } \phi \mid \phi \in \typ\}~. \]
\end{definition}
\noindent
As for the other operations, the $x$-extension of a type is compositional in the sense that it coincides with computing the type of a state with an already extended stack:
\begin{lemma}\label{lem:extend}
 For every state $\SH$, variable $x$ with $\S(x) \not\in \locs{\H}$ and $\eqclasses{\S}(x) = \set{x}$,
    %we have
 \[
     \Extend{x}{\sidtypeof{\pinst{\S}{x}{\bot}, \H}} = \sidtypeof{\S,\H}
 .\]
\end{lemma}
\begin{proof}
   See \Cref{app:extend}.
\end{proof}

\noindent
Rather than extending a type by a single variable, it will be convenient to extend it by all variables in an \emph{aliasing constraint} that are not aliases of an existing variable.

\begin{definition}[Extension of a type with regard to an aliasing constraint]
\label{def:type-ac-extension}
Let $\scls \subseteq \scls'$ be aliasing constraints.
Let $\vec{y}$ be a repetition-free sequence of all maximal variables in $\dom(\scls)$,
and let $\vec{y'}$ be the sequence obtained by replacing every variable in $y \in \vec{y}$ by the corresponding maximal variable in $\scls'$, i.e., by $\eqclass{\scls'}{y}$.
\footnote{We recall that we need maximal variables for maintaining canonic projections, i.e., type representations.}
Moreover, let $\vec{z} = \tuple{z_1, \ldots, z_n}$, $n \geq 0$, be a repetition-free sequence of all maximal variables in $\dom(\scls')$ that are not aliases of variables in $\dom(\scls)$.
\footnote{I.e., $z \in \vec{z}$ iff $z \in \dom(\scls')$, $z = \eqclass{\scls'}{z}$
and $z \not\in \scls'(y)$ for all $y \in \dom(\scls)$.}
Then the \emph{$\scls'$-extension} of a $\Sid$-type $\typ$ w.r.t. aliasing constraint $\scls$ is defined as $\Extend{\scls'}{\typ} \defn \typ_n$, where
\begin{align*}
    \typ_k = \begin{cases}
        \{\pinst{\phi}{\vec{y}}{\vec{y'}} \mid \phi\in\typ\}, & \text{if}~k = 0 \\
        \Extend{z_k}{\typ_{k-1}}, & \text{if}~0 < k \leq n~.
    \end{cases}
\end{align*}
\end{definition}

%\begin{definition}[Extension of a type with regard to an alias-constraint]
%Let $\scls \subseteq \scls'$ be alias-constraints and let $\typ$ be a type over variables $\dom(\scls)$.
%Let $\scls' \setminus \scls = \eqclass{\scls'}{x_1} \cup \cdots \eqclass{\scls'}{x_k}$ be the equivalence classes that only belong to $\scls'$, and where the representatives $x_i$ are chosen such that $x_i = \max \eqclass{\scls}{x_i}$\footnote{The choice of the representatives is only important for maintaining canonic type representations.}.
%Let $\typ_0 = \typ$ and let $\typ_{i+1}$ be the $x_i$-extension of $\typ_i$.
%We then call $\Extend{\scls'}{\typ} \defn \typ_k$ the $\scls'$-extension of $\typ$.
%\end{definition}

\noindent
The above operation preserves compositionality as it boils down to multiple type extensions:
%Since the above operation boils down to multiple applications of our type extension operator, it preserves compositionality:
%
\begin{lemma}\label{lem:extend-lifted}
Let $\SH$ be a state and $\scls$ be an aliasing constraint with $\scls \subseteq \eqclasses{\S}$.
Let $\S'$ be the restriction of $\S$ to the domain $\dom(\scls)$.
If $\S(x) \notin \locs{\H}$ for every variable $x \in \dom(\S)$ that is not an alias of a variable in $\dom(\scls)$, then $\Extend{\eqclasses{\S}}{\sidtypeof{\S',\H}} = \sidtypeof{\S,\H}$.
\end{lemma}
\begin{proof}
    Let $k \geq 0$ be the number of variables in $\dom(\S)$ that are no aliases of variables in $\dom(\scls)$.
    By \cref{def:type-ac-extension}, we have $\Extend{\eqclasses{\S}}{\sidtypeof{\S',\H}} = \typ_k$, i.e., we need to apply $k$ type extensions.
    The claim then follows from \cref{lem:extend} by induction on $k$.
\end{proof}

%%% Local Variables:
%%% mode: latex
%%% TeX-master: "../Thesis"
%%% End:

\subsection{Type Refinement}\label{sec:using-types:refinement}
The main insight required for effectively deciding whether a type $\typ$ belongs to $\phitypes{\scls}{\fa}$ is that
states with identical $\Sid$-types satisfy the same $\SLIDguarded$ formulas---a statement we formalize below.
This property is perhaps surprising, as types only contain formulas from the
DUSH fragment, which is largely orthogonal to $\SLIDguarded$. For
example, $\SLIDguarded$ formulas allow guarded negation and guarded
septraction, but neither quantifiers nor unguarded magic wands,
whereas DUSHs allow limited use of guarded quantifiers and unguarded
magic wands, but neither Boolean structure nor septraction.

\begin{theorem}[Refinement theorem]\label{thm:types-capture-sat}
  For all stacks $\S$, heaps $\H_1$, $\H_2$, and $\SLIDguarded$ formulas $\fa$,
  \[
      \sidtypeof{\S,\H_1} = \sidtypeof{\S,\H_2}
      \qquad\text{implies}\qquad
      \SHi{1} \sidmodels \phi ~~\text{iff}~~ \SHi{2} \sidmodels \phi~.
  \]
\end{theorem}
\begin{proof}
  See \Cref{app:types-capture-sat}.
\end{proof}
\noindent
\Cref{thm:types-capture-sat} immediately implies that, if the type of a state $\SH$ is equal to some already-known type of some other state $\SHpprime$ satisfying formula $\fa$, then
$\SH$ satisfies $\fa$.%---a crucial property for the correctness of our approach.
\begin{corollary}\label{cor:phi-typ-satisfies-phi}
    %Let $\SH$ be a state.
    If there is a type %$\typ$ such that 
    $\typ \in \phitypes{\eqclasses{\S}}{\fa}$
    with %and 
    $\sidtypeof{\S,\H} = \typ$, then $\SH \sidmodels \fa$.
\end{corollary}
%\begin{proof}
%  Since $\typ \in \phitypes{\eqclasses{\S}}{\fa}$, there exists a state $\SHpprime$ such that
%  (1) $\typ = \sidtypeof{\S',\H'}$,
%  (2) $\eqclasses{\S'} = \eqclasses{\S}$, and
%  (3) $\SHpprime \sidmodels \fa$.
%  Due to (2), we can assume w.l.o.g. that $\S = \S'$ (otherwise we can consider an isomorphic state $\SHpprime$ with this property).
%  Thus, $\SHprime \sidmodels \fa$ and $\sidtypeof{\S,\H'} = \typ = \sidtypeof{\S,\H}$.
%  Hence, by \cref{thm:types-capture-sat}, $\SH \sidmodels \fa$.
%\end{proof}
%
\noindent
Moreover, recall that $\SLIDguarded$ formulas are quantifier-free (although quantifiers may appear in predicate definitions).
As demonstrated below, this limitation is crucial for upholding \cref{thm:types-capture-sat}.
\begin{example}
  Recall $\SidLs$ from \cref{ex:sids}. Moreover, let $\SHi{k}$,
  $k \in \N$, be a list of length $k$ from $x_1$ to $x_2$. It then
  holds for all $i,j\geq 2$ that $\sidtypeof{\S,\H_i} =
  \sidtypeof{\S,\H_j}$. However,
  \begin{align*}
      \SHi{2} \not\sidmodels & \SHEX{\tuple{y_1,y_2}}
    \lseg(x_1,y_1)\sep\lseg(y_1,y_2)\sep\lseg(y_2,x_2),
      \text{ whereas, for all $j \geq 3$, } \\
  %\end{align*}
  %\begin{align*}
  \SHi{j} \sidmodels & \SHEX{\tuple{y_1,y_2}}
  \lseg(x_1,y_1)\sep\lseg(y_1,y_2)\sep\lseg(y_2,x_2).
  \end{align*}
  %holds for all $j \geq 3$.
  Hence, the refinement theorem
  does not hold if we admit quantifiers in $\SLIDguarded$ formulas.
\end{example}
\section{Algorithms for Computing Types}\label{ch:deciding-btw}
%\section{Reducing Satisfiability for GSL to Non-Emptiness of Types}\label{sec:deciding-btw:types-capture-sat}
%
As discussed in \cref{sec:types:understanding}, deciding whether a $\SLIDguarded$ formula $\fa$ is decidable boils down to computing finite sets of types $\phitypes{\scls}{\fa}$ for suitable aliasing constraints $\scls$.
We now present two algorithms for effectively computing $\phitypes{\scls}{\fa}$:
\cref{sec:deciding-btw:fixedpoint} deals with computing types of predicate calls defined by SIDs
and \cref{sec:deciding-btw:toplevel-types} shows how to compute types of $\SLIDguarded$ formulas, respectively.\footnote{Recall that the formulas in SIDs are symbolic heaps and \emph{not} $\SLIDguarded$ formulas; for example, they may contain quantifiers.}

%\subsection{The Refinement Theorem for Guarded Formulas}\label{sec:deciding-btw:types-capture-sat}
%\input{PartC/DecidingBTW-SatCapture}

\subsection{Computing the Types of Predicate Calls}\label{sec:deciding-btw:fixedpoint}
We first aim to compute, for every predicate $\pred \in \Preds{\Sid}$, the set of all $\scls$-types of $\pred$.
Specifically, for every aliasing constraint $\scls \in \EqClasses{\xxfvs{\pred}}$, where $\xx \subseteq \Var$ finite, we will compute
\[
  \PTypes{\scls}{\pred} \defn \phitypes{\scls}{\pred(\fvs{\pred})}.
\]
Once we have a way to compute these types, we can also compute types for
any $\SLIDguarded$ formula with free variables $\xx$, as we will see in
\cref{sec:deciding-btw:toplevel-types}.

\subsubsection{Assumptions}
Throughout this section, we fix a \ptrclosed{} SID $\Sid\in\IDbtw$ and a finite set of variables $\xx$; we assume w.l.o.g.~that $\xx \cap \fvs{\pred}=\emptyset$ for all predicates $\pred\in\Preds{\Sid}$.

\subsubsection{A Fixed-Point Algorithm for Computing the Types of   Predicates}\label{sec:deciding-btw:fixed-point}\index{type computation!fixed-point
  computation}

% This assumption will simplify reasoning about
% parameter instantiation by allowing us to disregard the possibility of
% double capture.
%
\begin{figure}[tb!]
  \centering
  \input{figtbl/PartC/fixedpoint-step-function}
  \caption[Algorithm: Computing the $\Sid$-types of SID rules]{Computing
    (a subset of) the $\Sid$-types of existentially-quantified symbolic heap $\phi \in \SymHeap$ for stacks with aliasing constraint $\scls$ under the assumption that $\thefun$ maps every predicate symbol $\pred$ and every aliasing constraint to (a subset of) the types $\PTypes{\scls}{\pred}$.

    Here, $\pinst{\scls}{\vec{u}}{\vec{v}}^{-1}$ denotes the addition of the variables $\vec{u}$ into the aliasing constraint $\scls$ such that the variables $\vec{u}$ are aliases of the variables $\vec{v}$ respectively; see \cref{def:reverse-ac} for details.
    Moreover, 
    we denote by $\eqrestr{\scls}{\vec{y}}$ the restriction of $\scls$ to the variables in $\vec{y}$, i.e.,
    $\eqrestr{\scls}{\vec{y}} \defn \scls \cap (\vec{y} \times \vec{y})$.
    }
  \label{fig:ptype-computation}
\end{figure}
We compute $\PTypes{\scls}{\pred}$ for all choices of $\scls$ and $\pred$ by a simultaneous fixed-point computation.
Specifically, our goal is to compute a (partial) function
$\thefun\colon \thefuntype$ that maps every predicate $\pred$
and every aliasing constraint $\scls \in \EqClasses{\xxfvs{\pred}}$ to the set of types $\PTypes{\scls}{\pred}$.
We start off the fixed-point computation with
$\thefun(\pred,\scls)=\emptyset$ for all $\pred$ and $\scls$;
each iteration adds to $\thefun$ some more types such that $\thefun(\pred,\scls) \subseteq \PTypes{\scls}{\pred}$;
and when we reach the fixed point, $\thefun(\pred,\scls) = \PTypes{\scls}{\pred}$ will hold for all
$\pred$ and $\scls$.
Each iteration amounts to applying the
function $\gls{ptypes-phi-Sigma}$ defined in
\cref{fig:ptype-computation} to all rule bodies $\phi\in\SymHeap$ of the SID $\Sid$ and all aliasing constraints $\scls$.
Here, $\thefun$ is the pre-fixed point from the previous iteration.
%We note that 
The function $\ptypesofKW$ operates on sets of types.
Hence, we need to lift $\ACompose$,
$\tinst{\cdot}{\cdot}{\cdot}{\cdot}$, $\ForgetKW$ and $\ExtendKW$ from types to sets of types in a point-wise manner, i.e.,
\begin{align*}
  & \set{\typi{1},\ldots,\typi{m}} \ACompose \set{\typi{1}',\ldots,\typi{n}'}
  \defn
    \set{\typi{i}
    \ACompose \typi{j}' \mid 1 \leq i \leq m, 1 \leq j \neq n, \typi{i}
  \ACompose \typi{j}' \neq \bot}, \\
  & \tinst{\set{\typi{1},\ldots,\typi{m}}}{\scls}{\vec{x}}{\vec{y}} \defn
    \set{\tinst{\typi{1}}{\scls}{\vec{x}}{\vec{y}},\ldots,\tinst{\typi{m}}{\scls}{\vec{x}}{\vec{y}}}, \\
  & \Forget{\scls,y}{\set{\typi{1},\ldots,\typi{m}}} \defn \set{\Forget{\scls,y}{\typi{1}},\ldots,\Forget{\scls,y}{\typi{m}}},     \text{ and } \\
  & \Extend{\scls}{\set{\typi{1},\ldots,\typi{m}}} \defn \set{\Extend{\scls}{\typi{1}},\ldots,\Extend{\scls}{\typi{m}}}.
\end{align*}
Further, $\ptypesofKW$ uses the following operation on aliasing constraints:
\begin{definition}[Reverse renaming of aliasing constraints]\label{def:reverse-ac}
  Let $\vec{x}$ be a sequence of pairwise distinct variables and let $\vec{y}$ be a sequence of (not necessarily pairwise distinct) variables with $|\vec{y}| = |\vec{x}|$.
  Moreover, let $\scls$ be an aliasing constraint with $\vec{x} \cap \dom(\scls) = \emptyset$ and $\vec{y} \subseteq \dom(\scls)$.
  Then, the \emph{reverse renaming} $\vec{x}$ to $\vec{y}$ in $\scls$ by is given by the aliasing constraint $\pinst{\scls}{\vec{x}}{\vec{y}}^{-1} \in \EqClasses{\dom(\scls)\cup\vec{x}}$
  defined by
  $$\pinst{\scls}{\vec{x}}{\vec{y}}^{-1} \defn \{(a_1,a_2) \mid \text{ there is } (b_1,b_2) \in \scls \text{ with } b_1 = \pinst{a_1}{\vec{x}}{\vec{y}} \text{ and } b_2 = \pinst{a_2}{\vec{x}}{\vec{y}}\}.
  $$
\end{definition}
Informally, the function $\ptypesof{\phi}{\scls}$ works as follows:
\begin{itemize}
\item If $\phi=\sleq{x}{y}$ or $\phi=\slneq{x}{y}$, we use the aliasing constraint $\scls$ to check whether the (dis)equality $\phi$ holds and then return either the type of the empty model or no type.
    This is justified because our semantics enforces that (dis)equalities only hold in the empty heap.
\item If $\phi=\pto{a}{\vec{b}}$, there is---up to isomorphism---only one state with aliasing constraint $\scls$ that satisfies $\phi$.
    We denote this state by $\ptrmodel{\scls}{\pto{a}{\vec{b}}}$ and return its type.
\item If $\phi = \pred(\vec{y})$, we look up the types of $\pred(\fvs{\pred})$ in the pre-fixed point $\thefun$ and then appropriately rename the formal parameters $\fvs{\pred}$ to the actual arguments $\vec{y}$: %. In more details:
    \begin{itemize}
      \item  For the look-up we use the aliasing constraint $\pinst{\scls}{\vec{z}}{\vec{y}}^{-1}$, which is obtained from the aliasing constraint $\scls$ by adding the formal parameters $\vec{z} \defn \fvs{\pred}$ to $\scls$ such that the $\vec{z}$ are aliases of the variables $\vec{y}$ respectively; see \cref{def:reverse-ac} for details.
      \item Crucially, we restrict $\pinst{\scls}{\vec{z}}{\vec{y}}^{-1}$ to the variables $\vec{x} \cup \vec{z}$ before we look up the types of $\pred(\fvs{\pred})$.
          This restriction guarantees that the computation of $\ptypesof{\phi}{\scls}$ does not diverge by considering larger and larger aliasing constraints in recursive calls.
          (An illustration of the problem as well as an argument why our solution does not lead to divergence can be found in \cref{sec:fixedpoint:complete}).
      \item After the loop-up we extend the types over aliasing constraint $\eqrestr{\pinst{\scls}{\vec{z}}{\vec{y}}^{-1}}{\vec{x}\cup\vec{z}}$ to types over aliasing constraint $\pinst{\scls}{\vec{z}}{\vec{y}}^{-1}$, undoing the earlier restriction.
      \item Finally, we rename the formal parameters $\vec{z}$ of the recursive call with the actual parameters $\vec{y}$ and obtain types over aliasing constraint $\scls$.
    \end{itemize}

\item If $\phi = \phi_1\sep\phi_2$, we apply the type composition operator developed in previous sections.
\item If $\phi = \SHEX{y}\phi'$, we consider all ways to extend the aliasing constraint $\scls$ with $y$ and recurse.
    Our treatment of predicate calls outlined above guarantees that this does not lead to divergence.
\end{itemize}

\paragraph*{Fixed Point Computation.}
We use the following wrapper for $\ptypesofKW$:
\begin{align*}
  & \theop\colon (\thefuntype)
  %&\phantom{\theop\colon}\;
    \to(\thefuntype), \\
  & \theop(\thefun) = \lambda (\pred, \scls) \ldotp \thefun(\pred,\scls) \cup
    \bigcup_{(\pred(\vec{y}) \Rule \phi) \in
    \Sid}\ptypesof{\scls}{\phi}.
\end{align*}

We observe that $\theop$ is a monotone function defined over a finite complete lattice:
  \begin{enumerate}
  \item The considered order $\subfun$ of $\thefuntype$ is the point-wise comparison of functions:
    \[f \mathrel{\gls{--rel-subfun}} g \defn \forall \pred \forall \scls
    \ldotp f(\pred,\scls) \subseteq g(\pred,\scls).\]
  \item $\thefuntype$ is a finite lattice because the image $\fpimg$ and the domain
    $\set{ \tuple{\pred,\scls} \mid \pred \in \Preds{\Sid},  \scls\in\EqClasses{\xxfvs{\pred}}}$ of the considered functions are finite.
  \item $\thefuntype$ is complete because the image $\fpimg$ of the considered functions is a complete lattice (the subset lattice over $\allsidtypes{\Sid}$).
\end{enumerate}
\noindent
Hence, $\theop$ has a least fixed point, which can be obtained by Kleene iteration:
$$
 \thelfp \defn \lim_{n \in \N} \prefixi{n}
$$
Moreover, since the lattice is finite, finitely many iterations suffice to reach the least fixed point.

\paragraph*{Correctness and Complexity.}
We analyze the correctness of our construction, i.e.,
\[ \text{for all}~\pred \in \Preds{\Sid}~\text{and}~\scls\in \EqClasses{\xxfvs{\pred}}.\quad \thelfp(\pred, \scls) =  \PTypes{\scls}{\pred}, \]
as well as its complexity in three steps, which can be found in~\ref{subec:correctness-fixed-point}:
\begin{enumerate}
\item We show $\thelfp(\pred, \scls) \subseteq
  \PTypes{\scls}{\pred}$.
   %in \cref{sec:fixedpoint:sound}.
\item We show $\thelfp(\pred, \scls) \supseteq
  \PTypes{\scls}{\pred}$.
   %in \cref{sec:fixedpoint:complete}.
\item We show that $\thelfp$ is computable in %doubly-exponential time, i.e., in 
    $2^{2^{\bigO(n^2 \log(n))}}$, where $n \defn \size{\Sid}+\size{\vec{x}}$.  
    %in \cref{sec:fixedpoint:correct}.
\end{enumerate}
%%% Local Variables:
%%% mode: latex
%%% TeX-master: "../Thesis"
%%% End:

% \section{Computing the Types of Arbitrary Guarded, Quantifier-Free
%   Formulas}\label{sec:deciding-btw:toplevel-types}
\subsection{Computing the Types of Guarded
  Formulas}\label{sec:deciding-btw:toplevel-types}
\index{type computation!of SLIDg formula@of $\SLIDguarded$ formula}

After we have established how to compute the types of predicate calls, we are now ready to
define a function $\gls{types-phi-Sigma}$ that computes the types of arbitrary $\SLIDguarded$ formulas $\phi$---i.e., quantifier-free guarded formulas---for some fixed stack-aliasing constraint $\scls$; the function is defined in~\cref{fig:ttype-computation}.

%on p.~\pageref{fig:ttype-computation}.

% The function $\ttypesofKW$ makes use of the auxiliary function
% $\extendscls{\scls}{\vec{y}}{\vec{z}}$, which extends $\scls$ by
% adding the $i$-th variable in $\vec{z}$ to the equivalence class of
% the $i$-th variable in $\vec{y}$ for all $i$.

% \begin{definition}[Extension of stack-aliasing
%   constraints]\index{extension!of stack-aliasing constraint}
%   Let $\vec{y}=\tuple{y_1,\ldots,y_k}$,
%   $\vec{z}=\tuple{z_1,\ldots,z_k}$ be (not necessarily
%   repetition-free) sequences of variables.
%   %
%   Let $\scls$ be a stack-aliasing constraint with
%   $\dom(\scls)\supseteq\vec{y}$ and
%   $\dom(\scls)\cap\vec{z}=\emptyset$.
%   %
%   The \emph{extension of $\scls$ with $\vec{y}=\vec{z}$},
%   $\extendscls{\scls}{\vec{y}}{\vec{z}}$, is the smallest
%   stack-aliasing constraint $\scls'$ with (1)
%   $\dom(\scls')=\dom(\scls)\cup\vec{z}$ and (2) $(y_i,z_i)\in\scls'$
%   for all $1\leq i \leq k$.
%   %
%   % \TODO{This is still quite ugly. Not sure yet if I want to spend time
%   % to improve this.}
% \end{definition}
%

\begin{figure}[tb!]
  \centering
  \input{figtbl/PartC/toplevel-function}
  \caption[Algorithm: Computing the $\Sid$-types of $\SLIDguarded$
  formulas]{Computation of $\Sid$-types for quantifier-free $\SLIDguarded$
    formula $\phi$ and stacks with aliasing
    constraint $\scls$.}
  \label{fig:ttype-computation}
\end{figure}

% \TODO{Add type computation for a simple example like
%   $(\ls(x_1,x_3)\sep\ls(x_3,x_2)) \wedge \ls(x_1,x_2)$? That nicely
%   illustrates the importance of the extra variables in the type
%   computation.}

\begin{theorem}[Correctness and Complexity of the Type Computation]
  \label{thm:ttypes-correct-complexity}
  %\label{thm:ttypes-correct}
  %\index{correctness!of Phi-type computation@of $\Sid$-type computation}
  %
  Let $\fa \in \SLIDguarded$ with $\fvs{\fa}=\vec{x}$ and
  $\values{\fa}\subseteq\{\nil\}$. Further, let $\scls\in\EqClasses{\vec{x}}$.
  Then, $\phitypes{\scls}{\fa} = \ttypesof{\fa}{\scls}$.
  Moreover, $\ttypesof{\fa}{\scls}$ can be computed in $2^{2^{\bigO(n^2 \log(n))}}$, where $n \defn \size{\Sid}+\size{\fa}$.
\end{theorem}

%
%\noindent
%Unsurprisingly, the asymptotic complexity of  $\ttypesof{\fa}{\scls}$ coincides with the asymptotic complexity of the fixed-point computation.

%\begin{theorem}[Complexity of type computation]\label{lem:ttypes-complexity}\index{complexity!of
%    Phi-type computation@of $\Sid$-type computation}
%  Let $\fa \in \SLIDguarded$ and let $\scls \in \EqClasses{\fvs{\fa}}$.
%  Moreover, let $n \defn \size{\Sid}+\size{\fa}$.
%  Then, $\ttypesof{\fa}{\scls}$ can be computed in $2^{2^{\bigO(n^2 \log(n))}}$.
%\end{theorem}
%
We now state the main result of this article:
\begin{theorem}[Decidability of $\SLIDguarded$]\label{thm:sat-decidable}
  \index{decidability!of SLIDg@of $\SLIDguarded$}
  Let $\fa \in \SLIDguarded$ and $n \defn \size{\Sid}+\size{\fa}$.
  It is decidable in time $2^{2^{\bigO(n^2 \log(n))}}$ whether $\fa$ is satisfiable.
\end{theorem}
\begin{proof}
  Let $\vec{x}\defn\fvs{\fa}$.
  Note that $\size{\vec{x}} \leq n$.
  The formula $\fa$ is satisfiable iff %if and only if 
    there exists a state $\SH$ with $\SH \sidmodels \fa$.
  %By~\cref{lem:pos-model-nonempty-type}
  By~\cref{cor:guarded-formula-non-empty-type},
  %we have 
  $\sidtypeof{\S,\H}\neq\emptyset$. % for such models $\SH$.
  Hence, it is sufficient to compute $\phitypes{\scls}{\phi}$ for all aliasing constraints $\scls$ with $\dom(\scls) = \vec{x}$ and check whether  $\phitypes{\scls}{\phi} \neq \emptyset$.

  By~\cref{thm:ttypes-correct-complexity} we can compute $\phitypes{\scls}{\fa} = \ttypesof{\fa}{\scls}$
  in $2^{2^{\bigO(n^2 \log(n))}}$ for a fixed aliasing constraints $\scls$.
  Since there are at most $n^n \in \bigO(2^{n \log(n)})$ stack-aliasing constraints, we can
  conclude that we can perform the satisfiability check in time
  $%\begin{align*}
    \bigO(2^{n \log(n)}) \cdot 2^{2^{\bigO(n^2 \log(n))}} ~=~ 2^{2^{\bigO(n^2 \log(n))}}%.
    %\tag*{\qedhere}
  $.%\end{align*}
\end{proof}
\noindent
Since the entailment query $\fa \sidmodels \fb$ is equivalent to checking the unsatisfiability of $\fa \wedge \neg \fb$, and the negation in $\fa \wedge \neg \fb$ is guarded, we obtain an entailment
checker with the same complexity:

\begin{corollary}[Decidability of entailment for
  $\SLIDguarded$]
\label{cor:gsl-entailment}
  %\index{decidability!of entailment for SLIDg@of entailment for $\SLIDguarded$}
  %
  Let $\fa,\fb \in \SLIDguarded$ and $n \defn
  \size{\Sid}+\size{\fa}+\size{\fb}$.
  The entailment problem $\fa \sidmodels \fb$ is decidable in time
  $2^{2^{\bigO(n^2 \log(n))}}$.
\end{corollary}
\begin{proof}
  If $\fa,\fb\in \SLIDguarded$, then $\fa \wedge \neg \fb \in \SLIDguarded$.
  The entailment $\fa \sidmodels \fb$ is valid iff $\fa \wedge \neg
  \fb$ is unsatisfiable.
  Since $\TwoExpTime$ is closed under complement, the claim follows from \cref{thm:sat-decidable}.
\end{proof}

\begin{example}
The entailments in \cref{ex:composition-projection} can be proven using our decision procedure.
    %\cref{cor:gsl-entailment}.
\end{example}

%\begin{example}
%The three entailments from \cref{ex:composition-projection} can be proven with the decision procedure from \cref{cor:gsl-entailment}.
%\end{example}
%
\noindent
Finally, our decision procedure is also applicable to (quantifier-free) symbolic heaps over inductive predicate definitions of bounded treewidth, because these formulas are always guarded.\footnote{Notice that only the formulas in the entailment query need to be quantifier-free; quantifiers are permitted in inductive definitions. Since we can use an arbitrary number of free variables at the top-level, this is a mild restriction.}
Hence, we also obtain a tighter complexity bound for the original decidability result of \citet{iosif2013tree}:
\begin{corollary}[Decidability of entailment for
  $\SLIDbtw$]
  Let $\fa,\fb \in \SLIDbtw$ be quantifier-free and $n \defn
  \size{\Sid}+\size{\fa}+\size{\fb}$.
  The entailment problem $\fa \sidmodels \fb$ is decidable in time
  $2^{2^{\bigO(n^2 \log(n))}}$.
\end{corollary}
\begin{proof}
  Follows from \cref{cor:gsl-entailment}, since
  every quantifier-free $\SLIDbtw$ formula is in $\SLIDguarded$.
  %Since every quantifier-free $\SLIDbtw$ formula is also a $\SLIDguarded$ formula,
  %the claim follows directly from 
  %  \cref{cor:gsl-entailment}.
\end{proof}

%%% Local Variables:
%%% mode: latex
%%% TeX-master: "../Thesis"
%%% End:

%%% Local Variables:
%%% mode: latex
%%% TeX-master: "../Thesis"
%%% End:

\section{Conclusion}\label{sec:conclusion}
We have given a unified and revised presentation of the decision procedures developed in \cite{katelaan2019effective,katelaan2020beyond}
covering
(1) the satisfiability of quantifier-free guarded separation logic and
(2) the entailment problem of (quantifier-free) symbolic heaps over SIDs of bounded treewidth.
%
%In particular, we have fixed the incompleteness issues of \cite{katelaan2019effective} and
%extended the original result beyond symbolic heaps while keeping the same (doubly-exponential) asymptotic complexity.
%As sketched in \cite{katelaan2020beyond}, we generalized the \emph{$\Sid$-profile} abstraction of \cite{katelaan2019effective} to the \emph{$\Sid$-type} abstraction by adding (limited) support for existential quantifiers
%(see \cref{ch:types}).
%We then used $\Sid$-types to develop a
%sound and complete, doubly-exponential decision procedure for guarded separation logic over the bounded treewidth fragment of separation
%logic (\cref{ch:deciding-btw}).
In particular, we have established a $\TwoExpTime$ upper bound for both problems.
A corresponding lower bound has been proven recently~\cite{echenim2019lower}.
Hence, we can conclude that our decision procedures have optimal computational complexity.

To the best of our knowledge, our decision procedure for $\SLIDguarded$ is the first decision procedure to support an SL fragment combining user-supplied inductive definitions, Boolean structure, and magic wands.
We obtained an almost tight delineation between decidability and undecidability:
We showed that any extension of $\SLIDguarded$ in which one of the guards is dropped, leads to 
undecidability.
%an undecidable fragment. %(\cref{sec:unguarded-undec}).

%One question remains: What is the fundamental difference between $\SLIDguarded$ and its
%less restrictive versions that explains the former's decidability and the latter's undecidability?
%%
%A close inspection of our undecidability proofs in \cref{sec:unguarded-undec} reveals that
%each proof ultimately relies specifying models that admit decompositions into two parts with unboundedly many dangling pointers.
%Specifically, we can split the CFG encoding into the induced word $w$ and the remainder; the word $w$ then introduces $\size{w}$ dangling pointers.
%%
%In contrast, we have shown in \cref{lem:pos-formula-pos-model}
%that every model of a $\SLIDguarded$ formula is itself guarded;
%by \cref{def:positive-model}, the number of dangling pointers is thus always bounded by the number of free variables of the formula.
%
%An interesting question for future research is whether the above observation allows generalizing $\SLIDbtw$ to a larger, yet decidable, SL fragment, where all models still have bounded treewidth.
We mention that recent follow-up work by~\citet{hal-03052687} generalizes the decidability of the entailment problem for $\SLIDbtw$ by weakening the establishment requirement. % of $\SLIDbtw$.
The result employs an abstraction inspired by the type abstraction presented in this article.
It is an interesting question whether this result can be lifted to guarded separation logic as well. 
Further, our undecidability results require an unbounded number of dangling pointers.
While~\cite{hal-03052687} supports classes of structures with unbounded treewidth, the entailment needs only to be checked for so-called normal structures of bounded treewidth.
It would be interesting to interpret this result in terms of the number of dangling pointers that need to be considered in order to understand whether a bounded number of dangling pointers in fundamental for decidability.

% Finally, it would be interesting to try and integrate the results of
% \cref{pt:ssl,pt:btw}, i.e., to investigate whether we can remove the
% guardedness restrictions if we assume strong-separation semantics.
% %
% This is not as trivial as adding the ``garbage components'' of the AMS
% abstraction to the $\Sid$-type abstraction, because there is, in
% general, no unique way to split a model of an $\IDbtw$ predicate into
% positive chunks and negative chunks (as defined in
% \cref{pt:ssl}).

%%% Local Variables:
%%% mode: latex
%%% TeX-master: "../Journal"
%%% End:
%

%\input{ArticleConclusion}
%\input{Journal/deposit}

%\bibliographystyle{plainnat}
\bibliographystyle{ACM-Reference-Format}
\bibliography{Bibliography,BibliographySepLog,MyPublications,Proceedings}


\begin{thebibliography}{52}

%%% the Web address.

\ifx \showCODEN    \undefined \def \showCODEN     #1{\unskip}     \fi
\ifx \showDOI      \undefined \def \showDOI       #1{#1}\fi
\ifx \showISBNx    \undefined \def \showISBNx     #1{\unskip}     \fi
\ifx \showISBNxiii \undefined \def \showISBNxiii  #1{\unskip}     \fi
\ifx \showISSN     \undefined \def \showISSN      #1{\unskip}     \fi
\ifx \showLCCN     \undefined \def \showLCCN      #1{\unskip}     \fi
\ifx \shownote     \undefined \def \shownote      #1{#1}          \fi
\ifx \showarticletitle \undefined \def \showarticletitle #1{#1}   \fi
\ifx \showURL      \undefined \def \showURL       {\relax}        \fi
% invisible to TeX
\providecommand\bibfield[2]{#2}
\providecommand\bibinfo[2]{#2}
\providecommand\natexlab[1]{#1}
\providecommand\showeprint[2][]{arXiv:#2}

\bibitem[\protect\citeauthoryear{Antonopoulos, Gorogiannis, Haa\-se, Kanovich,
  and Ouaknine}{Antonopoulos et~al\mbox{.}}{2014}]%
        {antonopoulos2014foundations}
\bibfield{author}{\bibinfo{person}{Timos Antonopoulos}, \bibinfo{person}{Nikos
  Gorogiannis}, \bibinfo{person}{Christoph Haa\-se}, \bibinfo{person}{Max~I.
  Kanovich}, {and} \bibinfo{person}{Jo{\"{e}}l Ouaknine}.}
  \bibinfo{year}{2014}\natexlab{}.
\newblock \showarticletitle{Foundations for Decision Problems in Separation
  Logic with General Inductive Predicates}. In
  \bibinfo{booktitle}{\emph{FOSSACS}}. \bibinfo{pages}{411--425}.
\newblock


\bibitem[\protect\citeauthoryear{Appel}{Appel}{2014}]%
        {appel2014program}
\bibfield{author}{\bibinfo{person}{Andrew~W. Appel}.}
  \bibinfo{year}{2014}\natexlab{}.
\newblock \bibinfo{booktitle}{\emph{Program Logics - for Certified Compilers}}.
\newblock \bibinfo{publisher}{Cambridge University Press}.
\newblock
\showISBNx{978-1-10-704801-0}


\bibitem[\protect\citeauthoryear{Bar-Hillel, Perles, and Shamir}{Bar-Hillel
  et~al\mbox{.}}{1961}]%
        {bar1961formal}
\bibfield{author}{\bibinfo{person}{Yehoshua Bar-Hillel}, \bibinfo{person}{Micha
  Perles}, {and} \bibinfo{person}{Eli Shamir}.}
  \bibinfo{year}{1961}\natexlab{}.
\newblock \showarticletitle{On formal properties of simple phrase structure
  grammars}.
\newblock \bibinfo{journal}{\emph{Sprachtypologie und Universalienforschung}}
  \bibinfo{volume}{14} (\bibinfo{year}{1961}), \bibinfo{pages}{143--172}.
\newblock


\bibitem[\protect\citeauthoryear{Batz, Kaminski, Katoen, Matheja, and
  Noll}{Batz et~al\mbox{.}}{2019}]%
        {batz2019quantitative}
\bibfield{author}{\bibinfo{person}{Kevin Batz},
  \bibinfo{person}{Benjamin~Lucien Kaminski}, \bibinfo{person}{Joost{-}Pieter
  Katoen}, \bibinfo{person}{Christoph Matheja}, {and} \bibinfo{person}{Thomas
  Noll}.} \bibinfo{year}{2019}\natexlab{}.
\newblock \showarticletitle{Quantitative separation logic: a logic for
  reasoning about probabilistic pointer programs}.
\newblock \bibinfo{journal}{\emph{{PACMPL}}} \bibinfo{volume}{3},
  \bibinfo{number}{{POPL}} (\bibinfo{year}{2019}),
  \bibinfo{pages}{34:1--34:29}.
\newblock


\bibitem[\protect\citeauthoryear{Berdine, Calcagno, Cook, Distefano, O'Hearn,
  Wies, and Yang}{Berdine et~al\mbox{.}}{2007}]%
        {berdine2007shape}
\bibfield{author}{\bibinfo{person}{Josh Berdine}, \bibinfo{person}{Cristiano
  Calcagno}, \bibinfo{person}{Byron Cook}, \bibinfo{person}{Dino Distefano},
  \bibinfo{person}{Peter~W. O'Hearn}, \bibinfo{person}{Thomas Wies}, {and}
  \bibinfo{person}{Hongseok Yang}.} \bibinfo{year}{2007}\natexlab{}.
\newblock \showarticletitle{Shape Analysis for Composite Data Structures}. In
  \bibinfo{booktitle}{\emph{CAV}}. \bibinfo{pages}{178--192}.
\newblock


\bibitem[\protect\citeauthoryear{Berdine, Calcagno, and O'Hearn}{Berdine
  et~al\mbox{.}}{2004}]%
        {berdine2004decidable}
\bibfield{author}{\bibinfo{person}{Josh Berdine}, \bibinfo{person}{Cristiano
  Calcagno}, {and} \bibinfo{person}{Peter~W. O'Hearn}.}
  \bibinfo{year}{2004}\natexlab{}.
\newblock \showarticletitle{A Decidable Fragment of Separation Logic}. In
  \bibinfo{booktitle}{\emph{{FSTTCS}}}. \bibinfo{pages}{97--109}.
\newblock


\bibitem[\protect\citeauthoryear{Berdine, Calcagno, and O'Hearn}{Berdine
  et~al\mbox{.}}{2005a}]%
        {berdine2005smallfoot}
\bibfield{author}{\bibinfo{person}{Josh Berdine}, \bibinfo{person}{Cristiano
  Calcagno}, {and} \bibinfo{person}{Peter~W. O'Hearn}.}
  \bibinfo{year}{2005}\natexlab{a}.
\newblock \showarticletitle{Smallfoot: Modular Automatic Assertion Checking
  with Separation Logic}. In \bibinfo{booktitle}{\emph{FMCO}}.
  \bibinfo{pages}{115--137}.
\newblock


\bibitem[\protect\citeauthoryear{Berdine, Calcagno, and O'Hearn}{Berdine
  et~al\mbox{.}}{2005b}]%
        {berdine2005symbolic}
\bibfield{author}{\bibinfo{person}{Josh Berdine}, \bibinfo{person}{Cristiano
  Calcagno}, {and} \bibinfo{person}{Peter~W. O'Hearn}.}
  \bibinfo{year}{2005}\natexlab{b}.
\newblock \showarticletitle{Symbolic Execution with Separation Logic}. In
  \bibinfo{booktitle}{\emph{APLAS}}. \bibinfo{pages}{52--68}.
\newblock


\bibitem[\protect\citeauthoryear{Berdine, Cook, and Ishtiaq}{Berdine
  et~al\mbox{.}}{2011}]%
        {berdine2011slayer}
\bibfield{author}{\bibinfo{person}{Josh Berdine}, \bibinfo{person}{Byron Cook},
  {and} \bibinfo{person}{Samin Ishtiaq}.} \bibinfo{year}{2011}\natexlab{}.
\newblock \showarticletitle{SLAyer: Memory Safety for Systems-Level Code}. In
  \bibinfo{booktitle}{\emph{CAV}}. \bibinfo{pages}{178--183}.
\newblock


\bibitem[\protect\citeauthoryear{Blom and Huisman}{Blom and Huisman}{2015}]%
        {blom2015witnessing}
\bibfield{author}{\bibinfo{person}{Stefan Blom} {and} \bibinfo{person}{Marieke
  Huisman}.} \bibinfo{year}{2015}\natexlab{}.
\newblock \showarticletitle{Witnessing the elimination of magic wands}.
\newblock \bibinfo{journal}{\emph{Int. J. Softw. Tools Technol. Transf.}}
  \bibinfo{volume}{17}, \bibinfo{number}{6} (\bibinfo{year}{2015}),
  \bibinfo{pages}{757--781}.
\newblock


\bibitem[\protect\citeauthoryear{Brochenin, Demri, and Lozes}{Brochenin
  et~al\mbox{.}}{2012}]%
        {brochenin2012almighty}
\bibfield{author}{\bibinfo{person}{R{\'{e}}mi Brochenin},
  \bibinfo{person}{St{\'{e}}phane Demri}, {and} \bibinfo{person}{{\'{E}}tienne
  Lozes}.} \bibinfo{year}{2012}\natexlab{}.
\newblock \showarticletitle{On the almighty wand}.
\newblock \bibinfo{journal}{\emph{Inf. Comput.}}  \bibinfo{volume}{211}
  (\bibinfo{year}{2012}), \bibinfo{pages}{106--137}.
\newblock


\bibitem[\protect\citeauthoryear{Brotherston}{Brotherston}{2007}]%
        {brotherston2007formalised}
\bibfield{author}{\bibinfo{person}{James Brotherston}.}
  \bibinfo{year}{2007}\natexlab{}.
\newblock \showarticletitle{Formalised Inductive Reasoning in the Logic of
  Bunched Implications}. In \bibinfo{booktitle}{\emph{SAS}}
  \emph{(\bibinfo{series}{LNCS})},
  \bibfield{editor}{\bibinfo{person}{Hanne~Riis Nielson} {and}
  \bibinfo{person}{Gilberto Fil{\'{e}}}} (Eds.), Vol.~\bibinfo{volume}{4634}.
  \bibinfo{publisher}{Springer}, \bibinfo{pages}{87--103}.
\newblock


\bibitem[\protect\citeauthoryear{Brotherston, Distefano, and dahl
  Petersen}{Brotherston et~al\mbox{.}}{2011}]%
        {brotherston2011automated}
\bibfield{author}{\bibinfo{person}{James Brotherston}, \bibinfo{person}{Dino
  Distefano}, {and} \bibinfo{person}{Rasmus~Lerche\ dahl Petersen}.}
  \bibinfo{year}{2011}\natexlab{}.
\newblock \showarticletitle{Automated Cyclic Entailment Proofs in Separation
  Logic}. In \bibinfo{booktitle}{\emph{{CADE-23}}}. \bibinfo{pages}{131--146}.
\newblock


\bibitem[\protect\citeauthoryear{Brotherston, Fuhs, P{\'{e}}rez, and
  Gorogiannis}{Brotherston et~al\mbox{.}}{2014}]%
        {brotherston2014decision}
\bibfield{author}{\bibinfo{person}{James Brotherston}, \bibinfo{person}{Carsten
  Fuhs}, \bibinfo{person}{Juan Antonio~Navarro P{\'{e}}rez}, {and}
  \bibinfo{person}{Nikos Gorogiannis}.} \bibinfo{year}{2014}\natexlab{}.
\newblock \showarticletitle{A decision procedure for satisfiability in
  separation logic with inductive predicates}. In
  \bibinfo{booktitle}{\emph{CSL-LICS}}. \bibinfo{pages}{25:1--25:10}.
\newblock


\bibitem[\protect\citeauthoryear{Calcagno and Distefano}{Calcagno and
  Distefano}{2011}]%
        {calcagno2011infer}
\bibfield{author}{\bibinfo{person}{Cristiano Calcagno} {and}
  \bibinfo{person}{Dino Distefano}.} \bibinfo{year}{2011}\natexlab{}.
\newblock \showarticletitle{Infer: An Automatic Program Verifier for Memory
  Safety of {C} Programs}. In \bibinfo{booktitle}{\emph{{NFM}}}.
  \bibinfo{pages}{459--465}.
\newblock


\bibitem[\protect\citeauthoryear{Calcagno, Distefano, Dubreil, Gabi,
  Hooimeijer, Luca, O'Hearn, Papakonstantinou, Purbrick, and
  ma~Rodriguez}{Calcagno et~al\mbox{.}}{2015}]%
        {calcagno2015moving}
\bibfield{author}{\bibinfo{person}{Cristiano Calcagno}, \bibinfo{person}{Dino
  Distefano}, \bibinfo{person}{J{\'{e}}r{\'{e}}my Dubreil},
  \bibinfo{person}{Dominik Gabi}, \bibinfo{person}{Pieter Hooimeijer},
  \bibinfo{person}{Martino Luca}, \bibinfo{person}{Peter~W. O'Hearn},
  \bibinfo{person}{Irene Papakonstantinou}, \bibinfo{person}{Jim Purbrick},
  {and} \bibinfo{person}{Dul\ ma Rodriguez}.} \bibinfo{year}{2015}\natexlab{}.
\newblock \showarticletitle{Moving Fast with Software Verification}. In
  \bibinfo{booktitle}{\emph{{NFM}}}. \bibinfo{pages}{3--11}.
\newblock


\bibitem[\protect\citeauthoryear{Calcagno, Distefano, O'Hearn, and
  Yang}{Calcagno et~al\mbox{.}}{2006}]%
        {calcagno2006beyond}
\bibfield{author}{\bibinfo{person}{Cristiano Calcagno}, \bibinfo{person}{Dino
  Distefano}, \bibinfo{person}{Peter~W. O'Hearn}, {and}
  \bibinfo{person}{Hongseok Yang}.} \bibinfo{year}{2006}\natexlab{}.
\newblock \showarticletitle{Beyond Reachability: Shape Abstraction in the
  Presence of Pointer Arithmetic}. In \bibinfo{booktitle}{\emph{SAS}}.
  \bibinfo{pages}{182--203}.
\newblock


\bibitem[\protect\citeauthoryear{Calcagno, Distefano, O'Hearn, and
  Yang}{Calcagno et~al\mbox{.}}{2011}]%
        {calcagno2011compositional}
\bibfield{author}{\bibinfo{person}{Cristiano Calcagno}, \bibinfo{person}{Dino
  Distefano}, \bibinfo{person}{Peter~W. O'Hearn}, {and}
  \bibinfo{person}{Hongseok Yang}.} \bibinfo{year}{2011}\natexlab{}.
\newblock \showarticletitle{Compositional Shape Analysis by Means of
  Bi-Abduction}.
\newblock \bibinfo{journal}{\emph{J. {ACM}}} \bibinfo{volume}{58},
  \bibinfo{number}{6} (\bibinfo{year}{2011}), \bibinfo{pages}{26:1--26:66}.
\newblock


\bibitem[\protect\citeauthoryear{Calcagno, O'Hearn, and Yang}{Calcagno
  et~al\mbox{.}}{2007}]%
        {calcagno2007local}
\bibfield{author}{\bibinfo{person}{Cristiano Calcagno},
  \bibinfo{person}{Peter~W. O'Hearn}, {and} \bibinfo{person}{Hongseok Yang}.}
  \bibinfo{year}{2007}\natexlab{}.
\newblock \showarticletitle{Local Action and Abstract Separation Logic}. In
  \bibinfo{booktitle}{\emph{LICS}}. \bibinfo{pages}{366--378}.
\newblock


\bibitem[\protect\citeauthoryear{Calcagno, Yang, and O'He\-arn}{Calcagno
  et~al\mbox{.}}{2001}]%
        {calcagno2001computability}
\bibfield{author}{\bibinfo{person}{Cristiano Calcagno},
  \bibinfo{person}{Hongseok Yang}, {and} \bibinfo{person}{Peter~W. O'He\-arn}.}
  \bibinfo{year}{2001}\natexlab{}.
\newblock \showarticletitle{Computability and Complexity Results for a Spatial
  Assertion Language for Data Structures}. In
  \bibinfo{booktitle}{\emph{APLAS}}. \bibinfo{pages}{289--300}.
\newblock


\bibitem[\protect\citeauthoryear{Chin, David, Nguyen, and Qin}{Chin
  et~al\mbox{.}}{2012}]%
        {chin2012automated}
\bibfield{author}{\bibinfo{person}{Wei{-}Ngan Chin}, \bibinfo{person}{Cristina
  David}, \bibinfo{person}{Huu~Hai Nguyen}, {and} \bibinfo{person}{Shengchao
  Qin}.} \bibinfo{year}{2012}\natexlab{}.
\newblock \showarticletitle{Automated verification of shape, size and bag
  properties via user-defined predicates in separation logic}.
\newblock \bibinfo{journal}{\emph{Sci. Comput. Program.}} \bibinfo{volume}{77},
  \bibinfo{number}{9} (\bibinfo{year}{2012}), \bibinfo{pages}{1006--1036}.
\newblock


\bibitem[\protect\citeauthoryear{Cook, Haase, Ouaknine, Parkinson, and
  Worrell}{Cook et~al\mbox{.}}{2011}]%
        {cook2011tractable}
\bibfield{author}{\bibinfo{person}{Byron Cook}, \bibinfo{person}{Christoph
  Haase}, \bibinfo{person}{Jo{\"{e}}l Ouaknine}, \bibinfo{person}{Matthew~J.
  Parkinson}, {and} \bibinfo{person}{James Worrell}.}
  \bibinfo{year}{2011}\natexlab{}.
\newblock \showarticletitle{Tractable Reasoning in a Fragment of Separation
  Logic}. In \bibinfo{booktitle}{\emph{{CONCUR}}}. \bibinfo{pages}{235--249}.
\newblock


\bibitem[\protect\citeauthoryear{Courcelle and Engelfriet}{Courcelle and
  Engelfriet}{2012}]%
        {courcelle2012graph}
\bibfield{author}{\bibinfo{person}{Bruno Courcelle} {and}
  \bibinfo{person}{Joost Engelfriet}.} \bibinfo{year}{2012}\natexlab{}.
\newblock \bibinfo{booktitle}{\emph{Graph Structure and Monadic Second-Order
  Logic - {A} Language-Theoretic Approach}}. \bibinfo{series}{Encyclopedia of
  mathematics and its applications}, Vol.~\bibinfo{volume}{138}.
\newblock \bibinfo{publisher}{Cambridge University Press}.
\newblock


\bibitem[\protect\citeauthoryear{Diestel}{Diestel}{2016}]%
        {diestel2016graph}
\bibfield{author}{\bibinfo{person}{Reinhard Diestel}.}
  \bibinfo{year}{2016}\natexlab{}.
\newblock \bibinfo{booktitle}{\emph{Graph Theory, 5th Edition}}.
  \bibinfo{series}{Graduate texts in mathematics}, Vol.~\bibinfo{volume}{173}.
\newblock \bibinfo{publisher}{Springer}.
\newblock
\showISBNx{978-3-662-53621-6}


\bibitem[\protect\citeauthoryear{Echenim, Iosif, and Peltier}{Echenim
  et~al\mbox{.}}{2020a}]%
        {echenim2020bernays}
\bibfield{author}{\bibinfo{person}{Mnacho Echenim}, \bibinfo{person}{Radu
  Iosif}, {and} \bibinfo{person}{Nicolas Peltier}.}
  \bibinfo{year}{2020}\natexlab{a}.
\newblock \showarticletitle{The Bernays-Sch{\"{o}}nfinkel-Ramsey Class of
  Separation Logic with Uninterpreted Predicates}.
\newblock \bibinfo{journal}{\emph{{ACM} Trans. Comput. Log.}}
  \bibinfo{volume}{21}, \bibinfo{number}{3} (\bibinfo{year}{2020}),
  \bibinfo{pages}{19:1--19:46}.
\newblock


\bibitem[\protect\citeauthoryear{Echenim, Iosif, and Peltier}{Echenim
  et~al\mbox{.}}{2020b}]%
        {echenim2019lower}
\bibfield{author}{\bibinfo{person}{Mnacho Echenim}, \bibinfo{person}{Radu
  Iosif}, {and} \bibinfo{person}{Nicolas Peltier}.}
  \bibinfo{year}{2020}\natexlab{b}.
\newblock \showarticletitle{Entailment Checking in Separation Logic with
  Inductive Definitions is 2-EXPTIME hard}.
\newblock   \bibinfo{volume}{73} (\bibinfo{year}{2020}),
  \bibinfo{pages}{191--211}.
\newblock


\bibitem[\protect\citeauthoryear{Echenim, Iosif, and Peltier}{Echenim
  et~al\mbox{.}}{2021}]%
        {hal-03052687}
\bibfield{author}{\bibinfo{person}{Mnacho Echenim}, \bibinfo{person}{Radu
  Iosif}, {and} \bibinfo{person}{Nicolas Peltier}.}
  \bibinfo{year}{2021}\natexlab{}.
\newblock \showarticletitle{{Decidable Entailments in Separation Logic with
  Inductive Definitions: Beyond Establishment}}. In
  \bibinfo{booktitle}{\emph{CSL}} \emph{(\bibinfo{series}{LIPICS})}.
\newblock


\bibitem[\protect\citeauthoryear{Enea, Leng{\'{a}}l, Sighireanu, and
  Vojnar}{Enea et~al\mbox{.}}{2017}]%
        {enea2017spen}
\bibfield{author}{\bibinfo{person}{Constantin Enea}, \bibinfo{person}{Ondrej
  Leng{\'{a}}l}, \bibinfo{person}{Mihaela Sighireanu}, {and}
  \bibinfo{person}{Tom{\'{a}}s Vojnar}.} \bibinfo{year}{2017}\natexlab{}.
\newblock \showarticletitle{{SPEN:} {A} Solver for Separation Logic}. In
  \bibinfo{booktitle}{\emph{{NFM}}}. \bibinfo{pages}{302--309}.
\newblock


\bibitem[\protect\citeauthoryear{et~al.}{et~al.}{2019}]%
        {sighireanu2019slcomp}
\bibfield{author}{\bibinfo{person}{Mihaela~Sighireanu et al.}}
  \bibinfo{year}{2019}\natexlab{}.
\newblock \showarticletitle{{SL-COMP:} Competition of Solvers for Separation
  Logic}. In \bibinfo{booktitle}{\emph{TACAS}}. \bibinfo{pages}{116--132}.
\newblock


\bibitem[\protect\citeauthoryear{Gotsman, Berdine, Cook, and Sagiv}{Gotsman
  et~al\mbox{.}}{2007}]%
        {gotsman2007threadmodular}
\bibfield{author}{\bibinfo{person}{Alexey Gotsman}, \bibinfo{person}{Josh
  Berdine}, \bibinfo{person}{Byron Cook}, {and} \bibinfo{person}{Mooly Sagiv}.}
  \bibinfo{year}{2007}\natexlab{}.
\newblock \showarticletitle{Thread-modular shape analysis}. In
  \bibinfo{booktitle}{\emph{PLDI}}. \bibinfo{pages}{266--277}.
\newblock


\bibitem[\protect\citeauthoryear{Hopcroft, Motwani, and Ullman}{Hopcroft
  et~al\mbox{.}}{2007}]%
        {hopcroft2007introduction}
\bibfield{author}{\bibinfo{person}{John~E. Hopcroft}, \bibinfo{person}{Rajeev
  Motwani}, {and} \bibinfo{person}{Jeffrey~D. Ullman}.}
  \bibinfo{year}{2007}\natexlab{}.
\newblock \bibinfo{booktitle}{\emph{Introduction to automata theory, languages,
  and computation, 3rd Edition}}.
\newblock \bibinfo{publisher}{Addison-Wesley}.
\newblock


\bibitem[\protect\citeauthoryear{Iosif, Rogalewicz, and Sim{\'{a}}cek}{Iosif
  et~al\mbox{.}}{2013}]%
        {iosif2013tree}
\bibfield{author}{\bibinfo{person}{Radu Iosif}, \bibinfo{person}{Adam
  Rogalewicz}, {and} \bibinfo{person}{Jir{\'i} Sim{\'{a}}cek}.}
  \bibinfo{year}{2013}\natexlab{}.
\newblock \showarticletitle{The Tree Width of Separation Logic with Recursive
  Definitions}. In \bibinfo{booktitle}{\emph{{CADE-24}}}.
  \bibinfo{pages}{21--38}.
\newblock


\bibitem[\protect\citeauthoryear{Iosif, Rogalewicz, and Vojnar}{Iosif
  et~al\mbox{.}}{2014}]%
        {iosif2014deciding}
\bibfield{author}{\bibinfo{person}{Radu Iosif}, \bibinfo{person}{Adam
  Rogalewicz}, {and} \bibinfo{person}{Tom{\'{a}}s Vojnar}.}
  \bibinfo{year}{2014}\natexlab{}.
\newblock \showarticletitle{Deciding Entailments in Inductive Separation Logic
  with Tree Automata}. In \bibinfo{booktitle}{\emph{ATVA}}.
  \bibinfo{pages}{201--218}.
\newblock


\bibitem[\protect\citeauthoryear{Ishtiaq and O'Hearn}{Ishtiaq and
  O'Hearn}{2001}]%
        {ishtiaq2001bi}
\bibfield{author}{\bibinfo{person}{Samin~S. Ishtiaq} {and}
  \bibinfo{person}{Peter~W. O'Hearn}.} \bibinfo{year}{2001}\natexlab{}.
\newblock \showarticletitle{{BI} as an Assertion Language for Mutable Data
  Structures}. In \bibinfo{booktitle}{\emph{{POPL}}}. \bibinfo{pages}{14--26}.
\newblock


\bibitem[\protect\citeauthoryear{Jacobs, Smans, Philippaerts, Vogels,
  Penninckx, and Piessens}{Jacobs et~al\mbox{.}}{2011}]%
        {jacobs2011verifast}
\bibfield{author}{\bibinfo{person}{Bart Jacobs}, \bibinfo{person}{Jan Smans},
  \bibinfo{person}{Pieter Philippaerts}, \bibinfo{person}{Fr{\'{e}}d{\'{e}}ric
  Vogels}, \bibinfo{person}{Willem Penninckx}, {and} \bibinfo{person}{Frank
  Piessens}.} \bibinfo{year}{2011}\natexlab{}.
\newblock \showarticletitle{VeriFast: {A} Powerful, Sound, Predictable, Fast
  Verifier for {C} and Java}. In \bibinfo{booktitle}{\emph{{NFM}}}.
  \bibinfo{pages}{41--55}.
\newblock


\bibitem[\protect\citeauthoryear{Jansen, Katelaan, Matheja, mas Noll, and
  Zuleger}{Jansen et~al\mbox{.}}{2017}]%
        {jansen2017unified}
\bibfield{author}{\bibinfo{person}{Christina Jansen}, \bibinfo{person}{Jens
  Katelaan}, \bibinfo{person}{Christoph Matheja}, \bibinfo{person}{Tho\ mas
  Noll}, {and} \bibinfo{person}{Florian Zuleger}.}
  \bibinfo{year}{2017}\natexlab{}.
\newblock \showarticletitle{Unified Reasoning About Robustness Properties of
  Symbolic-Heap Separation Lo\-gic}. In \bibinfo{booktitle}{\emph{ESOP}}.
  \bibinfo{pages}{611--638}.
\newblock


\bibitem[\protect\citeauthoryear{Jung, Krebbers, Jourdan, Bizjak, Birkedal, and
  Dreyer}{Jung et~al\mbox{.}}{2018}]%
        {jung2018iris}
\bibfield{author}{\bibinfo{person}{Ralf Jung}, \bibinfo{person}{Robbert
  Krebbers}, \bibinfo{person}{Jacques{-}Henri Jourdan}, \bibinfo{person}{Ales
  Bizjak}, \bibinfo{person}{Lars Birkedal}, {and} \bibinfo{person}{Derek
  Dreyer}.} \bibinfo{year}{2018}\natexlab{}.
\newblock \showarticletitle{Iris from the ground up: {A} modular foundation for
  higher-order concurrent separation logic}.
\newblock \bibinfo{journal}{\emph{J. Funct. Program.}}  \bibinfo{volume}{28}
  (\bibinfo{year}{2018}), \bibinfo{pages}{e20}.
\newblock


\bibitem[\protect\citeauthoryear{Katelaan, Matheja, and Zuleger}{Katelaan
  et~al\mbox{.}}{2019}]%
        {katelaan2019effective}
\bibfield{author}{\bibinfo{person}{Jens Katelaan}, \bibinfo{person}{Christoph
  Matheja}, {and} \bibinfo{person}{Florian Zuleger}.}
  \bibinfo{year}{2019}\natexlab{}.
\newblock \showarticletitle{Effective Entailment Checking for Separation Logic
  with Inductive Definitions}. In \bibinfo{booktitle}{\emph{TACAS}}.
  \bibinfo{pages}{319--336}.
\newblock


\bibitem[\protect\citeauthoryear{Katelaan and Zuleger}{Katelaan and
  Zuleger}{2020}]%
        {katelaan2020beyond}
\bibfield{author}{\bibinfo{person}{Jens Katelaan} {and}
  \bibinfo{person}{Florian Zuleger}.} \bibinfo{year}{2020}\natexlab{}.
\newblock \showarticletitle{Beyond Symbolic Heaps: Deciding Separation Logic
  With Inductive Definitions}. In \bibinfo{booktitle}{\emph{{LPAR}}}
  \emph{(\bibinfo{series}{EPiC Series in Computing})},
  Vol.~\bibinfo{volume}{73}. \bibinfo{publisher}{EasyChair},
  \bibinfo{pages}{390--408}.
\newblock


\bibitem[\protect\citeauthoryear{Le, Tatsuta, Sun, and Chin}{Le
  et~al\mbox{.}}{2017}]%
        {le2017decidable}
\bibfield{author}{\bibinfo{person}{Quang~Loc Le}, \bibinfo{person}{Makoto
  Tatsuta}, \bibinfo{person}{Jun Sun}, {and} \bibinfo{person}{Wei{-}Ngan
  Chin}.} \bibinfo{year}{2017}\natexlab{}.
\newblock \showarticletitle{A Decidable Fragment in Separation Logic with
  Inductive Predicates and Arithmetic}. In \bibinfo{booktitle}{\emph{CAV}}.
  \bibinfo{pages}{495--517}.
\newblock


\bibitem[\protect\citeauthoryear{Matheja}{Matheja}{2020}]%
        {matheja2020automated}
\bibfield{author}{\bibinfo{person}{Christoph Matheja}.}
  \bibinfo{year}{2020}\natexlab{}.
\newblock \emph{\bibinfo{title}{{A}utomated reasoning and randomization in
  separation logic}}.
\newblock Dissertation. \bibinfo{school}{RWTH Aachen University}.
\newblock


\bibitem[\protect\citeauthoryear{M{\"{u}}ller, Schwerhoff, and
  Summers}{M{\"{u}}ller et~al\mbox{.}}{2017}]%
        {muller2017viper}
\bibfield{author}{\bibinfo{person}{Peter M{\"{u}}ller}, \bibinfo{person}{Malte
  Schwerhoff}, {and} \bibinfo{person}{Alexander~J. Summers}.}
  \bibinfo{year}{2017}\natexlab{}.
\newblock \showarticletitle{Viper: {A} Verification Infrastructure for
  Permission-Based Reasoning}.
\newblock In \bibinfo{booktitle}{\emph{Dependable Software Systems
  Engineering}}. \bibinfo{pages}{104--125}.
\newblock


\bibitem[\protect\citeauthoryear{Pagel, Matheja, and Zuleger}{Pagel
  et~al\mbox{.}}{2020}]%
        {journals/corr/abs-2002-01202}
\bibfield{author}{\bibinfo{person}{Jens Pagel}, \bibinfo{person}{Christoph
  Matheja}, {and} \bibinfo{person}{Florian Zuleger}.}
  \bibinfo{year}{2020}\natexlab{}.
\newblock \showarticletitle{Complete Entailment Checking for Separation Logic
  with Inductive Definitions}.
\newblock \bibinfo{journal}{\emph{CoRR}}  \bibinfo{volume}{abs/2002.01202}
  (\bibinfo{year}{2020}).
\newblock
\showeprint[arxiv]{2002.01202}


\bibitem[\protect\citeauthoryear{Piskac, Wies, and Zufferey}{Piskac
  et~al\mbox{.}}{2013}]%
        {piskac2013automating}
\bibfield{author}{\bibinfo{person}{Ruzica Piskac}, \bibinfo{person}{Thomas
  Wies}, {and} \bibinfo{person}{Damien Zufferey}.}
  \bibinfo{year}{2013}\natexlab{}.
\newblock \showarticletitle{Automating Separation Logic Using {SMT}}. In
  \bibinfo{booktitle}{\emph{CAV}}. \bibinfo{pages}{773--789}.
\newblock


\bibitem[\protect\citeauthoryear{Piskac, Wies, and Zufferey}{Piskac
  et~al\mbox{.}}{2014a}]%
        {piskac2014automating}
\bibfield{author}{\bibinfo{person}{Ruzica Piskac}, \bibinfo{person}{Thomas
  Wies}, {and} \bibinfo{person}{Damien Zufferey}.}
  \bibinfo{year}{2014}\natexlab{a}.
\newblock \showarticletitle{Automating Separation Logic with Trees and Data}.
  In \bibinfo{booktitle}{\emph{CAV}}. \bibinfo{pages}{711--728}.
\newblock


\bibitem[\protect\citeauthoryear{Piskac, Wies, and Zufferey}{Piskac
  et~al\mbox{.}}{2014b}]%
        {piskac2014grasshopper}
\bibfield{author}{\bibinfo{person}{Ruzica Piskac}, \bibinfo{person}{Thomas
  Wies}, {and} \bibinfo{person}{Damien Zufferey}.}
  \bibinfo{year}{2014}\natexlab{b}.
\newblock \showarticletitle{G\-R\-A\-S\-S\-hopper - Complete Heap Verification
  with Mixed Specifications}. In \bibinfo{booktitle}{\emph{TACAS}}.
  \bibinfo{pages}{124--139}.
\newblock


\bibitem[\protect\citeauthoryear{Reynolds}{Reynolds}{2002}]%
        {reynolds2002separation}
\bibfield{author}{\bibinfo{person}{John~C. Reynolds}.}
  \bibinfo{year}{2002}\natexlab{}.
\newblock \showarticletitle{Separation Logic: {A} Logic for Shared Mutable Data
  Structures}. In \bibinfo{booktitle}{\emph{LICS}}. \bibinfo{pages}{55--74}.
\newblock


\bibitem[\protect\citeauthoryear{Schwerhoff and Summers}{Schwerhoff and
  Summers}{2015}]%
        {schwerhoff2015lightweight}
\bibfield{author}{\bibinfo{person}{Malte Schwerhoff} {and}
  \bibinfo{person}{Alexander~J. Summers}.} \bibinfo{year}{2015}\natexlab{}.
\newblock \showarticletitle{Light\-weight Support for Magic Wands in an
  Automatic Verifier}. In \bibinfo{booktitle}{\emph{ECOOP}}.
  \bibinfo{pages}{614--638}.
\newblock


\bibitem[\protect\citeauthoryear{Ta, Le, Khoo, and Chin}{Ta
  et~al\mbox{.}}{2016}]%
        {ta2016automated}
\bibfield{author}{\bibinfo{person}{Quang{-}Trung Ta},
  \bibinfo{person}{Ton~Chanh Le}, \bibinfo{person}{Siau{-}Cheng Khoo}, {and}
  \bibinfo{person}{Wei{-}Ngan Chin}.} \bibinfo{year}{2016}\natexlab{}.
\newblock \showarticletitle{Automated Mutual Explicit Induction Proof in
  Separation Logic}. In \bibinfo{booktitle}{\emph{{FM}}}
  \emph{(\bibinfo{series}{LNCS})}, Vol.~\bibinfo{volume}{9995}.
  \bibinfo{pages}{659--676}.
\newblock


\bibitem[\protect\citeauthoryear{Ta, Le, Khoo, and Chin}{Ta
  et~al\mbox{.}}{2018}]%
        {ta2018automated}
\bibfield{author}{\bibinfo{person}{Quang{-}Trung Ta},
  \bibinfo{person}{Ton~Chanh Le}, \bibinfo{person}{Siau{-}Cheng Khoo}, {and}
  \bibinfo{person}{Wei{-}Ngan Chin}.} \bibinfo{year}{2018}\natexlab{}.
\newblock \showarticletitle{Automated lemma synthesis in sym\-bo\-lic-heap
  separation logic}.
\newblock \bibinfo{journal}{\emph{{PACMPL}}} \bibinfo{volume}{2},
  \bibinfo{number}{{POPL}} (\bibinfo{year}{2018}), \bibinfo{pages}{9:1--9:29}.
\newblock


\bibitem[\protect\citeauthoryear{Thakur, Breck, and Reps}{Thakur
  et~al\mbox{.}}{2014}]%
        {thakur2014satisfiability}
\bibfield{author}{\bibinfo{person}{Aditya~V. Thakur}, \bibinfo{person}{Jason
  Breck}, {and} \bibinfo{person}{Thomas~W. Reps}.}
  \bibinfo{year}{2014}\natexlab{}.
\newblock \showarticletitle{Satisfiability modulo abstraction for separation
  logic with linked lists}. In \bibinfo{booktitle}{\emph{SPIN}}.
  \bibinfo{pages}{58--67}.
\newblock


\bibitem[\protect\citeauthoryear{Yang}{Yang}{2001}]%
        {yang2001local}
\bibfield{author}{\bibinfo{person}{Hongseok Yang}.}
  \bibinfo{year}{2001}\natexlab{}.
\newblock \emph{\bibinfo{title}{Local Reasoning for Stateful Programs}}.
\newblock \bibinfo{thesistype}{Ph.D. Dissertation}. \bibinfo{school}{University
  of Illinois at Urbana-Champaign}, \bibinfo{address}{Champaign, IL, USA}.
\newblock Advisor(s) Reddy, Uday S.
\newblock
\showISBNx{0-493-35008-X}
\newblock
\shownote{AAI3023240.}


\end{thebibliography}

\appendix
\section{Appendix}
\label{sec:appendix}

\subsection{Formal definition of substitution}
\label{app:substitution}

For $\vec{y} = \tuple{y_1,\ldots,y_k}$ and $\vec{z} = \tuple{z_1,\ldots,z_k}$, the substitution $\pinst{\phi}{\vec{y}}{\vec{z}}$ is defined by the table below.
Since quantified variables can be renamed before performing a substitution, we assume w.l.o.g. that $\vec{y}$ contains no variables that are bound by a quantifier in $\phi$.

\begin{tabular}{l@{\qquad}l@{\qquad}r}
    \hline\hline
    $\phi$ & $\pinst{\phi}{\vec{y}}{\vec{z}}$ \\
    \hline\hline
    $y_i$ & $z_i$  & $1 \leq i \leq k$ \\
    $u$ & $u$  & $u \notin \vec{y}$ \\
    $\tuple{v_1,\ldots,v_n}$ & $\tuple{\pinst{v_1}{\vec{y}}{\vec{z}}, \ldots, \pinst{v_n}{\vec{y}}{\vec{z}}}$ \\
    $\emp$ & $\emp$ \\
    $\sleq{u}{v}$ & $\sleq{\pinst{u}{\vec{y}}{\vec{z}}}{\pinst{v}{\vec{y}}{\vec{z}}}$ \\
    $\slneq{u}{v}$ & $\slneq{\pinst{u}{\vec{y}}{\vec{z}}}{\pinst{v}{\vec{y}}{\vec{z}}}$ \\
    $\pto{u}{\vec{v}}$ & $\pto{\pinst{u}{\vec{y}}{\vec{z}}}{\pinst{\vec{v}}{\vec{y}}{\vec{z}}}$ \\
    $\pred(\vec{x})$ & $\pred(\pinst{\vec{x}}{\vec{y}}{\vec{z}})$ \\
    $\neg \psi$ & $\neg (\pinst{\psi}{\vec{y}}{\vec{z}})$ \\
    $\psi \oplus \theta$
    & $\pinst{\psi}{\vec{y}}{\vec{z}} \oplus \pinst{\theta}{\vec{y}}{\vec{z}}$
    & $\oplus \in \set{\wedge,\vee,\sep,\mw}$ \\
    $\SHEX{x} \psi$ & $\SHEX{x} \pinst{\psi}{\vec{y}}{\vec{z}}$ & $x \notin \vec{y}$ \\
    $\FAO{x} \psi$ & $\FAO{x} \pinst{\psi}{\vec{y}}{\vec{z}}$ & $x \notin \vec{y}$ \\
    %$\SHEX{x} \psi$ & $\SHEX{\pinst{x}{\vec{y}}{\vec{z}}} \pinst{\psi}{\vec{y}}{\vec{z}}$ & $\pinst{x}{\vec{y}}{\vec{z}} \in \Var$ \\
    %$\FAO{x} \psi$ & $\FAO{\pinst{x}{\vec{y}}{\vec{z}}} \pinst{\psi}{\vec{y}}{\vec{z}}$ & $\pinst{x}{\vec{y}}{\vec{z}} \in \Var$ \\
    \hline
    \hline
%    %
%    %
%    \pinst{\SHEX{e} \phi}{\vec{y}}{\vec{z}}  & \defn &
%                                               \SHEX{\pinst{e}{\vec{y}}{\vec{z}}} \pinst{\phi}{\vec{y}}{\vec{z}}
\end{tabular}

\subsection{Proof of \cref{lem:predicate-pos-model}}
\label{app:predicate-pos-model}

\paragraph{Claim}
For all predicates $\pred\in\Preds{\Sid}$ and all states $\SH$, we have
  \[ \SH \sidmodels \pred(\vec{x})  \quad\text{implies}\quad \SH \in \Mpos{\Sid}. \]

\begin{proof}
  Let $\SH$ be a state such that $\SH \sidmodels \pred(\vec{x})$ for some predicate   $\pred\in\Preds{\Sid}$.
  The proof proceeds by strong mathematical induction on the number of rule applications needed to establish $\SH \sidmodels \pred(\vec{x})$:

  According to the semantics,
  there is a rule $(\pred(\vec{x}) \Rule \fa) \in \Sid$, for some $\fa = \SHEX{\vec{e}} \fa'$ with $\fa'=\ppto{y}{\vec{z}} \sep \pred_1(\vec{z_1}) \sep \cdots \sep \pred_k(\vec{z_k}) \sep \Pure$,
  $\Pure$ pure (note that for $k=0$ there are no recursive rule applications).
  Because of $\SH \sidmodels \pred(\vec{x})$,
  there exists a stack
  $\S' = \pinst{\S}{\vec{e}}{\vec{v}}$ for some values $\vec{v}$ such that $\SHpair{\S'}{\H} \models \fa'$.
  Thus, there are heaps $\H_0,\H_1,\ldots,\H_k$ with
  $\H = \H_0 \stdunion \H_1 \stdunion \cdots \stdunion \H_k$, $\H_0 \models \ppto{y}{\vec{z}}$ and
  $\SHpair{\S'}{\H_i} \models \pred_i(\vec{z_i})$ for all $1 \le i \le k$.
  By I.H., we have
  $\SHpair{\S'}{\H_i} \in \Mpos{\Sid}$.
  Hence, $\danglinglocs{\H_0} \subseteq \img(\S')$ for all $0 \le i \le k$.
    By the definition of establishment (see \cref{sec:sl-basics:btw}), we have
  $\S'(\vec{e}) \subseteq \dom(\H) \cup \S(\vec{x}) \cup \set{0}$.
    Because of $\H_i \subseteq \H$ and $\danglinglocs{\H} = \values{\H} \setminus (\dom(\H) \cup \{\nil\})$ we get that
  $\danglinglocs{\H_i} \subseteq \S(\vec{x})$.
    Hence, $\SH \in \Mpos{\Sid}$ because
  \begin{align*}
    \danglinglocs{\H} ~=~ &
    \danglinglocs{\H_0 \stdunion \H_1 \stdunion \cdots \stdunion \H_k} \\
      ~\subseteq~ &
  \danglinglocs{\H_0} \cup \danglinglocs{\H_1} \cup \cdots \cup \danglinglocs{\H_k} \\
      ~\subseteq~ & \S(\vec{x})  = \img(\S)~.
      \tag*{\qedhere}
  \end{align*}
\end{proof}

\subsection{Proof of \cref{lem:pos-formula-pos-model}}\label{app:pos-formula-pos-model}
\paragraph{Claim}
  For all $\phi \in \SLIDguarded$ and all states $\SH$, we have
  \[ \SH \sidmodels \phi \quad\text{implies}\quad \SH \in \Mpos{\Sid}. \]
\begin{proof}
  Let $\phi \in \SLIDguarded$ be a guarded formula and let $\SH$ be a state with $\SH \sidmodels \phi$.
  The proof proceeds by structural induction on $\phi$:

  \begin{description}
  \item[Case $\phi=\emp$, $\phi=\sleq{x}{y}$,$\slneq{x}{y}$:]
      Clearly, there are no dangling pointers in the empty heap.
  \item[Case $\phi=\pto{x}{\vec{y}}$.]
      Immediate because of $\danglinglocs{\H} \subseteq \S(\vec{y}) \subseteq \img(\S)$.
  \item[Case $\phi=\pred(\vec{x})$.] By~\cref{lem:predicate-pos-model}.
  \item[Case $\phi= \phi_1\sep\phi_2$.]
    Since $\SH \sidmodels \phi$, there exist heaps $\H_1, \H_2$ with $\H = \H_1 \stdunion \H_2$
    and $\SHi{i}\sidmodels\phi_i$.
    By I.H., we have $\SHi{i} \in \Mpos{\Sid}$, i.e.,
    $\danglinglocs{\H_i} \subseteq \img(\S)$.
    Thus,
    $\danglinglocs{\H} =
    \danglinglocs{\H_1 \stdunion \H_2}
    \subseteq
    \danglinglocs{\H_1} \cup \danglinglocs{\H_2} \subseteq \img(\S)$.
    Hence, $\SH \in \Mpos{\Sid}$.
  \item[Case $\phi = \phi_1\wedge\phi_2$.]
    %, where $\phi_2$ is either a
    %guarded formula or of the form $\neg \psi$, $\psi_1\sept\psi_2$,
    %or $\psi_1\mw\psi_2$.
    %In all these cases, $\SH \sidmodels \phi_1$,
    By the semantics of $\wedge$, this in particular means $\SH \sidmodels \phi_1$.
    By I.H., it then follows that $\SH \in \Mpos{\Sid}$.
    Notice that this case covers all standard conjunctions including the guarded negation, the guarded magic wand, and the guarded septraction.
  \item[Case $\phi = \phi_1 \vee \phi_2$.]
    Assume w.l.o.g.~that $\SH \sidmodels \phi_1$. By I.H., we have $\SH \in \Mpos{\Sid}$.
    \qedhere
  \end{description}
\end{proof}

\subsection{Proof of \cref{lem:guarded-iterated-star-predicates}}
\label{app:guarded-iterated-star-predicates}
\paragraph{Claim}
  Let $\phi \in \SLIDguarded$ be a guarded formula with $\fvs{\phi} = \xx$.
  Then, for every state $\SH \sidmodels \phi$, there are predicates $\pred_i \in \Preds{\Sid}$ and variables $\vec{z_i} \subseteq \xx$ such that $\SH \sidmodels \IteratedStar_{1\leq i\leq k}\pred_i(\vec{z_i})$.

\begin{proof}
    By structural induction on $\phi$.
    %For the remaining cases, consider the following:

  \begin{description}
  \item[Case $\phi=\emp$.]
      This case is immediate because, for $k = 0$,
      $\emp$ coincides with $\IteratedStar_{1 \leq i \leq k} \ldots$.
  \item[Case $\phi=\sleq{x}{y}$.]
      Since, in our semantics, the equality $\sleq{x}{y}$ entails $\emp$,
      we immediately obtain $\SH \in \Mpos{\Sid}$.
      The case for disequalities $\slneq{x}{y}$ is analogous.
  \item[Case $\phi=\pto{x}{\vec{y}}$.] Since $\Sid$ is \ptrclosed{},
      there exists a predicate $\pred \in \Sid$ such that $\SH \sidmodels \pred(x \concat \vec{y})$.
  \item[Case $\phi=\pred(\vec{x})$.] Clearly, the claim holds.
  \item[Case $\phi= \phi_1\sep\phi_2$.]
    Since $\SH \sidmodels \phi$, there exist domain-disjoint heaps $\H_1, \H_2$ with $\H = \H_1 \cup \H_2$
    such that $\SHi{1}\sidmodels\phi_1$ and $\SHi{2}\sidmodels\phi_2$.
    By the induction hypothesis, there exist predicate calls
    $\pred_{1,1}(\vec{x_1}),\ldots,\pred_{1,m}(\vec{x_m})$ and
    $\pred_{2,1}(\vec{y_1}),\ldots,\pred_{2,n}(\vec{y_n})$ such that
    \begin{align*}
      \SHi{1}\sidmodels\IteratedStar_{1\leq i \leq
      m}\pred_{1,i}(\vec{x_i}) \quad\text{and}\quad
    \SHi{2}\sidmodels\IteratedStar_{1\leq j \leq
      n}\pred_{2,j}(\vec{y_j}).
    \end{align*}
    The semantics of the separating conjunction $\sep$ and the fact $\H = \H_1 \cup \H_2$ then yield
    \[\SH \sidmodels \IteratedStar_{1\leq i \leq
        m}\pred_{1,i}(\vec{x_i}) \sep \IteratedStar_{1\leq j \leq
        n}\pred_{2,j}(\vec{y_j}).\]
  \item[Case $\phi = \phi_1\wedge\phi_2$.]
    %, where $\phi_2$ is either a
    %guarded formula or of the form $\neg \psi$, $\psi_1\sept\psi_2$,
    %or $\psi_1\mw\psi_2$.
    %In all these cases, $\SH \sidmodels \phi_1$,
    By the semantics of $\wedge$, this in particular means $\SH \sidmodels \phi_1$.
   By the induction hypothesis, the claim then holds for $\SH$ and $\phi_1$.
    Notice that this case covers all standard conjunctions including the guarded negation, the guarded magic wand, and the guarded septraction.
  \item[Case $\phi = \phi_1 \vee \phi_2$.]
    Assume w.l.o.g.~that $\SH \sidmodels \phi_1$. By the induction hypothesis, the claim then holds for $\SH$ and $\phi_1$. \qedhere
  \end{description}
\end{proof}

\subsection{Proof of \Cref{lem:g-to-sid}}
\label{app:g-to-sid}

\paragraph{Claim (Completeness of the encoding)}
  Let $\G=\theG$ and let $\Sid$ be the corresponding SID encoding. Let
  $1 \leq i \leq 2$, $x_1,x_2,x_3\in\Var$, and let
  $w \in \Lang{\G}$. Then there exists a model $\SH$ of
  $\Start(x_1,x_2,x_3)$ with
  $\inducedN{\Start}{\S,\H,x_2,x_3}=w$.

\begin{proof}
  We show the stronger claim that, for all $x_1,x_2,x_3\in\Var$,
  $w \in \Lang{\G}$, and $N \in \NTerm$, if $N \RuleSteps w$, then
  there exists a model $\SH$ of $N(x_1,x_2,x_3)$ with
  $\inducedN{N}{\S,\H,x_2,x_3}=w$.
  We proceed by mathematical induction on the number $m$ of
  $\RuleStep$ steps in a (minimal-length) derivation $N \RuleSteps w$.

  If $m=1$, $w=a_i$ for some $1 \leq i \leq n$ and there exists a rule
  $\GRule{N}{a_i}$.
  Let $\SH$ be a model of $\SHEX{a}\ppto{x_1}{\tuple{x_3,a}} \sep
  \letter{i}(a) \sep \sleq{x_1}{x_2}$.
  Note that this is a rule of the predicate $N$, so it holds that
  $\SH \sidmodels N(x_1,x_2,x_3)$.
  Moreover, $\inducedN{N}{\S,\H,x_2,x_3}=a_i$.
  If $m>1$, there exists a rule $\GRule{N}{AB}$ such that $N \RuleStep
  AB \RuleSteps w$.
  Then, there exist words $w_A,w_B$ with $w=w_A\concat
  w_B$, $A \RuleSteps w_A$, and $B \RuleSteps w_B$.

  Observe that both of the above derivations consist of strictly fewer
  than $m$ steps.
  Now, fix some variables $l,m,r \in \Var$.
  By I.H., there exist states $(\S_1,\H_1)$ and $(\S_2,\H_2)$ such that
  \begin{itemize}
  \item $\SHpair{\S_1}{\H_1} \sidmodels A(l,x_2,m)$ and
    $\inducedN{A}{\S_1,\H_1,x_2,m}=w_A$ as well as
  \item $\SHpair{\S_2}{\H_2} \sidmodels B(r,m,x_3)$ and $\inducedN{B}{\S_2,\H_2,m,x_3}=w_B$.
  \end{itemize}
  %
  %Let $v \in \Val$.
  %
  Assume w.l.o.g.~that (1) $\dom(\S_1)\cap\dom(\S_2)=m$, (2)
  $\S_1(m)=\S_2(m)$, (3) and $\H_1 \stdunion \H_2 \neq \bot$; if this is not the case,
  simply replace $\SHpair{\S_1}{\H_1}$ and $\SHpair{\S_2}{\H_2}$ with
  appropriate isomorphic models.
  We choose some location $k \in \Loc$ such that
  $k \notin \values{\H_1\stdunion\H_2}$.

  Let $\S \defn \S_1\cup\S_2\cup\set{x_1 \mapsto k}$ and $\H \defn \H_1 \cup
  \H_2 \cup \set{ k \mapsto \tuple{\S(l),\S(r)} }$.
  We obtain
  $\SH \sidmodels \ppto{x_1}{\tuple{l,r}} \sep A(l,x_2,m) \sep
  B(r,m,x_3)$ and thus also
  $\SH \sidmodels \SHEX{\tuple{l,r,m}} \ppto{x_1}{\tuple{l,r}} \sep
  A(l,x_2,m) \sep B(r,m,x_3)$. By definition of $\Sid$, we conclude
  $\SH\sidmodels N(x_1,x_2,x_3)$.
  Furthermore, observe that
  \begin{align*}
 \inducedN{N}{\S,\H,x_2,x_3}=&\inducedN{A}{\S,\H_1,x_2,m}\concat\inducedN{N}{\S,\H_2,m,x_3}\\=&w_A\concat
  w_B = w. \qedhere
  \end{align*}
\end{proof}

\subsection{Proof of \Cref{lem:sid-to-g}}
\label{app:sid-to-g}

\paragraph{Claim (Soundness of the encoding)}
  Let $\G=\theG$ and let $\Sid$ be the corresponding SID encoding. Let
  $x_1,x_2,x_3\in\Var$ and let
  $\SH\sidmodels \Start(x_1,x_2,x_3)$. Then
  $\inducedN{\Start}{\S,\H,x_2,x_3} \in \Lang{\G}$.

\begin{proof}
  We show the stronger claim that for all $x_1,x_2,x_3\in\Var$, all
  models $\SH$, and all $N \in \NTerm$, if
  $\SH\sidmodels N(x_1,x_2,x_3)$ then
  $N \RuleSteps \inducedN{N}{\S,\H,x_2,x_3}$.
  Observe that $\H$ is a tree overlaid with a linked list. We
  proceed by mathematical induction on the height $h$ of the tree in
  $\H$.

  If $h=0$, then $\Sid$ contains a rule
    \[ N(x_1,x_2,x_3) \Rule \SHEX{a}\ppto{x_1}{\tuple{x_3,a}} \sep
    \letter{k}(a) \sep \sleq{x_1}{x_2} \]
  whose right-hand side is satisfied by $\SH$. %, i.e.,
  %$\SH \sidmodels \SHEX{a}\ppto{x_1}{\tuple{x_3,a}} \sep \letter{k}(a)
  %\sep \sleq{x_1}{x_2}$.
  %
  Then $\inducedN{N}{\S,\H,x_2,x_3}=a_k$.
  By definition of $\Sid$, this implies
  $\GRule{N}{a_k}\in\Rules_1\cup\Rules_2$ and, consequently,
  $N \RuleStep a_k$. Hence, $N \RuleSteps a_k = \inducedN{N}{\S,\H,x_2,x_3}$.

  If $h>0$, there exists a rule $(N(x_1,x_2,x_3) \Rule \psi)\in\Sid$
  such that $\SH \sidmodels \psi$, where $\psi$ is of the form
  \[ \SHEX{l,r,m} \ppto{x_1}{\tuple{l,r}} \sep A(l,x_2,m) \sep
    B(r,m,x_3). \]
  Recall that by definition of $\Sid$, we have $\GRule{N}{AB}\in\Rules_1\cup\Rules_2$ \qquad \tagA.

  By the semantics of $\exists$ and $\sep$ there then are a stack
  $\S'$ with $\dom(\S')=\dom(\S)\cup\set{l,r,m}$ and heaps
  $\H_0,\H_A,\H_B$ such that $\H=\H_0\stdunion\H_A\stdunion\H_B$,
  $\SHpair{\S'}{\H_0}\sidmodels \ppto{x_1}{\tuple{l,r}}$,
  $\SHpair{\S'}{\H_A}\sidmodels A(l,x_2,m)$, and $\SHpair{\S'}{\H_B}\sidmodels B(r,m,x_3)$.
  Note that the height of the trees in $\H_A$ and $\H_B$ is at most
  $h-1$, so we can apply the induction hypotheses for these models to
  obtain
    \[ A \RuleSteps \inducedN{A}{\S',\H_1,x_2,m} \qquad\text{and}\qquad
    B \RuleSteps \inducedN{B}{\S',\H_2,m,x_3}. \]
  Together with \tagA, we derive
  \begin{align*}
  N &\RuleStep AB \\ &\RuleSteps \inducedN{A}{\S',\H_1,x_2,m} \concat
  \inducedN{B}{\S',\H_2,m,x_3} \\&=
  \inducedN{N}{\S',\H,x_2,x_3}\\&=\inducedN{N}{\S,\H,x_2,x_3}.\qedhere
  \end{align*}
\end{proof}

\subsection{Proof of \Cref{thm:undec-true}}
\label{app:undec-true}

\paragraph{Claim}
  The satisfiability problem for the fragment $\SLop{\wedge,\sep,\true}$ is undecidable.

\begin{proof}
  Let $\Sid$ be the encoding of the CFGs
  $\GA=\theGi{1}$ and $\GB=\theGi{2}$ as described in \cref{sec:unguarded-undec}.
  Moreover, consider the $\SLop{\wedge,\sep,\true}$ formula
  \[
     \phi \defn (\cfgA(a,x,y) \sep \true) \wedge (\cfgB(b,x,y) \sep \true).
  \]
  We claim that $\phi$ is satisfiable iff $\Lang{\GA}\cap\Lang{\GB}\neq\emptyset$; both implications are proven separately:

    If $\phi$ is satisfiable, there exists a state $\SH$ with $\SH \sidmodels \phi$.
    By \cref{lem:g-heap-has-word-heap},
    there exist heaps $\H_{w_1}, \H_{w_2} \subseteq \H$ such that
    $\inducedN{\cfgi}{\S,\H,x,y} = \inducedW{\S,\H_{w_i},x,y}$ for $i \in \{1,2\}$.

    Observe that both $\SHi{w_1}\sidmodels\word(x,y)$ and
    $\SHi{w_2}\sidmodels\word(x,y)$.
    Hence, $\H_{w_1} = \H_{w_2}$ and thus
    \[ w\defn\inducedN{\cfgB}{\S,\H,x,y} = \inducedN{\cfgA}{\S,\H,x,y}.\]
    By \cref{lem:sid-to-g}, we have $w \in \Lang{\GA}$ and $w
    \in \Lang{\GB}$, i.e., $w \in \Lang{\GA}\cap\Lang{\GB}$.

    Conversely, assume $\Lang{\GA}\cap\Lang{\GB}\neq\emptyset$.
    Then there exists a word $w \in \Lang{\GA}\cap\Lang{\GB}$.
    By \cref{lem:g-to-sid}, there exist states $\SHi{1},\SHi{2}$ with
    $\SHi{1}\sidmodels\cfgA(a,x,y)$ and
    $\SHi{2}\sidmodels\cfgB(b,x,y)$.
    Let $\H_{w_1}\subseteq\H_1,\H_{w_2}\subseteq\H_2$ be the unique
    heaps with
    $\inducedN{\cfgA}{\S,\H_1,x,y} = \inducedW{\S,\H_{w_1},x,y} = w
          = \inducedW{\S,\H_{w_2},x,y} = \inducedN{\cfgB}{\S,\H_2,x,y}$.

        Observe that $\SHi{w_1} \Iso \SHi{w_2}$ (see \cref{def:isomorphic-states}).
        Consequently, we can reason about an isomorphic state
        in which we replace $\H_2$ by a heap
        that contains $\H_{w_1}$ (as opposed to $\H_{w_2}$) as
        sub-heap and is otherwise disjoint from $\H_1$.
        That is, there
        exists a heap $\H_2'$ such that $(\S,\H_2)\Iso(\S,\H_2')$,
        $\values{\H_2'}\cap\values{\H_1}=\values{\H_{w_1}}$, and
        $\inducedN{\cfgB}{\S,\H_2',x,y}=\inducedW{\S,\H_{w_1},x,y} =
        w$.
        In particular,
        $\SHpair{\S}{\H_2'}\sidmodels\cfgB(b,x,y)$, because isomorphic states
        satisfy the same formulas (\cref{lem:sid:iso-models-same-formulas}).
    Now let $\H \defn \H_1 \cup \H_2'$ be the (non-disjoint) union of
    $\H_1$ and $\H_2'$.
    Since $\H_1\subseteq\H$ and $\SHi{1}\sidmodels\cfgA(a,x,y)$, we
    have $\SH \sidmodels \cfgA(a,x,y) \sep \true$; and similarly,
    since $\H_2'\subseteq\H$ and
    $\SHpair{\S}{\H_2'}\sidmodels\cfgB(a,x,y)$, we have that
    $\SH \sidmodels \cfgB(b,x,y)$. Consequently,
    $\SH \sidmodels \phi$.
\end{proof}

\subsection{Proof of \cref{lem:wordb-correct}}
\label{app:wordb-correct}

\paragraph{Claim}
  Let $\G_2=\theGi{2}$ be the CFG fixed in \cref{sec:unguarded-undec}.
  Moreover, let $\Sid$ be the corresponding SID encoding,
  $\wordB(x,y) \defn (\word(x,y) \sept \cfgB(a,x,y)) \sept \cfgB(a,x,y)$,
  and let $\SH$ be a state. Then  $\SH \sidmodels \wordB(x,y)$ iff
  $\SH\sidmodels\word(x,y)$ and $\inducedW{\S,\H,x,y} \in \Lang{\GB}$.

\begin{proof}
  Assume $\SH \sidmodels \wordB(x,y)$.
  By the semantics of $\sept$, there exists a heap $\H_1$ with
  $\SHi{1} \sidmodels \word(x,y) \sept \cfgB(a,x,y)$ such that
  $\SHpair{\S}{\H\stdunion\H_1} \sidmodels \cfgB(a,x,y)$.
  Observe that $\H_1$ contains precisely the inner nodes of
  $\SHpair{\S}{\H\stdunion\H_1}$, i.e., everything \emph{except} the
  part of the state that induces the word.
  Consequently, $\H$ is the part of the state that induces the word,
  i.e.,
  $\inducedN{\cfgB}{\S,\H\stdunion\H_1,x,y}=\inducedW{\S,\H,x,y}$ and
  $\SH \sidmodels \word(x,y)$.
  \cref{lem:sid-to-g} then yields
  $\inducedW{\S,\H,x,y}\in\Lang{\GB}$.

  Conversely, assume a state $\SH$ be such that
  $w \defn \inducedW{\S,\H,x,y} \in \Lang{\GB}$.
  As a consequence of \cref{lem:g-to-sid}, there exists a heap $\H_1$
  with $\SHpair{\S}{\H\stdunion\H_1} \sidmodels \cfgB(a,x,y)$.
  Because $\SH \sidmodels\word(x,y)$
  by assumption, the semantics of $\sept$ yields that
  $\SHi{1}\sidmodels \word(x,y) \sept \cfgB(a,x,y)$.
  Because $\SHpair{\S}{\H\stdunion\H_1} \sidmodels \cfgB(a,x,y)$, we
  obtain by the semantics of $\sept$ that
  $\SH \sidmodels (\word(x,y) \sept \cfgB(a,x,y)) \sept \cfgB(a,x,y)$.
\end{proof}

\subsection{Proof of \Cref{thm:undec-sept}}
\label{app:undec-sept}

\paragraph{Claim}
The satisfiability problem of $\SLop{\sept}$ is undecidable.

\begin{proof}
  We claim that
    $\psi \defn \wordB(x,y) \sept \cfgA(a,x,y)$ is satisfiable iff
    $\Lang{\GA}\cap\Lang{\GB}\neq\emptyset$.
%  %
%    Intuitively, this holds because $\psi$ is satisfiable iff it is
%    possible to replace the ``word part'' of a model of $\cfgA(a,x,y)$
%    with the ``word part'' of a model of $\cfgB(b,x,y)$.  Let us
%    formalize this intuition.

  Assume $\psi$ is satisfiable, i.e., there exists a state $\SH \sidmodels \psi$.
  By the semantics of $\sept$, there exists a
  heap $\H_0\subseteq\H$ with $\H_0 \sidmodels \wordB(x,y)$ and
  $\SHpair{\S}{\H\stdunion\H_0}\sidmodels \cfgA(a,x,y)$.
  As $\inducedW{\S,\H_0,x,y} \in \Lang{\GB}$, by
  \cref{lem:wordb-correct}, we have that
  $\SHi{0}\sidmodels\word(x,y)$.
  It follows that $\H_0$ is the unique sub-heap of
  $\H\stdunion\H_0$ with
  $\inducedN{\cfgA}{\S,\H\stdunion\H_0}=\inducedW{\S,\H_0,x,y}$.
  By \cref{lem:sid-to-g}, $\inducedW{\S,\H_0,x,y} \in
  \Lang{\GA}$. Together with \cref{lem:wordb-correct}, we thus have
  that $\inducedW{\S,\H_0,x,y} \in \Lang{\GA}\cap\Lang{\GB}$.

  Conversely, assume there exists a word
  $w\in\Lang{\GA}\cap\Lang{\GB}$.
  As shown in the proof of \cref{thm:undec-true}, there exist states $\SHi{1},\SHi{2},\SH$ with   $\SHi{1}\sidmodels\cfgA(a,x,y)$,  $\SHi{2}\sidmodels\cfgB(b,x,y)$
  and $\values{\H_1}\cap\values{\H_2}=\values{\H}$ such that
    \[\inducedN{\cfgA}{\S,\H_1,x,y} = \inducedN{\cfgB}{\S,\H_2,x,y} = \inducedW{\S,\H,x,y} = w~.\]

  %
%  By \cref{lem:g-to-sid}, there exist heaps $\H,\H_0,\H',\H_0'$ with
%  \begin{align*}
%  &\SH\sidmodels\cfgA(a,x,y),\\
%  &\inducedN{\cfgA}{\S,\H,x,y}=\inducedW{\S,\H_0,x,y},\\
%  &\SHprime\sidmodels\cfgB(a,x,y), \text{ and}\\
%  &\inducedN{\cfgB}{\S,\H',x,y}=\inducedW{\S,\H_0',x,y}.
%  \end{align*}
%  %
%  Since $\inducedW{\S,\H_0,x,y}=\inducedW{\S,\H_0',x,y}$, it holds
%  that $\H_0\Iso\H_0'$, so we can assume w.l.o.g.~that
%  $\H_0=\H_0'$---if the states are not isomorphic, simply replace
%  $\H'$ with an appropriate isomorphic heap to establish this
%  property.
%
  Let $\H_0 \subseteq \H_1$ be the sub-heap of $\H_1$ with $\H\stdunion\H_0=\H_1$.
  By \cref{lem:wordb-correct}, $\SH \sidmodels \wordB(x,y)$.
  Consequently, $\SHi{0}\sidmodels \psi$, i.e., $\psi$ is satisfiable.
\end{proof}

\subsection{Proof of \cref{lem:fsplit-unique}}
\label{app:fsplit-unique}
\paragraph{Claim}
For all $\vec{l} \subseteq \Loc$, every $\Sid$-forest has a unique $\vec{l}$-split $\gls{splitfl}$.
    \begin{proof}
      Let $\frst$ be a $\Sid$-forest with $\fgraph{\frst} = \tuple{\VDG,\EDG}$.
      Moreover, consider the graph
      \[
        \DG\defn \tuple{\VDG, \EDG \setminus \set{ \tuple{a,b} \mid a \in \Loc, b
          \in \vec{l}}}.
      \]
      Since $\tgraph{\frst}$ is a forest and
      $\mathcal{G} \subseteq \tgraph{\frst}$, $\mathcal{G}$ is a forest,
        i.e., all connected components $\mathcal{C}_1,\ldots,\mathcal{C}_k$ of $\mathcal{G}$ are trees.
      %
      %Hence, the connected components induce a $\Sid$-forest $\barfrst$:
      Formally, let $\locs{\mathcal{C}_i}$ be all locations in
      $\mathcal{C}_i$ and let $\tsucc{\mathcal{C}_i}{a}$ be
      the largest set of locations such that every edge in
      $\set{a} \times \tsucc{\mathcal{C}_i}{a}$ appears in $\mathcal{C}_i$.
      We then define:
      \begin{align*}
        &\ftree_i \defn \set{a \mapsto
        \tuple{\tsucc{\mathcal{C}_i}{a},\fruleinst{\frst}{a}} \mid a \in
          \locs{\mathcal{C}_i}},
          \tag{tree induced by component $\mathcal{C}_i$} \\
        &\barfrst \defn \set{\ftree_1, \ldots, \ftree_n}.
          \tag{$\Sid$-forest induced by the connected components}
      \end{align*}
      By construction, the forest $\barfrst$ is an $\vec{l}$-split of $\frst$.
      Moreover, since every $\vec{l}$-split must have the same domain
      and the same rule instances as $\frst$ and because every connected
      component gives rise to a single $\Sid$-tree, the
      $\vec{l}$-split
      $\fsplit{\frst}{\set{l}} = \barfrst$ is unique.
    \end{proof}

\subsection{Proof of \cref{lem:guarded-universal-intro}}
\label{app:guarded-universal-intro}
  Let %$\Sid\in\IDbtw$ and
  $\SH \in \Mpos{\Sid}$
  and $\fa$ be a quantifier free $\SLgeneric$ formula with
  $\SH \sidmodels \fa$.
  Moreover, let
  $\vec{v} \in {(\Loc \setminus (\dom(\H)\cup\img(\S)))}^{*}$ be
  a repetition-free sequence of locations.
  Then, for every set
  $\vec{a}\defn\set{a_1,\ldots,a_{\size{\vec{v}}}}$
  of fresh variables (i.e., $\vec{a}\cap\dom(\S)=\emptyset$),
  we have
  $\SH \sidmodels \FFA{\vec{a}} \pinst{\fa}{\vec{v}}{\vec{a}}$.
\begin{proof}[Proof sketch]
  Let $\vec{w}\in{(\Loc \setminus (\dom(\H)\cup\img(\S)))}^{*}$ be a
  repetition-free sequence of locations with
  $\size{\vec{v}}=\size{\vec{w}}$.
  We note that
  $\danglinglocs{\H} \subseteq \img(\S)$ because of $\SH \in \Mpos{\Sid}$.
  Hence, neither $\vec{v}$ nor $\vec{w}$ intersect with $\locs{\H}$ or $\img(\S)$.
  Thus, it follows that
  $\SH \sidmodels \pinst{\fa}{\vec{v}}{\vec{w}}$.
  Since $\vec{w}$ was arbitrary, $\SH \sidmodels \FFA{\vec{a}}
  \pinst{\fa}{\vec{v}}{\vec{a}}$ by the semantics of $\fforall$.
\end{proof}

%\begin{proof}[Proof sketch]
%  Let $\vec{w}\in{(\Loc \setminus (\dom(\H)\cup\img(\S)))}^{*}$ be a
%  repetition-free sequence of locations with
%  $\size{\vec{v}}=\size{\vec{w}}$.
%  We note that
%  $\danglinglocs{\H} \subseteq \img(\S)$ because of $\SH \in \Mpos{\Sid}$.
%  Hence, neither $\vec{v}$ nor $\vec{w}$ intersect with $\locs{\H}$ or $\img(\S)$.
%  Thus, it follows that
%  $\SH \sidmodels \pinst{\fa}{\vec{v}}{\vec{w}}$.
%  Since $\vec{w}$ was arbitrary, $\SH \sidmodels \FFA{\vec{a}}
%  \pinst{\fa}{\vec{v}}{\vec{a}}$ by the semantics of $\fforall$.
%\end{proof}

\subsection{Proof of \cref{lem:t-projection-sound}}\label{app:t-projection-sound}
\paragraph{Claim} Let $\ftree$ be a $\Sid$-tree- with $\fheapof{\ftree} = \H$.
  Then, $\SHpair{\_}{\fheapof{\ftree}} \sidmodels \ltproj{\img(\S)}{\ftree}$ (where $\_$ denotes an arbitrary stack).
\begin{proof}
  We prove the claim by mathematical induction on the height of $\ftree$.

  By construction, $\ftree$ has a root $r = \troot{\ftree}$ with $m \geq 0$ successors that are the root of subtrees $\ftree_1,\ldots,\ftree_m$.
  Hence, there is a rule instance $\truleinst{\ftree}{r}$ (up to applying commutativity of $\sep$) of the form\footnote{For $\theight{\ftree} = 0$, there are no successors, i.e., $\IteratedStar_{1 \leq i \leq m} \trootpred{\ftree_i}$ is equivalent to $\emp$.}
    \[
      \trootpred{\ftree}
      \Rule
      \ppto{a}{\vec{b}}
      \sep
      \big(
        \IteratedStar_{1 \leq i \leq m} \trootpred{\ftree_i}
      \big)
      \sep \IteratedStar\tholepreds{\ftree}{r},
    \]
    By the semantics of $\sep$ and $\mw$, we have
    \begin{align*}
        \SHpair{\_}{\set{\pto{a}{\vec{b}}}}
        \sidmodels
        \big(
            (\IteratedStar_{1 \leq i \leq m} \trootpred{\ftree_i})
            \sep
            \IteratedStar\tholepreds{\ftree}{r}\big)
        \mw \trootpred{\ftree}
    \end{align*}
  We apply the I.H. for each tree $\ftree_i$ and obtain
  \begin{align*}
    \SHpair{\_}{\fheapof{\ftree_i}} ~\sidmodels~
            (\IteratedStar\tallholepreds{\ftree_i}) \mw
            \trootpred{\ftree_i}.
  \end{align*}

  On the level of heaps, we have
  $\fheapof{\ftree} = \{ \pto{a}{\vec{b}} \} \stdunion \fheapof{\ftree_1} \stdunion \ldots \stdunion \fheapof{\ftree_m}$.
  Applying the semantics of $\sep$ and the definition of $\psi_i$ then yields
  \begin{align*}
    \SHpair{\_}{\fheapof{\ftree}} \sidmodels
        & \bigg(\big(
            (\IteratedStar_{1 \leq i \leq m} \trootpred{\ftree_i})
            \sep
            \IteratedStar\tholepreds{\ftree}{r}\big)
        \mw \trootpred{\ftree}\bigg) \\
        & \qquad
          \sep \IteratedStar_{1 \leq i \leq m}
            ((\IteratedStar\tallholepreds{\ftree_i}) \mw
            \trootpred{\ftree_i}).
  \end{align*}
  Applying \cref{lem:gmp} $m$ times, we then obtain
  \begin{align*}
        \SHpair{\_}{\fheapof{\ftree}} ~\sidmodels~ &
        \big((\IteratedStar_{1 \leq i \leq m} (\IteratedStar\tallholepreds{\ftree_i})) \sep
              \IteratedStar\tholepreds{\ftree}{r}\big) \mw \trootpred{\ftree} \\
        \qquad=\quad &
        \big(\IteratedStar\tallholepreds{\ftree}\big) \mw \trootpred{\ftree}.
        \tag{Def. of $\tallholepreds{\ftree}$}
  \end{align*}

%  We now prove the claim using (*).
%  Let
%    \[ \vec{w} ~=~ \locs{
%         \IteratedStar\tallholepreds{\ftree} \mw \trootpred{\ftree}
%       } \setminus (\dom(\H) \cup \img(\S))
%    \]
%  be all those locations that occur in the above formula but are neither allocated in $\fheapof{\ftree}$ nor in $\img(\S)$, and let $\vec{a} \defn \tuple{a_1,\ldots,a_{\size{\vec{w}}}}$ be fresh variables.
%
%  By \cref{lem:guarded-universal-intro} we then get
%  \begin{align*}
%      \SHpair{\S}{\fheapof{\ftree}} ~\sidmodels~
%        \FFA{\vec{w}}
%        \pinst{
%            \big(\IteratedStar\tallholepreds{\ftree}\big) \mw \trootpred{\ftree}
%        }{\vec{w}}{\vec{a}}
%      ~~=~~  \ltproj{\img(\S)}{\ftree}.
%  \end{align*}
\end{proof}

\subsection{Proof of \Cref{lem:sf-projection-sound}}
\label{app:sf-projection-sound}
\paragraph{Claim (Soundness of stack-projection)}
  Let %$\Sid\in\IDbtw$ and
  $\SH \in \Mpos{\Sid}$.
  Moreover, let $\frst$ be a $\Sid$-forest with $\fheapof{\frst} = \H$.
  Then, we have $\SH \sidmodels \sfproj{\S}{\frst}$.
\begin{proof}
  Let $\frst = \ktrees$ and $\phi = \IteratedStar_{1 \leq i \leq k}\; \ltproj{\vec{v}}{\ftree_i}$.
  By \cref{lem:t-projection-sound}, we know for each $i$ that
  $\SHpair{\S}{\theapof{\ftree_i}} \sidmodels \ltproj{\img(\S)}{\ftree_i}$.
  By definition,
  $\fheapof{\frst}=\theapof{\ftree_1}\stdunion \cdots \stdunion
  \theapof{\ftree_k}$.
  Applying the semantics of $\sep$ then yields $\SH \sidmodels\phi$ \qquad\tagA.

  Let $\vec{w} = \locs{\phi} \cap (\dom(\frst) \setminus \img(\S))$ be the locations that occur in $\phi$ and are allocated in $\fheapof{\frst}$ but are not the value of any stack variable,
  and let $\vec{v} = (\locs{\phi} \cap \Loc) \setminus (\img(\S) \cup \dom(\frst))$ be the locations that occur in the formula $\phi$ and are neither allocated nor the value of any stack variable.

  Then, we have
  \begin{displaymath}
      \sfproj{\S}{\frst} = \EEX{\vec{e}} \FFA{\vec{a}} \pinst{\phi}{\dom(\stkchc)\concat\vec{v}\concat\vec{w}}{\img(\stkchc)\concat\vec{a}\concat\vec{e}},
  \end{displaymath}
  where $\vec{e} \defn \tuple{e_1,e_2,\ldots,e_{\size{\vec{w}}}}$ and $\vec{a} \defn \tuple{a_1,a_2,\ldots,a_{\size{\vec{v}}}}$ denote some disjoint sets of fresh variables.

  The claim then follows by the implications below:
  %in \cref{fig:sfproj-correct}.
\begin{align*}
  & \SH \sidmodels \phi \tag*{(by \tagA)}\\
  \implies & \SH \sidmodels \phi\pinstINST{\dom(\stkchc)}{\img(\stkchc)} \tag*{(stack--heap semantics)}\\
  \implies& \SH \sidmodels \phi\pinstINST{\dom(\stkchc)}{\img(\stkchc)}\pinstINST{\vec{v} \concat \vec{w}}{\vec{a} \concat \vec{e}}\pinstINST{\vec{a} \concat \vec{e}}{\vec{v} \concat \vec{w}}
           \tag*{($\vec{a}$ and $\vec{e}$ are disjoint sets of fresh variables
           %, so $\pinstINST{\vec{a} \concat \vec{e} }{\vec{v} \concat \vec{w}}\pinstINST{\vec{v} \concat \vec{w}}{\vec{a} \concat \vec{e}}$ is the identity
           )}\\
  \implies & \SH \sidmodels %
  \phi\pinstINST{\dom(\stkchc)\concat\vec{v}\concat\vec{w}}{\img(\stkchc)\concat\vec{a} \concat \vec{e}} \pinstINST{\vec{a} \concat \vec{e}}{\vec{v}\concat\vec{w}} \tag*{($\vec{v}\cap\img(\S)=\emptyset$ and $\vec{w}\cap\img(\S)=\emptyset$)}\\
  \implies& \SH \sidmodels \FFA{\vec{a}} \phi\pinstINST{\dom(\stkchc)\concat\vec{v}\concat\vec{w}}{\img(\stkchc)\concat\vec{a} \concat \vec{e}}\pinstINST{\vec{e}}{\vec{w}}
              \tag*{(by \cref{lem:guarded-universal-intro})}\\
  \implies& \SH \sidmodels \EEX{\vec{e}} \FFA{\vec{a}} \phi\pinstINST{\dom(\stkchc)\concat\vec{v}\concat\vec{w}}{\img(\stkchc)\concat\vec{a} \concat \vec{e}}\tag*{(semantics of $\eexists$)}\\
  \implies & \SH \sidmodels \sfproj{\S}{\frst}
    \tag*{\qedhere}
\end{align*}
%  In the above calculations, we exploit that even if $\S$ is
%  \emph{not} injective, $\S\funcomp\stkchc$ is the identity
%  function---unlike $\stkchc \funcomp\S$. %, which is only the identity
%  %if $\S$ is injective.
\end{proof}

\subsection{Proof of~\cref{cor:fcompose-to-pcompose}}\label{app:fcompose-to-pcompose}

Before we prove \cref{cor:fcompose-to-pcompose}, we need two auxiliary results. 

\begin{lemma}\label{lem:funion-to-rescope}
  Let $\S$ be a stack and let $\frst_1,\frst_2$ be $\Sid$-forests with   %$\SHpair{\S}{\fheapof{\frst_1}},\SHpair{\S}{\fheapof{\frst_2}}\in\Mpos{\Sid}$.
  $\frst_1\funion\frst_2\neq \bot$.
  Then, $\sfproj{\S}{\frst_1\funion\frst_2} \in \sfproj{\S}{\frst_1} \PCompose \sfproj{\S}{\frst_2}$.
\end{lemma}
\begin{proof}
%  We assume $\frst_1\funion\frst_2\neq \bot$ and set $\frst_0 \defn \frst_1\funion\frst_2$.
%  We note that $\SHpair{\S}{\fheapof{\frst_0}}\in\Mpos{\Sid}$ by~\cref{lem:union-pos-model}.
%
  We set $\frst_0 \defn \frst_1\funion\frst_2$.
  For $i \in \{0,1,2\}$, 
  let $\phi_i = \IteratedStar_{\ftree \in \frst_i}\; \ltproj{\vec{v}}{\ftree}$,  
  let $\vec{w_i} = \locs{\phi_i} \cap (\dom(\frst_i) \setminus \img(\S))$ be the locations that occur in the formula $\phi$ and are allocated in $\fheapof{\frst_i}$ but are not the value of any stack variable,
  and let $\vec{v_i} = (\locs{\phi_i} \cap \Loc) \setminus (\img(\S) \cup \dom(\frst_i))$ be the locations that occur in the formula $\phi_i$ and are neither allocated nor the value of any stack variable.
  Then, we have
  \begin{displaymath}
      \sfproj{\S}{\frst_i} = \EEX{\vec{e_i}} \FFA{\vec{a_i}} \pinst{\phi_i}{\dom(\stkchc)\concat\vec{v_i}\concat\vec{w_i}}{\img(\stkchc)\concat\vec{a_i}\concat\vec{e_i}},
  \end{displaymath}
  where $\vec{e_i} \defn \tuple{e_1,e_2,\ldots,e_{\size{\vec{w_i}}}}$ and $\vec{a_i} \defn \tuple{a_1,a_2,\ldots,a_{\size{\vec{v_i}}}}$ denote some disjoint sets of fresh variables.
  Because of $\frst_0 = \frst_1\funion\frst_2$ we
  have $\vec{w_0} = \vec{w_1} \concat \vec{w_2}$ and hence
  can choose $\vec{e_0}$ such that $\vec{e_0} = \vec{e_1} \concat \vec{e_2}$.

  We now argue that we can find sequences of variables
  $\vec{u_i} \subseteq \vec{a_0} \cup \vec{e_{3-i}}$,
  for  $i=1,2$, such that
  \begin{multline*}
    \pinst{\phi_0}{\dom(\stkchc)\concat\vec{v_0}\concat\vec{w_0}}{\img(\stkchc)\concat\vec{a_0}\concat\vec{e_0}} \rewreq  \\
    \pinst{\phi_1}{\dom(\stkchc)\concat\vec{v_1}\concat\vec{w_2}}{\img(\stkchc)\concat\vec{a_1}\concat\vec{e_1}} \pinstINST{\vec{a_1}}{\vec{u_1}} \sep \\ \pinst{\phi_2}{\dom(\stkchc)\concat\vec{v_2}\concat\vec{w_2}}{\img(\stkchc)\concat\vec{a_2}\concat\vec{e_2}} \pinstINST{\vec{a_2}}{\vec{u_2}} (*)
  \end{multline*}

  We consider a location $l \in \vec{v_i}$ for $i \in \{1,2\}$.
  If $l \in \dom(\frst_{3-i})$,
  then there is a variable $e \in \vec{e_{3-i}}$ which replaces $l$ in the projection  $\sfproj{\S}{\frst_0}$.
  If $l \not\in \dom(\fheapof{\frst_{3-i}})$,
  then there is a variable $a \in \vec{a_0}$ which replaces $l$ in the projection $\sfproj{\S}{\frst_0}$.
  Hence, we can choose sequences of variables
  $\vec{u_i} \subseteq \vec{a_0} \cup \vec{e_{3-i}}$, for $i \in \{1,2\}$, such that the following holds for all $l \in \vec{v_i}$ and $i \in \{1,2\}$:
  \begin{displaymath}
    \pinst{l}{\vec{v_0}\concat\vec{w_0}}{\vec{a_0}\concat\vec{e_0}} = \pinst{l}{\vec{v_i}}{\vec{a_i}} \pinstINST{\vec{a_i}}{\vec{u_i}}
  \end{displaymath}
  The above then implies (*).
\end{proof}

\begin{lemma}\label{lem:fderive-to-derive}
  Let $\S$ be a stack and let $\frst_1,\frst_2$ be $\Sid$-forests with   %$\SHpair{\S}{\fheapof{\frst_1}}\in\Mpos{\Sid}$.
  $\frst_1\fderive\frst_2$.
  Then, $\sfproj{\S}{\frst_1} \deriveqf \sfproj{\S}{\frst_2}$.
\end{lemma}
\begin{proof}
  Since $\frst_1 \fderive \frst_2$, there exists a forest $\frst$ and trees $\ftree_1,\ftree_2,\ftree$ such that
  \begin{enumerate}
  \item $\frst_1 = \frst \cup \{\ftree_1,\ftree_2\}$,
  \item $\frst_2 = \frst \cup \{\ftree\}$,
  \item $\trootpred{\ftree_1} \in \tallholepreds{\ftree_2}$,
  \item $\trootpred{\ftree} = \trootpred{\ftree_2}$, and
  \item $\tallholepreds{\ftree} = \tallholepreds{\ftree_1} \cup (\tallholepreds{\ftree_2} \setminus \{\trootpred{\ftree_1}\})$.
  \end{enumerate}
  Intuitively, this implies that the projections of $\ftree_1$ and $\ftree_2$ can be merged into the projection of $\ftree$ via the generalized modus ponens
  (see \cref{lem:gmp}).
  In the following we make this claim formal.

  For $i \in \{1,2\}$, 
  let $\phi_i = \IteratedStar_{\ftree \in \frst_i}\; \ltproj{\vec{v}}{\ftree}$,
  let $\vec{w_i} = \locs{\phi_i} \cap \dom(\frst_i) \setminus \img(\S)$ be the locations that occur in the formula $\phi$ and are allocated in $\fheapof{\frst_i}$ but are not the value of any stack variable,
  and let $\vec{v_i} = (\locs{\phi_i} \cap \Loc) \setminus (\img(\S) \cup \dom(\frst_i))$ be the locations that occur in the formula $\phi_i$ and are neither allocated nor the value of any stack variable.
  Then, we have
  \begin{displaymath}
      \sfproj{\S}{\frst_i} = \EEX{\vec{e_i}} \FFA{\vec{a_i}} \pinst{\phi_i}{\dom(\stkchc)\concat\vec{v_i}\concat\vec{w_i}}{\img(\stkchc)\concat\vec{a_i}\concat\vec{e_i}},
  \end{displaymath}
  where $\vec{e_i} \defn \tuple{e_1,e_2,\ldots,e_{\size{\vec{w_i}}}}$ and $\vec{a_i} \defn \tuple{a_1,a_2,\ldots,a_{\size{\vec{v_i}}}}$ denote some disjoint sets of fresh variables.
  We note that
    $$\phi_1 \rewreq \IteratedStar_{\ftree' \in \frst}\; \ltproj{\vec{v}}{\ftree'} \sep \left(\IteratedStar\tallholepreds{\ftree_1}\right) \mw \trootpred{\ftree_1} \sep    \left(\IteratedStar\tallholepreds{\ftree_2}\right) \mw \trootpred{\ftree_2}$$
  and
  $$\phi_2 \rewreq \IteratedStar_{\ftree' \in \frst}\; \ltproj{\vec{v}}{\ftree'} \sep (\IteratedStar\tallholepreds{\ftree_1}\sep
    \IteratedStar(\tallholepreds{\ftree_2} \setminus \{\trootpred{\ftree_1})) \mw \trootpred{\ftree_2}.$$
  In particular, we have $\locs{\phi_2} \subseteq \locs{\phi_1}$.
  Hence, we can assume without loss of generality that
  $\vec{e_2} \subseteq \vec{e_1}$ and $\vec{a_2} \subseteq \vec{a_1}$.
  By (*), we have
  \begin{multline*}
    \sfproj{\S}{\frst_1} \rewreq \EEX{\vec{e_1}} \FFA{\vec{a_1}}
    \IteratedStar_{\ftree' \in \frst}\; \ltproj{\vec{v}}{\ftree'} \sep \left(\IteratedStar\tallholepreds{\ftree_1}\right) \mw \trootpred{\ftree_1} \sep\\
    \left(\IteratedStar\tallholepreds{\ftree_2}\right) \mw \trootpred{\ftree_2}
      \pinstINST{\dom(\stkchc)\concat\vec{v_i}\concat\vec{w_i}}{\img(\stkchc)\concat\vec{a_i}\concat\vec{e_i}},
  \end{multline*}
  and
  \begin{multline*}
    \sfproj{\S}{\frst_2} \rewreq \EEX{\vec{e_1}} \FFA{\vec{a_1}}
    \IteratedStar_{\ftree' \in \frst}\; \ltproj{\vec{v}}{\ftree'} \sep (\IteratedStar\tallholepreds{\ftree_1}\sep\\
    \IteratedStar(\tallholepreds{\ftree_2} \setminus \{\trootpred{\ftree_1})) \mw \trootpred{\ftree_2}
    \pinstINST{\dom(\stkchc)\concat\vec{v_i}\concat\vec{w_i}}{\img(\stkchc)\concat\vec{a_i}\concat\vec{e_i}},
  \end{multline*}
  We now recognize that $\sfproj{\S}{\frst_2}$ can be obtained from $\sfproj{\S}{\frst_1}$ by applying the generalized modus ponens rule and dropping the quantified variables $\vec{e_1} \setminus \vec{e_2}$ and $\vec{a_1} \setminus \vec{a_2}$, which is supported by our rewriting rules (see~\cref{fig:sl:rewreq}) because these variables do not appear in $\phi_2$.
\end{proof}

\paragraph{Claim (\cref{cor:fcompose-to-pcompose})}
  Let $\S$ be a stack and let $\frst_1,\frst_2$ be $\Sid$-forests such that $\frst_1\funion\frst_2\neq\bot$.
  Then,
  \[
    \frst \in \frst_1\FCompose\frst_2
    \quad\text{implies}\quad
    \sfproj{\S}{\frst} \in \sfproj{\S}{\frst_1} \PCompose
    \sfproj{\S}{\frst_2}.
  \]

\begin{proof}
  The claim is an immediate consequence of \cref{lem:funion-to-rescope,lem:fderive-to-derive}.
\end{proof}

%\begin{lemma}\label{lem:derive-to-fderive}
%  Let $\frst_1$ be a $\Sid$-forest and let $\phi$ be such that $\sfproj{\S}{\frst_1}\deriveqf \phi$.
%  Then there exist forests
%  $\frst_1',\frst_2$ with $\frst_1\AEQ\frst_1'$,
%  $\frst_1'\fderive\frst_2$ and $\sfproj{\S}{\frst_2}=\phi$.
%\end{lemma}

\subsection{Proof of~\cref{thm:fcompose-eq-pcompose}}\label{app:fcompose-eq-pcompose}

Before we prove \cref{thm:fcompose-eq-pcompose}, we need two auxiliary results. 

\begin{lemma}\label{lem:rescope-to-funion}
  Let $\S$ be a stack, let $\frst_1,\frst_2$ be $\Sid$-forests with $\dom(\frst_1) \cap \dom(\frst_2) \cap \img(\S) = \emptyset$, and let $\chi \in \sfproj{\S}{\frst_1} \PCompose \sfproj{\S}{\frst_2}$.
  Then, there exist forests $\frst_1',\frst_2'$ with $\frst_1\AEQ\frst_1'$, $\frst_2\AEQ\frst_2'$ and
  $\sfproj{\S}{\frst_1'\funion\frst_2'} \rewreq \chi$.
\end{lemma}
\begin{proof}
  For $i \in \{1,2\}$, 
  let $\phi_i = \IteratedStar_{\ftree \in \frst_i}\; \ltproj{\vec{v}}{\ftree}$,
  let $\vec{w_i} = \locs{\phi_i} \cap \dom(\frst_i) \setminus \img(\S)$ be the locations that occur in the formula $\phi$ and are allocated in $\fheapof{\frst_i}$ but are not the value of any stack variable,
    and let $\vec{v_i} = (\locs{\phi_i} \cap \Loc) \setminus (\img(\S) \cup \dom(\frst_i))$ be the locations that occur in the formula $\phi_i$ and are neither allocated nor the value of any stack variable.
  Then, we have
  \begin{displaymath}
      \sfproj{\S}{\frst_i} = \EEX{\vec{e_i}} \FFA{\vec{a_i}} \pinst{\phi_i}{\dom(\stkchc)\concat\vec{v_i}\concat\vec{w_i}}{\img(\stkchc)\concat\vec{a_i}\concat\vec{e_i}},
  \end{displaymath}
  where $\vec{e_i} \defn \tuple{e_1,e_2,\ldots,e_{\size{\vec{w_i}}}}$ and $\vec{a_i} \defn \tuple{a_1,a_2,\ldots,a_{\size{\vec{v_i}}}}$ denote some disjoint sets of fresh variables.

  By the definition of the re-scoping operation,
  we have $\chi = \EEX{\vec{e}} \FFA{\vec{a}} \phi$, where
  \begin{enumerate}
    \item $\vec{e} = \vec{e_1} \concat \vec{e_2}$, and
    \item $\phi = \pinst{\phi_1}{\dom(\stkchc)\concat\vec{v_1}\concat\vec{w_1}}{\img(\stkchc)\concat\vec{a_1}\concat\vec{e_1}}\pinstINST{\vec{a_1}}{\vec{u_1}} \sep \pinst{\phi_2}{\dom(\stkchc)\concat\vec{v_2}\concat\vec{w_2}}{\img(\stkchc)\concat\vec{a_2}\concat\vec{e_2}}\pinstINST{\vec{a_2}}{\vec{u_2}}$ for some sequences $\vec{u_i} \subseteq \vec{a} \cup \vec{e_{3-i}}$.
  \end{enumerate}
  We can now choose bijective functions $\renfun_1 \colon \Val \to \Val$ and $\renfun_2 \colon \Val\to \Val$ such that
  \begin{itemize}
    \item $\renfun_1(l) = \renfun_2(l) = l$ for all $l\in \img(\S)$,
    \item $\renfun_1(\nil) = \renfun_2(\nil) = \nil$,
    \item $\renfun_1(l) = \renfun_2(k)$ if and only if $\pinst{l}{\vec{v_1}\concat\vec{w_1}}{\vec{a_1}\concat\vec{e_1}}\pinstINST{\vec{a_1}}{\vec{u_1}} =         \pinst{k}{\vec{v_2}\concat\vec{w_2}}{\vec{a_2}\concat\vec{e_2}}\pinstINST{\vec{a_2}}{\vec{u_2}}$
        for all $l \in \vec{v_1}\concat\vec{w_1}, k \in \vec{v_2}\concat\vec{w_2}$, and
    \item $\dom(\renfun_1(\frst_1)) \cap \dom(\renfun_1(\frst_2)) = \emptyset$.
  \end{itemize}
  We set $\frst_1' \defn \renfun(\frst_1)$ and $\frst_2' \defn \renfun(\frst_2)$.
  By the above we have $\frst_1\AEQ\frst_1'$, $\frst_2\AEQ\frst_2'$ and $\frst_1'\funion\frst_2' \neq \bot$.
  Further, we get that $\lfproj{\img(\S)}{\frst_1'\funion\frst_2'} \rewreq
  \pinst{\lfproj{\img(\S)}{\frst_1}}{\dom(\renfun_1)}{\img(\renfun_1)}
  \sep \pinst{\lfproj{\img(\S)}{\frst_2}}{\dom(\renfun_2)}{\img(\renfun_2)}$.
  Finally, we get that $\sfproj{\S}{\frst_1'\funion\frst_2'} \rewreq \EEX{\vec{e}} \FFA{\vec{a}} \phi$ because we can appropriately rename the quantified variables by rewrite equivalence $\rewreq$.
\end{proof}

We note that the below lemma does not require the notion of $\S$-equivalence:

\begin{lemma}\label{lem:derive-to-fderive}
  Let $\frst_1$ be a $\Sid$-forest and let $\chi$ be a formula such that $\sfproj{\S}{\frst_1}\deriveqf \chi$.
  Then, there exist a forest $\frst_2$ with  $\frst_1\fderive\frst_2$ and $\sfproj{\S}{\frst_2} \rewreq \chi$.
\end{lemma}
\begin{proof}
  Let $\phi_1 = \IteratedStar_{\ftree \in \frst_1}\; \ltproj{\vec{v}}{\ftree}$, 
  let $\vec{w_1} = \locs{\phi_1} \cap (\dom(\frst_1) \setminus \img(\S))$ be the locations that occur in the formula $\phi$ and are allocated in $\fheapof{\frst_1}$ but are not the value of any stack variable,
    and let $\vec{v_1} = (\locs{\phi_1} \cap \Loc) \setminus (\img(\S) \cup \dom(\frst_1))$ be the locations that occur in the formula $\phi_1$ and are neither allocated nor the value of any stack variable.
  Then, we have
  \begin{displaymath}
      \sfproj{\S}{\frst_1} = \EEX{\vec{e_1}} \FFA{\vec{a_1}} \pinst{\phi_1}{\dom(\stkchc)\concat\vec{v_1}\concat\vec{w_1}}{\img(\stkchc)\concat\vec{a_1}\concat\vec{e_1}},
  \end{displaymath}
  where $\vec{e_1} \defn \tuple{e_1,e_2,\ldots,e_{\size{\vec{w_1}}}}$ and $\vec{a_1} \defn \tuple{a_1,a_2,\ldots,a_{\size{\vec{v_1}}}}$ denote some disjoint sets of fresh variables.
  By definition of the projection of forests, we have
    $$\phi_1 \rewreq \IteratedStar_{\ftree\in\frst} \left( \left(\IteratedStar \tallholepreds{\ftree}\right) \mw \trootpred{\ftree}\right).$$
  By the definition of $\deriveqf$ we have that there are predicates $\pred_1(\vec{x_1}), \pred_2(\vec{x_2})$, and formulae $\psi,\psi',\zeta$ such that
  \begin{enumerate}
  \item $\phi_1 \rewreq (\pred_2(\vec{x_2}) \sep \psi) \mw \pred_1(\vec{x_1}))) \sep (\psi' \mw \pred_2(\vec{x_2})) \sep \zeta$, and
  \item $\chi \rewreq \EEX{\vec{e_1}} \FFA{\vec{a_1}} (\psi \sep \psi') \mw \pred_1(\vec{x_1}) \sep \zeta$.
  \end{enumerate}
  Hence, there must be a forest $\frst$ and trees $\ftree_1,\ftree_2$ such that
  \begin{enumerate}
  \item $\frst_1 = \frst \cup \{\ftree_1,\ftree_2\}$,
  \item $\trootpred{\ftree_2} \in \tallholepreds{\ftree_1}$, and
  \item $\pinst{\trootpred{\ftree_2}}{\dom(\stkchc)\concat\vec{v_1}\concat\vec{w_1}}{\img(\stkchc)\concat\vec{a_1}\concat\vec{e_1}} = \pred_2(\vec{x_2})$.
%
%  \item $\trootpred{\ftree} = \trootpred{\ftree_2}$, and
%  \item $\tallholepreds{\ftree} = \tallholepreds{\ftree_1} \cup (\tallholepreds{\ftree_2} \setminus \{\trootpred{\ftree_1}\})$.
  \end{enumerate}
  Let $l = \troot{\ftree_2}$.
  Then, there is a tree $\ftree$ with
  $\{\ftree_1,\ftree_2\}= \fsplit{\{\ftree\}}{\set{l}}$.
  We set $\frst_2 = \frst \cup \{\ftree\}$.
  We note that $\frst_1\fderive\frst_2$.
  It remains to argue that   $\sfproj{\S}{\frst_2}\rewreq\chi$.

    Let $\phi_2 = \IteratedStar_{\ftree \in \frst_2}\; \ltproj{\vec{v}}{\ftree}$, let $\vec{w_2} = \locs{\phi_2} \cap (\dom(\frst_2) \setminus \img(\S))$ be the locations that occur in the formula $\phi$ and are allocated in $\fheapof{\frst_2}$ but are not the value of any stack variable,
    and let $\vec{v_2} = (\locs{\phi_2} \cap \Loc) \setminus (\img(\S) \cup \dom(\frst_2))$ be the locations that occur in the formula $\phi_2$ and are neither allocated nor the value of any stack variable.
  Then, we have
  \begin{displaymath}
      \sfproj{\S}{\frst_2} = \EEX{\vec{e_2}} \FFA{\vec{a_2}} \pinst{\phi_2}{\dom(\stkchc)\concat\vec{v_2}\concat\vec{w_2}}{\img(\stkchc)\concat\vec{a_2}\concat\vec{e_2}},
  \end{displaymath}
  where $\vec{e_2} \defn \tuple{e_1,e_2,\ldots,e_{\size{\vec{w_2}}}}$ and $\vec{a_2} \defn \tuple{a_1,a_2,\ldots,a_{\size{\vec{v_2}}}}$ denote some disjoint sets of fresh variables.
  We now note that $\locs{\phi_2} \cup \{l\} = \locs{\phi_1}$.
  Hence, we can assume without loss of generality that
  $\vec{e_2} \subseteq \vec{e_1}$ and $\vec{a_2} \subseteq \vec{a_1}$.
  Thus, we have $(\psi \sep \psi') \mw \pred_1(\vec{x_1}) \sep \zeta \rewreq \pinst{\phi_2}{\dom(\stkchc)\concat\vec{v_2}\concat\vec{w_2}}{\img(\stkchc)\concat\vec{a_2}\concat\vec{e_2}}$.
  Finally, we note that
  \begin{multline*}
  \chi \rewreq
  \EEX{\vec{e_1}} \FFA{\vec{a_1}} \pinst{\phi_2}{\dom(\stkchc)\concat\vec{v_2}\concat\vec{w_2}}{\img(\stkchc)\concat\vec{a_2}\concat\vec{e_2}} \rewreq \\
  \EEX{\vec{e_2}} \FFA{\vec{a_2}} \pinst{\phi_2}{\dom(\stkchc)\concat\vec{v_2}\concat\vec{w_2}}{\img(\stkchc)\concat\vec{a_2}\concat\vec{e_2}},
  \end{multline*}
  because we can drop the quantified variables $\vec{e_1} \setminus \vec{e_2}$ and $\vec{a_1} \setminus \vec{a_2}$, which is supported by our rewriting rules (see~\cref{fig:sl:rewreq}) because these variables do not appear in $\phi_2$.
\end{proof}

\paragraph{Claim (\cref{thm:fcompose-eq-pcompose})}
  If $\frst_1,\frst_2$ be $\Sid$-forests with $\frst_1\AEQ\frst_2$, then
  \begin{align*}
  \sfproj{\S}{\frst_1}\PCompose\sfproj{\S}{\frst_2} ~=~
  \set{\sfproj{\S}{\frst} \mid
  \frst \in \barfrst_1 \FCompose \barfrst_2,~
  \barfrst_1\AEQ\frst_1,~
  \barfrst_2\AEQ\frst_2
  }.
  \end{align*}

\begin{proof}
The claim is an immediate consequence of~\cref{cor:fcompose-to-pcompose,lem:rescope-to-funion,lem:derive-to-fderive}.
\end{proof}

\subsection{Proof of \Cref{lem:dush-finite}}
\label{app:dush-finite}

\paragraph{Claim}
  Let $n \defn \size{\Sid}+\size{\vec{x}}$, where $\vec{x}$ is a finite set of variables.
  Then %the number of DUSHs over $\Sid$ and $\vec{x}$
  %is exponentially bounded in $n$; more precisely,
  $\size{\DUSHx{\Sid}{\vec{x}}} \in 2^{\bigO(n^2 \log(n))}$.
\begin{proof}
  \newcommand{\Alphabet}{Z}
  We first show the following claim \tagA:
  every element of $\DUSHx{\Sid}{\vec{x}}$ can be
  encoded as a string of length $\bigO(n^2)$ over the alphabet
  $\Alphabet \defn \Preds{\Sid} \cup \vec{x} \cup \set{e_1,\ldots,e_{n^2}}
  \cup \set{a_1,\ldots,a_{n^2}} \cup \set{\emp,\sep,\mw,(,)}$ %\tagA,
  where $e_1,\ldots,e_{n^2}$ and $a_1,\ldots,a_{n^2}$
  are fresh variables.

  By definition, every DUSH $\fa \in \DUSHx{\Sid}{\vec{x}}$
  is of the form
\begin{align*}
  \phi =& \EEX{\vec{e}} \FFA{\vec{a}} \psi_1 \sep \cdots \sep \psi_m, \\
  \psi_i =& \zeta_i \mw \pred_i(\vec{z_i}) \text{
            for } 1 \leq i \leq m.
\end{align*}
Since $\fa$ is delimited, $\proot{\pred_i(\vec{z_i})} \in \vec{x}$.
Moreover, $\fa$ is the projection of a $\Sid$-forest $\frst$. Hence, every
variable $x\in\vec{x}$ can appear as a root parameter in at most one subformula $\psi_i$---otherwise, the value corresponding to $x$ would be
in the domain of two trees in $\frst$, which contradicts the fact
that $\frst$ is a $\Sid$-forest.
Consequently, the number $m$ of subformulas $\psi_i$ is bounded by $\size{\vec{x}} \leq n$.

Next, consider the subformulas $\zeta_i$ appearing on the left-hand side
of magic wands.
For every predicate call
$\pred'(\vec{z}')$ in $\zeta_i$,
$\proot{\pred'(\vec{z}')}$ is a hole.
Since the forest $\frst$ is delimited, it follows that
$\proot{\pred'(\vec{z}')} \in \vec{x}$.
Since no hole may occur more than once in a delimited USH,
the \emph{total} number of predicate calls across all
$\zeta_i$ is also bounded by $\size{\vec{x}} \leq n$.

Overall, $\phi$ thus contains at most $2n \in \bigO(n)$ predicate calls.
Each predicate call takes at most
$\size{\Sid} \leq n$ parameters.
Since there are no superfluous quantified variables,
this means that $\phi$ contains at most $n^2 - \size{\vec{x}} \leq n^2$ different
variables.
We can thus assume w.l.o.g.~that all existentially-quantified variables in $\fa$ are among the variables $e_1,\ldots,e_{n^2}$ and all universally-quantified variables are
among $a_1,\ldots,a_{n^2}$.
There then is no need to include the quantifiers explicitly in the
string encoding. After dropping the quantifiers, we obtain a formula
$\fa'$ that consists exclusively of letters from the alphabet $\Alphabet$.
Moreover, this formula consists of at most $\bigO(n^2)$ letters.
This concludes the proof of \tagA.

Now observe that $\size{\Alphabet} \in \bigO(n^2)$. Consequently, every
letter of $\Alphabet$ can be encoded by
$\bigO(\log(n^2)) = \bigO(\log(n))$ bits.
Therefore, every $\phi \in \DUSHx{\Sid}{\vec{x}}$ can be encoded by a
bit string of length $\bigO(n^2 \log(n))$.
Since there are $2^{\bigO(n^2 \log(n))}$ such strings, the claim follows.
\end{proof}

\subsection{Proof of \cref{lem:delimited-forest--delimited-projection}}
\label{app:delimited-forest--delimited-projection}

\paragraph{Claim}
  Let $\frst$ be a forest and let $\S$ be a stack. Then $\frst$ is
  $\S$-delimited iff $\sfproj{\S}{\frst}$ is delimited.

\begin{proof}
  Recall that the projection contains predicate calls corresponding to
  the roots and holes of the forest.  It thus holds for all forests
  that
  \[\finterface{\frst} = \set{\proot{\pred(\vec{z})} \mid
      \pred(\vec{z}) \in \fproj{\frst}}. \tag*{\tagA}\]
  We show that if $\frst$ is $\S$-delimited then $\sfproj{\S}{\frst}$
  is delimited. The proof of the other direction is completely
  analogous.

  If $\frst$ is $\S$-delimited then
  $\finterface{\frst} \subseteq \img(\S)$ and thus, by \tagA{},
  \begin{align*}
  \set{\proot{\pred(\vec{z})} \mid \pred(\vec{z}) \in \fproj{\frst}}
    &\subseteq \img(\S).
  \end{align*}
  Trivially, the set of root locations in the projection is a subset
  of the set of \emph{all} locations in the projection.
  \begin{align*}
  \set{\proot{\pred(\vec{z})} \mid \pred(\vec{z}) \in \fproj{\frst}}
    &\subseteq \locs{\fproj{\frst}}.
  \end{align*}
  Combining the above two observations, we conclude
  \begin{align*}
  &\set{\proot{\pred(\vec{z})} \mid \pred(\vec{z}) \in \fproj{\frst}}\\
    \subseteq &\img(\S) \cap \locs{\fproj{\frst}}.
  \end{align*}

  We apply $\stkchc$ on both sides to obtain that
  \begin{align*}
    %\sfproj{\S}{\frst}} \\=&
  &\stkchc(\set{\proot{\pred(\vec{z})} \mid
    \pred(\vec{z} \in \fproj{\frst})})\\
    \subseteq& \dom(\S) \cap
                              \underbrace{(\fvs{\sfproj{\S}{\frst}}}_{\set{\stkchc(l)
                                                    \mid l \in
                                                    \img(\S) \cap \locs{\fproj{\frst}}}} % \cup \set{l
               %                \in \locs{\fproj{\frst}} \mid l \notin
               % \img(\S)\cap \locs{\fproj{\frst}}} )
    %\\
    %                        =& \dom(\S) \cap \fvs{\sfproj{\S}{\frst}} \\
    \subseteq \fvs{\sfproj{\S}{\frst}}.
  \end{align*}

  Moreover, since there are no duplicate holes in $\frst$, and the
  holes of $\frst$ are mapped to the predicate calls on the left-hand
  side of magic wands in $\sfproj{\S}{\frst}$, no variable can
  occur twice as root parameter on the left-hand side of magic wands
  in $\sfproj{\S}{\frst}$.

  Consequently, $\sfproj{\S}{\frst}$ is delimited.
\end{proof}

\subsection{Proof of \cref{lem:delimited-forest-delimited-sub-forests}}
\label{app:delimited-forest-delimited-sub-forests}

We first some auxiliary definitions and results.
Recall that we described how $\Sid$-forests are merged in terms of splitting them at suitable locations (cf. \cref{def:forest:fsplit}).
Every split adds these locations to the interface of the resulting forest---provided they did not appear in the forest to begin with.
\begin{lemma}\label{lem:fsplit-interface}
  Let $\frst$ be a forest and $\vec{l}\subseteq\Loc$.
  Then, $\finterface{\fsplit{\frst}{\vec{l}}} = \finterface{\frst} \cup
  (\vec{l}\cap\dom(\frst))$.
\end{lemma}
\begin{proof}
  In the following, let $\values{\fgraph{\frst}}$ denote all those values that occur in the relation $\fgraph{\frst}$.

  \begingroup
  \allowdisplaybreaks
  \begin{align*}
    &\froots{\fsplit{\frst}{\vec{l}}} \\
    =& \froots{\frst}\cup \set{b \in \vec{l} \mid
    \EXO{a} (a,b)\in\fgraph{\frst}} \\
    =& \froots{\frst}
    \cup \set{b \in \vec{l}\cap\dom(\frst) \mid
      \EXO{a} (a,b)\in\fgraph{\frst}} \tag*{$(\values{\fgraph{\frst}}
      \subseteq \dom(\frst))$} \\
    =& \set{b \in \dom(\frst) \mid \FAO{a} (a,b)\notin \fgraph{\frst}}
    \\&\quad\cup \set{b \in \vec{l}\cap\dom(\frst) \mid
      \EXO{a} (a,b)\in\fgraph{\frst}} \tag*{(all and only roots have no predecessor)}\\
    =& \set{b \in \dom(\frst) \mid \FAO{a} (a,b)\notin \fgraph{\frst}}
    \\&\quad  \cup \set{b \in \vec{l}\cap\dom(\frst) \mid \FAO{a} (a,b)\notin \fgraph{\frst}}
    \\&\quad\cup \set{b \in \vec{l}\cap\dom(\frst) \mid
      \EXO{a} (a,b)\in\fgraph{\frst}}  \tag*{(second set subset of
    first set)}\\
    =& \froots{\frst}
      \cup \set{b \in \vec{l}\cap\dom(\frst) \mid \FAO{a} (a,b)\notin \fgraph{\frst}}
    \\&\quad\cup \set{b \in \vec{l}\cap\dom(\frst) \mid
    \EXO{a} (a,b)\in\fgraph{\frst}} \\
    =& \froots{\frst}\cup \set{b \in \vec{l}\cap\dom(\frst)}
  \end{align*}
  \endgroup
  Similarly,
  \begin{align*}
    \fallholes{\fsplit{\frst}{\vec{l}}} =&\fallholes{\frst} \cup \set{b \in \vec{l} \mid
    \EXO{a} (a,b)\in\fgraph{\frst}} \\
    =& \fallholes{\frst} \cup \set{b \in \vec{l}\cap\dom(\frst)}.
  \end{align*}
  By definition of interfaces, we thus obtain
  \begin{align*}
  \finterface{\fsplit{\frst}{\vec{l}}} =&
  \froots{\fsplit{\frst}{\vec{l}}} \cup
  \fallholes{\fsplit{\frst}{\vec{l}}}\\=&
                                          \froots{\frst}\cup \set{t \in
                                          \vec{l}\cap\dom(\frst)}\\&\,\cup\fallholes{\frst}
                                          \cup \set{b \in
                                          \vec{l}\cap\dom(\frst)}\\=&
                                                                      \finterface{\frst}
                                                                      \cup
                                                                      \set{b
                                                                      \in
                                                                      \vec{l}\cap\dom(\frst)}. \qedhere
  \end{align*}
\end{proof}

\begin{definition}\label{def:types:s-decomposition}
  \index{S-decomposition@$\S$-decomposition}%\index{S-decomposed@$\S$-decomposed}
  Let $\frst$ be an $\S$-delimited forest.
  We call $\fsplit{\frst}{\img(\S)}$ the \emph{$\S$-decompo\-sition} of $\frst$.
\end{definition}

\begin{lemma}%[Decompositions are delimited]
  \label{lem:s-decomposition-delimited}
  The $\S$-decomposition of an $\S$-delimited forest is $\S$-delimited.
  %Let $\barfrst$ be the $\S$-decomposition of $\S$-delimited forest
  %$\frst$. Then $\barfrst$ is $\S$-delimited.
\end{lemma}
\begin{proof}
  Let $\barfrst$ be the $\S$-decomposition of an $\S$-delimited forest $\frst$.
  By definition, $\barfrst=\fsplit{\frst}{\img(\S)}$.
  By~\cref{lem:fsplit-interface}, we have $\finterface{\barfrst} \subseteq \finterface{\frst} \cup \img(\S)$.
  Since $\frst$ is $\S$-delimited, $\finterface{\frst} \subseteq\img(\S)$.
  Overall, we thus obtain $\finterface{\barfrst} \subseteq \img(\S)$, i.e., $\barfrst$ is $\S$-delimited.
\end{proof}

We observe that, since the $\S$-decomposition of a forest is obtained by splitting the trees of the forest at all locations in $\img(\S)$, only the roots of the trees in an $\S$-decomposition of forest $\frst$ can be locations in $\img(\S)$:

\begin{lemma}\label{lem:s-decomposition-only-roots-free}
  For every $\Sid$-tree $\bartree$ in an $\S$-decomposition, we have  $\img(\S)\cap\dom(\bartree) = \set{\troot{\bartree}}$.
  %Let $\barfrst$ be the $\S$-decomposition of an $\S$-delimited forest
  %$\frst$ and let $\bartree \in \barfrst$.
  %
  %Then $\img(\S)\cap\dom(\bartree) = \set{\troot{\bartree}}$.
\end{lemma}
\begin{proof}
  Let $\barfrst$ be an $\S$-decomposition %of an $\S$-delimited forest
  and let $\bartree \in \barfrst$.
  Since $\barfrst$ is $\S$-delimited by
  \cref{lem:s-decomposition-delimited}, we have
  $\set{\troot{\bartree}} \subseteq \img(\S)$.
  Since $\troot{\bartree} \in \dom(\bartree)$,
  $\set{\troot{\bartree}} \subseteq \img(\S)\cap\dom(\bartree)$.

  Conversely, since $\barfrst = \fsplit{\frst}{\img(\S)}$, we have  $\froots{\barfrst} = \froots{\frst} \cup (\img(\S) \cap \dom(\frst))$, i.e., every location in  $\img(\S) \cap \dom(\barfrst)$ is a root of $\barfrst$.
  Consequently, $\img(\S)\cap\dom(\bartree) \subseteq \set{\troot{\bartree}}$.
\end{proof}

\begin{lemma}\label{lem:s-decomposition-splittable}
  Let $\SHi{1},\SHi{2}\in\Mpos{\Sid}$ be guarded states, and let $\frst$ be a $\S$-delimited forest with $\frst \in \sidfrstsof{\H_1\stdunion\H_2}$.
  Then, there exist forests $\frst_1,\frst_2$ with
  $\frst_1\funion\frst_2=\barfrst$ and $\fheapof{\frst_i}=\H_i$,
  where $\barfrst$ is the $\S$-decomposition of $\frst$.
\end{lemma}
\begin{proof}
  We let $\frst_i \defn \set{\bartree \in \barfrst \mid \troot{\bartree}\in \dom(\H_i)}$.
  Since $\frst_1 \funion \frst_2 = \frst$ and thus $\fheapof{\frst_1}\stdunion\fheapof{\frst_2}=\fheapof{\frst}$ by \cref{lem:forests:funion-heap},
  it suffices to show that for every tree $\bartree$ in $\frst_i$ that $\fheapof{\bartree}\subseteq \H_i$.

  To this end, let $\bartree \in \frst_i$.
  Assume towards a contradiction that $\dom(\bartree)\cap\dom(\H_{3-i}) \neq \emptyset$.
  Then there exist locations $\la_1\in\dom(\bartree)\cap\dom(\H_i)$ and $\la_2\in\dom(\bartree)\cap\dom(\H_{3-i})$ with $\la_2 \in \tsucc{\bartree}{\la_1}$.
  In particular, $\la_2\in\img(\H_i)$ and $\la_2\in\dom(\H_{3-i})$,
  implying that $\la_2 \in \danglinglocs{\H_i}$.
  However, since $\SHi{1},\SHi{2}\in\Mpos{\Sid}$, we have that $\la_2 \in \img(\S)$.
  Since $\la_2 \neq \troot{\ftree}$, this contradicts \cref{lem:s-decomposition-only-roots-free}.
\end{proof}

\paragraph*{We restate the claim of \cref{lem:delimited-forest-delimited-sub-forests}:}
Let $\SHi{1},\SHi{2}\in\Mpos{\Sid}$ be guarded states, and let $\frst$ be a $\S$-delimited forest with $\frst \in \sidfrstsof{\H_1\stdunion\H_2}$.
Then there exist $\S$-delimited forests $\frst_1,\frst_2$ with $\fheapof{\frst_i}=\H_i$ and  $\frst \in \frst_1 \FCompose \frst_2$.
\begin{proof}
  Let $\barfrst$ be the $\S$-decomposition of $\frst$.
  In particular, we then have $\barfrst\fderivestar\frst$ by definition of $\fderivestar$.
  Let $\frst_1,\frst_2$ be such that $\frst_1\funion\frst_2=\barfrst$
  and $\fheapof{\frst_i}=\H_i$. Such forests exist by
  \cref{lem:s-decomposition-splittable}.
  Then
  $\frst_1 \funion \frst_2 = \barfrst \fderivestar \frst$, i.e.,
  $\frst \in \frst_1 \FCompose \frst_2$. Since $\barfrst$ is
  $\S$-delimited (by \cref{lem:s-decomposition-delimited}), so are
  $\frst_1$ and $\frst_2$.
\end{proof}

\subsection{Proof of \Cref{thm:oursunion-to-derive}}
\label{app:oursunion-to-derive}

\paragraph{Claim}
  For all guarded states $\SHi{1}$ and $\SHi{2}$ with
  $\H_1\stdunion\H_2\neq\bot$,
  $\sidtypeof{\S, \H_1 \stdunion \H_2}$ can be computed from
  $\sidtypeof{\S, \H_1}$ and $\sidtypeof{\S,\H_2}$ as follows:
  \begin{align*}
    \sidtypeof{\S,\H_1 \stdunion \H_2} = \{
        \fa \in \DUSH{\Sid} \mid
        \text{ex. }
        \fb_1 \in \sidtypeof{\S,\H_1},
        \fb_2 \in \sidtypeof{\S,\H_2}
        \text{ such that }
        \fa\in\fb_1 \PCompose \fb_2
      \}.
  \end{align*}
\begin{proof}
  \begin{description}
  \item[$\subseteq$]
    Let $\fa \in \sidtypeof{\S,\H_1 \stdunion \H_2}$.
    By \cref{def:sid-type}, we know that
    (1) $\fa = \sfproj{\S}{\frst}$ for some forest $\frst \in \sidfrstsof{\H_1 \stdunion \H_2}$ and
    (2) $\fa$ is delimited.
    By (2) and \cref{lem:delimited-forest--delimited-projection}, $\frst$ is delimited as well.
    Moreover, by \cref{lem:delimited-forest-delimited-sub-forests}, there exist $\S$-delimited forests $\frst_1$ and $\frst_2$ with $\frst\in\frst_1\FCompose\frst_2$ and, for $i \in \{1,2\}$, $\fheapof{\frst_i}=\H_i$.
    By \cref{lem:delimited-forest--delimited-projection}, both
    $\fb_1 \defn \sfproj{\S}{\frst_1}$ and $\fb_2 \defn \sfproj{\S}{\frst_2}$
    are delimited---hence, $\fb_1 \in \sidtypeof{\S,\H_1}$ and $\fb_2 \in \sidtypeof{\S,\H_2}$.
    Furthermore, by \cref{thm:fcompose-eq-pcompose}, we have
    $\fa = \sfproj{\S}{\frst} \in \fb_1\PCompose\fb_2$.
  \item[$\supseteq$]
    Assume there exist formulas $\fa \in \DUSH{\Sid}$, $\fb_1 \in \sidtypeof{\S,\H_1}$, and $\fb_2 \in \sidtypeof{\S,\H_2}$ such that $\fa\in\fb_1 \PCompose \fb_2$.
    By \cref{def:sid-type}, there exist forests $\frst,\frst_1,\frst_2$ such that, for $i \in \{1,2\}$,
    we have $\fb_i = \sfproj{\S}{\frst_i}$ and
    $\fheapof{\frst_i} = \H_i$.
    Then, by \Cref{thm:fcompose-eq-pcompose},  there exist forests
    $\barfrst_1,\barfrst_2$ such that $\frst_1\AEQ\barfrst_1$,
    $\frst_2\AEQ\barfrst_2$,
    $\frst \in \barfrst_1 \FCompose \barfrst_2$, and
    $\sfproj{\S}{\frst}=\fa$.
    By \cref{def:alpha-eq}, we have, for $i \in \{1,2\}$,
          $\fheapof{\barfrst_i}=\fheapof{\frst_i} = \H_i$.
    Moreover, by \cref{lem:forests:funion-heap}, we have
    $\H_1\stdunion\H_2 = \fheapof{\barfrst_1\funion\barfrst_2}$.
    Since
    $\frst \in \barfrst_1 \FCompose \barfrst_2$,
    \cref{lem:fderive:samemodel}
    and \cref{def:forest-composition}
    yield that
    $\H_1 \stdunion \H_2 =\fheapof{\frst}$.
    Hence, $\frst \in \sidfrstsof{\S,\H_1\stdunion\H_2}$
    and thus also $\fa \in \sidtypeof{\S,\H_1\stdunion\H_2}$.
    \qedhere
  \end{description}
\end{proof}

\subsection{Proof of \cref{lem:typealloc-definable}}
\label{app:typealloc-definable}

We first show an auxiliary result, namely that the stack-allocated variables of a state $\SH$ correspond precisely to the roots of the $\S$-decomposed forests of $\H$:

\begin{lemma}\label{lem:s-decomp:valloc-are-roots}
  Let $\SH$ be a state and let $\frst \in \sidfrstsof{\H}$ an $\S$-delimited forest.
  Then, we have $\valloc{\S,\H} = \set{x \mid \S(x) \in \froots{\barfrst}}$, where $\barfrst$ is the $\S$-decomposition of $\frst$.
\end{lemma}
\begin{proof}
  By \cref{lem:fderive:samemodel}, $\fheapof{\barfrst}=\H$ and thus,
  in particular, $\dom(\barfrst)=\dom(\H)$. Consequently,
  \[\S(\valloc{\S,\H}) = \img(\S)\cap\dom(\barfrst).\]
  By \cref{lem:s-decomposition-only-roots-free},
  we have $\img(\S)\cap\dom(\bartree) = \set{\troot{\bartree}}$ for all $\bartree\in\barfrst$.
  Hence,
  \[\img(\S)\cap\dom(\barfrst) = \froots{\barfrst}.\]
  Overall, we thus have $\S(\valloc{\S,\H}) = \froots{\barfrst}$.
  By taking the inverse $\S^{-1}$ on both sides of the equation, we obtain that $\valloc{\S,\H} = \set{x \mid \S(x) \in \froots{\barfrst}}$.
\end{proof}

\paragraph*{We restate the claim of \cref{lem:typealloc-definable}:}
Let $\SH$ be a state with $\sidtypeof{\S,\H} \neq \emptyset$.
Then, $\valloc{\S,\H} = \typealloc{\sidtypeof{\S,\H}}$.
\begin{proof}
  By definition of DUSHs, all root parameters of all DUSHs in
  $\typealloc{\sidtypeof{\S,\H}}$ are in $\img(\S)$. Consequently,
  $\valloc{\S,\H} \supseteq \typealloc{\sidtypeof{\S,\H}}$

  For the other implication, let $\frst$ be a forest with
  $\fheapof{\frst}=\H$ and $\sfproj{\S}{\frst} \in
  \sidtypeof{\S,\H}$. Such a forest must exist, as
  $\sidtypeof{\S,\H} \neq \emptyset$ by assumption.
  Let $\barfrst$ be the $\S$-decomposition of $\frst$.
  By \cref{lem:s-decomposition-delimited}, $\barfrst$ is delimited and
  by \cref{lem:fderive:samemodel},
  $\fheapof{\barfrst}=\H$, implying
  $\sfproj{\S}{\barfrst} \in \sidtypeof{\S,\H}$
  and we can apply \cref{lem:s-decomp:valloc-are-roots} to obtain that
  \[\valloc{\S,\H} = \set{x \mid \S(x) \in \froots{\barfrst}}.\]
  Consequently, all variables in $\valloc{\S,\H}$ occur as root
  parameters on the right-hand side of magic wands in
  $\sfproj{\S}{\barfrst}$. Therefore,
  $\valloc{\S,\H} \subseteq \typealloc{\sidtypeof{\S,\H}}$.
\end{proof}

\subsection{Proof of \cref{cor:guarded-formula-non-empty-type}}
\label{app:guarded-formula-non-empty-type}

\paragraph{Claim}
  Let $\phi \in \SLIDguarded$ and let $\SH$ be a state with $\SH \sidmodels \phi$.
  Then, $\sidtypeof{\S,\H} \neq \emptyset$.

\begin{proof}
  By \cref{lem:pos-formula-pos-model}, $\SH$ is a guarded state.
  \cref{lem:guarded-iterated-star-predicates} then yields that
  there exist $k \geq 1$ predicate calls
  such that
  $\SH \sidmodels \IteratedStar_{1\leq i\leq k}\pred_i(\vec{x_i})$.

  We split the heap $\H$ into disjoint heaps
  $\H_1\stdunion\cdots\stdunion\H_k$ such that
  $\SHi{i}\sidmodels\pred_i(\vec{x_i})$ for each $i \in [1,k]$.
  Next, consider the forest
  $\frst\defn\set{\ftree_1,\ldots,\ftree_k}$,
  where each $\ftree_i$ is a
  $\Sid$-tree with $\fheapof{\ftree_i}=\H_i$; such trees
  exist by \cref{lem:model-of-pred-to-tree}. Observe further that
  each of these trees is delimited, because they do not have holes and
  their root is in $\S(\vec{x_i})$.
  By \Cref{lem:forests:funion-heap}, we have $\fheapof{\frst}=\H$.
  Finally, \cref{lem:sf-projection-sound} yields
  $\SH \sidmodels \sfproj{\S}{\frst}$ and thus
  $\sfproj{\S}{\frst}\in\sidtypeof{\S,\H}$.
  Hence, $\sidtypeof{\S,\H} \neq \emptyset$.
\end{proof}

\subsection{Proof of \cref{lem:types:homo:compose}}
\label{app:types:homo:compose}
\paragraph{Claim}
  For guarded states $\SHi{1}$ and $\SHi{2}$ with $\H_1 \stdunion \H_2 \neq \bot$, we have
  $\sidtypeof{\S, \H_1 \stdunion \H_2} = \sidtypeof{\S,\H_1} \ACompose \sidtypeof{\S,\H_2}$.
\begin{proof}
  We need to show that $\sidtypeof{\S,\H_1} \ACompose \sidtypeof{\S,\H_2} \neq \bot$.
  Assume that $\sidtypeof{\S,\H_i} = \emptyset$ for $i=1$ or $i=2$.
  Then, $\typealloc{\typi{i}} = \emptyset$ and we get that $\typealloc{\typi{1}} \cap \typealloc{\typi{2}} = \emptyset$,
  Otherwise, we have $\sidtypeof{\S,\H_i} \neq \emptyset$ for $i=1,2$.
  Then, $\valloc{\S,\H_i} = \typealloc{\sidtypeof{\S,\H_i}}$.
  $\H_1\stdunion\H_2\neq\bot$ then implies that
  $\typealloc{\typi{1}} \cap \typealloc{\typi{2}} = \emptyset$.
  The claim then follows from~\cref{thm:oursunion-to-derive}.
\end{proof}

\subsection{Proof of \Cref{lem:types:homo:existence}}
\label{app:types:homo:existence}

\paragraph{Claim}
For $i \in \{1,2\}$, let $\SHi{i}$ be states with $\sidtypeof{\S,\H_i} = \typi{i} \neq \emptyset$ and $\typi{1} \ACompose \typi{2} \neq \bot$.
  Then, there are states $\SHpair{\S}{\H'_i}$ such that $\sidtypeof{\S,\H'_i} = \typi{i}$ and
  $\sidtypeof{\S, \H_1' \stdunion \H'_2} = \typi{1} \ACompose \typi{2}$.

\begin{proof}
  We choose some states $\SHpair{\S}{\H'_i}$ that are isomorphic to $\SHi{i}$ such that $\locs{\H'_1} \cap \locs{\H'_2} \subseteq \img(\S)$.
  We have that $\SHpair{\S}{\H'_i} = \sidtypeof{\S,\H_i} = \typi{i}$ because isomorphic states have the same types (observe that the stack-projection replaces location that are not in the image of the stack by quantified variables).
  By~\cref{lem:typealloc-definable}, we have    $\typealloc{\typi{i}} = \typealloc{\SHpair{\S}{\H'_i}} = \valloc{\S,\H'_i}$.
  Thus, we get $\H'_1\stdunion\H'_2 \neq \bot$ from    $\typi{1} \ACompose \typi{2} \neq \bot$ and $\locs{\H'_1} \cap \locs{\H'_2} \subseteq \img(\S)$.
  Then, \cref{thm:oursunion-to-derive} yields that $\sidtypeof{\S, \H'_1 \stdunion \H'_2} = \sidtypeof{\S,\H'_1} \ACompose \sidtypeof{\S,\H'_2}$.
\end{proof}

\subsection{Proof of \Cref{lem:types:homo:rename}}
\label{app:types:homo:rename}

\paragraph{Claim}
  For $\vec{x}$, $\vec{y}$ as above and a stack $\S$
  with $\vec{y} \subseteq \dom(\S)$ and $\vec{x} \cap \dom(\S) = \emptyset$, we have
  \[
    \tinst{\sidtypeof{\pinst{\S}{\vec{x}}{\vec{y}},\H}}{\eqclasses{\S}}{\vec{x}}{\vec{y}}
    =
    \sidtypeof{\S,\H}
    .
  \]
\begin{proof}
  Let $\vec{y'}$ be the sequence obtained by replacing every variable in $y \in \vec{y}$ by the maximal variable $y'$ with $\S(y') = \S(y)$.
  We consider some $\fa \in  \sidtypeof{\pinst{\S}{\vec{x}}{\vec{y}},\H}$.
  Then, there exists a forest $\frst \in \sidfrstsof{\H}$ such that
    $\fa = \sfproj{
        \pinst{\S}{\vec{x}}{\vec{y}}
    }{\frst}$.
  By construction of stack-projections (cf. \cref{def:types:sf-projection}), we obtain that
  \[
    \pinst{\fa}{\vec{x}}{\vec{y'}}
    =
    \pinst{\sfproj{\pinst{\S}{\vec{x}}{\vec{y}}}{\frst}}{\vec{x}}{\vec{y'}}
    =
    \sfproj{\S}{\frst} \in \sidtypeof{\S,\H}.
  \]
  The converse direction is analogous.
\end{proof}

\subsection{Proof of \Cref{lem:types:homo:forget}}
\label{app:types:homo:forget}

\paragraph{Claim}
  Let $\SH$ be a guarded state such that $\S(x) \in \dom(\H)$ holds for some variable $x$.
  Then,
  \[ \Forget{\eqclasses{\S},x}{\sidtypeof{\S,\H}} = \sidtypeof{\pinst{\S}{x}{\bot},\H}~. \]
\begin{proof}
  We will use the following fact $(\dag)$ based on the construction of projections (cf. \cref{def:types:sf-projection}):
  If $\frst \in \sidfrstsof{\H}$, then
  $\S(x)$ is replaced in $\sfproj{\pinst{\S}{x}{\bot}}{\frst}$ by an existentially-quantified variable iff $x$ is replaced in $\Forget{\eqclasses{\S},x}{\sfproj{\S}{\frst}}$   by an existentially-quantified variable.

  Now, consider some $\fa \in \sidtypeof{\pinst{\S}{x}{\bot},\H}$.
  Then, there is some $\frst \in \sidfrstsof{\H}$ such that $\sfproj{\pinst{\S}{x}{\bot}}{\frst} = \fa \in  \DUSH{\Sid}$.
  Clearly, we have $\sfproj{\S}{\frst} \in \DUSH{\Sid}$ and hence
  $\sfproj{\S}{\frst} \in \sidtypeof{\S,\H}$.
  Applying $(\dag)$, we conclude that
  \[ \fa = \Forget{\eqclasses{\S},x}{\sfproj{\S}{\frst}} \in \Forget{\eqclasses{\S},x}{\sidtypeof{\S,\H}}~. \]
  Conversely, let $\fa \in \Forget{\eqclasses{\S},x}{\sidtypeof{\S,\H}}$.
  By construction, $\fa \in \DUSH{\Sid}$ and there is some $\frst \in \sidfrstsof{\H}$ such that $\Forget{\eqclasses{\S},x}{\sfproj{\S}{\frst}} = \fa$.
  Applying $(\dag)$, we can conclude that $\fa = \sfproj{\pinst{\S}{x}{\bot}}{\frst} \in \sidtypeof{\pinst{\S}{x}{\bot},\H}$.
\end{proof}

\subsection{Proof of~\Cref{lem:extend}}
\label{app:extend}

\paragraph{Claim}
 For every state $\SH$, variable $x$ with $\S(x) \not\in \locs{\H}$ and $\eqclasses{\S}(x) = \set{x}$,
    %we have
 \[
     \Extend{x}{\sidtypeof{\pinst{\S}{x}{\bot}, \H}} = \sidtypeof{\S,\H}
 .\]
\begin{proof}
  Assume $\phi$ is the $x$-extension of some $\phi' \in \sidtypeof{\pinst{\S}{x}{\bot},\H}$.
  Then there exists some $\frst \in \sidfrstsof{\H}$ such that $\phi' = \sfproj{\pinst{\S}{x}{\bot}}{\frst} \in \DUSH{\Sid}$.
  We can now choose a forest $\frst'$ with  $\frst\AEQ\frst'$ such that $\phi = \sfproj{\S}{\frst}$.
  Because of $\frst\AEQ\frst'$ and $\sfproj{\pinst{\S}{x}{\bot}}{\frst} \in \DUSH{\Sid}$ we get that $\sfproj{\S}{\frst'} \in \DUSH{\Sid}$.
  Hence, $\phi \in \sidtypeof{\S,\H}$.

  Conversely, let $\phi \in \sidtypeof{\S,\H}$.
  Then there exists some $\frst \in \sidfrstsof{\H}$ such that $\phi=\sfproj{\S}{\frst} \in \DUSH{\Sid}$.
  We note that $\S(x)\not\in \finterface{\frst}$ since $\S(x)\not\in \locs{\H}$.
  Hence, $\sfproj{\pinst{\S}{x}{\bot}}{\frst} \in \sidtypeof{\pinst{\S}{x}{\bot},\H}$.
  We distinguish two cases:
  First, if $x \not\in \fvs{\phi}$, then $\phi = \sfproj{\S}{\frst} = \sfproj{\pinst{\S}{x}{\bot}}{\frst} \in \sidtypeof{\pinst{\S}{x}{\bot}, \H}$.
  Second, if $x \in \fvs{\phi}$, then $\S(x)$ corresponds to a universally quantified variable in $\sfproj{\pinst{\S}{x}{\bot}}{\frst}$.
  Hence, $\phi$ is an $x$-instantiation of $\sfproj{\pinst{\S}{x}{\bot}}{\frst}$.
  By \cref{def:x-extension}, this means $\phi \in \Extend{x}{\sidtypeof{\pinst{\S}{x}{\bot}, \H}}$.
\end{proof}

\subsection{Proof of \Cref{thm:types-capture-sat}}
\label{app:types-capture-sat}

For a concise formalization, we assume---in addition to our global assumptions stated in \cref{sec:sl-basics:sid-assumptions}---that
%all SIDs considered in the remainder of this section are (a) in the bounded treewidth fragment $\IDbtw$ and (b) pointer-closed, i.e.,
%contain a dedicated predicate simulating each points-to assertion (cf. \cref{def:ptrclosed}).
%Moreover, since we allowed concrete locations in $\SLIDguarded$ formulas only as an auxiliary construct, we assume that
all formulas $\phi$ under consideration are $\SLIDguarded$ formulas without constant values except the null pointer, i.e., $\values{\phi} \subseteq \{\nil\}$. %%\emptyset$ (note that the null pointer $\nil$ is still permitted as it is not a location).

We will prove \cref{thm:types-capture-sat} by structural induction on the syntax of $\SLIDguarded$ formulas.
For most base cases---those that involve the heap---we rely on the fact that a state satisfies all
formulas in its type.
\begin{lemma}\label{lem:forms-in-type-hold}
  If $\phi \in \sidtypeof{\S,\H}$ for some state $\SH$, then $\SH \sidmodels \phi$.
\end{lemma}
\begin{proof}
  Since $\phi \in \sidtypeof{\S,\H}$, there exists a $\Sid$-forest
  $\frst$ with $\fheapof{\frst}=\H$ and $\sfproj{\S}{\frst}=\phi$.
  By \cref{lem:sf-projection-sound}, we have
  $\SHpair{\S}{\fheapof{\frst}}\sidmodels \sfproj{\S}{\frst}$
  and thus also $\SH \sidmodels \phi$.
\end{proof}
\noindent
Finally, to deal with the separating conjunction, we need another auxiliary result.
In \cref{lem:types:homo:compose}, we showed how two types can be composed into a single one, i.e.,
\[ \sidtypeof{\S,\H_1}\Compose\sidtypeof{\S,\H_2}=\sidtypeof{\S,\H_1\stdunion\H_2}~. \]
To prove \cref{thm:types-capture-sat}, we need the reverse:
Given a composed type, say $\sidtypeof{\S,\H}=\typ_1\Compose\typ_2$,
we need to \emph{decompose} $\H$ into two heaps whose types (in conjunction with stack $\S$) are $\typ_1$ and $\typ_2$.
\begin{lemma}[Type decomposability]\label{lem:types:decompose}
Let $\SH$ be a state with $\emptyset \neq \sidtypeof{\S,\H}=\typ_1\ACompose\typ_2$.
Then, there exist $\H_1,\H_2$ such that $\H=\H_1\stdunion\H_2$, $\typ_1=\sidtypeof{\S,\H_1}$, and $\typ_2=\sidtypeof{\S,\H_2}$.
\end{lemma}
Since proving type decomposability involves a bit more technical machinery, we refer the interested reader to \cref{app:types:decompose} for a detailed proof.

With the above three lemmas at hand, we can now prove the refinement theorem.

\paragraph{Claim (Refinement theorem)}
  For all stacks $\S$, heaps $\H_1$, $\H_2$, and $\SLIDguarded$ formulas $\fa$,
  \[
      \sidtypeof{\S,\H_1} = \sidtypeof{\S,\H_2}
      \qquad\text{implies}\qquad
      \SHi{1} \sidmodels \phi ~~\text{iff}~~ \SHi{2} \sidmodels \phi~.
  \]
\begin{proof}%[Proof (of \cref{thm:types-capture-sat})]
  We only show that if $\SHi{1} \sidmodels \phi$ then $\SHi{2} \sidmodels \phi$;
  the converse direction is symmetric.
  We proceed by induction on the structure of the $\SLIDguarded$ formula $\phi$.
  Let us assume that $\SHi{1} \sidmodels \phi$.

%  If $\SHi{1}$ or $\SHi{2}$ is not a guarded state in $\Mpos{\Sid}$,
%  then $\sidtypeof{\S,\H_1}$ is undefined and neither model satisfies $\phi$ by
%  \cref{lem:pos-formula-pos-model}.
%
%  Let us thus assume $\SHi{1},\SHi{2}\in \Mpos{\Sid}$.

  \begin{description}
  \item[Case $\phi=\emp$.]
    By the semantics of $\emp$, we have $\H_1 = \emptyset$.
    Let $\frst$ be the empty forest.
    Then
    $\frst \in \sidfrstsof{\H_1}$ and thus
    $\emp=\sfproj{\S}{\frst} \in \sidtypeof{\S,\H_1} =
    \sidtypeof{\S,\H_2}$.
    Hence, by \cref{lem:forms-in-type-hold}, we have $\SHi{2} \sidmodels \emp$.
  \item[Cases $\phi = \sleq{x}{y}$, $\phi=\slneq{x}{y}$.]
      We observe that the states $\SHi{1}$ and $\SHi{2}$ have the same stack.
      Then we proceed as in the case for $\phi=\emp$.
  \item[Case $\phi = \pto{x}{\tuple{y_1,\ldots,y_k}}$.]
    By assumption, $\Sid$ is \ptrclosed{} (see \cref{def:ptrclosed}), i.e.,
    $\SHi{1} \sidmodels \ptrpred_{k}(x,y_1,\ldots,y_k)$.
    We define a $\Sid$-forest $\frst = \set{\ftree}$, where $\ftree$ is
    \begin{align*}
    \ftree = \{ \S(x) \mapsto \langle \emptyset,
        \ptrpred_{k}(\S(x),\S(y_1),\ldots,\S(y_k))\; \Rule
        \pto{\S(x)}{\tuple{\S(y_1),\ldots,\S(y_k)}} \rangle\}~.
    \end{align*}
    Observe that $\frst \in \sidfrstsof{\H_1}$ and
    \begin{align*}
\ptrpred_{k}(x,y_1,\ldots,y_k)=\sfproj{\S}{\frst} \in
    \sidtypeof{\S,\H_1} = \sidtypeof{\S,\H_2}.
    \end{align*}
    Hence, by \cref{lem:forms-in-type-hold}, we have
    $\SHi{2} \sidmodels \ptrpred_{k}(x,y_1,\ldots,y_k)$.
    By definition of the predicate $\ptrpred_k$, we conclude that
    $\SHi{2} \sidmodels \pto{x}{\tuple{y_1,\ldots,y_k}}$.
  \item[Case $\phi = \pred(z_1,\ldots,z_k)$.]
    By \cref{lem:model-of-pred-to-tree}, there exists a $\Sid$-tree
    $\ftree$ such that  $\trootpred{\ftree}=\pred(\S(z_1),\ldots,\S(z_k))$,
    $\tallholepreds{\ftree}=\emptyset$, and
    $\fheapof{\set{\ftree}}=\H_1$.
    Let
    \[
      \psi \defn
      \pinst{\pred(\S(z_1),\ldots,\S(z_k))}{\dom(\stkchc)}{\img(\stkchc)}
      = \sfproj{\S}{\set{\ftree}}
    .\]
    Then, $\psi \in \sidtypeof{\S,\H_1} = \sidtypeof{\S,\H_2}$
    and, by
    \cref{lem:forms-in-type-hold}, $\SHi{2} \sidmodels \psi$.
    Observe that while $\psi \neq \pred(\vec{z})$ is possible, we have
    by definition of $\stkchc$ that the parameters of the predicate
    call in $\psi$ evaluate to the same locations as the parameters
    $\vec{z}$.  Hence, $\SHi{2}\sidmodels\pred(z_1,\ldots,z_k)$.
  \item[Case $\phi = \phi_1 \wedge \phi_2$.]
    We then have $\SHi{1} \sidmodels \phi_1$ and $\SHi{1} \sidmodels \phi_2$.
    By I.H.,
    $\SHi{2} \sidmodels \phi_1$ and $\SHi{2} \sidmodels \phi_2$.
    Hence,
    $\SHi{2} \sidmodels \phi_1 \wedge \phi_2$.
  \item[Cases $\phi = \phi_1 \vee \phi_2$, $\phi = \phi_1 \wedge\neg \phi_2$.]
    Analogous to previous case.
  \item[Case $\phi = \phi_1 \sep \phi_2$.]
    By the semantics of $\sep$, there exist heaps $\Hoo$ and $\Hot$ such
    that $\SHpair{\S}{\H_{1,i}} \sidmodels \phi_i$ for $1 \leq i \leq 2$.
    Let $\typ_i \defn \sidtypeof{\S,\H_{1,i}}$.
    By \cref{lem:pos-formula-pos-model}, we have that $\SHpair{\S}{\H_{1,i}}\in\Mpos{\Sid}$ for $1 \leq i \leq 2$.
    By \cref{lem:types:homo:compose} we have that $\sidtypeof{\S, \H_1 \stdunion \H_2} = \typ_1 \ACompose \typ_2$.
    By \cref{cor:guarded-formula-non-empty-type}, we have that $\sidtypeof{\S,\H_1} \neq \emptyset$.
    We can then apply \cref{lem:types:decompose} to $\SHi{2}$, $\typ_1$ and $\typ_2$ in order to obtain states $(\S,\Hto)$ and $(\S,\Htt)$ with $\H_2=\Hto\stdunion\Htt$,
    $\sidtypeof{\S,\Hto}=\typ_1$, and $\sidtypeof{\S,\Htt}=\typ_2$.

    We can thus apply the I.H. to both
    $\Hoo$, $\Hot$, $\phi_1$ and $\Hto$, $\Htt$, $\phi_2$
    to conclude that
    $\SHpair{\S}{\H_{2,1}} \sidmodels \phi_1$
    and $\SHpair{\S}{\H_{2,2}} \sidmodels \phi_2$.
    Since $\Hto \stdunion \Htt = \H_2$, the semantics of $\sep$ then yields $\SHi{2} \sidmodels \phi$.
  \item[Case $\phi = \phi_0 \wedge (\phi_1 \protect\sept \phi_2)$.]
    Then there exists a heap $\H_0$ with $\SHi{0} \sidmodels \phi_1$
    and $\SHpair{\S}{\H_1 \stdunion \H_0} \sidmodels \phi_2$.

    Since $\SHi{1}$ and $\SHi{2}$ have the same type, we have
    $\valloc{\S,\H_1}=\valloc{\S,\H_2}$.  We can therefore
    assume w.l.o.g.~that $\H_2 \stdunion \H_0$ is defined---if this
    is not the case, simply replace $\H_0$ with a heap $\H_0'$ such
    that $\SHi{0} \Iso \SHpair{\S}{\H_0'}$ and both
    $\H_1 \stdunion \H_0'$ and $\H_2 \stdunion \H_0'$ are defined.
    Then, by \cref{lem:sid:iso-models-same-formulas}, we can conclude that $(\S, \H_1\stdunion
    \H_0')\sidmodels \phi$.

    By \cref{lem:pos-formula-pos-model} we have $\SHi{0},\SHi{1}\in\Mpos{\Sid}$. \cref{lem:types:homo:compose} then yields that
    \begin{align*}
    \sidtypeof{\S,\H_1\stdunion\H_0} ~=~& \sidtypeof{\S,\H_1} \Compose
    \sidtypeof{\S,\H_0}\\ ~=~& \sidtypeof{\S,\H_2} \Compose
    \sidtypeof{\S,\H_0}\\ ~=~& \sidtypeof{\S,\H_2\stdunion\H_0}.
    \end{align*}

    Now, we apply the I.H. for $\phi_0$, $\H_1$ and $\H_2$ to conclude that $\SHi{2} \sidmodels \phi_0$, as well as for $\phi_2$, $\SHpair{\S}{\H_1\stdunion\H_0}$ and
    $\SHpair{\S}{\H_2\stdunion\H_0}$ to conclude that
    $\SHpair{\S}{\H_2\stdunion\H_0} \sidmodels \phi_2$.
    Hence, by the semantics of $\sept$ and $\wedge$, we have
    $\SHi{2}\sidmodels \phi_0 \wedge (\phi_1 \sept \phi_2)$.
  \item[Case $\phi = \phi_0 \wedge (\phi_1 \mw \phi_2)$.]
    Analogous to the previous case for guarded septraction, except that we must consider
    \emph{arbitrary} models $\H_0$ with $\SHi{0} \sidmodels \phi_1$
    and $\SHpair{\S}{\H_1 \stdunion \H_0} \sidmodels
    \phi_2$.
    \qedhere
  \end{description}
\end{proof}

\subsection{Proof of \cref{lem:types:decompose} (type decomposability)}
\label{app:types:decompose}

We need a couple of auxiliary definitions and lemmata before we can show this result in \cref{lem:types:decompose} at the end of this section.
\begin{definition}
We call $\frst$ \emph{$\S$-decomposed} iff
$\frst=\fsplit{\frst}{\img(\S)}$.
\end{definition}

\begin{lemma}\label{lem:type-split-to-forest-split}
  Let $\SH$ be a state with $\emptyset \neq \sidtypeof{\S,\H}$ and let $\typ_1,\typ_2\in\ssidtypes{\eqclasses{\S}}{\Sid}$ be types with $\sidtypeof{\S,\H}=\typ_1\ACompose\typ_2$
  Then there exist $\S$-decomposed, $\S$-delimited $\Sid$-forests
  $\frst,\frst_1,\frst_2$ such that
  \begin{enumerate}
  \item $\sfproj{\S}{\frst_i}\in\typ_i$, $1\leq i \leq 2$,
  \item $\frst_1\funion\frst_2=\frst$, and
  \item $\fheapof{\frst}=\H$.
  \end{enumerate}
\end{lemma}
\begin{proof}
  By assumption we have $\emptyset \neq \sidtypeof{\S,\H}$.
  Take an arbitrary forest $\barfrst$ with
  $\sfproj{\S}{\barfrst}\in\sidtypeof{\S,\H}$.
  By definition, $\barfrst$ is $\S$-delimited.
  Let $\frst\defn\fsplit{\frst}{\img(\S)}$ be the $\S$-decomposition of $\barfrst$.
  By \cref{lem:s-decomposition-delimited}, $\frst$ is $\S$-delimited.
  By \cref{lem:fderivestar-is-split}, $\frst \fderivestar \barfrst$.
  \Cref{lem:fderive:samemodel} thus gives us that
  $\fheapof{\frst}=\H$.
  Hence, $\sfproj{\S}{\frst}\in\sidtypeof{\S,\H}$.

  By definition of $\ACompose$, there exist formulas
  $\psi_1\in\typ_1$, $\psi_2 \in \typ_2$
  with $\sfproj{\S}{\frst} \in \psi_1\PCompose\psi_2$.
  By definition, there exist $\Sid$-forests $\frstb_1,\frstb_2$ with $\sfproj{\S}{\frstb_i}=\psi_i$.
  Because $\typ_1\ACompose\typ_2$ is defined, we have $\typealloc{\typ_1}\cap\typealloc{\typ_2}=\emptyset$, allowing us to assume w.l.o.g.~that $\frstb_1\funion\frstb_2\neq\bot$.
  By \cref{thm:fcompose-eq-pcompose}, there then
  exist forests $\frst_1,\frst_2$ with $\sfproj{\S}{\frst_i}=\psi_i$ and $\frst \in
  \frst_1 \FCompose \frst_2$.
  Because $\frst$ is $\S$-decomposed, this implies that zero $\fderive$-steps were taken by $\FCompose$, i.e., $\frst= \frst_1 \funion \frst_2$.
  Moreover, because $\frst$ is $\S$-decomposed and $\S$-delimited, we get that $\frst_1$ and $\frst_2$ are $\S$-decomposed and $\S$-delimited as well.
\end{proof}
\begin{definition}[Roots of a DUSH]
  Let $\phi=\EEX{\vec{e}}\FFA{\vec{a}} \IteratedStar_{1\leq i\leq k} (\zeta_i \mw \pred_i(\vec{z_i}))$ be a DUSH.
  The \emph{roots} of $\psi$ are the set
  \[\gls{dushroots-phi} \defn
  \bigcup_{1 \leq i \leq k} \eqclasses{\S}(\proot{\pred_i(\vec{z_i})}).\]
\end{definition}

Clearly, the roots of a forest are connected to the roots of a DUSH via the stack:

\begin{lemma}\label{lem:froots-to-dushroots}
  Let $\frst$ be a $\Sid$-forest.
  Then,
  % $\dushroots{\phi}=\set{x \mid \S(x) \in \froots{\frst}}$
  $\dushroots{\sfproj{\S}{\frst}}=\set{x \mid \S(x) \in
    \froots{\frst}}$.
  %$\dushroots{\phi}=\bigcup\stkinv(\froots{\frst})$.
\end{lemma}
\begin{proof}
  Let $\phi\defn\sfproj{\S}{\frst}$.
  By definition of DUSHs, $\froots{\frst}\subseteq\img(\S)$. Every root $l\in\froots{\frst}$ is therefore replaced by a variable in $\stkchc(l)$ by stack--forest projection.
  Since $\dushroots{\phi}$ closes the set of roots under all $\eqclasses{\S}(\cdot)$, we obtain that $\dushroots{\phi}$ contains \emph{all} variables $x$ with  $\S(x)\in\froots{\frst}$.
\end{proof}

\begin{lemma}\label{lem:type-closed-under-decomp}
  Let $\typ$ be a $\Sid$-type and $\psi \in \typ$.
  There exists a formula $\psi' \in \typ$ such that
  $\psi' \derivestar \psi$ and
  $\typealloc{\typ}=\dushroots{\psi'}$.
\end{lemma}
\begin{proof}
% $\psi'=\EEX{\vec{e}}\IteratedStar_{1\leq i \leq k}\FFA{\vec{a_i}}(\zeta_i \mw
%   \pred_i(\vec{z_i}))$,
  Let $\SH$ be such that $\sidtypeof{\S,\H}=\typ$.
  By~\cref{lem:typealloc-definable} we then have that
  $\valloc{\S,\H}=\typealloc{\typ}$ \tagA.
  By definition of $\Sid$-types, there then exists a $\Sid$-forest $\frst$ with $\fheapof{\frst}=\H$ and $\sfproj{\S}{\frst}=\psi$.
  Let $\barfrst \defn \fsplit{\frst}{\img(\S)}$ be the
  $\S$-decomposition of $\frst$ and write
  $\psi'\defn\sfproj{\S}{\barfrst}$. We show that $\psi'$ has the
  desired properties.

  First, $\barfrst\fderivestar\frst$ by
  \cref{lem:fderivestar-is-split}.
  Observe that $\frst \in \barfrst \FCompose \emptyset$ (where $\emptyset$ is the empty forest).
  \Cref{cor:fcompose-to-pcompose} therefore guarantees that
  $\psi \in \psi' \PCompose \emp$ and thus $\psi' \derivestar \psi$.

  Second, by \cref{lem:fderive:samemodel},
  $\fheapof{\barfrst}=\fheapof{\frst}$ and thus $\psi'\in\typ$.
  Moreover, by \cref{lem:s-decomp:valloc-are-roots},
  $\valloc{\S,\H} = \set{x \mid \S(x) \in \froots{\barfrst}}$.
  We combine the above with \tagA{} to derive
  $\typealloc{\typ}=\set{x \mid \S(x) \in \froots{\barfrst}}$.
  \Cref{lem:froots-to-dushroots} then yields that
  $\typealloc{\typ}=\dushroots{\psi'}$.
\end{proof}

%\cmtodo{CMFIXME: this is central for refinement and correctness; not needed elsewhere}
%\begin{lemma}[Type decomposability]\label{lem:types:decompose}
%  \index{decomposability!of Phi-types@of $\Sid$-types}
%  %
%  Let $\SH\in\Mpos{\Sid}$ and assume that
%  $\sidtypeof{\S,\H}=\typ_1\ACompose\typ_2$.
%  %
%  Then there exist $\H_1,\H_2$ such that
%  %$\SHpair{\S}{\H_1},\SHpair{\S}{\H_2}\in\Mpos{\Sid}$,
%  $\H=\H_1\stdunion\H_2$, $\typ_1=\sidtypeof{\S,\H_1}$, and
%  $\typ_2=\sidtypeof{\S,\H_2}$.
%\end{lemma}

\paragraph{Claim (\cref{lem:types:decompose})}
Let $\SH$ be a state with $\emptyset \neq \sidtypeof{\S,\H}=\typ_1\ACompose\typ_2$.
Then, there exist $\H_1,\H_2$ such that $\H=\H_1\stdunion\H_2$, $\typ_1=\sidtypeof{\S,\H_1}$, and $\typ_2=\sidtypeof{\S,\H_2}$.

\begin{proof}
  Let $\typ \defn \sidtypeof{\S,\H}$.
  By \cref{lem:type-split-to-forest-split}, there exist $\S$-decomposed, $\S$-delimited forests $\frst,\frst_1,\frst_2$ with
  \begin{enumerate}
  \item $\sfproj{\S}{\frst_i}\in\typ_i$, $1\leq i \leq 2$,
  \item $\frst_1\funion\frst_2=\frst$, and
  \item $\fheapof{\frst}=\H$.
  \end{enumerate}
  Define $\H_1\defn \fheapof{\frst_1}$,
  $\H_2\defn\fheapof{\frst_2}$.
  Then, $\H_1 \stdunion \H_2=\H$ because of $\fheapof{\frst_1\funion\frst_2}=\H$.
  Because the $\frst_i$ are $\S$-delimited, we have
  $\danglinglocs{\H_i}\subseteq\img(\S)$.
  Hence, $\SHi{1},\SHi{2} \in \Mpos{\Sid}$.
  By \cref{lem:s-decomp:valloc-are-roots} and \cref{lem:typealloc-definable} we have
  \[
     \valloc{\S,\H} = \set{x \mid \S(x) \in \froots{\frst}} = \typealloc{\typ} \tag*{(*)}.
  \]
  Further, we have
  \[
    \valloc{\S,\H_i} = \set{x \mid \S(x) \in \froots{\frst_i}} =
    \typealloc{\typ_i}, \tag*{\tagA}
  \]
  where the first equality follows from
  \cref{lem:s-decomp:valloc-are-roots}, and the second equality holds by (*) and because the $\frst_i$ are $\S$-decomposed.

  We will show that $\typ_1=\sidtypeof{\S,\H_1}$; the argument for $\typ_2=\sidtypeof{\S,\H_2}$ is symmetrical.
  We prove the inclusions $\typ_1\subseteq\sidtypeof{\S,\H_1}$ and $\typ_1\supseteq\sidtypeof{\S,\H_1}$ separately.

  ``$\typ_1\subseteq\sidtypeof{\S,\H_1}$:''
  Let $\psi_1 \in \typ_1$.
  By \cref{lem:type-closed-under-decomp},
  there exists a formula $\psi_1' \in \typ_1$ such that $\psi_1' \derivestar \psi_1$, and $\typealloc{\typ_1}=\dushroots{\psi_1'}$ (\#).
  Let $\psi_2'\defn \sfproj{\S}{\frst_2} \in \typ_2$.
  By the definition of projections, we have $\psi_i'=\EEX{\vec{e_i}} \FFA{\vec{a_i}} \phi_i$ for some $\vec{e_i},\vec{a_i},\phi_i$,
  where we can assume w.l.o.g. that  $\vec{e_1},\vec{e_2},\vec{a_1},\vec{a_2}$ are pairwise disjoint.
  By definition of type composition, $\ACompose$,
  it follows that $\psi \defn \EEX{\vec{e_1},\vec{e_2}} \FFA{\vec{a_1},\vec{a_2}}\phi_1\sep\phi_2\in\typ$.
  Then, there is an $\Sid$-forest
  $\frstb \in \sidfrstsof{\S,\H}$ such that    $\sfproj{\S}{\frstb}=\psi$.
  From (\#), (*) and \tagA{} we obtain that $\valloc{\S,\H}=\dushroots{\psi}$.
  By~\cref{lem:delimited-forest-delimited-sub-forests} there exist $\Sid$-forests
  $\frstb_1,\frstb_2$ with   $\frstb=\frstb_1\funion\frstb_2$,
  $\fheapof{\frstb_i}=\H_i$ and $\frstb \in \frstb_1 \FCompose \frstb_2$.
  With $\sfproj{\S}{\frstb_1\funion\frstb_2} = \EEX{\vec{e_1},\vec{e_2}} \FFA{\vec{a_1},\vec{a_2}}\phi_1\sep\phi_2$ we then must have that $\sfproj{\S}{\frstb_i}= \psi_i'$.
  Therefore, $\psi_1'\in\sidtypeof{\S,\H_1}$.
  Since $\psi_1'\derivestar \psi_1$, \cref{lem:derive-to-fderive}
  then gives us a forest $\frstb_1'$
  s.t.~$\frstb_1 \fderivestar \frstb_1'$ and
  $\sfproj{\S}{\frstb_1'} = \psi_1$.
  Because also $\fheapof{\frstb_1'}=\fheapof{\frstb_1}$ by
  \cref{lem:fderive:samemodel}, $\frstb_1' \in \sidfrstsof{\H_1}$ and
  $\sfproj{\S}{\frstb_1'}\in\sidtypeof{\S,\H_1}$.
  Thus, $\psi_1 \in \sidtypeof{\S,\H_1}$.

  ``$\typ_1\supseteq\sidtypeof{\S,\H_1}$:''
  Let $\psi_1 \in \sidtypeof{\S,\H_1}$.
  By \cref{lem:type-closed-under-decomp},
  there exists a formula $\psi_1' \in \sidtypeof{\S,\H_1}$ such that $\psi_1' \derivestar \psi_1$, and $\typealloc{\typ_1}=\dushroots{\psi_1'}$ (\#).
  %Because of $\psi_1' \in \sidtypeof{\S,\H_1}$ there is a tree  $\frstb_1' \in \sidfrstsof{\S,\H}$ such that    $\sfproj{\S}{\frstb_1'}=\psi_1'$.
  Let $\psi_2'\defn \sfproj{\S}{\frst_2} \in \sidtypeof{\S,\H_2}$.
  By the definition of projections, we have $\psi_i'=\EEX{\vec{e_i}} \FFA{\vec{a_i}} \phi_i$ for some $\vec{e_i},\vec{a_i},\phi_i$,
  where we can assume w.l.o.g. that  $\vec{e_1},\vec{e_2},\vec{a_1},\vec{a_2}$ are pairwise disjoint.
  By \cref{lem:types:homo:compose} we then have that    $\psi \defn \EEX{\vec{e_1},\vec{e_2}} \FFA{\vec{a_1},\vec{a_2}}\phi_1\sep\phi_2 \in
  \sidtypeof{\S,\H_1} \ACompose \sidtypeof{\S,\H_2} = \sidtypeof{\S, \H_1 \stdunion \H_2} = \typ$.
  From (\#), (*) and \tagA{} we obtain that $\valloc{\S,\H}=\dushroots{\psi}$.
  Because of $\typ=\typ_1\ACompose\typ_2$ there are some $\psi_i''\in \typ_i$ such that $\psi \in \psi_1'' \PCompose \psi_2''$.
  Because of $\valloc{\S,\H_i} = \typealloc{\typ_i}$ we must have that $\psi_i'' = \psi_i'$.
  Therefore, $\psi_1'\in\typ_1$.
  We now consider some forest $\frstb_1$ with $\sidtypeof{\S,\fheapof{\frstb_1}} = \typ_1$ and  $\sfproj{\S}{\frstb_1'} = \psi_1'$.
  Since $\psi_1'\derivestar \psi_1$,   \cref{lem:derive-to-fderive}
  then gives us a forest $\frstb_1'$
  s.t.~$\frstb_1 \fderivestar \frstb_1'$ and
  $\sfproj{\S}{\frstb_1'} = \psi_1$.
  Because also $\fheapof{\frstb_1'}=\fheapof{\frstb_1}$ by
  \cref{lem:fderive:samemodel},
  we get that $\sfproj{\S}{\frstb_1}\in\sidtypeof{\S,\fheapof{\frstb_1}} = \typ_1$.
  Hence, $\psi_1 \in \typ_1$.
\end{proof}

\subsection{Correctness of the Fixed Point Algorithm For Computing Types of Predicate Calls}
\label{subec:correctness-fixed-point}

\subsubsection{Soundness of the Type
  Computation}\label{sec:fixedpoint:sound}\index{type computation!soundness}

We organize the soundness proof into a sequence of simple lemmata about the base cases of the fixed point algorithm and about the operations $\ACompose$, $\tinst{\cdot}{\cdot}{\cdot}{\cdot}$, $\ForgetKW$ and $\ExtendKW$.
The soundness of the overall algorithm is then be a direct consequence of these lemmata.
We first characterize the types of atomic formulas.
\begin{lemma}\label{lem:types-of-emp}
 For all aliasing constraints $\scls$, $\goaltypes{\emp} = \set{ \set{\emp} }$.
\end{lemma}
\begin{proof}
  Let $\SH$ be a state with $\sidtypeof{\S,\H}\in\goaltypes{\emp}$ and $\eqclasses{\S}=\scls$.
  By definition, $\SH \sidmodels \emp$ and thus $\H = \emptyset$.
  We now argue that $\sidtypeof{\S,\H} = \set{\emp}$.

  We note that $\emptyset \in \sidfrstsof{\H}$ ($\emptyset$ is the forest that does not contain any trees) and $\emp = \sfproj{\S}{\set{\emptyset}}$.
  Hence, $\emp \in \sidtypeof{\S,\H}$.

  Conversely, let $\psi \in \sidtypeof{\S,\H}$.
  By definition, there is a $\Sid$-forest $\frst = \set{\ftree_1,\ldots,\ftree_k}$ with    $\sfproj{\S}{\frst}=\psi$ and $\fheapof{\frst}=\H$.
  Since $\H = \emptyset$, we have $\fheapof{\frst}=\emptyset$ and thus $\theapof{\ftree_i}=\emptyset$ for all $i \in [1,k]$.
  Hence, $k = 0$.
  By \cref{def:types:sf-projection} (stack projections), we then have $\psi = \emp$.
  \qedhere
  \end{proof}

\begin{lemma}\label{lem:types-of-sleq}
  Let $\scls$ be a aliasing constraint and
  $x,y\in\dom(\scls)$.
  \begin{itemize}
      \item If $\tuple{x,y}\in\scls$, then $\goaltypes{\sleq{x}{y}} = \set{ \set{\emp} }$ and
                                   $\goaltypes{\slneq{x}{y}} = \emptyset$.
    \item Otherwise,
            $\goaltypes{\slneq{x}{y}} = \set{ \set{\emp} }$ and
            $\goaltypes{\sleq{x}{y}} = \emptyset$.
  \end{itemize}
\end{lemma}
\begin{proof}
We only consider the case $\sleq{x}{y}$ as the argument for $\slneq{x}{y}$ is completely analogous.
If $\tuple{x,y}\in\scls$, our semantics of equalities enforces
$\goaltypes{\emp}=\goaltypes{\sleq{x}{y}}$. The claim then follows
from \cref{lem:types-of-emp}.
If $\tuple{x,y}\notin\scls$, it holds for all $\S$ with
$\eqclasses{\S}=\scls$ that $\S(x)\neq\S(y)$.
The semantics of $\sleq{x}{y}$ then yields, for all heaps $\H$, that
$\SH \not\sidmodels\sleq{x}{y}$. Hence,
$\goaltypes{\sleq{x}{y}}=\emptyset$.
\end{proof}

\begin{lemma}\label{lem:types-of-pto}
  Let $\scls$ be an aliasing constraint, let $a \in \dom(\scls)$,
  and let $\vec{b}\in\Var^{*}$ with
  $\vec{b}\subseteq\dom(\scls)$.
  Then, $\goaltypes{\pto{a}{\vec{b}}} =
  \set{\sidtypeof{\ptrmodel{\scls}{\pto{a}{\vec{b}}}}}$.
\end{lemma}
\begin{proof}
  \begin{description}
  \item[``$\supseteq$'']
    Let $\SH\defn\ptrmodel{\scls}{\pto{a}{\vec{b}}}$.
    By definition, $\SH \sidmodels \pto{a}{\vec{b}}$.
    Hence, $\sidtypeof{\S,\H} \in \goaltypes{\pto{a}{\vec{b}}}$.
  \item[``$\subseteq$''] Let $\SH$ be a state such that
    $\sidtypeof{\S,\H}\in\goaltypes{\pto{a}{\vec{b}}}$ and
    $\eqclasses{\S}=\scls$. By definition,
    $\SH \sidmodels \pto{a}{\vec{b}}$ and thus, by the semantics of points-to assertions,
    $\H = \set{\S(a) \mapsto \S(\vec{b})}$.
    Consequently, $\SH \Iso \ptrmodel{\scls}{\pto{a}{\vec{b}}}$ and
    $\sidtypeof{\S,\H} =
    \sidtypeof{\ptrmodel{\scls}{\pto{a}{\vec{b}}}}$.
    Since $\SH$ was an arbitrary model of $\phi$ with
    $\sidtypeof{\S,\H}\in\goaltypes{\pto{a}{\vec{b}}}$ and
    $\eqclasses{\S}=\scls$, we have
    $\goaltypes{\pto{a}{\vec{b}}} \subseteq \set{
      \sidtypeof{\ptrmodel{\scls}{\pto{a}{\vec{b}}}} }$. \qedhere
  \end{description}
\end{proof}
%
%The lifted operators $\ACompose$ operator can be used to compute the
%types of $\phi_1 \sep \phi_2$ from the types of $\phi_1$
%and $\phi_2$.
%This is formalized in the following lemma.
%
\noindent
We next consider the operations $\ACompose$ (type composition), $\tinst{\cdot}{\cdot}{\cdot}{\cdot}$ (variable renaming), $\ForgetKW$ and $\ExtendKW$ (lifted to sets of types).
The lemmata for these operations below are more general than what is needed for the soundness of our fixed point algorithm because we will also use them for proving the correctness of our algorithm dealing with guarded formulas, see
\cref{sec:deciding-btw:toplevel-types}.
\begin{lemma}[Type composition]\label{lem:types:compose-lifted}
  For $\phi_1,\phi_2\in\SLIDguarded$,
  $\phitypes{\scls}{\phi_1 \sep \phi_2} = \phitypes{\scls}{\phi_1} \ACompose
  \phitypes{\scls}{\phi_2}$.
\end{lemma}
\begin{proof}
  We show each inclusion separately.
  \begin{itemize}
  \item Let $\typ \in \phitypes{\scls}{\phi_1 \sep \phi_2}$.
    Moreover, fix a state $\SH$ be such that $\SH \sidmodels \phi_1\sep\phi_2$ and $\typ = \sidtypeof{\S,\H}$.
    Then, there exist heaps $\H_1,\H_2$ such that
    $\SHi{i} \sidmodels \phi_i$ and $\H = \H_1\stdunion\H_2$.
    By~\cref{lem:pos-formula-pos-model}, we have
    $\SHi{1},\SHi{2}\in\Mpos{\Sid}$.
    By \Cref{lem:types:homo:compose},
    $\typ = \sidtypeof{\S,\H_1} \ACompose \sidtypeof{\S,\H_2}$.
    Since $\SHi{i} \sidmodels \phi_i$, we have $\sidtypeof{\S,\H_i} \in \phitypes{\scls}{\phi_i}$.
    Hence, $\typ \in \phitypes{\scls}{\phi_1} \ACompose \phitypes{\scls}{\phi_2}$.

  \item Let $\typ \in \phitypes{\scls}{\phi_1}
      \ACompose \phitypes{\scls}{\phi_2}$.
    Then, there are $\typi{1} \in \phitypes{\scls}{\phi_1}$ and $\typi{2} \in \phitypes{\scls}{\phi_2}$ such that~$\typ = \typi{1} \ACompose \typi{2}$.
    Moreover, there are states $\SHi{i}$ such that $\eqclasses{\S}=\scls$, $\sidtypeof{\S,\H_i} = \typi{i}$ and $\SHi{i} \sidmodels \phi_i$.
    By~\cref{lem:guarded-iterated-star-predicates} we have $\typi{i} \neq \emptyset$.
    By~\cref{lem:types:homo:existence}
    there are states $\SHpair{\S}{\H'_i}$ such that $\sidtypeof{\S,\H'_i} = \typi{i}$ and
    $\sidtypeof{\S, \H_1' \stdunion \H'_2} = \typi{1} \ACompose \typi{2}$.
    By~\cref{cor:phi-typ-satisfies-phi} we have
    $\SHpair{\S}{\H'_i} \sidmodels \phi_i$.
    By the semantics of $\sep$,
    $\SHpair{\S}{\H'_1\stdunion\H'_2} \sidmodels \phi_1\sep\phi_2$.
    Hence, $\typ \in \phitypes{\scls}{\phi_1 \sep \phi_2}$. \qedhere
  \end{itemize}
\end{proof}

\noindent
%To deal with variable renaming, we first formally define the reverse renaming operation $\pinst{\scls}{\vec{x}}{\vec{y}}^{-1}$ on aliasing constraints, which we already used in \cref{fig:ptype-computation}:
%
%\begin{definition}[Reverse renaming of aliasing constraints]\label{def:reverse-ac}
%  Let $\vec{x}$ be a sequence of pairwise distinct variables and let $\vec{y}$ be a sequence of (not necessarily pairwise distinct) variables with $|\vec{y}| = |\vec{x}|$.
%  Moreover, let $\scls$ be an aliasing constraint with $\vec{x} \cap \dom(\scls) = \emptyset$ and $\vec{y} \subseteq \dom(\scls)$.
%    Then, the \emph{reverse renaming} $\vec{x}$ to $\vec{y}$ in $\scls$ by is given by the aliasing constraint $\pinst{\scls}{\vec{x}}{\vec{y}}^{-1} \in \EqClasses{\dom(\scls)\cup\vec{x}}$
%  defined by
%  $$\pinst{\scls}{\vec{x}}{\vec{y}}^{-1} \defn \{(a_1,a_2) \mid \text{ there is } (b_1,b_2) \in \scls \text{ with } b_1 = \pinst{a_1}{\vec{x}}{\vec{y}} \text{ and } b_2 = \pinst{a_2}{\vec{x}}{\vec{y}}\}.
%  $$
%  %  $$\pinst{\scls}{\vec{x}}{\vec{y}} \defn \{(a_1,a_2) \mid \text{ there is } (b_1,b_2) \in \scls \text{ with } a_i = \pinst{b_i}{\vec{x}}{\vec{y}},a_1=y_i, b_2 = x_j,a_2=y_j \}.
%%  $$
%\end{definition}
%
\begin{lemma}[Renaming of type sets]\label{lem:types:instantiation-lifted}
  Let $\vec{x}$ and $\vec{y}$ be sequences of variables as in \cref{def:reverse-ac} from above.
  Then, for every $\SLIDguarded$ formula $\phi$,  we have
  $$\tinst{\phitypes{\pinst{\scls}{\vec{x}}{\vec{y}}^{-1}}{\phi}}{\scls}{\vec{x}}{\vec{y}} = \phitypes{\scls}{\pinst{\phi}{\vec{x}}{\vec{y}}}.
  $$
\end{lemma}
\begin{proof}
    Let $\sidtypeof{\S,\H} \in \phitypes{\scls}{\pinst{\phi}{\vec{x}}{\vec{y}}}$.
  By definition, this means $\eqclasses{\S}=\scls$.
  Moreover, we note that $\pinst{\scls}{\vec{x}}{\vec{y}}^{-1} = \eqclasses{\pinst{\S}{\vec{x}}{\vec{y}}}$.
  By \cref{lem:types:homo:rename}, we have
  $\tinst{\sidtypeof{\pinst{\S}{\vec{x}}{\vec{y}},\H}}{\scls}{\vec{x}}{\vec{y}}=\sidtypeof{\S,\H}$,
  i.e., $\sidtypeof{\S,\H} \in\tinst{\phitypes{\pinst{\scls}{\vec{x}}{\vec{y}}^{-1}}{\phi}}{\scls}{\vec{x}}{\vec{y}}$.
  The converse direction is analogous.
\end{proof}

\begin{lemma}[Forgetting a variable in type sets]\label{lem:types:forget-lifted}
  Let $\phi\in\SLIDguarded$ be a formula with free variables $\vec{x} \cup \{y\}$ such that $y \not\in \vec{x}$.
  Moreover, assume that, for every state $\SH$, $\SH \sidmodels \phi$ implies $\S(y) \in \dom(\H)$.
  Then, for every aliasing constraint $\scls$ with $\dom(\scls) = \vec{x}$, we have
  $$\phitypes{\scls}{\EXO{y}{\phi}} = \bigcup_{\scls' \in \EqClasses{\vec{x} \cup\{y\}} \text{ with } \scls'|_\vec{x} = \scls} \Forget{\scls',y}{\phitypes{\scls'}{\phi}}.$$
\end{lemma}
\begin{proof}
    Let $\typ \in \Forget{\scls',y}{\phitypes{\scls'}{\phi}}$, where $\scls' \in \EqClasses{\vec{x} \cup \{y\}}$ is an aliasing constraint satisfying $\scls'|_{\vec{x}} = \scls$.
    Then there exists a state $\SH$ such that
    $\typ = \Forget{\scls',y}{\sidtypeof{\S,\H}}$,
    $\SH \sidmodels \phi$,
    $\dom(\S) = \vec{x}$, and
    $\eqclasses{\S}|_{\dom(\scls)} = \scls$.
    By assumption, $\S(y) \in \dom(\H)$.
    \cref{lem:types:homo:forget} then yields
  \begin{align*}
      \sidtypeof{\pinst{\S}{y}{\bot},\H} = \Forget{\scls',y}{\sidtypeof{\S,\H}} = \typ.
  \end{align*}
  By the semantics of existential quantifiers, we have $\SHpair{\pinst{\S}{y}{\bot}}{\H}\sidmodels \EXO{y}{\phi}$.
  Hence, $\typ \in \phitypes{\scls}{\EXO{y}{\phi}}$.

  Conversely, let $\typ \in \phitypes{\scls}{\EXO{y}{\phi}}$.
  Then, there is a state $\SH$ such that
  $\SHpair{\S}{\H} \sidmodels \EXO{y}{\phi}$,
  $\typ = \sidtypeof{\S,\H}$,
  $\dom(\S)=\vec{x} \setminus \{y\}$ and
  $\eqclasses{\S} = \scls$.
    By the semantics of the existential quantifier, there is a value $v$ such that $(\pinst{\S}{y}{v}, \H) \sidmodels \phi$.
  By assumption we have $\S(y) \in \dom(\H)$.
    Then, for $\scls' = \eqclasses{\pinst{\S}{y}{v}} \in \EqClasses{\vec{x} \cup \{y\}}$, \cref{lem:types:homo:forget} yields
  \begin{align*}
      \typ = \Forget{\scls',y}{\sidtypeof{\pinst{\S}{y}{v},\H}}
           %= \sidtypeof{\pinst{\pinst{\S}{y}{v}}{y}{\bot},\H}
           = \sidtypeof{\S,\H}
      \in \Forget{\scls',y}{\phitypes{\scls'}{\phi}}.
      \tag*{\qedhere}
  \end{align*}
\end{proof}

\begin{lemma}[Extending type sets]
\label{lem:extend-lifted-types}
  Let $\phi\in\SLIDguarded$ be a formula with free variables $\vec{x}$.
  Moreover, let $\scls \subseteq \scls'$ be alias constraints
  with $\dom(\scls) = \xx$.
  Then, $\phitypes{\scls'}{\phi} \subseteq \Extend{\scls'}{\phitypes{\scls}{\phi}}$.
\end{lemma}
\begin{proof}
  We consider some $\typ \in \phitypes{\scls'}{\phi}$.
  Then, there is a state $\SH$ with $\dom(\S) = \xx = \dom(\scls)$ and $\sidtypeof{\S,\H} = \typ$.
  We now choose some extension $\S'$ of $\S$ to $\dom(\scls')$ such that $\S(x) \not\in \locs{\H}$ for every variable $x \in \dom(\S)$ that is not an alias of a variable in $\dom(\scls)$.
  By~\cref{lem:extend-lifted} we then have $\sidtypeof{\S',\H} = \Extend{\scls'}{\sidtypeof{\S,\H}}$.
  Hence, $\Extend{\scls'}{\typ} \in \Extend{\scls'}{\phitypes{\scls}{\phi}}$.
\end{proof}
\noindent
We are now ready to prove the soundness of the fixed point computation, i.e.,
\[ \thelfp(\pred, \scls) \subseteq \PTypes{\scls}{\pred}. \]
We first need to establish that  $\ptypesof{\scls}{\phi}$ is sound when $\phi$ is a rule of predicate $\pred$, i.e., that $\ptypesof{\scls}{\phi} \subseteq \phitypes{\scls}{\phi}$ holds, under the assumption that $\thefun$ maps every pair of predicate identifier and aliasing constraint to a subset of the corresponding types.
As SID rules are guaranteed to be existentially-quantified symbolic heaps, it suffices to prove this result for arbitrary $\phi \in \SymHeap$:

\begin{lemma}\label{lem:typesof:conditional-correct}
  Let $\phi \in \SymHeap$ and $\scls \in \EqClassesKW$ with $\dom(\scls)\supseteq\fvs{\phi}$.
  %
  % Moreover, let
  % \[\thefun\colon \thefuntype\] be such that
  % $\thefun(\pred,\scls')\subseteq\PTypes{\scls'}{\pred}$ for all
  % $\pred \in \Preds{\Sid}$ and all
  % $\scls'\in\EqClasses{\xxfvs{\pred}}$.
  Moreover, let
  \[\thefun\colon \thefuntype\] be such that for all
  $\pred \in \Preds{\Sid}$ and all
  $\scls'\in\EqClasses{\xxfvs{\pred}}$, it holds that
  $\thefun(\pred,\scls')\subseteq\PTypes{\scls'}{\pred}$.
  Then $\ptypesof{\phi}{\scls} \subseteq \goaltypes{\phi}$.
\end{lemma}
\begin{proof}
  We proceed by induction on the structure of $\phi$ and apply the lemmata from above.
  \begin{description}
  \item[Cases $\phi=\sleq{x}{y}$, $\phi=\slneq{x}{y}$.] The claim
    follows from \cref{lem:types-of-sleq}.
  \item[Case $\phi = \pto{a}{\vec{b}}$.] The claim
    follows from \cref{lem:types-of-pto}.
  \item[Case $\phi = \pred(\vec{y})$.]
    Let $\vec{z} = \fvs{\pred}$.
          By I.H., we have $$\thefun(\pred,\eqrestr{\pinst{\scls}{\vec{z}}{\vec{y}}^{-1}}{\vec{z}\cup\vec{x}})
          \subseteq \PTypes{\eqrestr{\pinst{\scls}{\vec{z}}{\vec{y}}^{-1}}{\vec{z}\cup\vec{x}}}{\pred}.$$
    By~\cref{lem:extend-lifted-types}, we have
    \begin{align*}
        \Extend{\pinst{\scls}{\vec{z}}{\vec{y}}^{-1}}{
        \phitypes{\eqrestr{\pinst{\scls}{\vec{z}}{\vec{y}}^{-1}}{
            \vec{z}\cup\vec{x}}
    }{\pred}} \subseteq
        \phitypes{\pinst{\scls}{\vec{z}}{\vec{y}}^{-1}}{\pred}.
    \end{align*}
    Moreover, by~\cref{lem:types:instantiation-lifted} we have
          $$\tinst{\phitypes{\pinst{\scls}{\vec{z}}{\vec{y}}^{-1}}{\pred}}{\scls}{\vec{z}}{\vec{y}} = \phitypes{\scls}{\pinst{\pred}{\vec{z}}{\vec{y}}}.$$
    Hence, we get
    \begin{align*}
        \tinst{\Extend{\scls}{\thefun(\pred,\eqrestr{\pinst{\scls}{\vec{z}}{\vec{y}}^{-1}}{\vec{z}\cup\vec{x}})}}
    {\scls}{\vec{z}}{\vec{y}} \subseteq  \phitypes{\scls}{\pinst{\pred}{\vec{z}}{\vec{y}}}.
    \end{align*}
  \item[Case $\phi = \phi_1 \sep \phi_2$.]
    The claim
    follows from~\cref{lem:types:compose-lifted} and the induction hypothesis.
  \item[Case $\phi = \SHEX{y} \phi$.]
    The claim
    follows from~\cref{lem:types:forget-lifted} and the induction hypothesis.
  \end{description}
\end{proof}

\begin{lemma}[Soundness of type computation] \label{lem:fp-sound}
  $\thelfp(\pred, \scls) \subseteq
  \PTypes{\scls}{\pred}$.
\end{lemma}
\begin{proof}
  A straightforward induction on top of
  \cref{lem:typesof:conditional-correct}.
\end{proof}

\subsubsection{Completeness of the Type
  Computation}\label{sec:fixedpoint:complete}

We now establish the completeness of the fixed point computation, i.e.,
$\thelfp(\pred, \scls) \supseteq \PTypes{\scls}{\pred}$.
The main challenge is our treatment of predicate calls $\ptypesof{\pred(\vec{y})}{\scls}$,
for which the recursive look-up $\thefun(\pred,\eqrestr{\pinst{\scls}{\vec{z}}{\vec{y}}^{-1}}{\vec{x}\cup\vec{z}})$
restricts the stack-aliasing constraint $\pinst{\scls}{\vec{z}}{\vec{y}}^{-1}$ to $\vec{x}\cup \vec{z}$, where
$\vec{z} \defn \fvs{\pred}$.
This restriction of the variables is necessary in order to avoid having to consider larger and larger sequences of variables, which would lead to divergence as we illustrate below.
Hence, our goal will be to establish that $\ptypesofKW$ discovers all types even though we restrict the variables in the recursive look-up.

We now illustrate the need for restricting the variables in the recursive look-up:
We assume a stack $\S$ with $\dom(\S)=\xx \cup
\vec{y}$ and pick a rule
\[\pred(\fvs{\pred}) \Rule \SHEX{\vec{e}} \ppto{a}{\vec{b}} \sep
  \pred_1(\vec{z_1}) \sep \cdots \sep \pred_k(\vec{z_k}).\]
We extend $\S$ to a stack $\S'$ with
$\dom(\S')=\dom(\S)\cup\vec{e}$ and are left with computing the types of
\[
    \pinst{\left(\ppto{a}{\vec{b}} \sep
  \pred_1(\vec{z_1}) \sep \cdots \sep \pred_k(\vec{z_k})\right)}{\fvs{\pred}}
  {\vec{y}}
  .
\]
At a first glance, this implies recursively computing the types of the calls, $\pred_i(\vec{z_i})$, w.r.t.~the variables $\xx'\defn\xx\cup\vec{y}\cup\vec{e}$, i.e., we additionally have to consider the existentially quantified variables $\vec{e}$.
We attempt to do so by picking a rule for each predicate, say we first pick a rule for predicate $\pred_i$.
Then, we need to consider an extension $\S''$  of $\S'$ with $\dom(\S'')=\dom(\S')\cup\vec{e_i}$ for the existentially quantified variables $\vec{e_i}$ on the right hand side of the picked rule.
Continuing in this fashion, the computation diverges as we have to extend the set of considered variables $\xx$ again and again.

However, a more careful analysis reveals that restricting the aliasing constraints to $\eqrestr{\pinst{\scls}{\vec{z}}{\vec{y}}^{-1}}{\vec{x}\cup\vec{z}}$ for the recursive look-up (followed by extending and renaming the obtained set of types) is sufficient.
This is a consequence of the \emph{establishment} property, which we require for all SIDs in $\IDbtw$.
We formalize this insight in the notion of a \emph{tree closure}, which restricts the locations a subtree can share with the variables appearing in the rule instance of its parent node:
\begin{definition}[Tree Closure]
   Let $\vec{u}$ be a set of locations and let $\ftree$ be a $\Sid$-tree.
   Moreover, let $\subtree$ be a proper subtree of $\ftree$ and let $\pred(\vec{w})= \trootpred{\subtree}$.
   Let $\loc \in \dom(\ftree)$ be the parent location of the root of $\subtree$ and let $\truleinst{\ftree}{\loc} = \pred(\vec{v}) \Rule  \pinst{\fa}{\fvs{\pred}\cdot\vec{e}}{\vec{v}\cdot\vec{m}}$ be the rule instance at location $\loc$.
   We say $\ftree$ is $\vec{u}$-closed for $\subtree$, if $\tptrlocs{\subtree} \cap (\vec{v}\cup\vec{m}) \subseteq \vec{w} \cup \vec{u} \cup \{\nil\}$.

   Furthermore, we say $\ftree$ is $\vec{u}$-closed,
   if $\ftree$ is $\vec{u}$-closed for all proper subtrees of $\ftree$.
\end{definition}

\begin{example}
The tree from~\cref{fig:ex:sid-tree:forest}
is $\emptyseq$-closed, where $\emptyseq$ is the empty sequence.
If we replace location $8$ everywhere in the tree with location $7$, then the resulting tree is not $\emptyseq$-closed anymore (consider the subtree rooted at location $2$), but $7$-closed.
\end{example}

\begin{lemma}\label{lem:establishment-helper}
   Let $\ftree$ be some $\Sid$-tree with $\tallholepreds{\ftree}=\emptyset$ and
   $\pred(\vec{u})= \trootpred{\ftree}$.
   Let $\loc \in \dom(\ftree)$ be some location and let $\truleinst{\ftree}{\loc} = \pred(\vec{v}) \Rule  \pinst{\fa}{\fvs{\pred}\cdot\vec{e}}{\vec{v}\cdot\vec{m}}$ be the rule instance at location $\loc$ of $\ftree$.
   Then, we have $\vec{v} \cup \vec{m} \subseteq \dom(\ftree) \cup \vec{u} \cup \{\nil\}$.
\end{lemma}
\begin{proof}
   A direct consequence of establishment.
\end{proof}
\noindent
Recall that, by~\cref{lem:model-of-pred-to-tree}, we have $\SH \sidmodels \pred(\fvs{\pred})$ if and only if there exists a $\Sid$-tree $\ftree$ with $\trootpred{\ftree}=\pred(\S(\fvs{\pred}))$,
$\tallholes{\ftree}=\emptyset$ and $\fheapof{\set{\ftree}}=\H$.
The completeness of our fixed point algorithm relies on the observation
that such trees $\ftree$ are $\S(\fvs{\pred})$-closed:
\begin{lemma}\label{lem:trees-with-no-holes-are-closed}
   Every $\Sid$-tree $\ftree$ with $\tallholepreds{\ftree}=\emptyset$ and
    $\pred(\vec{u})= \trootpred{\ftree}$ is $\vec{u}$-closed.
\end{lemma}
\begin{proof}
   Let $\subtree$ be a proper subtree of $\ftree$ and let $\pred(\vec{w})= \trootpred{\subtree}$.
   Let $\loc \in \dom(\ftree)$ be the parent location of the root of $\subtree$ and let $\truleinst{\ftree}{\loc} = \pred(\vec{v}) \Rule  \pinst{\fa}{\fvs{\pred}\cdot\vec{e}}{\vec{v}\cdot\vec{m}}$ be the rule instance at location $\loc$.
   We consider some $\locAlt \in \tptrlocs{\subtree} \cap (\vec{v}\cup\vec{m})$.
    Then $\locAlt$ is either equal to $\nil$ (and there is nothing to show) or $\locAlt$ is a dangling or an allocated location in $\fheapof{\subtree}$.

   Assume $\locAlt$ is dangling in $\fheapof{\subtree}$:
   We define the stack $\S: \fvs{\pred} \to \Loc$ by setting $\S(\fvs{\subpred}) = \vec{w}$.
    By~\cref{lem:model-of-pred-to-tree}, we have $\SHpair{\S}{\fheapof{\subtree}} \sidmodels \pred(\fvs{\pred})$.
   By~\cref{lem:predicate-pos-model} we have  $\SHpair{\S}{\fheapof{\subtree}} \in \Mpos{\Sid}$.
    Hence, $\locAlt \in \img(\S) = \vec{w}$. In other words, dangling locations do not invalidate that $\ftree$ is $\vec{u}$-closed.

   Assume $\locAlt$ is allocated in $\fheapof{\subtree}$, i.e., $\locAlt \in \dom(\subtree)$:
   To prove that $\ftree$ is $\vec{u}$-closed, it is sufficient that $\locAlt \notin \vec{w}$ implies that $\locAlt \in \vec{u}$.
   Hence, let us assume that $\locAlt \notin \vec{w}$.
   Let $\remtree = \ftree \setminus \subtree$ be the remainder of $\ftree$ after splitting off $\subtree$, i.e., $\fsplit{\set{\ftree}}{ \set{\troot{\subtree}}}= \set{\subtree,\remtree}$.
   We note that $\locAlt \notin \dom(\remtree)$ because a location cannot be allocated in two subtrees.
   We choose a fresh location $\locAlt' \in \Loc \setminus \tptrlocs{\ftree}$ and
   then create a tree $\subtree'$ as a copy of $\subtree$ except that we replace every occurrence of location $\locAlt$ in $\subtree$ with $\locAlt'$.
   We note that $\trootpred{\subtree'} = \subpred(\vec{w})= \trootpred{\subtree}$ because of $\locAlt \not\in \vec{w}$.
   Hence, there is a tree $\ftree'$ such that $\fsplit{\set{\ftree'}}{ \set{\loc}}= \set{\subtree',\remtree}$.
   We note that, by construction of $\ftree'$, we have that (1) $\locAlt' \notin \dom(\ftree')$, (2) $\trootpred{\ftree'}= \trootpred{\ftree} = \pred(\vec{u})$, (3) $\tallholepreds{\ftree'}=\emptyset$, (4) $\loc$ is the parent of $\subtree'$ in $\ftree'$, and (5) $\truleinst{\ftree'}{\loc} = \pred(\vec{v}) \Rule  \pinst{\fa}{\fvs{\pred}\cdot\vec{e}}{\vec{v}\cdot\vec{m}}$.
    \cref{lem:establishment-helper} then yields $\vec{v} \cup \vec{m} \subseteq \dom(\ftree') \cup \vec{u} \cup \{\nil\}$.
   Since $\locAlt$ is allocated, this means $\locAlt \in \dom(\ftree') \cup \vec{u}$.
   With (1) we then obtain $\locAlt \in \vec{u}$.
\end{proof}

\noindent
We are now ready to prove the completeness of our fixed-point algorithm for computing types.
We will show that the fixed-point algorithm discovers, for all
predicates $\pred$ and aliasing constraints $\scls \in \EqClasses{\xxfvs{\pred}}$, all types in the following set:
\begin{align*}
  \{ \sidtypeof{\S,\theapof{\ftree}} \mid &\;\eqclasses{\S}=\scls,\\
    &\;\trootpred{\ftree}=\pred(\S(\fvs{\pred})),\\&\;\tallholes{\ftree}=\emptyset \},
\end{align*}

where---as shown above---we can rely on the assumption that the considered trees $\ftree$ are $\S(\xx)$-closed.

\begin{lemma}\label{lem:lfp-contains-all-tree-types}
   Let $\S$ be a stack with $\dom(\S)=\xx\cup\fvs{\pred}$
   and $\S(\fvs{\pred}) = \vec{v}$.
   Let $\ftree$ be an $\S(\xx)$-closed $\Sid$-tree with $\trootpred{\ftree}=\pred(\vec{v})$ and $\tallholepreds{\ftree}=\emptyset$.
   Then,
   \[\sidtypeof{\S,\theapof{\ftree}} \in
   \prefixi{\theight{\ftree}+1}(\pred,\eqclasses{\S}).\]
\end{lemma}
\begin{proof}
We prove the claim by strong mathematical induction on $\theight{\ftree}$:

Let $\H \defn \theapof{\ftree}$, $r \defn \troot{\ftree}$, and $\scls \defn \eqclasses{\S}$.
Since $\ftree$ is an $\Sid$-tree with  $\trootpred{\ftree}=\pred(\vec{v})$,
there is a rule $(\pred(\vec{x}) \Rule \fa) \in \Sid$ with $\fa = \SHEX{\vec{e}} \fa'$, $\fa'=\ppto{y}{\vec{z}} \sep \pred_1(\vec{z_1}) \sep \cdots \sep \pred_k(\vec{z_k}) \sep \Pure$, $\Pure$ pure\footnote{In case of $\theight{\ftree}=0$, the rule $(\pred(\vec{x}) \Rule \fa)$ is non-recursive, and we have $k=0$ and there are no existentially quantified variables, i.e., $\vec{e} = \epsilon$.}, such that $\truleinst{\ftree}{r} = \pred(\vec{v}) \Rule \pinst{\fa'}{\fvs{\pred}\cdot\vec{e}}{\vec{v}\cdot\vec{m}}$ for some $\vec{m}\in\Loc^{*}$ (i.e., the root $r$ of $\ftree$ is labeled with an instance of the rule).

Let $\S' = \pinst{\pinst{\S}{\fvs{\pred}}{\vec{v}}}{\vec{e}}{\vec{m}}$.
Moreover, for $1 \le i \leq k$, let $\ftree_i$ be the subtree of $\ftree$ such that $\trootpred{\ftree_i} = \pinst{\pred_i(\vec{z_i})}{\fvs{\pred}\cdot\vec{e}}{\vec{v}\cdot\vec{m}}$; let $\H_i = \fheapof{\ftree_i}$.
By~\cref{lem:model-of-pred-to-tree}, we have
$\SHpair{\S'}{\H_i}\sidmodels\pred_i(\vec{z_i})$ and,
by~\cref{lem:predicate-pos-model}, we have $\SHpair{\S'}{\H_i} \in \Mpos{\Sid}$.
Finally, we denote by $\H_0$ the unique heap such that $\SHpair{\S'}{\H_0} \sidmodels \pto{y}{\vec{z}}$.
Clearly, $\SHpair{\S'}{\H_0} \in \Mpos{\Sid}$ and $\H=\H_0\stdunion\cdots\stdunion\H_k$.

We use the following abbreviations:
\begin{align*}
&  \scls'\defn\eqclasses{\S'} \\
&  \typ_0 \defn \sidtypeof{\ptrmodel{\scls'}{\pto{y}{\vec{z}}}} \\
  &  \typ_i \defn \sidtypeof{\S',\H_i}, i \geq 1
\end{align*}
Since $\ftree$ is $\S(\xx)$-closed we have that
  $\tptrlocs{\ftree_i} \cap (\vec{v}\cup\vec{m}) \subseteq \S'(\vec{z_i}) \cup \S(\xx) \cup \{\nil\} = \S'(\vec{z_i}) \cup \S'(\xx) \cup \{\nil\}$.
Furthermore, due to $\img(\S') = \vec{v}\cup\vec{m}$, we have $\S'(x) \not\in \tptrlocs{\ftree_i} \supseteq \locs{\H_i}$ for all $x \in \dom(\scls')$ for which there is no $y \in \vec{z_i} \cup \xx$ with $\S'(x) = \S'(y)$.
Let $\S_i$ be the restriction of $\S'$ to $\xx \cup \vec{z_i}$.
\cref{lem:extend-lifted} then yields
\[
\sidtypeof{\S',\H_i} = \Extend{\scls'}{\sidtypeof{\S_i,\H_i}}. \tag*{\tagA}
\]
We introduce some more abbreviations:
\begin{align*}
& \S_i' \defn \S \cup \set{\fvs{\pred_i} \mapsto \S'(\vec{z_i})},\\
& \typ_i' \defn \sidtypeof{\S_i', \H_i}\\
& \scls_i \defn \eqclasses{\S_i'}
\end{align*}
\newcommand{\fvsiINST}{\tinstINST{\fvs{\pred_i}}{\vec{z_i}}}
Observe that $\S_i = \S_i'\fvsiINST$.
By~\cref{lem:types:homo:rename},
$\typ_i'\fvsiINST=\sidtypeof{\S_i,\H_i}$.
We then apply $\tagA$ to obtain
\[
  \typ_i =
  \sidtypeof{\S',\H_i} = \Extend{\scls'}{\sidtypeof{\S_i,\H_i}} =
  \Extend{\scls'}{\typ_i'\fvsiINST}.
  \tag*{\tagB}
\]
Finally, we note that $\ftree_i$ is $\S_i'(\xx)$-closed,    $\trootpred{\ftree_i}=\pred(\S_i'(\fvs{\pred_i}))$ and $\tallholepreds{\ftree_i}=\emptyset$.
Since $\theight{\ftree_i} <
\theight{\ftree}$, we can then apply the induction hypothesis and conclude that
\begin{align*}
  \typ_i'~=~ &  \sidtypeof{\S_i',\H_i} \\~\in~& \theop^{\theight{\ftree_i}+1}(\lambda
           (\pred',\scls')\ldotp\emptyset)(\pred_i,\eqclasses{\S_i'}) \\
  ~\subfun~ & \theop^{\theight{\ftree}}(\lambda
           (\pred',\scls')\ldotp\emptyset)(\pred_i,\scls_i). \tag*{\tagC}
\end{align*}
To finish the proof, we set $\thefun \defn \prefixi{\theight{\ftree}}$ and proceed as follows:
\begingroup
  \allowdisplaybreaks
\begin{align*}
  & \prefixi{\theight{\ftree}+1}(\pred,\scls) \\
  \\
  ~=~ & \bigcup_{(\pred(\fvs{\pred}) \Rule \phi) \in
      \Sid}\ptypesof{\scls}{\phi} \tag{by definition} \\
    ~\supseteq~ & \ptypesof{\scls}{\SHEX{\vec{e}}\phi'} \tag{$\phi = \exists\vec{e}. \phi'$} \\
  ~\supseteq~ & \Forget{\scls',\vec{e}}{\ptypesof{\phi'}{\scls'}}
    \tag{\cref{lem:types:forget-lifted}} \\
  ~=~ & \Forget{\scls',\vec{e}}{\ptypesof{\pto{y}{\vec{z}}}{\scls'} \ACompose
      \ptypesof{\pred_1(\vec{z_1})}{\scls'}
      \tag{Def. of $\phi'$, \cref{lem:types:compose-lifted}}
      \\& \qquad\qquad \ACompose \cdots \ACompose
      \ptypesof{\pred_1(\vec{z_k})}{\scls'}}  \\
  ~=~ & \Forget{\scls',\vec{e}}{\set{\typ_0} \ACompose
    \Extend{\scls'}{\tinst{\thefun(\pred_1,\scls_1)}{\fvs{\pred_1}}{\vec{z_1}}}
    \tag{Def. of $\typ_0$, \cref{lem:extend-lifted-types}}
  \\& \qquad\qquad\ACompose \cdots \ACompose
              \Extend{\scls'}{\tinst{\thefun(\pred_k,\scls_k)}{\fvs{\pred_k}}{\vec{z_k}}}}
  \\
    ~\supseteq~ & \Forget{\scls',\vec{e}}{\set{\typ_0} \ACompose
    \Extend{\scls'}{\tinst{\set{\typ_1'}}{\fvs{\pred_1}}{\vec{z_1}}}
                \tag*{(by \tagC{})}
    \\& \qquad\qquad \ACompose \cdots \ACompose
              \Extend{\scls'}{\tinst{\set{\typ_k'}}{\fvs{\pred_k}}{\vec{z_k}}}} \\
  ~=~ & \Forget{\scls',\vec{e}}{ \set{\typ_0 \ACompose \typ_1 \cdots
              \ACompose \typ_k} } \tag*{(by \tagB{})}\\
  ~=~ & \set{ \Forget{\scls',\vec{e}}{ \sidtypeof{\S',\H} }  }
  \tag*{(by \cref{lem:types:homo:compose})} \\
  ~=~& \set{ \sidtypeof{\S,\H} }. \tag*{(by
     \cref{lem:types:homo:forget}, as $\S(\vec{e})\subseteq\dom(\H)$)}
\end{align*}
\endgroup
Read from bottom to top, we have
$\sidtypeof{\S,\H} \in \prefixi{\theight{\ftree}+1}(\pred,\scls)$.
\end{proof}

\noindent
Completeness then follows by exploiting the one-to-one correspondence between $\S(\xx)$-closed $\Sid$-trees without holes and the models of a predicate:

\begin{lemma}[Completeness of type computation]\label{lem:fp-complete}
  \index{type computation!completeness}
  Let $\pred \in \Preds{\Sid}$  such that
    $\fvs{\pred}=\vec{z}=\tuple{z_1,\ldots,z_k} \subseteq \vec{x}$.
    Moreover, let $\SH \sidmodels \pred(\vec{z})$ for some state $\SH$ with $\xx = \dom(\S)$.
  Then,
  \[\sidtypeof{\S,\H}\in \thelfp(\pred,\eqclasses{\S}).\]
\end{lemma}
\begin{proof}
By \cref{lem:model-of-pred-to-tree}, there exists a  $\Sid$-tree $\ftree$ such that
    $\trootpred{\ftree}=\pred(\S(\vec{z}))$,
$\tallholes{\ftree}=\emptyset$, and $\fheapof{\set{\ftree}}=\H$.
    By~\cref{lem:trees-with-no-holes-are-closed},  the tree $\ftree$ is $\S(\vec{z})$-closed.
    Since $\vec{z} \subseteq \vec{x}$, $\ftree$ is also $\S(\vec{x})$-closed.
By~\cref{lem:lfp-contains-all-tree-types}, we know that
\[\sidtypeof{\S,\theapof{\ftree}} \in
   \prefixi{\theight{\ftree}+1}(\pred,\eqclasses{\S}).\]
Recalling that $\thelfp = \lim_{n \in \N} \prefixi{n}$ and $\H=\theapof{\ftree}$, we conclude that
 \begin{align*}
     \sidtypeof{\S,\H} \in \thelfp(\pred,\eqclasses{\S}).\tag*{\qedhere}
 \end{align*}
\end{proof}

\subsubsection{Complexity of the Fixed-Point Computation}\label{sec:fixedpoint:correct}
%
%Together, the results of
%\cref{sec:fixedpoint:sound,sec:fixedpoint:complete} imply that the
%fixed point $\thelfp$ contains all and only the $\Sid$-types of all
%predicates and aliasing constraints.
%
%\begin{lemma}\label{lem:fp-correct}
%  For all $\pred\in\Preds{\Sid}$ and
%  $\scls \in \EqClasses{\xx\cup\fvs{\pred}}$, it holds that
%  $\thelfp(\pred, \scls) = \PTypes{\scls}{\pred}$.
%\end{lemma}
%\begin{proof}
%Immediate from \cref{lem:fp-sound,lem:fp-complete}.
%% By \cref{lem:fp-sound}, $\thelfp(\pred, \scls) \subseteq
%% \PTypes{\scls}{\pred}$ and by \cref{lem:fp-complete} (with $\vec{y} \defn
%% \fvs{\pred}$), $\thelfp(\pred, \scls) \supseteq
%% \PTypes{\scls}{\pred}$.
%\end{proof}

%Put together, \cref{lem:fp-sound,lem:fp-complete} establish that our algorithm for computing types is sound and complete, i.e., it indeed yields a decision procedure for single predicate calls.
%
We now establish that the types of predicates can be computed in \emph{doubly-exponential time}.
As a first step, we consider a special case: the complexity of computing the types of single points-to assertions $\pto{a}{\vec{b}}$. 
Intuitively, single points-to assertions correspond to $\Sid$-trees of size one.
To compute their types, we systematically enumerate all such trees and check for each tree whether the points-to assertion in the tree node coincides with $\pto{\S(a)}{\S(\vec{b})}$.

\begin{lemma}\label{lem:pto-types-computable}
  Let $\scls$ be an aliasing constraint, let $a \in \dom(\scls)$, and let $\vec{b}\in\Var^{*}$ with $\vec{b}\subseteq\dom(\scls)$.
  Let $n\defn \max\{\size{\Sid},\size{\dom(\scls)}\}$.
  Then, $\sidtypeof{\ptrmodel{\scls}{\pto{a}{\vec{b}}}}$ is computable in $2^{\bigO(n \log(n))}$.
\end{lemma}
\begin{proof}
  Let $\SH = \ptrmodel{\scls}{\pto{a}{\vec{b}}}$.
  W.l.o.g. we can assume that $\img(\S) \subseteq \{\nil,1,\ldots,n+1\}$ (otherwise we can select an isomorphic model with this property).
  We observe that a single location is allocated in $\H$.
  Hence, in order to compute the type of $\SH$ we need to considers exactly those forests that consists of a single tree with a single rule instance whose points-to assertion agrees with $\pto{a}{\vec{b}}$.
  We collect those rule instances in the set $\mathbf{R}$:
  \begin{multline*}
  \mathbf{R} \defn \{ \pred(\vec{l}) \Rule(\ppto{v}{\vec{w}} \sep \pred_1(\vec{z_1}) \sep \cdots \sep \pred_k(\vec{z_k}) \sep \Pure) \pinstINST{\fvs{\pred}\cdot\vec{e}}{\vec{l}\cdot\vec{m}} \in \RuleInst{\Sid} \mid\\
  \vec{l}\cdot\vec{m}\in\mathcal{L}^{*},
  \pinst{v}{\fvs{\pred}\cdot\vec{e}}{\vec{l}\cdot\vec{m}}=\S(a),
  \pinst{\vec{w}}{\fvs{\pred}\cdot\vec{e}}{\vec{l}\cdot\vec{m}}=\S(\vec{b}))\}
  \end{multline*}
  We note that $|\vec{l}\cdot\vec{m}| \le n$ for all rule instances.
  Then,
  $\sidtypeof{\S,\H}$ is given by the projections of the forests that consists of a single tree with a rule instance from $\mathbf{R}$:
  \[
    \sidtypeof{\S,\H} = \set{ \sfproj{\S}{ \set{\set{a \mapsto
    \tuple{\emptyset,\mathcal{R}}}}} \mid \mathcal{R} \in \mathbf{R}} \cap \DUSH{\Sid}.
  \]
  For computing $\sidtypeof{\S,\H}$, we only need those rules instances in
  $\mathbf{R}$ such that $\vec{l}\cdot\vec{m} \subseteq \{\nil,\ldots,n+1\}$:
  We have $\img(\S) \subseteq \{0,\ldots,n\}$ and we can rename values not in  $\{\nil,\ldots,n+1\}$ to obtain a $\S$-equivalent forest with the desired property;
  such forests have the same projections due to~\cref{lem:stack-equivalence}.
  Thus, we can compute $\sidtypeof{\S,\H}$ by considering $n \cdot n^n \in 2^{\bigO(n \log(n))}$ rule instances.
\end{proof}
\begin{theorem}[Complexity of type computation]\label{lem:fp-complexity}
  Let $n \defn \size{\Sid}+\size{\vec{x}}$.
  Then, one can compute the set $\thelfp$ assigning sets of types to predicates in
  $2^{2^{\bigO(n^2 \log(n))}}$.
\end{theorem}
\begin{proof}
\cref{thm:types-finite} gives us a bound on the
the size of all types over aliasing constraints in $\EqClasses{\vec{x}}$:
$\size{\ssidtypes{\vec{x}}{\Sid}} \in 2^{2^{\bigO(n^2 \log(n))}}$.
Moreover, the number of predicates of $\Sid$ is bounded by $n$.
Consequently, the number of functions with signature $\thefuntype$ is bounded by
\[n \cdot 2^{2^{\bigO(n^2 \log(n))}} =
  2^{2^{\bigO(n^2 \log(n)) + 1}} = 2^{2^{\bigO(n^2 \log(n))}}.\]
Since every iteration of the fixed-point computation discovers at
least one new type, the computation terminates after at most
$2^{2^{\bigO(n^2 \log(n))}}$ many iterations.
We will show that each iteration takes at most $2^{2^{\bigO(n^2 \log(n))}}$ steps.
This is sufficient to establish the claim because
\[
  \underbrace{2^{2^{\bigO(n^2 \log(n))}}}_{\text{number of iterations}} \cdot
  \underbrace{2^{2^{\bigO(n^2 \log(n))}}}_{\text{cost per iteration}}  =  2^{2^{\bigO(n^2 \log(n))}}.
\]
\noindent
We now study the time spent in each iteration:
Given some predicate $\pred \in \Preds{\Sid}$ and aliasing constraint $\scls \in \EqClasses{\vec{x}}$, we need to compute
\begin{itemize}
\item the function $\ptypesof{\phi}{\scls}$ for each rule $\pred(\fvs{\pred}) \Rule \phi \in \Sid$, where $\thefun$ is the pre-fixed point from the previous iteration, and
\item the union of the results of these function calls (note that we need to compute at most one union operation per rule $\phi \in \Sid$).
\end{itemize}
We argue below that each call $\ptypesof{\phi}{\scls}$ can be done in at most $2^{2^{\bigO(n^2 \log(n))}}$ many steps.
We further observe that the union over a set of types is linear in the number of types, i.e., linear in $2^{2^{\bigO(n^2 \log(n))}}$.
Hence, each iteration takes at most $2^{2^{\bigO(n^2 \log(n))}}$ many steps because
\[
  \underbrace{n}_{\text{number of rules}} \cdot
  \underbrace{\bigO(2^{n\log(n)})}_{\text{number of aliasing constraints}} \cdot \underbrace{2^{2^{\bigO(n^2 \log(n))}}}_{\text{cost for a fixed rule and aliasing constraint}} =  2^{2^{\bigO(n^2 \log(n))}}.
\]
To conclude the proof, we consider the cost of evaluating $\ptypesof{\phi}{\scls}$ for a fixed rule body $\phi$ and aliasing constraint $\scls$.
Since the type for each right-hand side of a rule is computed at most once,
we note that the recursive calls of $\ptypesofKW$ lead to at most $|\phi| \le n$ evaluations of base cases, i.e., (dis-)equalities and points-to assertions, and operations $\ACompose$,
$\tinst{\cdot}{\cdot}{\cdot}{\cdot}$, $\ForgetKW$ and $\ExtendKW$.
It remains to establish the cost of these operations:
\begin{enumerate}
\item Evaluating a (dis-)equality takes constant time.
\item The evaluation of a points-to assertions can be done in time $\bigO(2^{n \log(n)})$ by  \cref{lem:pto-types-computable} (observing that $|\dom(\scls)| \le \size{\Sid}+\size{\vec{x}} = n$).
\item The evaluation of the operations $\ACompose$, $\tinst{\cdot}{\cdot}{\cdot}{\cdot}$, $\ForgetKW$ and $\ExtendKW$ each takes time polynomial in the size of the types, i.e., $2^{\bigO(n^2 \log(n))}$ (see \cref{lem:dush-finite}).
    For $\tinst{\cdot}{\cdot}{\cdot}{\cdot}$, and $\ForgetKW$ this is trivial.
    For the composition operation, $\ACompose$, the polynomial bound follows because (1) the number of formulas that can be obtained by re-scoping is bounded by the number of types, and (2) the number of formulas that can be obtained by $\deriveqf$ steps is also bounded by the number of types.
    Similarly, the number of formulas that can be obtained by $\ExtendKW$ is bounded by the number of types.
    As the number of types to which each function is applied is bounded by $2^{2^{\bigO(n^2 \log(n))}}$ we obtain the following cost for each $\ACompose$, $\tinst{\cdot}{\cdot}{\cdot}{\cdot}$, $\ForgetKW$ and $\ExtendKW$:
    \[
    \underbrace{\mathit{poly}(2^{\bigO(n^2 \log(n))})}_\text{cost of operation for a single type} \cdot
    \underbrace{2^{2^{\bigO(n^2 \log(n))}}}_\text{number of types}
    = 2^{2^{\bigO(n^2 \log(n))}}.
    \]
\end{enumerate}
Hence, the cost of evaluating $\ptypesof{\phi}{\scls}$ for a fixed rule body $\phi$ and aliasing constraint $\scls$ is
\[
  \underbrace{n}_{\text{size of the rule $\phi$}} \cdot
\underbrace{2^{2^{\bigO(n^2 \log(n))}}}_{\text{cost of each of the } n \text{ operations}} =
  2^{2^{\bigO(n^2 \log(n))}}. \qedhere
\]
\end{proof}

\subsection{Correctness of the Algorithm For Computing the Types of Guarded Formulas} \label{subec:correctness-GSL} 

The correctness of $\ttypesofKW$ is almost immediate from our previous results established for computing types of predicates; we only need two additional lemmata which we state below:

\begin{lemma}
\label{lem:types-intersection-union}
   Let $\phi_1, \phi_2 \in\SLIDguarded$ be two formulas and let $\scls$ be a stack-aliasing constraint.
   Then, $\phitypes{\scls}{\phi_1 \wedge \phi_2} = \phitypes{\scls}{\phi_1}\cap \phitypes{\scls}{\phi_2}$, $\phitypes{\scls}{\phi_1 \vee \phi_2} = \phitypes{\scls}{\phi_1}\cup \phitypes{\scls}{\phi_2}$  and $\phitypes{\scls}{\phi_1 \wedge \neg \phi_2} = \phitypes{\scls}{\phi_1}\setminus \phitypes{\scls}{\phi_2}$
\end{lemma}
\begin{proof}
   We only show the first claim, the other two claims are shown analogously.

   By definition of types, the inclusion $\phitypes{\scls}{\phi_1 \wedge \phi_2} \subseteq \phitypes{\scls}{\phi_1}\cap \phitypes{\scls}{\phi_2}$ is straightforward. For the converse direction, we consider some $\typ \in \phitypes{\scls}{\phi_1}\cap \phitypes{\scls}{\phi_2}$.
   Because of $\typ \in \phitypes{\scls}{\phi_1}$ there is a state $\SH$ with $\typ=\sidtypeof{\S,\H}$ and $\SH \sidmodels \phi_1$.
   By~\cref{lem:pos-formula-pos-model}, we have $\SH\in\Mpos{\Sid}$.
   Thus, \cref{cor:phi-typ-satisfies-phi} yields $\SH\sidmodels \phi_2$.
   Hence, $\SH \sidmodels \phi_1 \wedge \phi_2$ and we obtain that $\typ \in \phitypes{\scls}{\phi_1 \wedge \phi_2}$.
\end{proof}

\begin{lemma}
\label{lem:types-mw-septraction}
   Let $\phi_0,\phi_1, \phi_2 \in\SLIDguarded$ be three formulas and let $\scls$ be a stack-aliasing constraint.
   Then,
   $$\phitypes{\scls}{\phi_0 \wedge (\phi_1 \sept \phi_2)} = \{\typ \in \phitypes{\scls}{\phi_0} \mid
   \exists \typ' \in \phitypes{\scls}{\phi_1}\ldotp
   \typ \ACompose \typ' \in \phitypes{\scls}{\phi_2}\},$$
   $$\phitypes{\scls}{\phi_0 \wedge (\phi_1 \mw \phi_2)} = \{\typ \in \phitypes{\scls}{\phi_0} \mid
   \forall \typ' \in \phitypes{\scls}{\phi_1}\ldotp
   \typ \ACompose \typ' \in \phitypes{\scls}{\phi_2}\}.$$
\end{lemma}
\begin{proof}
   We only show the first claim, the second claim is shown analogously.

   Let $\typ \in \phitypes{\scls}{\phi_0 \wedge (\phi_1 \sept \phi_2)}$.
   Then, there is a state $\SH$ with $\typ=\sidtypeof{\S,\H}$, $\SH \sidmodels \phi_0$, and
   $\SH \sidmodels \phi_1 \sept \phi_2$.
   By the semantics of $\sept$, there exists a heap $\H_1$ with $\SHi{1} \sidmodels \phi_1$ and $\SHpair{\S}{\H\stdunion\H_1} \sidmodels \phi_2$.
   Let $\typ_1 \defn \sidtypeof{\S,\H_1}$ and $\typ_2 \defn \sidtypeof{\S,\H\stdunion \H_2}$.
   By \cref{lem:types:homo:compose}, $\typ_2 = \typ \ACompose \typ_1$.
   Hence, $\typ \in \{\phitypes{\scls}{\phi_0} \mid  \exists \typ' \in \phitypes{\scls}{\phi_1}\ldotp \typ \ACompose \typ' \in \phitypes{\scls}{\phi_2}\}$.

   Conversely, let $\typ \in \phitypes{\scls}{\phi_0}$
   such that there is an $\typ' \in \phitypes{\scls}{\phi_1}$ with $\typ \ACompose \typ' \in \phitypes{\scls}{\phi_2}.$
   Then, there is a state $\SH$ with $\typ=\sidtypeof{\S,\H}$ and $\SH \sidmodels \phi_0$.
   Further, there is a state $\SHi{1}$ with $\sidtypeof{\S,\H_1}=\typ'$ and $\SHi{1} \sidmodels \phi_1$.
   We can assume w.l.o.g.~that $\H \stdunion \H_1 \neq \bot$---otherwise, replace $\H_1$ with an isomorphic heap that has this property.
   \Cref{lem:types:homo:compose} yields  $\sidtypeof{\S,\H\stdunion\H_1} = \typ \ACompose \typ_1 \in \goaltypes{\fa_2}$.
   Since $\phi_0,\phi_1\in\SLIDguarded$, we have $\SH\in\Mpos{\Sid}$ and $\SHi{1}\in\Mpos{\Sid}$ by \cref{lem:pos-formula-pos-model}.
   Thus, also $\SHpair{\S}{\H\stdunion\H_1} \in \Mpos{\Sid}$.
   \Cref{cor:phi-typ-satisfies-phi} then gives us that  $\SHpair{\S}{\H\stdunion\H_1} \sidmodels \phi_2$. Therefore, $\SH \sidmodels \phi_1 \sept \phi_2$, which implies that $\typ \in \phitypes{\scls}{\phi_1 \sept \phi_2}$.
   Hence, $\typ \in \phitypes{\scls}{\phi_0 \wedge (\phi_1 \sept \phi_2)}$.
\end{proof}

\paragraph*{We restate the claim of \cref{thm:ttypes-correct-complexity}:}
  Let $\fa \in \SLIDguarded$ with $\fvs{\fa}=\vec{x}$ and
  $\values{\fa}\subseteq\{\nil\}$. Further, let $\scls\in\EqClasses{\vec{x}}$.
  Then, $\phitypes{\scls}{\fa} = \ttypesof{\fa}{\scls}$.
  Moreover, $\ttypesof{\fa}{\scls}$ can be computed in $2^{2^{\bigO(n^2 \log(n))}}$, where $n \defn \size{\Sid}+\size{\fa}$.
\begin{proof}
  We first prove that $\phitypes{\scls}{\fa} = \ttypesof{\fa}{\scls}$.
  The proof proceeds by induction on $\fa$:
  \begin{description}
  \item[Case $\fa = \emp$.]
    By~\cref{lem:types-of-emp}.
  \item[Case] $\fa=\sleq{x}{y}$, $\fa=\slneq{x}{y}$.]
    By~\cref{lem:types-of-sleq}.
  \item[Case $\fa=\pto{a}{\vec{b}}$.]
    By~\cref{lem:types-of-pto}.
  \item[Case $\fa=\pred(\vec{y})$.]
    By~\cref{lem:types:instantiation-lifted}, \cref{lem:fp-sound} and \cref{lem:fp-complete}.

  \item[Case $\fa=\fa_1\sep\fa_2$.]
    By~\cref{lem:types:compose-lifted} and the I.H..
   % \begin{align*}
%      & \goaltypes{\fa_1 \sep \fa_2} \\
%      %
%      = & \phitypes{\scls}{\phi_1} \ACompose
%  \phitypes{\scls}{\phi_2} \tag*{(by \cref{lem:types:compose-lifted})}
%      \\
%      %
%      = & \ttypesof{\fa_1}{\scls} \ACompose \ttypesof{\fa_2}{\scls}
%          \tag*{(by the induction hypotheses)} \\
%      %
%      = & \ttypesof{\fa_1\sep\fa_2}{\scls}.
%    \end{align*}
  %
  \item[Case $\fa=\fa_1\wedge\fa_2$, $\fa=\fa_1\vee\fa_2$, $\fa=\fa_1\wedge\neg\fa_2$. ]
    By~\cref{lem:types-intersection-union} and the I.H.
  \item[Case $\fa=\fa_0\wedge(\fa_1\protect\sept\fa_2)$, $\fa=\fa_0\wedge(\fa_1\mw\fa_2)$.]
    By~\cref{lem:types-mw-septraction} and the I.H.  
  \end{description}
  
  We now turn to the complexity claim: 
  
  We recall that the number of types in  $\phitypes{\scls}{\phi}$
  is bounded by
  $2^{2^{\bigO(n^2 \log(n))}}$ (see~\cref{thm:types-finite}).
  The evaluation of $\ttypesof{\fa}{\scls}$ consists of at most $\size{\fa} \leq n$ invocations of the form $\ttypesof{\cdot}{\scls}$.
  We will show that each of these invocations can be evaluated in time at most $2^{2^{\bigO(n^2 \log(n))}}$; this is sufficient to establish the claim because of $n \cdot 2^{2^{\bigO(n^2 \log(n))}} = 2^{2^{\bigO(n^2 \log(n))}}$:
  \begin{itemize}
  \item For $\emp$ and (dis-)equalities, the evaluation time is constant.
  \item For points-to assertions, this follows from~\cref{lem:pto-types-computable}.
  \item For predicate calls, this follows from~\cref{lem:fp-complexity}.
  \item For $\wedge$, $\vee$, and $\neg$, the bound follows because each of these operations can be implemented in linear time in terms of the number of types.
  \item For $\sep$, this follows because (1) $\Compose$ is applied to at most $2^{2^{\bigO(n^2 \log(n))}} \cdot 2^{2^{\bigO(n^2 \log(n))}} = 2^{2^{\bigO(n^2 \log(n))}}$ many types and (2) the  composition $\typ_1 \Compose \typ_2$ takes time at most $\mathit{poly}(2^{\bigO(n^2 \log(n))})$, as argued in the proof of~\cref{lem:fp-complexity}.
      Hence, the cost of $\Compose$ is $\mathit{poly}(2^{\bigO(n^2 \log(n))}) \cdot 2^{2^{\bigO(n^2 \log(n))}} = 2^{2^{\bigO(n^2 \log(n))}}$.
\item For septraction and the magic wand, this is analogously to the cases for $\wedge$ resp. $\vee$ and $\sep$. \qedhere
\end{itemize}
\end{proof}

\end{document}